\documentclass[12pt]{article}
\usepackage{mathrsfs}
\makeatletter \@addtoreset{equation}{section} \makeatother
\renewcommand{\theequation}{\thesection.\arabic{equation}}
\newcommand{\ba}{\begin{array}}
\newcommand{\ea}{\end{array}}
\newcommand{\beq}{\begin{equation}}
\newcommand{\eeq}{\end{equation}}
\newcommand{\bea}{\begin{eqnarray}}
\newcommand{\eea}{\end{eqnarray}}
\def\bce{\begin{center}}
\def\ece{\end{center}}

\def\nonu{\nonumber}

\def\pa{\partial}
\def\al{\alpha}
\def\be{\beta}
\def\ga{\gamma}
\def\Ga{\Gamma}
\def\de{\delta}

\def\la{\lambda}

\def\eps6{{\displaystyle \mathop{\epsilon}^{6}}{}}
\def\g6{{\displaystyle \mathop{g}^{6}}{}}

\def\nab6{{\displaystyle \mathop{\nabla}^{6}}{}}
\def\0{{\sst{(0)}}}
\def\1{{\sst{(1)}}}
\def\2{{\sst{(2)}}}
\def\3{{\sst{(3)}}}
\def\4{{\sst{(4)}}}
\def\5{{\sst{(5)}}}
\def\6{{\sst{(6)}}}
\def\7{{\sst{(7)}}}
\def\8{{\sst{(8)}}}

\def\ba{\begin{array}}
\def\ea{\end{array}}
\def\beq{\begin{equation}}
\def\eeq{\end{equation}}
\def\be{\begin{equation}}
\def\ee{\end{equation}}

\def\la{\lambda}
\def\eps{\epsilon}

\def\ba{\begin{array}}
\def\ea{\end{array}}
\def\beq{\begin{equation}}
\def\eeq{\end{equation}}
\def\be{\begin{equation}}
\def\ee{\end{equation}}

\def\la{\lambda}
\def\eps{\epsilon}

\def\eps6{{\displaystyle \mathop{\epsilon}^{6}}{}}

\def\nab6{{\displaystyle \mathop{\nabla}^{6}}{}}

\newcommand{\bean}{\begin{eqnarray*}}
\newcommand{\eean}{\end{eqnarray*}}

\begin{document}
\thispagestyle{empty} \addtocounter{page}{-1}
   \begin{flushright}
%PUPT-2395 \\
%CALT-68-nnnn \\
%{\tt hep-th/yymmnnn}\\
\end{flushright}

\vspace*{1.3cm}
  
\centerline{ \Large \bf
  The Structure of}
\vspace*{0.5cm}
\centerline{ \Large \bf
the ${\cal N}=4$ Supersymmetric Linear $W_{\infty}[\la]$ Algebra
}
%\vspace*{0.3cm}
%\centerline{ \Large \bf
%in the Two-Dimensional  SYK Models} 
%and }
\vspace*{1.5cm}
\centerline{ {\bf  Changhyun Ahn}
\footnote{On leave from the Department of Physics, Kyungpook National University, Taegu 41566, Korea and 
address during Sept. 01, 2022-Feb. 28, 2023:
Department of Physics, New York University, 
726 Broadway, New York, NY 10003, USA}
} 
\vspace*{1.0cm} 
\centerline{\it 
 Department of Physics, Kyungpook National University, Taegu
41566, Korea} 
\vspace*{0.5cm}
\centerline{\it 
%Institut f$\ddot{u}$r Theoretische Physik,
%ETH Zurich, 8093 Z$\ddot{u}$rich, Switzerland
Department of Physics, New York University, 
726 Broadway, New York, NY 10003, USA
}
\vspace*{0.8cm} 
\centerline{\tt ahn@knu.ac.kr
  %\qquad  
} 
\vskip2cm

\centerline{\bf Abstract}
\vspace*{0.5cm}

For the vanishing deformation parameter $\la$,
the full structure of
the (anti)commutator relations in 
the ${\cal N}=4$ supersymmetric linear $W_{\infty}[\la=0]$ algebra
is obtained
for arbitrary weights $h_1$ and $h_2$ of the currents
appearing on the left hand sides in  these (anti)commutators.
The $w_{1+\infty}$ algebra can be seen from this by taking the vanishing
limit of other deformation parameter $q$ with
the proper contractions of the currents. 
For the nonzero $\la$,
the complete structure of the
${\cal N}=4$ supersymmetric linear $W_{\infty}[\la]$ algebra
is determined for the arbitrary weight $h_1$ together with
the constraint $h_1-3 \leq h_2 \leq h_1+1$.
The additional structures  on the right hand sides
in the (anti)commutators, compared to the
above $\la=0$ case, arise for the  arbitrary weights $h_1$
and $h_2$ where the weight $h_2$ is outside of above region.

\vspace*{2cm}
 \begin{flushright}
{\it On the occasion of my thirtieth Ph.D. anniversary}
\end{flushright}

\baselineskip=18pt
\newpage
\renewcommand{\theequation}
{\arabic{section}\mbox{.}\arabic{equation}}

\tableofcontents

%%%%%%%%%%%%%%%%%%%%%%%%%%%%%%%%%%%%%%%%%%%%%%%%%%%%%%%%%%%%%%%%%%%%%
%%%%%%%%%%%%%%%%%%%%%%%%%%%%%%%%%%%%%%%%%%%%%%%%%%%%%%%%%%%%%%%%%%%%%%
%%%%%%%%%%%%%%%%%%%%%%%%%%%%%%%%%%%%%%%%%%%%%%%%%%%%%%%%%%%%%%%%%%%%%
%%%%%%%%%%%%%%%%%%%%%%%%%%%%%%%%%%%%%%%%%%%%%%%%%%%%%%%%%%%%%%%%%%%%%%
\section{ Introduction and Outlook}
%1%%%%%%%%%%%%%%%%%%%%%%%%%%%%%%%%%%%%%%%%%%%%%%%%%%%%%%%%%%%%%%%%%%%%%
%%%%%%%%%%%%%%%%%%%%%%%%%%%%%%%%%%%%%%%%%%%%%%%%%%%%%%%%%%%%%%%%%%%%%

In the description of
\cite{BVd1,BVd2},
the conformal weights of the free bosonic and fermionic operators
in the two dimensional conformal field theory
do depend on the deformation parameter $\la$. See also the relevant work
\cite{FMS} for the role of this $\la$ in the similar two dimensional model.
The bosonic and fermionic currents made from the above free fields
quadratically by including the multiple derivatives have
integer or half integer weights.
%($h=\frac{1}{2}, 1, \frac{3}{2},
%\cdots$). 
The algebra from these currents has the $\la$ dependent structure
constants on the right hand sides of the (anti)commutator relations
because the relative coefficients
between the free fields in the expression of the currents reveal the
$\la$ dependence in nontrivial way.
By construction in \cite{BVd1,BVd2},
it is a new feature, compared to the previous construction
by using free fields (for example, \cite{BK,OS,Odake}),
that
there exist the bosonic current of weight-$1$
and the fermionic current of weight-$\frac{1}{2}$.
The multiple number of free bosonic and fermionic operators
can be introduced.
Then it is straightforward to write down the corresponding algebra
because the defining operator product expansions(OPEs)
between these multiple free fields
satisfy independently.
See also the relevant work in \cite{Ahn2202}
without a deformation parameter we mentioned above.

The structure of the ${\cal N}=2$
supersymmetric linear $W_{\infty}^{K,K}[\la]$ algebra
where $K$ is the number of complex bosons or the number of complex
fermions
is found in \cite{Ahn2203}.
The structure constants for vanishing
deformation parameter are generalized to the ones
for nonzero $\la$.
However, it turns out that
there are some constraints in the weights for the currents
on the left hand sides of the algebra
according to the result of this paper.
Of course, when we go
from the algebra i) at $\la=0$ to the algebra ii)
at $\la \neq 0$, once the corresponding
structure constants in the latter which lead to
the ones in the former when we take
$\la \rightarrow 0$ are determined, then definitely we expect to have
the generalized structure constants for nonzero $\la$ in the algebra
ii)
via above transition. This is the simplest deformed algebra.

On the other hand,
suppose that there are some ``additional'' structure constants
having the factor $\la$ in the (new or known) currents appearing in the
latter algebra ii).
Then
after we take the above $\la \rightarrow 0$ limit in this algebra ii),
we still have the same  former algebra i) at $\la=0$
because these ``additional'' terms vanish in this limit.
This is another deformed algebra.
Therefore, it is nontrivial to determine the ``additional''
current terms having
the $\la$ factor explicitly for general $h_1$ and $h_2$.
They can appear as either the new currents with the
coefficients having the $\la$ factor or
previously known currents with the structure constants
having the $\la$ factor explicitly.
%History of 2203

%History of 2205

Recently \cite{Ahn2205},
by considering the particular number $K=2$ of free fields above,
the ${\cal N}=4$ supersymmetric linear $W_{\infty}[\la]$ algebra is studied.
For low weights, the explicit OPEs between
the currents of ${\cal N}=4$ multiplet are obtained.
All the structure constants appearing on the right hand sides
of these OPEs are given in terms of the above deformation parameter
$\la$ explicitly but the general behavior of those on the weights is
not known.

In this paper,
we continue to study the structure of
the ${\cal N}=4$ supersymmetric linear $W_{\infty}[\la]$ algebra found in
\cite{Ahn2205}.
We try to determine the ${\cal N}=4$ multiplet for any weight $h$
in terms of above free field operators.
The corresponding five components are determined explicitly
for arbitrary weight $h$. 
After the explicit form of
the ${\cal N}=4$ multiplet is obtained, 
we perform their OPEs by using the defining OPEs between the
free fields. In doing this,
we should use the previous results in \cite{Ahn2203} for the
OPEs between the nonsinglet currents, as an intermediate step.
We will observe the appearance of the extra structures
(described before) on the
right hand sides of the (anti)commutators for the
particular weights $h_1$ and $h_2$.
Eventually we will determine the (anti)commutator relations
between the currents of the ${\cal N}=4$ multiplet, as in the abstract.
In the various footnotes, we emphasize that
the extra structures on the
right hand sides of the (anti)commutators
arise for the specific $h_1$ and $h_2$.

In section $2$,
we construct the ${\cal N}=4$ multiplet in terms of free fields
for general weight $h$.
In section $3$,
the explicit (anti)commutator relations between the nonsinglet currents
for the weights
$h_1$ and $h_2$ satisfying some constraints are obtained.
In section $4$,
the fundamental commutator relations between the
${\cal N}=4$ multiplets of $SO(4)$ singlets or nonsinglets are determined.
In Appendices,
the details appearing in sections $3$ and $4$
are described. In particular, the remaining ten (anti)commutators
for the ${\cal N}=4$ multiplet are given here.
We are using the Thielemans package \cite{Thielemans}
with a mathematica \cite{mathematica}.

We summarize what we have obtained as follows.
At the vanishing deformation parameter $\la=0$,
the complete structure
of the ${\cal N}=4$ supersymmetric linear $W_{\infty}[\la =0]$
algebra is given by
(\ref{Onerel}),
(\ref{Tworel}), (\ref{Threerel}), (\ref{Fourrel}) and (\ref{finalfinal})
in addition to Appendices
(\ref{sixcase}), (\ref{sevencase}), (\ref{eightcase}),
(\ref{ninecase}), (\ref{tencase}), (\ref{elevencase}), (\ref{twelvecase}),
(\ref{thirteencase}), (\ref{fourteencase}) and (\ref{fifteencase})
where all the $\la$ dependence appearing on the right hand sides
is gone by putting $\la=0$.
Compared to the previous work in \cite{AKK1910},
due to the presence of the weights $1$ and $\frac{1}{2}$ currents
mentioned before, in general, they do appear on the
right hand sides of above (anti)commutators at $\la=0$.
However, in the most of the examples, the structure constants
appearing in these currents are vanishing at $\la=0$. See also the footnotes
in sections $3$ and $4$.
It is an open problem whether
the present algebra at $\la=0$
can be reduced to the one of \cite{AKK1910}
by decoupling the above
weights $1$ and $\frac{1}{2}$ currents.
For nonzero $\la$, still the above (anti)commutator relations
between the currents can be used
for the weights $h_1$ and $h_2$ satisfying some constraints.
From the analysis in the footnotes of section $4$, this algebra
at nonzero $\la$ is different from the one in \cite{AK2009}.
In other words, they have common algebra at $\la =0$
(their structure constants are the same) and for nonzero
$\la$, one deformed algebra is given by \cite{AK2009}
and another one is the algebra obtained in this paper.

Then what happens for the generic weights $h_1$ and $h_2$?
First of all, 
the ${\cal N}=4$ supersymmetric linear $W_{\infty}[\la]$
algebra is linear in the sense that the right hand sides of the
OPEs between the currents contain all the possible
current terms which are linear.
Furthermore, the ${\cal N}=4$ multiplet is given by (\ref{one}),
(\ref{two}), (\ref{three}), (\ref{four}) and (\ref{last}).
Then we can perform any OPE inside the Thielemans package \cite{Thielemans}
using
the explicit forms of the ${\cal N}=4$ multiplet.
Of course, from the beginning we should fix the weights $h_1$
and $h_2$ within the package.
Then the possible poles are given by the highest singular
term which is $(h_1+h_2)$-th order pole, the next singular term
which is $(h_1+h_2-1)$-th order pole, and so on until the first order pole.
Then the next step is to rewrite all the singular terms in terms of
the components of the ${\cal N}=4$ multiplet
of $SO(4)$ nonsinglets or singlets. This is straightforward
because the possible current terms at the particular singular term
are known. The weight $h$ is fixed and the possible currents
can be determined. They consist of the current having that weight $h$,
the current having the weight $(h-1)$ with one derivative,
$\cdots$, and the current having the weight $1$ with
$(h-1)$ derivative.
See also Appendix $C$ for the specific examples.

Now we introduce the arbitrary coefficients on these
possible current terms and solve the linear equations for these
unknown coefficients by requiring that
the algebra is closed in the sense that the right hand side of the
OPE contains the components of ${\cal N}=4$ multiplet.
It will turn out that
they can be determined explicitly in terms of the deformation parameter
$\la$. The point is,
according to the result of this paper,
that there exists a critical singular term where the possible current
of weight $h_c$ is allowed. If $h \leq h_c$, then
we expect to have the additional structures
(either the presence of new currents or the different
$\la$ dependent structure constants in the previously known currents)
on the right hand side of the
OPE.
In this case, although the structure constants are known in terms of
the $\la$ for fixed $h_1$ and $h_2$, their explicit expressions
for generic $h_1$ and $h_2$ are not known so far.
On the other hand, if $h \geq h_c$, then still we can use
the previous (anti)commutator relations described in the
previous paragraph, for nonzero $\la$
without any modifications.
In this case, all the structure constants are known and they
are given by those in Appendix $A$.

Let us list some future directions along the line of  the present paper.

%%%%%%%%%%%%%%%%%%%%%%%%%%%%%%%%%%%%%%%%%%%%
$\bullet$ The ${\cal N}=4$ superspace OPE
%%%%%%%%%%%%%%%%%%%%%%%%%%%%%%%%%%%%%%%%%%%%

Although we have found $15$ (anti)commutator relations explicitly,
it is nice to observe its ${\cal N}=4$ superspace description.
In order to perform the ${\cal N}=4$ superspace approach,
we need to rewrite the above fundamental OPEs in
(\ref{Onerel}),
(\ref{Tworel}), (\ref{Threerel}), (\ref{Fourrel}) and (\ref{finalfinal})
such that the second element with the coordinate $w$
on the left hand side of the OPE should be the lowest component of
the ${\cal N}=4$ multiplet.
That is, they are given by
$\big[(\Phi^{(h_1)}_{0})_m,
  (\Phi^{(h_2)}_{0})_n \big]$,
$ \big[(\Phi^{(h_1),i}_{\frac{1}{2}})_r,
  (\Phi^{(h_2)}_{0})_n \big]$, $\big[(\Phi^{(h_1),ij}_{1})_m,
  (\Phi^{(h_2)}_{0})_n \big]$, $\big[(\Phi^{(h_1),i}_{\frac{3}{2}})_r,
  (\Phi^{(h_2)}_{0})_n \big]$, and $\big[(\Phi^{(h_1)}_{2})_m,
  (\Phi^{(h_2)}_{0})_n \big]$ in the commutators. 
After these are obtained, then it is straightforward to
express them in the ${\cal N}=4$ superspace.
For consistency check, it is obvious to extract the remaining
$10$ (anti)commutator relations (or its corresponding OPEs) from the above
${\cal N}=4$ superspace description.
In other words, we do not have to calculate the
remaining (anti)commutators separately and this is the power of
${\cal N}=4$ supersymmetry.

%%%%%%%%%%%%%%%%%%%%%%%%%%%%%%%%%%%%%%%%%%%%%%%%%%%%%%%%%%%%%%%%%%%%%
$\bullet$ The complete structure constants for any $h_1$ and $h_2$
for nonzero $\la$
%%%%%%%%%%%%%%%%%%%%%%%%%%%%%%%%%%%%%%%%%%%%%%%%%%%%%%%%%%%%%%%%%%%%%

One way to determine these is that it is better to
consider the modes of the currents in terms of
those in the free fields. By simplifying the (anti)commutator
relations in terms of the modes of the free fields,
we can express them in terms of several (anti)commutator relations
according to the decomposition of the (anti)commutator in
quantum mechanics. Then we can use the corresponding
(anti)commutator relations for (\ref{fundOPE}).
We can try to obtain the general structure by fixing
the weights $h_1$ and $h_2$. It is rather nontrivial to determine
the structure constants for generic $h_1$ and $h_2$ by varying them.

%%%%%%%%%%%%%%%%%%%%%%%%%%%%%%%%%%%%%%%%%%%%%%%%%%%%%
$\bullet$ Realization of the present algebra
in the celestial conformal field theory
%%%%%%%%%%%%%%%%%%%%%%%%%%%%%%%%%%%%%%%%%%%%%%%%%%%%

At the vanishing deformation parameter $\la=0$,
the algebra is known completely. In other words,
the structure constants are given in Appendix $A$
by inserting the $\la=0$. We are left with another deformation
parameter $q$.
As described before, we realize that
the $w_{1+\infty}$ algebra can be obtained from the
$SO(4)$ singlet currents via proper contractions of the
currents with vanishing $q$ limit at $\la=0$.
It would be interesting to observe whether there exists
any  realization of the present algebra
in the celestial conformal field theory,
along the line of
\cite{PPR,Strominger,Strominger1,GHPS,MRSV,Jiang,HPS}, or not.

%%%%%%%%%%%%%%%%%%%%%%%%%%%%%%%%%%%%%%%%%%%%%%%%%%%%%%%%%%%%%%%%%%%%%
%%%%%%%%%%%%%%%%%%%%%%%%%%%%%%%%%%%%%%%%%%%%%%%%%%%%%%%%%%%%%%%%%%%%%%
\section{ The ${\cal N}=4$ multiplet }
%1%%%%%%%%%%%%%%%%%%%%%%%%%%%%%%%%%%%%%%%%%%%%%%%%%%%%%%%%%%%%%%%%%%%%%
%%%%%%%%%%%%%%%%%%%%%%%%%%%%%%%%%%%%%%%%%%%%%%%%%%%%%%%%%%%%%%%%%%%%%

The ${\cal N}=4$ multiplet for any weight $h$
is described by using the free bosonic and fermionic fields.

%%%%%%%%%%%%%%%%%%%%%%%
\subsection{Review}
%%%%%%%%%%%%%%%%%%%%%%%

The $\beta \, \ga$ and $b \, c$ systems satisfy the
following operator product expansions 
\bea
\ga^{i,\bar{a}}(z)\, \beta^{\bar{j},b}(w) =
\frac{1}{(z-w)}\, \de^{i \bar{j}}\, \de^{\bar{a} b} + \cdots\, ,
\qquad
c^{i, \bar{a}}(z) \, b^{\bar{j},b}(w) =
\frac{1}{(z-w)}\, \de^{i \bar{j}}\, \de^{\bar{a} b} + \cdots\, .
\label{fundOPE}
\eea
The fundamental indices $a, b $ of $SU(2)$ run over $a, b =1,2$
while the antifundamental indices $\bar{a}, \bar{b}$
of $SU(2)$ run over
$\bar{a}, \bar{b}=1,2$.
Similarly the  fundamental indices $i, j $ of $SU(N)$
run over $i, j =1,2, \cdots, N$
and the antifundamental indices $\bar{i}, \bar{j}$ of
$SU(N)$ run over
$\bar{i}, \bar{j}=1,2, \cdots, N$.
The $(\beta, \ga)$ fields are bosonic operators
and the $(b, c)$ fields are fermionic operators. 

Then the $SU(N)$ singlet currents (the generalization of \cite{BVd1,BVd2})
can be obtained by
taking the bilinears of above free fields
with a summation over the (anti)fundamental indices
of $SU(N)$ as follows:
\bea
V_{\la,\bar{a} b}^{(h)+} & = & \sum_{i=0}^{h-1}\, a^i(h, \la)\,
\pa^{h-1-i}\,
(( \pa^i \, \beta^{\bar{l} b} ) \, \de_{l \bar{l}}  \,
\ga^{l \bar{a}}) +
 \sum_{i=0}^{h-1}\, a^i(h, \la+\frac{1}{2})\, \pa^{h-1-i}\,
 (( \pa^i \, b^{\bar{l} b} ) \,  \de_{l \bar{l}} \,
 c^{l \bar{a}} )\, ,
 \nonu \\
 V_{\la,\bar{a} b}^{(h)-} & = & -\frac{(h-1+2\la)}{(2h-1)}\,
 \sum_{i=0}^{h-1}\, a^i(h, \la)\, \pa^{h-1-i}\,
 (( \pa^i \, \beta^{\bar{l} b} ) \, \de_{l \bar{l}} \, 
 \ga^{l \bar{a}}) \nonu \\
 & + &
 \frac{(h-2\la)}{(2h-1)}\,
 \sum_{i=0}^{h-1}\, a^i(h, \la+\frac{1}{2})\, \pa^{h-1-i}\,
 (( \pa^i \, b^{\bar{l} b} ) \,  \de_{l \bar{l}} \,
 c^{l \bar{a}} )\, ,
\nonu \\
Q_{\la,\bar{a} b}^{(h)+} & = & \sum_{i=0}^{h-1}\, \al^i(h, \la)\,
\pa^{h-1-i}\,
(( \pa^i \, \beta^{\bar{l} b} ) \, \de_{l \bar{l}} \,
 c^{l \bar{a}}) -
 \sum_{i=0}^{h-2}\, \beta^i(h, \la)\, \pa^{h-2-i}\,
 (( \pa^i \, b^{\bar{l} b} ) \, \de_{l \bar{l}} \,
 \ga^{l \bar{a}} )\, ,
\nonu \\
Q_{\la,\bar{a} b}^{(h)-} & = & \sum_{i=0}^{h-1}\, \al^i(h, \la)\,
\pa^{h-1-i}\,
(( \pa^i \, \beta^{\bar{l} b} ) \, \de_{l \bar{l}} \,
 c^{l \bar{a}}) +
 \sum_{i=0}^{h-2}\, \beta^i(h, \la)\, \pa^{h-2-i}\,
 (( \pa^i \, b^{\bar{l} b} ) \, \de_{l \bar{l}} \,
 \ga^{l \bar{a}} )\, .
\label{VVQQla}
\eea
Note that the weights of these currents
are given by $h$, $h$, $(h-\frac{1}{2})$
and $(h-\frac{1}{2})$ respectively. The
conformal weights of $(\beta , \ga)$ fields are given by
$(\la,1-\la)$ while
conformal weights of $(b , c)$ fields are given by
$(\frac{1}{2}+\la,\frac{1}{2}-\la)$.
The above weights for the currents do not depend on
the deformation parameter $\la$ due to the particular
combinations of the free fields.
By counting the number of (anti)fundamental indices,
there exist four components labeled by $(\bar{a} b)=(11,12,21,22)$
in each current.

The relative coefficients appearing in (\ref{VVQQla})
depend on the conformal weight $h$ and deformation
parameter $\la$ explicitly and they are given by the binomial
coefficients and the rising Pochhammer symbols
where $(a)_n \equiv a(a+1) \cdots (a+n-1)$
as follows \cite{BVd1,BVd2}:
\bea 
 a^i(h, \la) \equiv \left(\begin{array}{c}
h-1 \\  i \\
 \end{array}\right) \, \frac{(-2\la-h+2)_{h-1-i}}{(h+i)_{h-1-i}}\, ,
 \qquad 0 \leq i \leq (h-1)\, ,
 \nonu \\
 \al^i(h, \la) \equiv \left(\begin{array}{c}
h-1 \\  i \\
 \end{array}\right) \, \frac{(-2\la-h+2)_{h-1-i}}{(h+i-1)_{h-1-i}}\, ,
 \qquad 0 \leq i \leq (h-1)\, ,
 \nonu \\
  \beta^i(h, \la) \equiv \left(\begin{array}{c}
h-2 \\  i \\
  \end{array}\right) \, \frac{(-2\la-h+2)_{h-2-i}}{(h+i)_{h-2-i}}\, ,
  \qquad 0 \leq i \leq (h-2)\, .
  \label{coeff}
  \eea

Let us consider   
the following currents consisting of
$(b , c)$ fields, $(\beta, \ga)$ fields,
$(\ga , b)$ fields and $(\beta, c)$ fields respectively 
by taking the linear combinations of (\ref{VVQQla})
with the help of (\ref{coeff})
\bea
W^{\la,\bar{a} b}_{F,h}(b,c) & = &
%\frac{n_{W_{F,h}}}{q^{h-2}}
\frac{2^{h-3}(h-1)!}{(2h-3)!!}\,
\,
\frac{(-1)^h}{\sum_{i=0}^{h-1}\, a^i( h, \frac{1}{2})}\,
\Bigg[\frac{(h-1+2\la)}{(2h-1)}\, V_{\la,\bar{a} b}^{(h)+} + V_{\la,\bar{a} b}^{(h)-}
\Bigg]\, ,
\nonu \\
W^{\la,\bar{a} b}_{B,h}(\beta,\ga) & = &
% \frac{n_{W_{B,h}}}{q^{h-2}}\,
\frac{2^{h-3}\,h!}{(2h-3)!!}\,
 \frac{(-1)^h}{\sum_{i=0}^{h-1}\, a^i( h, 0)}
\,
\Bigg[\frac{(h-2\la)}{(2h-1)}\, V_{\la,\bar{a} b}^{(h)+} - V_{\la,\bar{a} b}^{(h)-}
  \Bigg]\, ,
\nonu \\
Q^{\la, \bar{a} b}_{h+\frac{1}{2}}(\ga,b) & = & \frac{1}{2} \,
%\frac{n_{W_{Q,h+\frac{1}{2}}}}{q^{h-1}}
\frac{2^{h-\frac{1}{2}}h!}{(2h-1)!!}
\, \frac{(-1)^{h+1}  \, h }{
   \sum_{i=0}^{h-1} \, \beta^i( h+1, 0)}\, \Bigg[
  Q_{\la,\bar{a} b}^{(h+1)-} - Q_{\la,\bar{a} b}^{(h+1)+}\Bigg]\, ,
\nonu \\
\bar{Q}^{\la, b \bar{a}}_{h+\frac{1}{2}}(\beta,c) & = & \frac{1}{2} \,
%\frac{n_{W_{Q,h+\frac{1}{2}}}}{q^{h-1}} \,
\frac{2^{h-\frac{1}{2}}h!}{(2h-1)!!}\,
\frac{(-1)^{h+1}  }{
   \sum_{i=0}^{h} \, \al^i( h+1, 0)} \,
\Bigg[ Q_{\la,\bar{a} b}^{(h+1)-} +
  Q_{\la,\bar{a} b}^{(h+1)+}\Bigg]\, .
\label{WWQQnonzerola}
\eea
The overall coefficients do not depend on the
deformation parameter $\la$.
Then we have eight bosonic currents for the weight
$h=1,2, \cdots $ and eight
fermionic currents
for the weight
$h+\frac{1}{2}=\frac{3}{2}, \frac{5}{2}, \cdots$
as well as four fermionic currents 
$\bar{Q}^{\la, b \bar{a}}_{\frac{1}{2}}$
of the weight $\frac{1}{2}$ in
(\ref{WWQQnonzerola}).
Note that four fermionic currents
$Q^{\la,\bar{a} b}_{ \frac{1}{2}}$ of  the weight $\frac{1}{2}$
are identically zero.

The stress energy tensor of weight $2$ is given by
\bea
L & = &
\Big(W^{\la,\bar{a} a}_{\mathrm{B},2}+
W^{\la,\bar{a} a}_{\mathrm{F},2} \Big)\, ,
\label{Lterm}
\eea
which can be written as $V_{\la, \bar{a} a}^{(2)+}$. 
The corresponding central charge is 
\bea
c_{cen}= 6\,N\, (1-4\la)\, ,
\label{central}
\eea
which depends on the deformation parameter $\la$ explicitly.
The above bosonic and fermionic currents in
(\ref{WWQQnonzerola}) are quasiprimary operators
under the stress energy tensor (\ref{Lterm}) by using the defining
OPEs in (\ref{fundOPE}). The central charge (\ref{central}) becomes
$c_{cen}=6N$ at $\la=0$.

%%%%%%%%%%%%%%%%%%%
\subsection{The ${\cal N}=4$ multiplet}
%%%%%%%%%%%%%%%%%%%

%%%%%%%%%%%%%%%%%%%%%
\subsubsection{The lowest component}
%%%%%%%%%%%%%%%%%%%%%

It is known, in \cite{Ahn2205}, that the lowest components
$\Phi_0^{(h)}$ for the weights $h=1,2,3$ and $4$
have their explicit $\la$ dependences
$(h-2\la)$ and $(h-1+2\la)$ in their relative coefficients.
Then the question is how we determine these relative coefficients
for arbitrary weight $h$.
We realize that there exists an additional overall factor  $-4$
from the weight $h$ to the weight $(h+1)$ in (\ref{WWQQnonzerola}).
Moreover, the denominator of the overall factor can be extracted
easily and is given by $\frac{1}{(2h+1)}$
in terms of the weight $h$. We expect to have
the factor $(-4)^h$ from the above analysis.
The other numerical ($h$ independent) factor can appear
in general. This can be fixed only after we calculate the
OPE between this lowest component and itself and obtain the
central term. We will compute this central term later.
For the time being we simply write down the following form
\bea
\Phi_0^{(h)} & = &\frac{(-4)^{h-2}}{(2h-1)}\,
\Bigg[ -(h-2\la)\,
  W^{\la,\bar{a} a }_{\mathrm{F},h}+
  (h-1+2\la) \, W^{\la,\bar{a} a}_{\mathrm{B},h} \Bigg]\, .
\label{one}
\eea
For the weights $h=1,2,3,4$, we can observe that
the corresponding numerical values appearing in the relative
coefficients of (\ref{one}) can be seen from the ones in
\cite{Ahn2205}.
The normalization in (\ref{one}) is different from
the one in \cite{Ahn2205} where there appear
the additional numerical factors $16,8,12,24$
for the weights $h=1,2,3,4$ respectively. At the moment,
it is not easy to figure out the exact $h$ dependence
from these values. In other words, we take the 
$h$ dependence as in (\ref{one}) together with the additional numerical
factor $(-4)^{-2}$.
As explained before, the overall factor in (\ref{one})
can be determined by the normalization of the highest order
singular term in the OPE between the $\Phi_{0}^{(h)}$ and itself.

%%%%%%%%%%%%%%%%%%%%%
\subsubsection{The second components}
%%%%%%%%%%%%%%%%%%%%%

From the observation of \cite{Ahn2205} for the weights $h=1,2,3,4$,
the second components
contain the various fermionic currents
and their relative coefficients are common for any weights
$h=1,2,3,4$.
This implies that it is natural to take these relative coefficients
for any $h$ and the question is how we obtain
the overall numerical factor.
By taking the normalization of (\ref{one}) we can extract
the overall factors for
the weights $h=1,2,3,4$ and they are given by
$-\frac{1}{16} \times 1=-\frac{1}{16}$, $-\frac{1}{8}\times (-2)=
\frac{1}{4}$, $-\frac{1}{12}\times 12 =-1$ and
$-\frac{1}{24}\times 96=4$ respectively.
We can easily see that there exists $(-4)^h$ dependence when we increase
the weight by $1$. Therefore, the general expression for
the weight $h$ is given by $4(-4)^{h-4}$
which covers the above numerical values for the weights $h=1,2,3,4$.
Then we can write down the corresponding second components
as follows:
\bea
\Phi^{(h),1}_{\frac{1}{2}}
&
=& 4 \, (-4)^{h-4}\,
\Bigg[\frac{1}{2}\,\Big(
Q^{\la,11}_{h+\frac{1}{2}}
+i\sqrt{2}\,Q^{\la,12}_{h+\frac{1}{2}}
+2i \sqrt{2}\,Q^{\la,21}_{h+\frac{1}{2}}
-2\,Q^{\la,22}_{h+\frac{1}{2}}
\nonu \\
& + & 2\,\bar{Q}^{\la,11}_{h+\frac{1}{2}}
+2i \sqrt{2}\, \bar{Q}^{\la,12}_{h+\frac{1}{2}}
+i\sqrt{2}\,\bar{Q}^{\la,21}_{h+\frac{1}{2}}
-\bar{Q}^{\la,22}_{h+\frac{1}{2}}
\Big)\,\Bigg]\, ,
\nonu\\
\Phi^{(h),2}_{\frac{1}{2}}
&
=& 4 \, (-4)^{h-4}\,\Bigg[
-\frac{i}{2}\,\Big(
Q^{\la,11}_{h+\frac{1}{2}}
+2i\sqrt{2} \,Q^{\la,21}_{h+\frac{1}{2}}
-2 \,Q^{\la,22}_{\frac{3}{2}}
+2\, \bar{Q}^{\la,11}_{h+\frac{1}{2}}
+2i\sqrt{2} \, \bar{Q}^{\la,12}_{h+\frac{1}{2}}
-\bar{Q}^{\la,22}_{h+\frac{1}{2}}
\Big)\, \Bigg]\, ,
\nonu\\
\Phi^{(h),3}_{\frac{1}{2}}
&
=& 4 \, (-4)^{h-4}\,\Bigg[
-\frac{i}{2}\,\Big(
Q^{\la,11}_{h+\frac{1}{2}}
+i\sqrt{2} \,Q^{\la,12}_{h+\frac{1}{2}}
-2\,Q^{\la,22}_{h+\frac{1}{2}}
+2\, \bar{Q}^{\la,11}_{h+\frac{1}{2}}
+i \sqrt{2} \, \bar{Q}^{\la,21}_{h+\frac{1}{2}}
-\bar{Q}^{\la,22}_{h+\frac{1}{2}}
\Big)\, \Bigg]\, ,
\nonu\\
\Phi^{(h),4}_{\frac{1}{2}}
&
=& 4 \, (-4)^{h-4}\,\Bigg[
-\frac{1}{2}\,Q^{\la,11}_{h+\frac{1}{2}}
-Q^{\la,22}_{h+\frac{1}{2}}
+\bar{Q}^{\la,11}_{h+\frac{1}{2}}
+\frac{1}{2}\,\bar{Q}^{\la,22}_{h+\frac{1}{2}} \Bigg]
\,.
\label{two}
\eea
Or we can calculate the OPEs between the supersymmetry generators
and the lowest component and read off the first order pole
which will provide the second components in (\ref{two}).
We will calculate the central terms coming from the
highest order pole between the second components and itself
later. As described before, once we fix this central term, then
the overall factor in (\ref{two}) where we use the
normalization in (\ref{one}) can be determined.

%%%%%%%%%%%%%%%%%%%%%
\subsubsection{The third components}
%%%%%%%%%%%%%%%%%%%%%

The third components
contain the various bosonic currents
and their relative coefficients are equal to
each other for any weights $h=1,2,3,4$ in the analysis of \cite{Ahn2205}.
Then we obtain the following results
by replacing them with the corresponding expressions
for arbitrary weight $h$ with the above overall factors
in previous section 
\bea
\Phi^{(h),12}_{1}
&
=&  4 \, (-4)^{h-4}\,\Bigg[
2i\,W^{\la,11}_{\mathrm{B},h+1}
-\sqrt{2}\,W^{\la,12}_{\mathrm{B},h+1}
-2i\,\,W^{\la,22}_{\mathrm{B},h+1}
\nonu \\
& + & 2i\,W^{\la,11}_{\mathrm{F},h+1}
-2\sqrt{2}\,W^{\la,12}_{\mathrm{F},h+1}
-2i\,W^{\la,22}_{\mathrm{F},h+1}\, \Bigg]\, ,
\nonu\\
\Phi^{(h),13}_{1}
&
=& 
 4 \, (-4)^{h-4}\,\Bigg[
-2i\,W^{\la,11}_{\mathrm{B},h+1}
+4\sqrt{2}\,W^{\la,21}_{\mathrm{B},h+1}
+2i\,\,W^{\la,22}_{\mathrm{B},h+1}
\nonu \\
& - & 2i\,W^{\la,11}_{\mathrm{F},h+1}
+2\sqrt{2}\,W^{\la,21}_{\mathrm{F},h+1}
+2i\,W^{\la,22}_{\mathrm{F}+h+1}\, \Bigg] \, ,
\nonu\\
\Phi^{(h),14}_{1}
&
=&  4 \, (-4)^{h-4}\,\Bigg[
2\,W^{\la,11}_{\mathrm{B},h+1}
+i\sqrt{2}\,W^{\la,12}_{\mathrm{B},h+1}
+4i\sqrt{2}\,\,W^{\la,21}_{\mathrm{B},h+1}
-2\,W^{\la,22}_{\mathrm{B},h+1}
\nonu \\
& - & 2\,W^{\la,11}_{\mathrm{F},h+1}
-2i\sqrt{2}\,W^{\la,12}_{\mathrm{F},h+1}
 -  2i\sqrt{2}\,W^{\la,21}_{\mathrm{F},h+1}
+2\,W^{\la,22}_{\mathrm{F},h+1}\, \Bigg]\, ,
\nonu\\
\Phi^{(h),23}_{1}
&
=&  4 \, (-4)^{h-4}\,\Bigg[
-2\,W^{\la,11}_{\mathrm{B},h+1}
-i\sqrt{2}\,W^{\la,12}_{\mathrm{B},h+1}
-4i\sqrt{2}\,\,W^{\la,21}_{\mathrm{B},h+1}
+2\,W^{\la,22}_{\mathrm{B},h+1}
\nonu \\
& - & 2\,W^{\la,11}_{\mathrm{F},h+1}
-2i\sqrt{2}\,W^{\la,12}_{\mathrm{F},h+1}
-  2i\sqrt{2}\,W^{\la,21}_{\mathrm{F},h+1}
+2\,W^{\la,22}_{\mathrm{F},h+1}\, \Bigg]\, ,
\nonu\\
\Phi^{(h),24}_{1}
&
=&  4 \, (-4)^{h-4}\,\Bigg[
-2i\,W^{\la,11}_{\mathrm{B},h+1}
+4\sqrt{2}\,W^{\la,21}_{\mathrm{B},h+1}
+2i\,\,W^{\la,22}_{\mathrm{B},h+1}
\nonu \\
& + & 2i\,W^{\la,11}_{\mathrm{F},h+1}
-2\sqrt{2}\,W^{\la,21}_{\mathrm{F},h+1}
-2i\,W^{\la,22}_{\mathrm{F},h+1}\, \Bigg]\, ,
\nonu\\
\Phi^{(h),34}_{1}
&
=&  4 \, (-4)^{h-4}\,\Bigg[
-2i\,W^{\la,11}_{\mathrm{B},h+1}
+\sqrt{2}\,W^{\la,12}_{\mathrm{B},h+1}
+2i\,\,W^{\la,22}_{\mathrm{B},h+1}
\nonu \\
& + & 2i\,W^{\la,11}_{\mathrm{F},h+1}
-2\sqrt{2}\,W^{\la,12}_{\mathrm{F},h+1}
-2i\,W^{\la,22}_{\mathrm{F},h+1}\, \Bigg]\, .
\label{three}
\eea
In principle, 
the OPEs between the supersymmetry generators
of ${\cal N}=4$ superconformal algebra
and the second components and the first order pole
will provide the third components in (\ref{three}).
The central terms coming from the
highest order pole between the third components and itself
will be determined later.
Note that we can express the linear combination of
$W_{B,h+1}^{\la, \bar{a}b}$ and another linear combination
of $W_{F,h+1}^{\la, \bar{a}b}$ in terms of
$\Phi_1^{(h),ij}$ and $\frac{1}{2}\, \varepsilon^{ijkl}\,
\Phi_1^{(h),kl}$.
For example, the $\Phi_1^{(h),12}$ and the
$\Phi_1^{(h),34}$ look similar to each other in the sense that the
field contents are the same 
and half of them have opposite signs.
By adding or subtracting these two relations,
the two independent field contents can be written in terms of
the third components of
${\cal N}=4$ multiplet as above.

%%%%%%%%%%%%%%%%%%%%%
\subsubsection{The fourth components}
%%%%%%%%%%%%%%%%%%%%%

The fourth components
contain the various fermionic currents
and their relative coefficients are equal to
each other for any weights $h=1,2,3,4$ in the analysis of
\cite{Ahn2205}.
By replacing them with the corresponding expressions
for arbitrary weight $h$ with the above overall factors
in previous section we determine the following
results as follows:
\bea
\tilde{\Phi}^{(h),1}_{\frac{3}{2}}
&
\equiv &
\Phi^{(h),1}_{\frac{3}{2}} -\frac{1}{(2h+1)}\, (1-4\la)\,
\pa \,\Phi^{(h),1}_{\frac{1}{2}}
\nonu \\
&=&  4 \, (-4)^{h-4}\,\Bigg[
  -\frac{1}{2}\,\Big(
Q^{\la,11}_{h+\frac{3}{2}}
+i\sqrt{2}\,Q^{\la,12}_{h+\frac{3}{2}}
+2i\sqrt{2}\,Q^{\la,21}_{h+\frac{3}{2}}
-2\,Q^{\la,22}_{h+\frac{3}{2}}
\nonu \\
& - & 2\,\bar{Q}^{\la,11}_{h+\frac{3}{2}}
-2i\sqrt{2}\,\bar{Q}^{\la,12}_{h+\frac{3}{2}}
-i\sqrt{2}\,\bar{Q}^{\la,21}_{h+\frac{3}{2}}
+\bar{Q}^{\la,22}_{h+\frac{3}{2}}
\Big)\, \Bigg] \, ,
\nonu\\
\tilde{\Phi}^{(h),2}_{\frac{3}{2}}
&
\equiv &
\Phi^{(h),2}_{\frac{3}{2}} -\frac{1}{(2h+1)}\, (1-4\la)\,
\pa \,\Phi^{(h),2}_{\frac{1}{2}}
\nonu \\
&
=&  4 \, (-4)^{h-4}\,\Bigg[
\frac{i}{2}\,\Big(\,
Q^{\la,11}_{h+\frac{3}{2}}
+2i\sqrt{2}\,Q^{\la,21}_{h+\frac{3}{2}}
-2\,Q^{\la,22}_{h+\frac{3}{2}}
-2\,\bar{Q}^{\la,11}_{h+\frac{3}{2}}
-2i\sqrt{2}\,\bar{Q}^{\la,12}_{h+\frac{3}{2}}
+\bar{Q}^{\la,22}_{h+\frac{3}{2}}
\Big)\, \Bigg]\, ,
\nonu\\
\tilde{\Phi}^{(h),3}_{\frac{3}{2}}
&
\equiv &
\Phi^{(h),3}_{\frac{3}{2}} -\frac{1}{(2h+1)}\, (1-4\la)\,
\pa \,\Phi^{(h),3}_{\frac{1}{2}}
\nonu \\
&
=&  4 \, (-4)^{h-4}\,\Bigg[
\frac{i}{2}\,\Big(
Q^{\la,11}_{h+\frac{3}{2}}
+i\sqrt{2}\,Q^{\la,12}_{h+\frac{3}{2}}
-2\,Q^{\la,22}_{h+\frac{3}{2}}
-2\,\bar{Q}^{\la,11}_{h+\frac{3}{2}}
-i\sqrt{2}\,\bar{Q}^{\la,21}_{h+\frac{3}{2}}
+\bar{Q}^{\la,22}_{h+\frac{3}{2}}
\Big)\, \Bigg]\, ,
\nonu\\
\tilde{\Phi}^{(h),4}_{\frac{3}{2}}
&
\equiv &
\Phi^{(h),4}_{\frac{3}{2}} -\frac{1}{(2h+1)}\, (1-4\la)\,
\pa \,\Phi^{(h),4}_{\frac{1}{2}}
\nonu \\
&
=&  4 \, (-4)^{h-4}\,\Bigg[
\frac{1}{2}\,
\Big(
Q^{\la,11}_{h+\frac{3}{2}}
+2\,Q^{\la,22}_{h+\frac{3}{2}}
+2\,\bar{Q}^{\la,11}_{h+\frac{3}{2}}
+\bar{Q}^{\la,22}_{h+\frac{3}{2}}
\Big)\, \Bigg]\, .
\label{four}
\eea
Compared to the $\Phi_{\frac{3}{2}}^{(h),i}$
which belongs to the components of the ${\cal N}=4$ multiplet,
the $\tilde{\Phi}_{\frac{3}{2}}^{(h),i}$ in (\ref{four}) are quasiprimary
fields under the stress energy tensor (\ref{Lterm}).
The $\Phi_{\frac{3}{2}}^{(h),i}$ and the
$\Phi_{\frac{1}{2}}^{(h+1),i}$ by considering that the weight $h$
in (\ref{two}) is replaced with the weight $(h+1)$
look similar to each other in the sense that the
field contents are the same 
and half of them have opposite signs.
By adding or subtracting these two relations as before,
the two independent field contents can be written in terms of
the second components of
the $(h+1)$-th ${\cal N}=4$ multiplet
and the  fourth components of
the $h$-th ${\cal N}=4$ multiplet.
We expect that the OPEs between the supersymmetry generators
of ${\cal N}=4$ superconformal algebra
and the third components 
will provide the fourth components in (\ref{four}).

%%%%%%%%%%%%%%%%%%%%%
\subsubsection{The last component}
%%%%%%%%%%%%%%%%%%%%%

Finally we describe the last component for
arbitrary weight $h$ as follows:
\bea
\tilde{\Phi}^{(h)}_{2}
&
\equiv &
\Phi^{(h)}_{2} -\frac{1}{(2h+1)}\, (1-4\la)\,
\pa^2 \,\Phi^{(h)}_{0}
=
 4 \, (-4)^{h-4}\,\Bigg[
-2\,\Big(
W^{\la,\bar{a} a}_{\mathrm{B},h+2}
%+W^{\la,22}_{\mathrm{B},3}
+W^{\la, \bar{a} a}_{\mathrm{F},h+2}
%+W^{\la,22}_{\mathrm{F},3}
\Big) \Bigg].
\label{last}
\eea
Under the stress energy tensor (\ref{Lterm}),
this is a quasiprimary operator.
The OPEs between the supersymmetry generators
of ${\cal N}=4$ superconformal algebra
and the fourth components 
will provide the last component in (\ref{last}).
By replacing  $h$ with $(h+2)$ in (\ref{one}),
we can express
$W^{\la,\bar{a} a}_{\mathrm{B},h+2}$ and $W^{\la, \bar{a} a}_{\mathrm{F},h+2}$
in terms of $\Phi_0^{(h+2)}$ and $\tilde{\Phi}^{(h)}_{2}$
by simple linear combinations as before.

Therefore,
the ${\cal N}=4$ multiplet
is summarized by (\ref{one}), (\ref{two}), (\ref{three}),
(\ref{four}) and (\ref{last}) together with (\ref{VVQQla}),
(\ref{coeff}) and
(\ref{WWQQnonzerola}).
Their algebra will be obtained explicitly by using the
defining relations in (\ref{fundOPE}).

%%%%%%%%%%%%%%%%%%%%%%%%%%%%%%%%%%%%%%%%%%%%%%%%%%%%%%%%%%%%%%%%%%%%%
%%%%%%%%%%%%%%%%%%%%%%%%%%%%%%%%%%%%%%%%%%%%%%%%%%%%%%%%%%%%%%%%%%%%%%
\section{ The ${\cal N}=4$ supersymmetric linear $W_{\infty}^{2,2}$
  algebra between the adjoints and the bifundamentals under the
$U(2) \times U(2)$ symmetry}
%1%%%%%%%%%%%%%%%%%%%%%%%%%%%%%%%%%%%%%%%%%%%%%%%%%%%%%%%%%%%%%%%%%%%%%
%%%%%%%%%%%%%%%%%%%%%%%%%%%%%%%%%%%%%%%%%%%%%%%%%%%%%%%%%%%%%%%%%%%%%

In order to
obtain the algebra
between (\ref{one}), (\ref{two}), (\ref{three}), (\ref{four})
and (\ref{last}),
it is necessary to determine the algebra
between the currents in (\ref{WWQQnonzerola}).
In the footnotes, we present some examples
where there are extra structures (described in the
introduction) on the right hand sides of
the (anti)commutator relations for the specific weights
$h_1$ and $h_2$.

%%%%%%%%%%%%%%%%%%%%
\subsection{ The (anti)commutator relations between
  the nonsinglet currents}
%%%%%%%%%%%%%%%%%%%%

Let us consider
the algebra between
the currents consisting of $(b,c)$ fields in (\ref{WWQQnonzerola}).
By multiplying the Pauli matrix of $SU(2)$ with
the additional factor $\frac{1}{2}$ properly
and summing over the indices $\bar{a}$ and $b$ as in \cite{Ahn2205},
we can construct the three fundamentals of $SU(2)$.
By multiplying the Kronecker delta (or $2 \times 2$  identity
matrix) with the contractions of
the indices, we obtain the singlet of $SU(2)$.
First of all, in $SU(2)$, there is no symmetric
$d^{\hat{A}\hat{B}\hat{C}}$
symbols.

%%%%%%%%%%%%%%%%%%%
\subsubsection{The commutator relation with $h_1=h_2, h_2\pm 1$
for nonzero $\la$}
%%%%%%%%%%%%%%%%%%%

Then we can associate
$(W^{\la,12}_{\mathrm{F},h}+W^{\la,21}_{\mathrm{F},h})$,
$i \, (W^{\la,12}_{\mathrm{F},h}-W^{\la,21}_{\mathrm{F},h})$
and $(W^{\la,11}_{\mathrm{F},h}-W^{\la,22}_{\mathrm{F},h})$ 
with the triplets $W^{\la,\hat{A}=1}_{\mathrm{F},h}$,
$W^{\la,\hat{A}=2}_{\mathrm{F},h}$,
and
$W^{\la,\hat{A}=3}_{\mathrm{F},h}$ of $SU(2)$ respectively.
Moreover, 
the $(W^{\la,11}_{\mathrm{F},h}+W^{\la,22}_{\mathrm{F},h})=
W^{\la,\bar{a}a}_{\mathrm{F},h}$
plays the role of the singlet
$ W^{\la,\hat{A}=0}_{\mathrm{F},h} \equiv W^{\la}_{\mathrm{F},h}$
of $SU(2)$.

One of the commutator relations in \cite{Ahn2205} can be
written as the following commutator relation
\footnote{
\label{footnoteone}
  We can calculate the OPE between
  $W^{\la,\hat{A}=1}_{\mathrm{F},h_1=6}(z)$ and
  $W^{\la,\hat{B}=1}_{\mathrm{F},h_2=4}(w)$
  where $h_1=h_2+2$ and read off the ninth order
  pole which has the structure constant
  $\frac{131072}{5} (\la-1) \la (\la+1) (2 \la-3)
  (2 \la-1) (2 \la+1) (2 \la+3)$
  appearing in the current $ W^{\la}_{\mathrm{F},h_1+h_2-2-h=1}(w)$.
  In this case, the weight $h$ is given by odd number $h=7$.
  We do not see this term from the commutator (\ref{first})
  because the even $h$ appears in the last term of (\ref{first}).
  This implies that if the weights $h_1$ and $h_2$ do not satisfy the
  above constraint ($h_1=h_2$ or $h_1=h_2\pm 1$) for nonzero $\la$,
  then we cannot use the formula in (\ref{first}) fully
  and there appear the extra terms on the right hand sides of the
  corresponding OPE. Because the above extra factor contains
  $\la$, there will be no problem for vanishing $\la$ when we use
  (\ref{first}). It would be interesting to obtain the above commutator
  for generic $h_1$ and $h_2$. It seems that
  there is a critical singular term
  in the sense that we still have the structure of (\ref{first})
  for the poles less than this critical singular term. Of course,
  for the poles greater than the critical singular term there exist
other extra terms in general.}
\bea
\big[(W^{\la,\hat{A}}_{\mathrm{F},h_1})_m,(W^{\la,\hat{B}}_{\mathrm{F},h_2})_n\big] 
\!&=& \!
-\sum^{h_1+h_2-3}_{h= -1,
\mbox{\footnotesize odd}} \, q^h\,
p_{\mathrm{F}}^{h_1,h_2, h}(m,n,\la)
\,
%\frac{i}{2}\,
i \,
f^{\hat{A} \hat{B}  \hat{C}} \, (   W^{\la,\hat{C}}_{\mathrm{F},h_1+h_2-2-h} )_{m+n}
\nonu \\
\!& + \!& 
\left(\begin{array}{c}
m+h_1-1 \\  h_1+h_2-1 \\
 \end{array}\right) \,
c_{F} (h_1,h_2,\la) \,
\delta^{\hat{A} \hat{B}}\,
%\delta^{h_1 h_2}\,
q^{h_1+h_2-4}\,\delta_{m+n}
\nonu \\
\!&+\!& \sum^{h_1+h_2-3}_{h= 0, \mbox{\footnotesize even}} \, q^h\,
p_{\mathrm{F}}^{h_1,h_2, h}(m,n,\la)
\,
\, \delta^{\hat{A} \hat{B}}\,
(   W^{\la}_{\mathrm{F},h_1+h_2-2-h} )_{m+n}
%\Big)
\, .
\label{first}
\eea
The structure constant is given by (\ref{structla}).

Let us calculate the central term in (\ref{first}) explicitly.
One of the reasons why we are doing this is that
we have not seen any literatures which provides
all the details in the calculation and
it is useful to observe the general structure of
the computation of any OPEs including the free fields.
In order to calculate the highest singular term in the OPE $
V_{\la, \bar{a} b}^{(h_1)+}(z) \, V_{\la, \bar{c} d}^{(h_2)+}(w)$,
we need to calculate the central term in the OPE
between $V_{\la, \bar{a} b}^{(h_1)+}(z) $ and
$\de_{l \bar{l}} \,
\ga^{l \bar{c}}\, \pa^i \, \beta^{\bar{l} d}(w)$ coming from
$V_{\la, \bar{c} d}^{(h_2)+}(w)$.
It is known that
the following OPE satisfies
\bea
V_{\la,\bar{a} b}^{(h_1)+}(z)\, \ga^{j \bar{c}}(x) &=&
\delta_{b \bar{c}}\, \sum_{j=0}^{h_1-1}\, a^{j}(h_1,\la)\,
(-1)^{h_1} \, j! \,\sum_{t=0}^{j+1}\,
(j+1-t)_{h_1-1-j} \, \frac{1}{t!}
\, \frac{1}{(z-x)^{h_1-t}}\,
\pa^t \, \ga^{l \bar{a} }(x)
\nonu \\
& + &  \cdots.
\label{opeexp2}
\eea
The next step is to calculate the OPE between
$ \pa^t_x\, \ga^{l \bar{a} }(x)$
appearing in the last factor in (\ref{opeexp2})
and $\pa^i_w \, \beta^{\bar{l} d}(w)$.
We have the defining OPE relation in (\ref{fundOPE}).
The multiple derivative with respect to $x$
acting on $\frac{1}{(x-w)}$ can be obtained explicitly and
the similar  multiple derivative with respect to $w$
acting on $\frac{1}{(x-w)}$ can be rewritten
as the corresponding multiple derivative
with respect to $x$ acting on  $\frac{1}{(x-w)}$
with the number of minus signs.
Then we obtain the following result
\bea
\pa^t_x\, \ga^{l \bar{a} }(x)\, \pa^i_w \, \beta^{\bar{l} d}(w)
= \frac{1}{(x-w)^{t+i+1}}\, (-1)^t\, (t+i)! \, \delta^{l\bar{l}}\,
\de^{d \bar{a}} + \cdots \, .
\label{t+i+1}
\eea
Now we expand 
$ \frac{1}{(z-x)^{h_1-t}}$ appearing in the second factor from the
last in (\ref{opeexp2})
around $x=w$ by using the Taylor expansion.
Then we obtain that the coefficient of $(x-w)^{t+i+1}$ in the
$\frac{1}{(z-x)^{h_1+i+1}}$ evaluated at $x=w$
is given by $\frac{1}{(t+i+1)!}\, (h_1-t)_{t+i+1}$.
We are left, by collecting the contributions from
(\ref{opeexp2}) and (\ref{t+i+1}), with
\bea
&& \Bigg[\delta_{b \bar{c}}\, \sum_{j=0}^{h_1-1}\, a^{j}(h_1,\la)\,
(-1)^{h_1} \, j! \,\sum_{t=0}^{j+1}\,
(j+1-t)_{h_1-1-j} \, \frac{1}{t!}
\Bigg]\, \Bigg[ (-1)^t\, (t+i)! \, \delta^{l\bar{l}}\,
  \de^{d \bar{a}} \Bigg]\nonu \\
&& \times 
\Bigg[\frac{1}{(t+i+1)!}\, (h_1-t)_{t+i+1}
  \Bigg]\, ,
\label{threepieces}
\eea
in the $\frac{1}{(z-w)^{h_1+i+1}}$ term.
After acting the derivative
$\pa_w^{h_2-1-i}$ from the remaining factor in
the first part of
$ V_{\la, \bar{c} d}^{(h_2)+}(w)$
on the
$\frac{1}{(z-w)^{h_1+i+1}}$ term,
we obtain $(h_1+i+1)_{h_2-1-i}$.
By combining with (\ref{threepieces}), the final contribution
from the central terms in the OPE between
 $
V_{\la, \bar{a} b}^{(h_1)+}(z)$ and the
first part of $ V_{\la, \bar{c} d}^{(h_2)+}(w)$
can be written as 
\bea
&& N\, \de_{b \bar{c}}\, \de_{d \bar{a}}\,
\sum_{j=0}^{h_1-1}\, \sum_{i=0}^{h_2-1}\, \sum_{t=0}^{j+1}\,
a^j(h_1,\la)\, a^i(h_2,\la)
\nonu \\
&& \times \frac{ j! \,(t+i)!}{t! \, (t+i+1)!}
\,(-1)^{h_1+t}\, (j+1-t)_{h_1-1-j}\, (h_1-t)_{t+1+i}\,
(h_1+1+i)_{h_2-1-i} \, .
\label{cont1}
\eea
We can do the similar calculation for the
contribution from the
second part of $V_{\la, \bar{c} d}^{(h_2)+}(w)$.
In this case, we should use the following
intermediate result in \cite{Ahn2205}
\bea
V_{\la,\bar{a} b}^{(h_1)+}(z)\, c^{j \bar{c}}(x) &=&
\delta_{b \bar{c}} \sum_{j=0}^{h_1-1} a^{j}(h_1,\la+\frac{1}{2})
(-1)^{h_1}  j! \sum_{t=0}^{j+1} (j+1-t)_{h_1-1-j} 
\frac{1}{t!}
\frac{1}{(z-x)^{h_1-t}}
\pa^t  c^{l \bar{c} }(w)
\nonu \\
& + &  \cdots.
\label{opeexp3}
\eea

By starting with (\ref{opeexp3})
and following the procedures in (\ref{t+i+1}),
(\ref{threepieces}) and (\ref{cont1})
we have described above,
we obtain the following contribution
\bea
&& -N\, \de_{b \bar{c}}\, \de_{d \bar{a}}\,
\sum_{j=0}^{h_1-1}\, \sum_{i=0}^{h_2-1}\, \sum_{t=0}^{j+1}\, 
a^{j}(h_1,\la+\frac{1}{2})\,
a^i(h_2,\la+\frac{1}{2})
\nonu \\
&& \times \frac{ j! \,(t+i)!}{t! \, (t+i+1)!}
\,(-1)^{h_1+t}\, (j+1-t)_{h_1-1-j}\, (h_1-t)_{t+1+i}\,
(h_1+1+i)_{h_2-1-i} \, .
\label{cont2}
\eea

Therefore, we obtain the final central term,
by adding (\ref{cont1}) and (\ref{cont2}), as follows:
\bea
&& V_{\la, \bar{a} b}^{(h_1)+}(z) \, V_{\la, \bar{c} d}^{(h_2)+}(w)\Bigg|_{\frac{1}{
    (z-w)^{h_1+h_2}}} = N\, \de_{b \bar{c}}\, \de_{d \bar{a}}\,
\sum_{j=0}^{h_1-1}\, \sum_{i=0}^{h_2-1}\, \sum_{t=0}^{j+1}\,
\nonu \\
&& \times 
\Bigg( a^j(h_1,\la)\, a^i(h_2,\la)-a^{j}(h_1,\la+\frac{1}{2})\,
a^i(h_2,\la+\frac{1}{2})\Bigg)
\nonu \\
&& \times \frac{ j! \,(t+i)!}{t! \, (t+i+1)!}
\,(-1)^{h_1+t}\, (j+1-t)_{h_1-1-j}\, (h_1-t)_{t+1+i}\,
(h_1+1+i)_{h_2-1-i} \, .
\label{++}
\eea
Due to the behavior of two Kronecker deltas,
the central term is nonzero only for
the case where the second index of the first operator
should equal to the first index of the second operator
and
the first index of the first operator
should equal to the second index of the second operator
on the left hand side. In Appendix $B$, we present other central terms.

By realizing that the overall factor appearing in the first
current of (\ref{WWQQnonzerola})
is given by $(-4)^{h-2}$,
we can write down the central terms 
\bea
c_F(h_1,h_2,\la) & \equiv &
(-4)^{h_1+h_2-4}\, \de_{b \bar{a}} \, \de_{d \bar{c}} \, \Bigg[
  \frac{(h_1-1+2\la)}{(2h_1-1)} \, \frac{(h_2-1+2\la)}{(2h_2-1)}\,
V_{\la, \bar{a} b}^{(h_1)+}(z) \, V_{\la, \bar{c} d}^{(h_2)+}(w) \nonu \\
& + &
V_{\la, \bar{a} b}^{(h_1)-}(z) \, V_{\la, \bar{c} d}^{(h_2)-}(w)+
\frac{(h_1-1+2\la)}{(2h_1-1)}\,
V_{\la, \bar{a} b}^{(h_1)+}(z) \, V_{\la, \bar{c} d}^{(h_2)-}(w)
\nonu \\
&+& \frac{(h_2-1+2\la)}{(2h_2-1)}\,
V_{\la, \bar{a} b}^{(h_1)-}(z) \, V_{\la, \bar{c} d}^{(h_2)+}(w) \Bigg]_{\frac{1}{
    (z-w)^{h_1+h_2}}}\, ,
\label{cF}
\eea
where the relation (\ref{++}) and the relations in
Appendix (\ref{+---})
are used. In the last term of (\ref{cF}), we can use
the central term of $V_{\la, \bar{c} d}^{(h_2)+}(z) \,
V_{\la, \bar{a} b}^{(h_1)-}(w)$ with the extra factor $(-1)^{h_1+h_2}$.

In Appendices (\ref{ope-one}) and (\ref{ope-two}),
we present the corresponding OPEs for $h_1=h_2=4$
where the indices $\hat{A}$ and $\hat{B}$ are equal to each other
for the former while they are different from each other for the latter. 
Compared to the commutator relation in (\ref{first}), there exists
$(-1)^{h-1}$ factor.
Due to the Kronecker delta, the former corresponds to the last
two terms in (\ref{first}) while the latter
corresponds to the first term in (\ref{first}).
The other four cases between the nonsinglet currents are
checked explicitly and we do not present them in this paper.
The commutator relations between the nonsinglet currents are
given in \cite{Ahn2203}.
The corresponding commutator relations
between the nonsinglet currents and the singlet currents can be determined
similarly.

%%%%%%%%%%%%%%%%%%%
\subsubsection{The second commutator relation  with $h_1=h_2, h_2\pm 1$
for nonzero $\la$}
%%%%%%%%%%%%%%%%%%%

Similarly, we obtain the following commutator relation
for the currents consisting of $(\beta,\ga)$ fields
\footnote{
As in the footnote \ref{footnoteone}, the OPE between
  $W^{\la,\hat{A}=1}_{\mathrm{B},h_1=7}(z)$ and
$W^{\la,\hat{B}=1}_{\mathrm{B},h_2=3}(w)$ where $h_1=h_2+4$ can be
calculated and the ninth order
  pole contains the structure constant
  $-\frac{524288}{11}  (\la-1) \la (2 \la-3) (2 \la-1)
  (2 \la+1) (\la^2-\la+5)$
  appearing in the current $ W^{\la}_{\mathrm{B},h_1+h_2-2-h=1}(w)$
  and  the seventh order
  pole contains the structure constant
  $-6144 (\la-1) \la (2 \la-3)
  (2 \la-1) (2 \la+1)$
  appearing in the current $ W^{\la}_{\mathrm{B},h_1+h_2-2-h=3}(w)$.
  In this case, the weight $h$ is given by odd number $h=7$ or $h=5$
  while the corresponding dummy variable $h$ is given by even number
  in (\ref{BB}).
If the weights $h_1$ and $h_2$ do not satisfy the
  above constraint ($h_1=h_2$ or $h_1=h_2 \pm 1$) for nonzero $\la$,
  then we cannot use the formula in (\ref{BB}) exactly
  because the extra terms on the right hand sides of the
  corresponding OPE occur, compared to the $\la=0$ case.
  We observe that for the poles less than the
  critical singular term mentioned in the footnote \ref{footnoteone},
  the above commutator can be used precisely and for the poles
  greater than that singular term there exist extra terms on the right hand
  side of the OPE. Although these extra terms can be obtained for fixed
  $h_1$ and $h_2$, at the moment, those for general
  $h_1$ and $h_2$ are not known. It seems that as the difference between
  $h_1$ and $h_2$ increases, the pole corresponding to
  the critical singular term decreases.}
\bea
\big[(W^{\la,\hat{A}}_{\mathrm{B},h_1})_m,(W^{\la,\hat{B}}_{\mathrm{B},h_2})_n\big] 
\!&=& \!
-\sum^{h_1+h_2-3}_{h= -1, \mbox{\footnotesize odd}} \,
q^h\, p_{\mathrm{B}}^{h_1,h_2, h}(m,n,\la)
%\, \frac{i}{2}
i \, f^{\hat{A} \hat{B}  \hat{C}} \,
(   W^{\la,\hat{C}}_{\mathrm{B},h_1+h_2-2-h} )_{m+n}
\nonu \\
\!& + \!&
\left(\begin{array}{c}
m+h_1-1 \\  h_1+h_2-1 \\
 \end{array}\right) \,
c_{B}(h_1,h_2,\la) \,
\delta^{\hat{A} \hat{B}}\,
q^{h_1+h_2-4}\,\delta_{m+n} 
\nonu \\
\!&+\!& \sum^{h_1+h_2-3}_{h= 0, \mbox{\footnotesize even}} \, q^h\,
p_{\mathrm{B}}^{h_1,h_2, h}(m,n,\la)
\,
\delta^{\hat{A} \hat{B}}\,
(   W^{\la}_{\mathrm{B},h_1+h_2-2-h} )_{m+n} \, .
%\Bigg)\ 
\label{BB}
\eea
The central term appearing in (\ref{BB}), by recalling the
definition of the second relation of (\ref{WWQQnonzerola}),
can be described by
\bea
c_B(h_1,h_2,\la) & \equiv &
(-4)^{h_1+h_2-4}\, \de_{b \bar{a}} \, \de_{d \bar{c}} \, \Bigg[
  \frac{(h_1-2\la)}{(2h_1-1)} \, \frac{(h_2-2\la)}{(2h_2-1)}\,
V_{\la, \bar{a} b}^{(h_1)+}(z) \, V_{\la, \bar{c} d}^{(h_2)+}(w) \nonu \\
& + &
V_{\la, \bar{a} b}^{(h_1)-}(z) \, V_{\la, \bar{c} d}^{(h_2)-}(w)-
\frac{(h_1-2\la)}{(2h_1-1)}\,
V_{\la, \bar{a} b}^{(h_1)+}(z) \, V_{\la, \bar{c} d}^{(h_2)-}(w)
\nonu \\
&-& \frac{(h_2-2\la)}{(2h_2-1)}\,
V_{\la, \bar{a} b}^{(h_1)-}(z) \, V_{\la, \bar{c} d}^{(h_2)+}(w) \Bigg]_{\frac{1}{
    (z-w)^{h_1+h_2}}}\, .
\label{cB}
\eea
The previous relation (\ref{++}) and the previous relations in
Appendix (\ref{+---})
can be used. As before,
in the last term of (\ref{cB}), 
the central term of $V_{\la, \bar{c} d}^{(h_2)+}(z) \,
V_{\la, \bar{a} b}^{(h_1)-}(w)$ with the extra factor $(-1)^{h_1+h_2}$
can be used.
The relevant OPEs are given in Appendices (\ref{ope-three}) and
(\ref{ope-four}).
The additional 
$(-1)^{h-1}$ factor appears in the OPEs.

%%%%%%%%%%%%%%%%%%%%%%%
\subsubsection{Other commutator relations
 with $h_1=h_2, h_2+ 1$
for nonzero $\la$}
%%%%%%%%%%%%%%%%%%%%%%%

The remaining commutator relations
between the bosonic currents
and the fermionic currents
can be described as 
\bea
\big[(W^{\la,\hat{A}}_{\mathrm{F},h_1})_m,(Q^{\la,\hat{B}}_{h_2+\frac{1}{2}})_r\big] 
\!&=& \!
\sum^{h_1+h_2-3}_{h= -1} \, q^h\, q_{\mathrm{F}}^{h_1,h_2+\frac{1}{2}, h}(m,r,\la)
\,
%\frac{i}{2}
\, \Bigg( i \, f^{\hat{A} \hat{B}  \hat{C}} \, (
Q^{\la,\hat{C}}_{h_1+h_2-\frac{3}{2}-h} )_{m+r}
\nonu \\
\!& + \!&
\, \delta^{\hat{A} \hat{B}}\,
(   Q^{\la}_{h_1+h_2-\frac{3}{2}-h} )_{m+r} \Bigg)\, ,
\nonu \\
\big[(W^{\la,\hat{A}}_{\mathrm{B},h_1})_m,(Q^{\la,\hat{B}}_{h_2+\frac{1}{2}})_r\big] 
\!&=& \!
%-
\sum^{h_1+h_2-3}_{h= -1} \, q^h\,
q_{\mathrm{B}}^{h_1,h_2+\frac{1}{2}, h}(m,r,\la)
\,
%\frac{i}{2}\,
\Bigg( -i \, 
f^{\hat{A} \hat{B}  \hat{C}} \,
(   Q^{\la,\hat{C}}_{h_1+h_2-\frac{3}{2}-h} )_{m+r}
\nonu \\
\!& + \!&
%\frac{1}{N}\,
\delta^{\hat{A} \hat{B}}\,
(   Q^{\la}_{h_1+h_2-\frac{3}{2}-h} )_{m+r} \Bigg)\, ,
\nonu \\
\big[(W^{\la,\hat{A}}_{\mathrm{F},h_1})_m,
  (\bar{Q}^{\la,\hat{B}}_{h_2+\frac{1}{2}})_r\big] 
\!&=& \!
%-
\sum^{h_1+h_2-2}_{h=-1} \, q^h\,  (-1)^h\,
q_{\mathrm{F}}^{h_1,h_2+\frac{1}{2}, h}(m,r,\la)
\,
%\frac{i}{2}
\Bigg( - i \, f^{\hat{A} \hat{B}  \hat{C}} \,
( \bar{Q}^{\la,\hat{C}}_{h_1+h_2-\frac{3}{2}-h} )_{m+r}
\nonu \\
\!&+\!&
\, \delta^{\hat{A} \hat{B}}\,
(   \bar{Q}^{\la}_{h_1+h_2-\frac{3}{2}-h} )_{m+r} \Bigg) \, ,
\nonu \\
\big[(W^{\la,\hat{A}}_{\mathrm{B},h_1})_m,
  (\bar{Q}^{\la,\hat{B}}_{h_2+\frac{1}{2}})_r\big] 
\!&=& \!
\sum^{h_1+h_2-2}_{h= -1} \, q^h\,  (-1)^h \,
q_{\mathrm{B}}^{h_1,h_2+\frac{1}{2}, h}(m,r,\la)
\,
%\frac{i}{2}
\Bigg( i \, f^{\hat{A} \hat{B}  \hat{C}} \,
(   \bar{Q}^{\la,\hat{C}}_{h_1+h_2-\frac{3}{2}-h} )_{m+r}
\nonu \\
\!& +\! &
%\frac{1}{N}
\, \delta^{\hat{A} \hat{B}}\,
(   \bar{Q}^{\la}_{h_1+h_2-\frac{3}{2}-h} )_{m+r} \Bigg) \, . 
\label{WQ}
\eea
The corresponding OPEs for fixed $h_1$ and $h_2$
can be found in Appendices
(\ref{ope-five}), (\ref{ope-six}),(\ref{ope-seven}),
(\ref{ope-eight}),(\ref{ope-nine}), (\ref{ope-ten}),
(\ref{ope-eleven}) and (\ref{ope-twelve}).
In the last two commutator relations of (\ref{WQ}),
the upper limit of $h$ is given by $h=h_1+h_2-2$
due to the presence of the lowest fermionic current
$\bar{Q}^{\la,b \bar{a}}_{\frac{1}{2}}$.
The additional 
$(-1)^{h-1}$ factor appears when we change the commutator relations
into the corresponding OPEs
\footnote{
\label{foot}
  As in two previous examples, the OPE between
  $W^{\la,\hat{A}=1}_{\mathrm{F},h_1=5}(z)$ and
$Q^{\la,\hat{B}=1}_{\mathrm{B},h_2+\frac{1}{2}=\frac{7}{2}}(w)$
where $h_1=h_2+2$ can be
obtained and the seventh order
  pole contains the structure constant
  $-\frac{6144}{35}  (\la-1) (\la+1) (\la+2)
  (2 \la-3) (2 \la+1) (2 \la+3)$
  appearing in the current $ Q^{\la}_{h_1+h_2-\frac{3}{2}-h=\frac{3}{2}}(w)$.
  On the other hand, the 
  structure constant $ q_{\mathrm{F}}^{5,
    \frac{7}{2},5}(m,r,\la)$
  contains the $\la$ dependent factor
  $\frac{4}{1575} (\la-1) (\la+1) (2 \la-3)
    (2 \la+1) (4 \la^2-14 \la-9)$.
  By subtracting  the contribution
  $\frac{2}{225} (\la-1) \la
  (\la+1) (2 \la-3) (2 \la-1) (2 \la+1)$
  coming from the 
  structure constant $ q_{\mathrm{F}}^{4,
    \frac{7}{2},5}(m,r,\la)$, where $4$ is realized by
  $(h_1-1)$, from the above,
  we obtain
  $-\frac{2}{525}  (\la-1) (\la+1) (\la+2)
  (2 \la-3) (2 \la+1) (2 \la+3)$.
  Note that the additional term
  in the  $ q_{\mathrm{F}}^{4,
    \frac{7}{2},5}(m,r,\la)$ contains the factor
  $\la$.
  By considering the numerical factor $46080$
  when we move from the modes in the commutator
  to the differential operators in the OPE
  and multiplying this into the above factor,
  we obtain the previous structure constant in the current
  $ Q^{\la}_{h_1+h_2-\frac{3}{2}-h=\frac{3}{2}}(w)$.
This implies that if the weights $h_1$ and $h_2$ do not satisfy the
  above constraint ($h_1=h_2$ or $h_1=h_2+1$) for nonzero $\la$,
  then the formula in the first equation of (\ref{WQ})
  cannot be used fully because the extra contribution from the
  structure constant on the right hand sides of the
  corresponding OPE occur. As before, for the poles less than
  the critical singular term, still we can use some of terms in
  the corresponding expression of the first relation of (\ref{WQ}).

  Similarly,  let us consider the last equation of (\ref{WQ}).
  Then the OPE between
  $W^{\la,\hat{A}=1}_{\mathrm{B},h_1=5}(z)$ and
$\bar{Q}^{\la,\hat{B}=1}_{\mathrm{B},h_2+\frac{1}{2}=\frac{7}{2}}(w)$ provides
the seventh order pole with
$-\frac{2048}{35}  (\la-1) (\la+1) (2 \la-3)
(2 \la+1) (2 \la+3) (11 \la-10)$
appearing in the current $ \bar{Q}^{\la}_{\frac{3}{2}}(w)$.
This can be obtained by adding the
extra contribution from  $ q_{\mathrm{B}}^{4,
  \frac{7}{2},5}(m,r,\la)$  in addition to
the one from  $ q_{\mathrm{B}}^{5,
  \frac{7}{2},5}(m,r,\la)$ as in previous paragraph.
Note that 
the $ \bar{Q}^{\la}_{\frac{1}{2}}(w)$
term on the right hand side of the OPE can be
determined by the contribution from the
structure constant $ q_{\mathrm{B}}^{5,
  \frac{7}{2},6}(m,r,\la)$ only.
Therefore,
the last equation of (\ref{WQ}) should be modified
for nonzero $\la$ with more general $h_1$ and $h_2$
where the weight $h_2$ is outside of the above allowed region.
The similar behaviors corresponding to
the second and the third equations of (\ref{WQ})
we do not present in this paper can appear.}.

%%%%%%%%%%%%%%%%%%%%%%%%%
\subsubsection{The final anticommutator relation
 with $h_1=h_2$
for nonzero $\la$}
%%%%%%%%%%%%%%%%%%%%%%%%%%

The final anticommutator relation
between the fermionic currents is summarized by
\bea
\{(Q^{\la,\hat{A}}_{h_1+\frac{1}{2}})_r,(\bar{Q}^{\la,\hat{B}}_{h_2+\frac{1}{2}})_s\} 
\!&=& \!
\sum^{h_1+h_2-1}_{h= 0} \, q^h \,o_{\mathrm{F}}^{h_1+\frac{1}{2},h_2+
\frac{1}{2}, h}(r,s,\la)
\,
%\frac{i}{2}
\Bigg( i \, f^{\hat{A} \hat{B}  \hat{C}} \,
(   W^{\la,\hat{C}}_{F,h_1+h_2-h} )_{r+s}
\nonu \\
\!& + \!&
%\frac{1}{N}
\, \delta^{\hat{A} \hat{B}}\,
(   W^{\la}_{F,h_1+h_2-h} )_{r+s} \Bigg)
\nonu \\
\!&+\!& \sum^{h_1+h_2-1}_{h= 0} \, q^h \,
o_{\mathrm{B}}^{h_1+\frac{1}{2},h_2+\frac{1}{2}, h}(r,s,\la)
\,
%\frac{i}{2}
\Bigg( -i \, f^{\hat{A} \hat{B}  \hat{C}} \,
(   W^{\la,\hat{C}}_{B,h_1+h_2-h} )_{r+s}
\nonu \\
\!& +\! &
%\frac{1}{N}
\, \delta^{\hat{A} \hat{B}}\,
(   W^{\la}_{B,h_1+h_2-h} )_{r+s} \Bigg)
\nonu \\
\!&+\!&
\left(\begin{array}{c}
r+h_1-\frac{1}{2} \\  h_1+h_2 \\
 \end{array}\right) \,
c_{Q}(h_1,h_2,\la)  \, \delta^{\hat{A} \hat{B}}
\,  q^{h_1+ h_2-2}
\delta_{r+s} \, .
\label{Final}
  \eea
  The corresponding OPEs for fixed
  $h_1$ and $h_2$ can be found in Appendices (\ref{ope-thirteen})
  and (\ref{ope-fourteen})
\footnote{
As in the examples appearing in the previous footnotes, the OPE between
  $Q^{\la,\hat{A}=1}_{h_1+\frac{1}{2}=\frac{3}{2}}(z)$ and
$\bar{Q}^{\la,\hat{B}=1}_{h_2+\frac{1}{2}=\frac{7}{2}}(w)$
where $h_1=h_2-2$ can be
determined and the fourth order
  pole contains the structure constant
  $\frac{128}{5} (\la-1) (2 \la-3) (2 \la-1)$
  appearing in the current $ W^{\la}_{F,h_1+h_2-h=1}(w)$.
  On the other hand, the 
  structure constant $ o_{\mathrm{F}}^{\frac{3}{2},
    \frac{7}{2},3}(r,s,\la)$
  contains the $\la$ dependent factor
  $\frac{8}{15} (2 \la-1)(4 \la^2+2 \la+3)$.
  We can take the sum (due to the odd $h$)
  of these with proper numerical value
  $-48$ for the latter and obtain
  $-\frac{128}{5} \la (2 \la-1) (2 \la+7)$ which vanishes
  at $\la=0$ and this implies that this extra contribution
  should appear.
  Furthermore,
   the fourth order
  pole contains the structure constant
  $\frac{256}{5} \la (\la+1) (2 \la+1)$
  appearing in the current $ W^{\la}_{B,h_1+h_2-h=1}(w)$.
  On the other hand, the 
  structure constant $ o_{\mathrm{B}}^{\frac{3}{2},
    \frac{7}{2},3}(r,s,\la)$
  contains
  $\frac{16}{15} \la (4 \la^2-6 \la+5)$.
  Obviously they are different from each other and the extra
  term $-\frac{256}{5} (\la-4) \la (2 \la-1)$ should appear from similar
  analysis.
  Similarly,   
the third order
  pole contains the structure constant
  $\frac{96}{5} (\la-1) (2 \la-3)$
  appearing in the current $ W^{\la}_{F,h_1+h_2-h=2}(w)$.
  Moreover, the 
  structure constant $ o_{\mathrm{F}}^{\frac{3}{2},
    \frac{7}{2},2}(r,s,\la)$
  contains
  $-\frac{4}{5} (4 \la^2+10 \la-9)$.
  By taking the difference (due to the even $h$) between the latter
  with an additional numerical value $8$ and the former,
  we obtain $-32 \la (2 \la-1)$ which vanishes at $\la=0$
  and should appear.
  The third order
  pole also contains the structure constant
  $\frac{96}{5} (\la+1) (2 \la+1)$
 appearing in the current $ W^{\la}_{B,h_1+h_2-h=2}(w)$.
  On the other hand, the 
  structure constant $ o_{\mathrm{B}}^{\frac{3}{2},
    \frac{7}{2},2}(r,s,\la)$
  contains
  $-\frac{4}{5}  (4 \la^2-14 \la-3)$.
  We obtain $-32 \la (2 \la-1)$ from the difference
  between the latter and the former as before and this $\la$ dependent
  contribution should appear.  
  If the weights $h_1$ and $h_2$ are not equal to each other
  for nonzero $\la$,
  then the formula in the first equation of (\ref{Final})
  cannot be used fully because the extra contribution from the
  structure constant on the right hand sides of the
  corresponding OPE arises.
}.
The additional 
$(-1)^{h}$ factor appears when we change the anticommutator relations
into the corresponding OPEs.
Because the lowest weight of $W_{B,h}^{\la,\bar{a} b}$
is given by $1$, the upper limit of the second
summation is also
given by $h=h_1+h_2-1$ which is the same as the one in the first
summation of (\ref{Final}).
  Here the central term appearing in (\ref{Final}) can be described by
\bea
c_Q(h_1,h_2,\la) & \equiv &  8(-4)^{h_1+h_2-4} 
\de_{b \bar{a}}  \de_{d \bar{c}}  \Bigg[
-Q_{\la, \bar{a} b}^{(h_1+1)+}(z)  Q_{\la, \bar{c} d}^{(h_2+1)+}(w)  + 
Q_{\la, \bar{a} b}^{(h_1+1)-}(z)  Q_{\la, \bar{c} d}^{(h_2+1)-}(w)
\nonu \\
& - &
Q_{\la, \bar{a} b}^{(h_1+1)+}(z) \, Q_{\la, \bar{c} d}^{(h_2+1)-}(w)
+ Q_{\la, \bar{a} b}^{(h_1+1)-}(z) \, Q_{\la, \bar{c} d}^{(h_2+1)+}(w)
\nonu \\
& - & (-1)^{h_1+h_2}\,
Q_{\la, \bar{c} d}^{(h_2+1)+}(z) \, Q_{\la, \bar{a} b}^{(h_1+1)+}(w)  + 
 (-1)^{h_1+h_2}\,Q_{\la, \bar{c} d}^{(h_2+1)-}(z) \, Q_{\la, \bar{a} b}^{(h_1+1)-}(w)
\nonu \\
& - & (-1)^{h_1+h_2}\,
Q_{\la, \bar{c} d}^{(h_2+1)+}(z) \, Q_{\la, \bar{a} b}^{(h_1+1)-}(w)
\nonu \\
& + &  (-1)^{h_1+h_2}\,
Q_{\la, \bar{c} d}^{(h_2+1)-}(z) \, Q_{\la, \bar{a} b}^{(h_1+1)+}(w)
\Bigg]_{\frac{1}{
    (z-w)^{h_1+h_2+1}}}\, ,
\label{cQ}
\eea
where we can use Appendix (\ref{CEN}).

Therefore, the
seven (anti)commutator relations between
  the nonsinglet currents
  are given by (\ref{first}), (\ref{BB}), (\ref{WQ}) and
  (\ref{Final}).
  
%%%%%%%%%%%%%%%%%%%%
\subsection{The (anti)commutator relations between
  the nonsinglet currents in explicit forms
 with $h_1=h_2, h_2\pm 1$
for nonzero $\la$
}
%%%%%%%%%%%%%%%%%%%%

In order to construct the algebra from the
${\cal N}=4$ multiplets, we should rewrite the previous
(anti)commutator relations in the basis of four components
of $(\bar{a} b)=(11,12,21,22)$ in each current.
For example,
from (\ref{first}), the commutator relation
between $\hat{A}=1$ (sum of $(12)$ and $(21)$) and
$\hat{B}=1$ is known. 
Moreover, the commutator relation
between $\hat{A}=1$ and
$\hat{B}=2$ (difference of $(12)$ and $(21)$ up to an overall factor)
is known. Then we obtain the commutator relation between
$\hat{A}=1$ and the element $(12)$ current by adding the above two
commutator relations.
By realizing that there is no singular term in the commutator
relation between the element $(12)$ current and itself,
the above analysis leads to the commutator relation between
the element $(21)$ and the element $(12)$ as follows:
\bea
\big[(W^{\la,21}_{\mathrm{F},h_1})_m,(W^{\la,12}_{\mathrm{F},h_2})_n\big] 
\!&=& \!
\frac{1}{2}\, \sum^{h_1+h_2-3}_{h= 0,
\mbox{\footnotesize even}} \, q^h\,
p_{\mathrm{F}}^{h_1,h_2, h}(m,n,\la)
 \, ( W^{\la,11}_{\mathrm{F},h_1+h_2-2-h} + W^{\la,22}_{\mathrm{F},h_1+h_2-2-h} )_{m+n}
\nonu \\
\!& + \!&
\frac{1}{2}\,
\left(\begin{array}{c}
m+h_1-1 \\  h_1+h_2-1 \\
 \end{array}\right) \,
c_{F} (h_1,h_2,\la) \,
%\delta^{\hat{A} \hat{B}}\, \delta^{h_1 h_2}\,
q^{h_1+h_2-4}\,\delta_{m+n}
\label{FIRST}
\\
\!&-\!& \frac{1}{2}\,
\sum^{h_1+h_2-3}_{h= -1, \mbox{\footnotesize odd}} \, q^h\,
p_{\mathrm{F}}^{h_1,h_2, h}(m,n,\la)
\,
(   W^{\la,11}_{\mathrm{F},h_1+h_2-2-h}- W^{\la,22}_{\mathrm{F},h_1+h_2-2-h} )_{m+n}
\, .
\nonu
\eea

On the other hand,
by subtracting the previous two commutator relations
and by using the fact that
there is no singular term in the commutator
relation between the element $(21)$ current and itself, 
we  obtain the following commutator relation
between the element $(12)$ and the element $(21)$
as follows:
\bea
\big[(W^{\la,12}_{\mathrm{F},h_1})_m,(W^{\la,21}_{\mathrm{F},h_2})_n\big] 
\!&=& \!
\frac{1}{2}\, \sum^{h_1+h_2-3}_{h= 0,
\mbox{\footnotesize even}} \, q^h\,
p_{\mathrm{F}}^{h_1,h_2, h}(m,n,\la)
 \, ( W^{\la,11}_{\mathrm{F},h_1+h_2-2-h} + W^{\la,22}_{\mathrm{F},h_1+h_2-2-h} )_{m+n}
\nonu \\
\!& + \!&
\frac{1}{2}\,
\left(\begin{array}{c}
m+h_1-1 \\  h_1+h_2-1 \\
 \end{array}\right) \,
c_{F} (h_1,h_2,\la) \,
%\delta^{\hat{A} \hat{B}}\, \delta^{h_1 h_2}\,
q^{h_1+h_2-4}\,\delta_{m+n}
\label{SECOND}
\\
\!&-\!& \frac{1}{2}\,
\sum^{h_1+h_2-3}_{h= -1, \mbox{\footnotesize odd}} \, q^h\,
p_{\mathrm{F}}^{h_1,h_2, h}(m,n,\la)
\,
(   W^{\la,11}_{\mathrm{F},h_1+h_2-2-h}- W^{\la,22}_{\mathrm{F},h_1+h_2-2-h} )_{m+n}
\, .
\nonu
\eea

Similarly, the commutator relation
having both $\hat{A}=1$ and $\hat{B}=3$
and the commutator relation having $\hat{A}=2$
and $\hat{B}=3$ provide the commutator relation
between the element $(12)$ and the current having
$\hat{B}=3$ and
 the commutator relation
between the element $(21)$ and the current having
$\hat{B}=3$. Furthermore,
 the commutator relation
having both $\hat{A}=1$ and $\hat{B}=0$
and the commutator relation having $\hat{A}=2$
and $\hat{B}=0$ provide
 the commutator relation
between the element $(12)$ and the current having
$\hat{B}=0$ and
 the commutator relation
between the element $(21)$ and the current having
$\hat{B}=0$.
Then we obtain the following commutator relations by linear combination
of previous four commutators
\bea
\big[(W^{\la,12}_{\mathrm{F},h_1})_m,(W^{\la,11}_{\mathrm{F},h_2})_n\big] 
\!&=& \!
\frac{1}{2}\, (\sum^{h_1+h_2-3}_{h= 0,
  \mbox{\footnotesize even}} -\sum^{h_1+h_2-3}_{h= -1,
  \mbox{\footnotesize odd}})
\, q^h\,
p_{\mathrm{F}}^{h_1,h_2, h}(m,n,\la)
 \, ( W^{\la,12}_{\mathrm{F},h_1+h_2-2-h} )_{m+n} \, ,
 \nonu \\
\big[(W^{\la,12}_{\mathrm{F},h_1})_m,(W^{\la,22}_{\mathrm{F},h_2})_n\big] 
\!&=& \!
\frac{1}{2}\, (\sum^{h_1+h_2-3}_{h= 0,
  \mbox{\footnotesize even}} +\sum^{h_1+h_2-3}_{h= -1,
  \mbox{\footnotesize odd}})
\, q^h\,
p_{\mathrm{F}}^{h_1,h_2, h}(m,n,\la)
 \, ( W^{\la,12}_{\mathrm{F},h_1+h_2-2-h} )_{m+n} \, ,
 \nonu \\
 \big[(W^{\la,21}_{\mathrm{F},h_1})_m,(W^{\la,11}_{\mathrm{F},h_2})_n\big] 
\!&=& \!
\frac{1}{2}\, (\sum^{h_1+h_2-3}_{h= 0,
  \mbox{\footnotesize even}} +\sum^{h_1+h_2-3}_{h= -1,
  \mbox{\footnotesize odd}})
\, q^h\,
p_{\mathrm{F}}^{h_1,h_2, h}(m,n,\la)
 \, ( W^{\la,21}_{\mathrm{F},h_1+h_2-2-h} )_{m+n} \, ,
 \nonu \\
 \big[(W^{\la,21}_{\mathrm{F},h_1})_m,(W^{\la,22}_{\mathrm{F},h_2})_n\big] 
&=& 
\frac{1}{2} (\sum^{h_1+h_2-3}_{h= 0,
  \mbox{\footnotesize even}} -\sum^{h_1+h_2-3}_{h= -1,
  \mbox{\footnotesize odd}})
q^h
p_{\mathrm{F}}^{h_1,h_2, h}(m,n,\la)
  ( W^{\la,21}_{\mathrm{F},h_1+h_2-2-h} )_{m+n}.
 \label{THIRD}
 \eea

 From the results of  
 the commutator relation
 between the current having the index $\hat{A}=0$
 and the current having
 $\hat{B}=1$ and
  the commutator relation
 between the current having the index $\hat{A}=0$
 and the current having
 $\hat{B}=2$,
 the commutator relation between the
 current having the index $\hat{A}=0$
 and the element $(12)$
 and the one between  the
 current having the index $\hat{A}=0$
 and the element $(21)$ can be determined.
 Moreover, from the commutator relation
 having the index $\hat{A}=3$ and $\hat{B}=1$
and  the commutator relation
  having the index $\hat{A}=3$ and $\hat{B}=2$,
  the commutator relation between the
 current having the index $\hat{A}=3$
 and the element $(12)$
 and the one between  the
 current having the index $\hat{A}=3$
 and the element $(21)$ can be fixed.
 Therefore, by linear combinations of these results,
 we can determine the following commutator relations
 \bea
 \big[(W^{\la,11}_{\mathrm{F},h_1})_m,(W^{\la,12}_{\mathrm{F},h_2})_n\big] 
\!&=& \!
\frac{1}{2}\, (\sum^{h_1+h_2-3}_{h= 0,
  \mbox{\footnotesize even}} +\sum^{h_1+h_2-3}_{h= -1,
  \mbox{\footnotesize odd}})
\, q^h\,
p_{\mathrm{F}}^{h_1,h_2, h}(m,n,\la)
 \, ( W^{\la,12}_{\mathrm{F},h_1+h_2-2-h} )_{m+n} \, ,
 \nonu \\
 \big[(W^{\la,22}_{\mathrm{F},h_1})_m,(W^{\la,12}_{\mathrm{F},h_2})_n\big] 
\!&=& \!
\frac{1}{2}\, (\sum^{h_1+h_2-3}_{h= 0,
  \mbox{\footnotesize even}} -\sum^{h_1+h_2-3}_{h= -1,
  \mbox{\footnotesize odd}})
\, q^h\,
p_{\mathrm{F}}^{h_1,h_2, h}(m,n,\la)
 \, ( W^{\la,12}_{\mathrm{F},h_1+h_2-2-h} )_{m+n} \, ,
 \nonu \\
 \big[(W^{\la,11}_{\mathrm{F},h_1})_m,(W^{\la,21}_{\mathrm{F},h_2})_n\big] 
\!&=& \!
\frac{1}{2}\, (\sum^{h_1+h_2-3}_{h= 0,
  \mbox{\footnotesize even}} -\sum^{h_1+h_2-3}_{h= -1,
  \mbox{\footnotesize odd}})
\, q^h\,
p_{\mathrm{F}}^{h_1,h_2, h}(m,n,\la)
 \, ( W^{\la,21}_{\mathrm{F},h_1+h_2-2-h} )_{m+n} \, ,
 \nonu \\
 \big[(W^{\la,22}_{\mathrm{F},h_1})_m,(W^{\la,21}_{\mathrm{F},h_2})_n\big] 
&=& 
\frac{1}{2} (\sum^{h_1+h_2-3}_{h= 0,
  \mbox{\footnotesize even}} +\sum^{h_1+h_2-3}_{h= -1,
  \mbox{\footnotesize odd}})
q^h
p_{\mathrm{F}}^{h_1,h_2, h}(m,n,\la)
( W^{\la,21}_{\mathrm{F},h_1+h_2-2-h} )_{m+n}.
 \label{FOURTH}
 \eea

 Finally, from the results of
 the commutator relation between the current having the
 index $\hat{A}=0$ and the current having the index $\hat{B}=3$
 and  the commutator relation between the current having the
 index $\hat{A}=0$ and the current having the index $\hat{B}=0$,
 we can determine the following commutator relations
 by realizing that there is no nontrivial commutator relation
 between the element $(11)$ and the element $(22)$
 \bea
 \big[(W^{\la,11}_{\mathrm{F},h_1})_m,(W^{\la,11}_{\mathrm{F},h_2})_n\big] 
\!&=& \!
\sum^{h_1+h_2-3}_{h= 0,
  \mbox{\footnotesize even}} 
\, q^h\,
p_{\mathrm{F}}^{h_1,h_2, h}(m,n,\la)
\, ( W^{\la,11}_{\mathrm{F},h_1+h_2-2-h} )_{m+n} \, \nonu \\
& + &
\frac{1}{2}\,
\left(\begin{array}{c}
m+h_1-1 \\  h_1+h_2-1 \\
 \end{array}\right) \,
c_{F} (h_1,h_2,\la) \,
%\delta^{\hat{A} \hat{B}}\, \delta^{h_1 h_2}\,
q^{h_1+h_2-4}\,\delta_{m+n} \, ,
\nonu \\
 \big[(W^{\la,22}_{\mathrm{F},h_1})_m,(W^{\la,22}_{\mathrm{F},h_2})_n\big] 
\!&=& \!
\sum^{h_1+h_2-3}_{h= 0,
  \mbox{\footnotesize even}} 
\, q^h\,
p_{\mathrm{F}}^{h_1,h_2, h}(m,n,\la)
\, ( W^{\la,22}_{\mathrm{F},h_1+h_2-2-h} )_{m+n} \, \nonu \\
& + &
\frac{1}{2}\,
\left(\begin{array}{c}
m+h_1-1 \\  h_1+h_2-1 \\
 \end{array}\right) \,
c_{F} (h_1,h_2,\la) \,
%\delta^{\hat{A} \hat{B}}\, \delta^{h_1 h_2}\,
q^{h_1+h_2-4}\,\delta_{m+n} \, .
\label{FIFTH}
\eea

Therefore, there are 
(\ref{FIRST}), (\ref{SECOND}), (\ref{THIRD}),
(\ref{FOURTH}) and (\ref{FIFTH}).
Other complete relations are summarized by
Appendices (\ref{res1}), (\ref{res2}), (\ref{res3}),
(\ref{res4}), (\ref{res5}) and (\ref{res6}).
By adding the two equations of (\ref{FIFTH}),
we obtain the commutator between
the singlet currents and by taking the contractions of the
currents with the vanishing $q$ limit for $\la=0$
we obtain the $w_{1+\infty}$ algebra \cite{Bakas,Ahn2111,Ahn2202}.

%%%%%%%%%%%%%%%%%%%%%%%%%%%%%%%%%%%%%%%%%%%%%%%%%%%%%%%%%%%%%%%%%%%%%
%%%%%%%%%%%%%%%%%%%%%%%%%%%%%%%%%%%%%%%%%%%%%%%%%%%%%%%%%%%%%%%%%%%%%%
\section{ The ${\cal N}=4$ supersymmetric linear $W_{\infty}^{2,2}$
algebra between the ${\cal N}=4$ multiplets}
%1%%%%%%%%%%%%%%%%%%%%%%%%%%%%%%%%%%%%%%%%%%%%%%%%%%%%%%%%%%%%%%%%%%%%%
%%%%%%%%%%%%%%%%%%%%%%%%%%%%%%%%%%%%%%%%%%%%%%%%%%%%%%%%%%%%%%%%%%%%%

Based on the results in previous sections,
we present the first five commutator relations
between the ${\cal N}=4$ multiplets. In the footnotes,
some particular examples for the specific weights
$h_1$ and $h_2$ show that there are extra terms on the right hand sides
of the commutators for the nonzero $\la$ we explained in the introduction.

%%%%%%%%%%%%%%%%%%
\subsection{The commutator relation between the lowest components
with $h_1=h_2, h_2 \pm 1$ for nonzero $\la$}
%%%%%%%%%%%%%%%%%%

Let us consider the
commutator relation between the lowest components $\Phi_0^{(h_1)}(z)$
and $\Phi_0^{(h_2)}(w)$ of the
${\cal N}=4$ multiplet. Because
there are no singular terms in the OPE of
$W_{F,h_1}^{\la, \bar{a} a}(z)$ and $W_{B,h_2}^{\la, \bar{b} b}(w)$
according to (\ref{fundOPE}),
the commutator relation consists of two parts.
That is, they are given by
$\big[(W_{F,h_1}^{\la, \bar{a} a})_m,
  (W_{F,h_2}^{\la, \bar{b} b})_n \big]$
and $\big[(W_{B,h_1}^{\la, \bar{a} a})_m,
  (W_{B,h_2}^{\la, \bar{b} b})_n \big]$. Then we can use
the relation (\ref{FIFTH}) and the last two relations of
Appendix (\ref{res1}). We need to express the right hand sides
of these commutator relations in terms of the components of
the ${\cal N}=4$ multiplet in order to present the complete
algebra between them.
The two relations in (\ref{FIFTH}) and the last two
relations of Appendix (\ref{res1}) can be added respectively because
the structure constants are common at each expression. 

As described before, from the relations (\ref{one}) and (\ref{last}),
we obtain the following relations 
\bea
W_{F,h}^{\la, \bar{a} a} &= & -\frac{1}{(-4)^{h-2}}\, \Phi_0^{(h)}
-\frac{(h-1+2\la)}{8(2h-1)(-4)^{h-6}}\, \tilde{\Phi}_2^{(h-2)}\, ,
\nonu \\
W_{B,h}^{\la, \bar{a} a} &= & \frac{1}{(-4)^{h-2}}\, \Phi_0^{(h)}
-\frac{(h-2\la)}{8(2h-1)(-4)^{h-6}}\, \tilde{\Phi}_2^{(h-2)}\, .
\label{rel1}
\eea
Once we observe the currents on the left hand sides, then
we should rewrite them in terms of the components of
${\cal N}=4$ multiplet according to (\ref{rel1}).

On the other hand, there exist the following relations
between the lowest bosonic currents and
the lowest component $\Phi_0^{(1)}$  of the
first ${\cal N}=4$ multiplet from (\ref{one}) and the weight-$1$ current
$U = 2 (W_{F,1}^{\la, \bar{a} a}+W_{B,1}^{\la, \bar{a} a})$
of the ${\cal N}=4$ stress energy tensor
\bea
W_{F,1}^{\la, \bar{a} a} &= & 4 \, \Phi_0^{(1)} +\la \, U\, ,
\nonu \\
W_{B,1}^{\la, \bar{a} a} &= & - 4 \, \Phi_0^{(1)} +\frac{1}{2}\, (1-2\la)\, U
\, .
\label{rel2}
\eea
This implies that by comparing (\ref{rel1}) with (\ref{rel2}),
we can identify the previous
weight-$1$ current as 
$\tilde{\Phi}_2^{(-1)} \equiv 256 \, U$.

Furthermore, there are relations 
between 
the lowest component $\Phi_0^{(2)}$
of the second ${\cal N}=4$ multiplet and the weight-$2$
stress energy tensor
$L$ of the ${\cal N}=4$ stress energy tensor
together with (\ref{one}) and (\ref{Lterm}) as follows:
\bea
W_{F,2}^{\la, \bar{a} a} &= & - \Phi_0^{(2)} +\frac{1}{3}\, (1+2\la) \, L\, ,
\nonu \\
W_{B,2}^{\la, \bar{a} a} &= & \Phi_0^{(2)} +\frac{2}{3}\, (1-\la)\, L
\, .
\label{rel3}
\eea
Again, from the relations (\ref{rel1}) and (\ref{rel3}),
we identify the stress energy tensor in terms of the
component of ${\cal N}=4$ multiplet as
$\tilde{\Phi}_2^{(0)} \equiv -32 \, L$.

Therefore, the commutator relation
between the lowest component of the $h_1$-th ${\cal N}=4$ multiplet
and
the lowest component of the $h_2$-th ${\cal N}=4$ multiplet
  can be described as, by substituting the relation (\ref{rel1}) into the
  above right hand sides of the relevant commutator relations described
  before,
  \bea
&& \big[(\Phi^{(h_1)}_{0})_m,
  (\Phi^{(h_2)}_{0})_n \big]  = 
\frac{(-4)^{h_1-2}}{(2h_1-1)} \, \frac{(-4)^{h_2-2}}{(2h_2-1)}
\Bigg[ 
\left(\begin{array}{c}
m+h_1-1 \\  h_1+h_2-1 \\
\end{array}\right) \, q^{h_1+h_2-4}
\nonu \\
&&
  \times \Bigg(
  (h_1-2\la)(h_2-2\la) \, c_F(h_1,h_2,\la) +
  (h_1-1+2\la)(h_2-1+2\la)\, c_B(h_1,h_2,\la) \Bigg)\,
  \de_{m+n}
  \nonu \\
  && +   \sum_{h=0,\mbox{\footnotesize even}}^{h_1+h_2-3} \, \Bigg( -
  (h_1-2\la)(h_2-2\la)\, q^h\, 
p_{\mathrm{F}}^{h_1,h_2, h}(m,n,\la)
\nonu \\
&& +  
 %\sum_{h=0,\mbox{\footnotesize even}}^{h_1+h_2-3} \,
 (h_1-1+2\la)(h_2-1+2\la)\, q^h\, 
p_{\mathrm{B}}^{h_1,h_2, h}(m,n,\la)
\,  \Bigg)\,  \frac{1}{(-4)^{h_1+h_2-h-4}}\,
(\Phi_{0}^{(h_1+h_2-2-h)})_{m+n}
\nonu \\
&& - \sum_{h=0,\mbox{\footnotesize even}}^{h_1+h_2-3} \, \Bigg(
(h_1-2\la)(h_2-2\la)(h_1+h_2-h-3+2\la)\, q^h\, 
p_{\mathrm{F}}^{h_1,h_2, h}(m,n,\la)
\nonu \\
&& +
% \sum_{h=0,\mbox{\footnotesize even}}^{h_1+h_2-3} \,
(h_1-1+2\la)(h_2-1+2\la)(h_1+h_2-h-2-2\la)\, q^h\, 
p_{\mathrm{B}}^{h_1,h_2, h}(m,n,\la)
\,
\Bigg)\nonu \\
&& \times \frac{1}{
  8(2h_1+2h_2-2h-5)(-4)^{h_1+h_2-h-8}}
\, (\tilde{\Phi}_{2}^{(h_1+h_2-h-4)})_{m+n}
  \Bigg] \, .
\label{Onerel}
\eea
Each central term is given by (\ref{cF}) and (\ref{cB}).
Due to the even weight $h$ in the summation,
the currents of even (or odd) weights can occur depending on
the weights $h_1$ and $h_2$.
When we change the
above commutator relation (\ref{Onerel})
into the corresponding OPE,
due to the factor $(-1)^{h-1}$ for even $h$, then there
exists an extra minus sign on the right hand side of the OPE.
We can easily observe that for the maximum value of
dummy variable $h$, $h=h_1+h_2-3$ with odd $(h_1+h_2)$,
on the right hand side of (\ref{Onerel}), there appear
the currents
$\Phi_0^{(1)}$ and $\tilde{\Phi}_2^{-1}$ which is related to
the previous current $U$ of ${\cal N}=4$ stress energy tensor.
Note that we have vanishing structure constant
$p_B^{h_1,h_2,h_1+h_2-3}$
at $\la=0$ \cite{Ahn2203}.
On the other hand, for even $(h_1+h_2)$,
the maximum value of the weight $h$ is given by $h=h_1+h_2-4$
because the weight $h$ should be even. For this value,
 there appear
 the currents
 $\Phi_0^{(2)}$ and $\tilde{\Phi}_2^{(0)}$, which is related to
 the previous current $L$, with proper
 $\la$ dependent
 structure constants on the right hand side of (\ref{Onerel}).
 Let us emphasize that for nonzero $\la$, the above commutator
 holds for the arbitrary weight $h_1$ under the
 condition $h_1=h_2$ or $h_1=h_2\pm 1$. We take the particular example
 which shows that if the weights $h_1$ and $h_2$
 do not satisfy this condition, then there exist other terms on the right
 hand of the above commutator
 \footnote{
Let us take the OPE between $\Phi^{(h_1=1)}_{0}(z)$ and $
\Phi^{(h_2=3)}_{0}(w) $ where the weights satisfy
$h_1=h_2-2$. The third order pole of this OPE
provides the structure constant $2048 \la (2 \la-1)$
appearing in the  current $\Phi^{(1)}_{0}(w)$ and
the structure constant $\frac{4}{5} \la (2 \la-1)
(4 \la-1)$
appearing in the  current $U(w)$.
According to (\ref{Onerel}), the exponent of the first current
implies $h_1+h_2-2-h=2-h$. Moreover, $h$ is given by $h=0$
from the summation and the exponent is $2$. Therefore, the current
$\Phi^{(1)}_{0}(w)$ cannot appear from the (\ref{Onerel}).
 Furthermore, the exponent in the second current
 implies $h_1+h_2-4-h=-h=-1$ from the presence of $U$.
 Moreover, $h$ is given by $h=0$
 from the summation and there is a contradiction.
 In this case also, the current
$U(w)$ cannot appear from the corresponding term in (\ref{Onerel}).
Note that the two structure constants contain the $\la$ factor
and they become zero at $\la=0$.
From the above analysis, there exist extra terms on the right hand side
of the commutator (or corresponding OPE)
for the weights which do not satisfy the constraint
$h_1=h_2$ or $h_1=h_2 \pm 1$ we mentioned before.
As in previous footnotes in section $3$, this feature
also arises for the singlet currents.
 \label{footnotefirst}}.

%%%%%%%%%%%%%%%%%%%%%%%%%%%%%%
%\subsubsection{The central term}
%%%%%%%%%%%%%%%%%%%%%%%%%%%%%%

%%%%%%%%%%%%%%%%%%
\subsection{The commutator relation between the lowest component and the
  second component with $h_1=h_2, h_2+1$  for nonzero $\la$}
%%%%%%%%%%%%%%%%%%

Let us consider the
commutator relation between the lowest component $\Phi_0^{(h_1)}(z)$ of the
$h_1$-th ${\cal N}=4$ multiplet in (\ref{one}) and
the second component $\Phi_{\frac{1}{2}}^{(h_2),i}(w)$
of $h_2$-th ${\cal N}=4$ multiplet in (\ref{two}).

We expect to have the fermionic currents
$Q_{h+\frac{1}{2}}^{\la, \bar{a} b}$ and
$\bar{Q}_{h+\frac{1}{2}}^{\la, b \bar{a}}$ on the right hand side of the
commutator relation because the product of bosonic and fermionic
operators produce the fermionic ones.
As before, we need to rewrite them in terms of
the components of the ${\cal N}=4$ multiplet
in order to complete the algebra we are considering.
For the index $i=1$ of these components of the ${\cal N}=4$ multiplets,
the following relations are satisfied 
\bea
\frac{1}{2}\,\Big(
Q^{\la,11}_{h+\frac{1}{2}}
+i\sqrt{2}\,Q^{\la,12}_{h+\frac{1}{2}}
+2i \sqrt{2}\,Q^{\la,21}_{h+\frac{1}{2}}
-2\,Q^{\la,22}_{h+\frac{1}{2}}\Big) & = &
\frac{1}{2}\, \Bigg[ \frac{1}{4(-4)^{h-4}}\, \Phi^{(h),1}_{\frac{1}{2}}
  - \frac{1}{4(-4)^{h-5}}\, \tilde{\Phi}^{(h-1),1}_{\frac{3}{2}} \Bigg]\, ,
\nonu \\
\frac{1}{2}\Big( 2\,\bar{Q}^{\la,11}_{h+\frac{1}{2}}
+2i \sqrt{2} \bar{Q}^{\la,12}_{h+\frac{1}{2}}
+i\sqrt{2}\bar{Q}^{\la,21}_{h+\frac{1}{2}}
-\bar{Q}^{\la,22}_{h+\frac{1}{2}}
\Big) & = &\frac{1}{2} \Bigg[ \frac{1}{4(-4)^{h-4}} \Phi^{(h),1}_{\frac{1}{2}}
  + \frac{1}{4(-4)^{h-5}} \tilde{\Phi}^{(h-1),1}_{\frac{3}{2}} \Bigg].
\nonu \\
\label{twotwo}
\eea
Note that the weight on both sides is given by $(h+\frac{1}{2})$.
On the right hand side, the $(h-1)$-th component of ${\cal N}=4$ multiplet
appears also.
Other relations for $i=2,3,4$ appear in Appendix (\ref{RELL}).
For $h=0$, the above relation (\ref{twotwo}) with others in
Appendix (\ref{RELL}) implies that
there exists the relation between the currents
$\tilde{\Phi}^{(-1),i}_{\frac{3}{2}} = -\frac{1}{4}\, \Phi^{(0),i}_{\frac{1}{2}}$
together with (\ref{two}) and (\ref{four})
because the left hand side of the first relation of (\ref{twotwo})
is identically zero
and moreover, the weight-$\frac{1}{2}$ current of
the ${\cal N}=4$ stress energy tensor has the
following relation $-i \, \Ga^i = 64 \,  \Phi^{(0),i}_{\frac{1}{2}}$
with (\ref{two}).
For $h=1$, the corresponding weight-$\frac{3}{2}$
currents
of ${\cal N}=4$ stress energy tensor
can be written as $G^i= 64\, \tilde{\Phi}^{(0),i}_{\frac{3}{2}}$
with (\ref{four})
by subtracting the two relation of (\ref{twotwo}).
Therefore, the components, $U, \Ga^i, G^i$ and $L$
of ${\cal N}=4$ stress energy tensor,
except the weight-$1$ currents $T^{ij}$ which will appear in next
subsection, 
are given by the components of the ${\cal N}=4$ multiplet,
$\tilde{\Phi}_2^{(-1)}, \Phi_{\frac{1}{2}}^{(0),i},
\tilde{\Phi}_{\frac{3}{2}}^{(0),i}$ and $\tilde{\Phi}_2^{(0)}$ up to
normalization respectively.

The result of the commutator relation we are considering, by replacing
all the fermionic currents with the
${\cal N}=4$ components appearing in
(\ref{twotwo}) and Appendix (\ref{RELL}), can be summarized by
\bea
&& \big[(\Phi^{(h_1)}_{0})_m,
  (\Phi^{(h_2),i}_{\frac{1}{2}})_r \big]  = 
5\frac{(-4)^{h_1-2}}{(2h_1-1)} \, 4(-4)^{h_2-4}
\Bigg[ 
  \Bigg( -(h_1-2\la)\,
  \sum_{h=-1}^{h_1+h_2-3} \, q^h\, 
  q_{\mathrm{F}}^{h_1,h_2+\frac{1}{2}, h}(m,r,\la)
  \nonu \\
  && -  (h_1-2\la)\, \sum_{h=-1}^{h_1+h_2-2} 
 q^h\, (-1)^h 
q_{\mathrm{F}}^{h_1,h_2+\frac{1}{2}, h}(m,r,\la)
 + (h_1-1+2\la)
\sum_{h=-1}^{h_1+h_2-3} 
 q^h
q_{\mathrm{B}}^{h_1,h_2+\frac{1}{2}, h}(m,r,\la)
\nonu \\
&& +  (h_1-1+2\la)\,
\sum_{h=-1}^{h_1+h_2-2} \,
 q^h\, (-1)^h\,
q_{\mathrm{B}}^{h_1,h_2+\frac{1}{2}, h}(m,r,\la)
\Bigg)\,
\frac{1}{8(-4)^{h_1+h_2-h-6}}\,
(\Phi_{\frac{1}{2}}^{(h_1+h_2-2-h)})_{m+r}
\nonu \\
 &&   +
  \Bigg( 
  \sum_{h=-1}^{h_1+h_2-3} \,(h_1-2\la)\, q^h\, 
  q_{\mathrm{F}}^{h_1,h_2+\frac{1}{2}, h}(m,r,\la)
 -  \sum_{h=-1}^{h_1+h_2-2} \,
 (h_1-2\la)\, q^h\, (-1)^h\, 
q_{\mathrm{F}}^{h_1,h_2+\frac{1}{2}, h}(m,r,\la)
\nonu \\
&& -  
\sum_{h=-1}^{h_1+h_2-3} \,
(h_1-1+2\la)\, q^h\, 
q_{\mathrm{B}}^{h_1,h_2+\frac{1}{2}, h}(m,r,\la)
\label{Tworel}
 \\
&& +  
\sum_{h=-1}^{h_1+h_2-2} \,
(h_1-1+2\la)\, q^h\, (-1)^h\,
q_{\mathrm{B}}^{h_1,h_2+\frac{1}{2}, h}(m,r,\la)
\Bigg)\,
\frac{1}{8(-4)^{h_1+h_2-h-7}}\,
(\tilde{\Phi}_{\frac{3}{2}}^{(h_1+h_2-3-h)})_{m+r} \,
\Bigg]\, .
\nonu
\eea
Note that the upper limit of $h$ having the term with the
factor $(-1)^h$ is given by $h=h_1+h_2-2$.
Except these four terms,
the range of $h$ in all the other terms
is the same.
Then these four terms occur if their structure constants are nonvanishing.
Depending on the odd $h$ or even $h$,
some of terms in (\ref{Tworel}) can cancel each other
because four kinds of each two terms have the same structure
except the factor $(-1)^h$.
At $h=h_1+h_2-2$, 
there appear the current $\Phi_{\frac{1}{2}}^{(0),i}$ and
the current $\tilde{\Phi}_{\frac{3}{2}}^{(-1),i}$ which are related to
the previous weight-$\frac{1}{2}$ current $\Ga^i$ as above, with
the relevant structure constants.
Moreover, we have vanishing structure constants
$q_F^{h_1,h_2+\frac{1}{2},h_1+h_2-2}=0=q_B^{h_1,h_2+\frac{1}{2},h_1+h_2-2}$
at $\la=0$ \cite{Ahn2203}.
For the $h \leq h_1+h_2-3$ with odd $h$ in (\ref{Tworel}),
there appear the current $\tilde{\Phi}_{\frac{3}{2}}^{(h_1+h_2-3-h),i}$ terms
because the coefficients of the current $\Phi_{\frac{1}{2}}^{(h_1+h_2-2-h),i}$
terms are identically vanishing.
Note that there are relative signs in (\ref{Tworel}).
On the other hand, for
the $h \leq h_1+h_2-3$ with even $h$,
there appear the current $\Phi_{\frac{1}{2}}^{(h_1+h_2-2-h),i}$ terms
because the coefficients of the current
$\tilde{\Phi}_{\frac{3}{2}}^{(h_1+h_2-3-h),i}$
terms are identically zero. We also present some example
where the weights $h_1$ and $h_2$ do not satisfy
the condition $h_1=h_2$ or $h_1=h_2+1$
\footnote{Along the line of the footnote
  \ref{footnotefirst}, we  consider the OPE
  between
$\Phi^{(h_1=4)}_{0}(z)$ and $
  \Phi^{(h_2=2),i}_{\frac{1}{2}}(w)$ where $h_1=h_2+2$.
  The sixth order pole of this OPE gives us
   the structure constant $\frac{2048}{7} (\la-1) \la (2 \la-1)
  (2 \la+1) (4 \la-1)$ appearing in the current
  $\Phi_{\frac{1}{2}}^{(h_1+h_2-2-h=0),i}(w)$ with the weight $h=4$.
   By substituting the various expressions in the corresponding terms of
   (\ref{Tworel}),
  we can check that we obtain the above structure constant correctly
  where $q_F^{4,\frac{5}{2},4}$ term corresponds to
  $\frac{256}{3} (\la-1) \la (2 \la-3) (2 \la-1) (2 \la+1)$
  while  $q_B^{4,\frac{5}{2},4}$ term
  corresponds to $-\frac{512}{3}  (\la-1) \la (\la+1)
  (2 \la-1) (2 \la+1)$. There is no
  current $\tilde{\Phi}_{\frac{3}{2}}^{(-1),i}(w)$ term in the sixth order pole
  from our calculation
  and this is consistent with (\ref{Tworel}) because
  all the coefficients vanish for the even $h=4$.

  Now we can move to the fifth order pole of this OPE. The
  current $G^i(w)$ term of ${\cal N}=4$ stress energy tensor
  has the structure constant
  $-\frac{512}{7}  (\la-1) (2 \la+1) (6 \la^2-3 \la-4)$
  which can be identified with the coefficient appearing in
  the second current $\tilde{\Phi}_{\frac{3}{2}}^{(h_1+h_2-3-h=0),i}(w)$
  of (\ref{Tworel}) by substituting $q_F^{4,\frac{5}{2},3}$
  and $q_B^{4,\frac{5}{2},3}$ correctly as before.
  On the other hand, there exists the current
  $\Phi_{\frac{1}{2}}^{(h_1+h_2-2-h=1),i}(w)$ term with
  the structure constant
  $\frac{8192}{3} (\la-1) \la (2 \la-1) (2 \la+1)$ having
  the $\la$ factor. We can check that this structure constant
  is equal to the one in the first current terms in (\ref{Tworel})
  where the previous
$q_F^{4,\frac{5}{2},3}$
  and $q_B^{4,\frac{5}{2},3}$ are replaced by
$q_F^{3,\frac{5}{2},3}$
  and $q_B^{3,\frac{5}{2},3}$ respectively. That is, the $h_1$
  is replaced by $(h_1-1)$.
  Therefore, we cannot use the equation (\ref{Tworel})
  for this particular pole fully.
  We can check that
  the remaining lower order poles
can be described by (\ref{Tworel}) without any extra terms precisely.}.

%%%%%%%%%%%%%%%%%%
\subsection{The commutator relation between the lowest component and the
  third component with $h_1=h_2,h_2+1,h_2+2$  for nonzero $\la$}
%%%%%%%%%%%%%%%%%%

Let us consider the
commutator relation between the lowest component $\Phi_0^{(h_1)}(z)$ of the
$h_1$-th ${\cal N}=4$ multiplet and
 the
 third component $\Phi_1^{(h_2),ij}(w)$
 of the $h_2$-th ${\cal N}=4$ multiplet. Because
there are no singular terms in the OPE of
$W_{F,h_1}^{\la, \bar{a} b}(z)$ and $W_{B,h_2}^{\la, \bar{c} d}(w)$,
the commutator relation consists of two parts.
That is, there are 
$\big[(W_{F,h_1}^{\la, \bar{a} a})_m,
  (W_{F,h_2}^{\la, \bar{c} d})_n \big]$
and $\big[(W_{B,h_1}^{\la, \bar{a} a})_m,
  (W_{B,h_2}^{\la, \bar{c} d})_n \big]$. Then we can use
the previous relations, (\ref{FIRST}), (\ref{SECOND}), (\ref{THIRD}),
(\ref{FOURTH}) and (\ref{FIFTH}) and 
Appendix (\ref{res1}). We need to express the right hand sides
of these commutator relations in terms of the components of
the ${\cal N}=4$ multiplet as before.

We obtain the following relations
\bea
2i\,W^{\la,11}_{\mathrm{B},h+1}
-\sqrt{2}\,W^{\la,12}_{\mathrm{B},h+1}
-2i\,\,W^{\la,22}_{\mathrm{B},h+1}
&= & \frac{1}{8(-4)^{h-4}} \Bigg[ \Phi_1^{(h),12}-  \Phi_1^{(h),34}\Bigg]\, ,
\nonu \\
 2i\,W^{\la,11}_{\mathrm{F},h+1}
-2\sqrt{2}\,W^{\la,12}_{\mathrm{F},h+1}
-2i\,W^{\la,22}_{\mathrm{F},h+1} & = &
\frac{1}{8(-4)^{h-4}} \Bigg[ \Phi_1^{(h),12}+  \Phi_1^{(h),34}\Bigg] \, .
\label{rell}
\eea
For other relations, 
we present them in Appendix (\ref{RELL1}).
By using (\ref{rell}) and Appendix (\ref{RELL1}), we can express
the weight-$1$ current $T^{ij}$ of the ${\cal N}=4$ stress
energy tensor as $i\, T^{ij} = 64 \,\Phi_1^{(0),ij}$ with (\ref{three}).
On the right hand sides of (\ref{rell}), there are only the components of
$h$-th ${\cal N}=4$ multiplet. By linear combinations,
any component of the $h$-th ${\cal N}=4$ multiplet can be written
in terms of $W_{F,h+1}^{\la, \bar{a} b}$ and
$W_{B,h+1}^{\la, \bar{c} d}$ explicitly.

It turns out that the corresponding commutator relation,
after substituting all the bosonic currents
in terms of the components of
the ${\cal N}=4$ multiplet described above, can be written as
\bea
&& \big[(\Phi^{(h_1)}_{0})_m,
  (\Phi^{(h_2),ij}_1)_n \big]  = 
\frac{(-4)^{h_1-2}}{(2h_1-1)} \, 4(-4)^{h_2-4}
\Bigg[ 
 -(h_1-2\la)\,
  \sum_{h=-1,\mbox{\footnotesize even}}^{h_1+h_2-2} \, q^h\, 
  p_{\mathrm{F}}^{h_1,h_2+1, h}(m,n,\la)
  \nonu \\
  && \times
 \frac{1}{8(-4)^{h_1+h_2-h-6}}\,
  \Bigg(\Phi_{1}^{(h_1+h_2-2-h),ij} + \frac{1}{2}\,
  \varepsilon^{ijkl} \,\Phi_{1}^{(h_1+h_2-2-h),kl} \Bigg)_{m+n}
\nonu \\
 &&   +(h_1-1+2\la)\,
  \sum_{h=-1,\mbox{\footnotesize even}}^{h_1+h_2-2} \, q^h\, 
  p_{\mathrm{B}}^{h_1,h_2+1, h}(m,n,\la)
  \nonu \\
  && \times
\frac{1}{8(-4)^{h_1+h_2-h-6}}\,
  \Bigg(\Phi_{1}^{(h_1+h_2-2-h),ij} - \frac{1}{2}\,
  \varepsilon^{ijkl} \,\Phi_{1}^{(h_1+h_2-2-h),kl}\Bigg)_{m+n} \,
\Bigg]\, .
\label{Threerel}
\eea
Due to the even weight $h$, on the right hand side of (\ref{Threerel}),
the currents of odd (or even) weights can appear
depending on the weights $h_1$ and $h_2$.
Due to the $SO(4)$ index $ij$ appearing on the
left hand side of (\ref{Threerel}), the field contents
on the right hand side are different from the ones in (\ref{Onerel}).
At the maximum value of $h$, $h=h_1+h_2-2$,
the current $\Phi_{1}^{(0),ij}$ term and the
current $ \varepsilon^{ijkl} \,
\Phi_{1}^{(0),kl}$ term,
which are related to
the previous weight-$1$ current of
the ${\cal N}=4$ stress energy tensor,
occur on the right hand side of (\ref{Threerel}).
Note that as before we have vanishing structure constant
$p_B^{h_1,h_2+1,h_1+h_2-2}$
at $\la=0$ \cite{Ahn2203}.
In this case also, there some constraints on the weight $h_2$
for nonzero $\la$
\footnote{For the OPE between $\Phi^{(h_1=4)}_{0}(z)$ and
  $\Phi^{(h_2=1),ij}_1(w)$ where the weights satisfy
  $h_1=h_2+3$, the fifth order pole of this OPE
  has the structure constant $-\frac{8192}{7}  \la
  (2 \la-1) (4 \la^2-2 \la-5)$ in the current $\Phi_1^{(0),ij}(w)$ term
  and the structure constant $\frac{8192}{7} \la (2 \la-1)
  (4 \la-1)$ in the  current $\frac{1}{2}\,
  \varepsilon^{ijkl} \, \Phi_1^{(0),kl}(w)$ term from our calculation.
  However,
  the corresponding structure constants $p_F^{h_1=4,h_2+1=2,h=2}$
  and $p_B^{h_1=4,h_2+1=2,h=2}$
  do not produce these $\la$ dependent
  structure constants respectively. This implies that there are extra
  contributions in the structure constants
  for the weights which do not satisfy
the condition $h_1=h_2$, $h_1=h_2+1$, or $h_1=h_2+2$.}.

%%%%%%%%%%%%%%%%%%
\subsection{The commutator relation between the lowest component and the
  fourth component with $h_1=h_2+1,h_2+2$  for nonzero $\la$}
%%%%%%%%%%%%%%%%%%

In this case, we can use the previous relation (\ref{twotwo})
and Appendix (\ref{RELL}) by simply
substituting $h$ with $(h+1)$
in order to rewrite the right hand side of
this commutator relation in terms of the components of
the ${\cal N}=4$ multiplet.

Then we can determine the following result for the
commutator relation
between the lowest component $\Phi_0^{(h_1)}(z)$ of the
$h_1$-th ${\cal N}=4$ multiplet and
 the
 fourth component (and derivative term)
 $\tilde{\Phi}_{\frac{3}{2}}^{(h_2),i}(w)$
 of the $h_2$-th ${\cal N}=4$ multiplet
\bea
&& \big[(\Phi^{(h_1)}_{0})_m,
  (\tilde{\Phi}^{(h_2),i}_{\frac{3}{2}})_r \big]  = 
\frac{(-4)^{h_1-2}}{(2h_1-1)} \, 4(-4)^{h_2-4}
\Bigg[ 
  \Bigg( 
 (h_1-2\la)\, \sum_{h=-1}^{h_1+h_2-2} \, q^h\, 
  q_{\mathrm{F}}^{h_1,h_2+\frac{3}{2}, h}(m,r,\la)
  \nonu \\
  && - (h_1-2\la) \sum_{h=-1}^{h_1+h_2-1} 
  q^h\, (-1)^h 
q_{\mathrm{F}}^{h_1,h_2+\frac{3}{2}, h}(m,r,\la)
 -  (h_1-1+2\la)
\sum_{h=-1}^{h_1+h_2-2} 
 q^h
q_{\mathrm{B}}^{h_1,h_2+\frac{3}{2}, h}(m,r,\la)
\nonu \\
&& +  (h_1-1+2\la)\,
\sum_{h=-1}^{h_1+h_2-1} \,
 q^h\, (-1)^h\,
q_{\mathrm{B}}^{h_1,h_2+\frac{3}{2}, h}(m,r,\la)
\Bigg)\,
 \frac{1}{8(-4)^{h_1+h_2-h-5}}\,
(\Phi_{\frac{1}{2}}^{(h_1+h_2-1-h),i})_{m+r}
\nonu \\
 &&   +
  \Bigg( - (h_1-2\la)\,
  \sum_{h=-1}^{h_1+h_2-2} \, q^h\, 
  q_{\mathrm{F}}^{h_1,h_2+\frac{3}{2}, h}(m,r,\la)
  \label{Fourrel} \\
  && - (h_1-2\la) \sum_{h=-1}^{h_1+h_2-1} 
  q^h\, (-1)^h 
q_{\mathrm{F}}^{h_1,h_2+\frac{3}{2}, h}(m,r,\la)
 +  (h_1-1+2\la)
\sum_{h=-1}^{h_1+h_2-2} 
 q^h 
q_{\mathrm{B}}^{h_1,h_2+\frac{3}{2}, h}(m,r,\la)
\nonu \\
&& + (h_1-1+2\la)\, 
\sum_{h=-1}^{h_1+h_2-1} \,
 q^h\, (-1)^h\,
q_{\mathrm{B}}^{h_1,h_2+\frac{3}{2}, h}(m,r,\la)
\Bigg)\,
\frac{1}{8(-4)^{h_1+h_2-h-6}}\,
(\tilde{\Phi}_{\frac{3}{2}}^{(h_1+h_2-2-h),i})_{m+r} \,
\Bigg]\, .
\nonu
\eea
The form of this (\ref{Fourrel}) looks similar to
the previous one in (\ref{Tworel}) because the field contents
are the same.
The relative signs are different from each other and we observe that
after replacing $h_2$ with $(h_2+1)$ appearing in (\ref{Tworel}),
the corresponding expressions occur in (\ref{Fourrel}).
At $h=h_1+h_2-1$, 
there appear the $\Phi_{\frac{1}{2}}^{(0),i}$ and
the $\tilde{\Phi}_{\frac{3}{2}}^{(-1),i}$ which are related to
the weight-$\frac{1}{2}$ current $\Ga^i$.
Except four terms having the factor $(-1)^h$
with $h=h_1+h_2-1$, there some cancellation between the currents.
Furthermore, we have vanishing structure constants
$q_F^{h_1,h_2+\frac{3}{2},h_1+h_2-1}=0=q_B^{h_1,h_2+\frac{3}{2},h_1+h_2-1}$
at $\la=0$ \cite{Ahn2203}.
For the $h \leq h_1+h_2-2$ with odd $h$ in (\ref{Fourrel}),
there appear the current $\Phi_{\frac{1}{2}}^{(h_1+h_2-1-h),i}$ terms
because the coefficients of the
current $\tilde{\Phi}_{\frac{3}{2}}^{(h_1+h_2-2-h),i}$
terms are identically vanishing.
On the other hand, for
the $h \leq h_1+h_2-2$ with even $h$,
there appear the current
$\tilde{\Phi}_{\frac{3}{2}}^{(h_1+h_2-2-h),i}$ terms
because the coefficients of the current $\Phi_{\frac{1}{2}}^{(h_1+h_2-2-h),i}$
terms are identically zero. This behavior
is different from the one in (\ref{Tworel}) because
the appearance of signs behaves differently. We also comment on the
possible ranges for the weight $h_2$ as described before \footnote{
Similar to the previous examples, we consider the OPE
  between
$\Phi^{(h_1=4)}_{0}(z)$ and $
  \Phi^{(h_2=1),i}_{\frac{3}{2}}(w)$ where the weights
  satisfy $h_1=h_2+3$.
  The sixth order pole of this OPE provides 
  the structure constant $
\frac{32768}{7} (\la-1) \la (2 \la-1) (2 \la+1) (4 \la-1)
$ appearing in the current
  $\tilde{\Phi}_{\frac{3}{2}}^{(h_1+h_2-2-h=-1),i}(w)$ with the weight $h=4$.
By substituting the various expressions in the corresponding
terms of (\ref{Fourrel}),
  we can check that we obtain the above structure constant correctly
  where $q_F^{4,\frac{5}{2},4}$ term corresponds to
  the $\la$ dependent factor
  $\frac{256}{3} (\la-1) \la (2 \la-3) (2 \la-1) (2 \la+1)$
  while  $q_B^{4,\frac{5}{2},4}$ term
  corresponds to the $\la$ dependent factor
  $-\frac{512}{3}  (\la-1) \la (\la+1)
  (2 \la-1) (2 \la+1)$. There is no
  $\Phi_{\frac{1}{2}}^{(1),i}(w)$ term in the sixth order pole
  from our calculation 
  and this is consistent with (\ref{Fourrel}) because
  all the coefficients vanish for the even weight $h=4$.

  Now we can move to the fifth order pole of above OPE. The
 $\Phi_{\frac{1}{2}}^{(h_1+h_2-2-h=1),i}(w)$ term  has the structure constant
  $-\frac{2048}{7}  (\la-1) (2 \la+1) (6 \la^2-3 \la-4)$
  which can be identified with the coefficient appearing in
  the second current $\Phi_{\frac{1}{2}}^{(h_1+h_2-2-h=1),i}(w)$
  of (\ref{Fourrel}) by substituting $q_F^{4,\frac{5}{2},3}$
  and $q_B^{4,\frac{5}{2},3}$ correctly.
  On the other hand, there exists the current
 $G^i(w)$ term of ${\cal N}=4$ stress energy tensor  with
  the structure constant
  $\frac{128}{3} (\la-1) \la (2 \la-1) (2 \la+1)$ having
  the $\la$ factor. We can check that this structure constant
  is equal to the one in the first current terms in (\ref{Fourrel})
  where the previous
$q_F^{4,\frac{5}{2},3}$
  and $q_B^{4,\frac{5}{2},3}$ are replaced by
$q_F^{3,\frac{5}{2},3}$
  and $q_B^{3,\frac{5}{2},3}$ respectively. The weight $h_1$ is replaced by
  the weight $(h_1-1)$ and this reflects the fact that the weight
  $h_1$ is increased by $1$ from $h_1=h_2+2$ to $h_1=h_2+3$.
  Now
  we have seen the extra structures on the right hand side of this
particular pole for the weights we are considering.}.

%%%%%%%%%%%%%%%%%%
\subsection{The commutator relation between the lowest component and the
last component with $h_1=h_2+1,h_2+2,h_2+3$  for nonzero $\la$}
%%%%%%%%%%%%%%%%%%

Finally, the commutator relation
between the lowest component $\Phi_0^{(h_1)}(z)$ of the
$h_1$-th ${\cal N}=4$ multiplet and
 the
 last component (and derivative term) $\tilde{\Phi}_{2}^{(h_2)}(w)$
 of the $h_2$-th ${\cal N}=4$ multiplet
 can be determined by using (\ref{rel1})
\bea
&& \big[(\Phi^{(h_1)}_{0})_m,
  (\tilde{\Phi}^{(h_2)}_{2})_n \big]  = 
\frac{(-4)^{h_1-2}}{(2h_1-1)} \, 4(-4)^{h_2-4}
\Bigg[ \nonu \\
  & &
\left(\begin{array}{c}
m+h_1-1 \\  h_1+h_2-1 \\
 \end{array}\right) \, q^{h_1+h_2-2}\,
  \Bigg( 2(h_1-2\la) \, c_F -2(h_1-1+2\la)\, c_B \Bigg) \, \de_{m+n}
  \nonu \\
  && +
  \sum_{h=0,\mbox{\footnotesize even}}^{h_1+h_2-1}
\frac{2\, q^h}{(-4)^{h_1+h_2-h-2}}
  \, \Bigg( -
 (h_1-2\la)\, 
p_{\mathrm{F}}^{h_1,h_2+2, h}(m,n,\la)
\nonu \\
&& -  
 %\sum_{h=0,\mbox{\footnotesize even}}^{h_1+h_2-3} \,
(h_1-1+2\la)\, 
p_{\mathrm{B}}^{h_1,h_2+2, h}(m,n,\la)
\,  \Bigg)\, (\Phi_{0}^{(h_1+h_2-h)})_{m+n}
\nonu \\
&& +   \sum_{h=0,\mbox{\footnotesize even}}^{h_1+h_2-1} \,
\frac{q^h}{4
(2h_1+2h_2-2h-1) 
  (-4)^{h_1+h_2-h-6}} \,
\Bigg( \nonu \\
&& 
-(h_1-2\la)(h_1+h_2-h-1+2\la)\, 
p_{\mathrm{F}}^{h_1,h_2+2, h}(m,n,\la)
\nonu \\
&& +  
 %\sum_{h=0,\mbox{\footnotesize even}}^{h_1+h_2-3} \,
(h_1-1+2\la)(h_1+h_2-h-2\la)\, 
p_{\mathrm{B}}^{h_1,h_2+2, h}(m,n,\la)
\,  \Bigg)\, (\tilde{\Phi}_{2}^{(h_1+h_2-h-2)})_{m+n}
  \Bigg]\, .
\label{finalfinal}
\eea
The field contents appearing in (\ref{finalfinal}) are the same as
the one in (\ref{Onerel}).
As before, we observe that for the maximum value of
dummy variable $h$, $h=h_1+h_2-1$ with odd $(h_1+h_2)$,
on the right hand side of (\ref{finalfinal}), there appear
the currents
$\Phi_0^{(1)}$ and $\tilde{\Phi}_2^{-1}$ which is related to
the previous current $U$ of ${\cal N}=4$ stress energy tensor.
On the other hand, for even $(h_1+h_2)$,
the maximum value of the weight $h$ is given by $h=h_1+h_2-2$
because the weight $h$ should be even. For this value,
 there appear
 the currents
 $\Phi_0^{(2)}$ and $\tilde{\Phi}_2^{(0)}$, which is related to
 the previous current $L$, with proper
 $\la$ dependent
 structure constants on the right hand side of (\ref{finalfinal}).
 Note that as before we have vanishing structure constant
$p_B^{h_1,h_2+2,h_1+h_2-1}$
at $\la=0$ \cite{Ahn2203}.
We comment on the other possible cases for the weights
\footnote{
From the OPE between $\Phi^{(h_1=4)}_{0}(z)$ and $
\Phi^{(h_2=0)}_{2}(w) $ where the weights satisfy
$h_1=h_2+4$, the fifth order pole of this OPE
provides the structure constant $\frac{8192}{7} \la (2 \la-1)
(4 \la-1)$
appearing in the  current $\Phi^{(1)}_{0}(w)$ and
the structure constant $\frac{256}{7} \la (2 \la-1)
(10 \la^2-5 \la-2)$
appearing in the current $U(w)$.
In (\ref{finalfinal}), the exponent of the first current
implies the weight
$h_1+h_2-h=4-h$. Moreover, the weight $h$ is given by $h=0$ or $h=2$
from the summation. Therefore, the current
$\Phi^{(1)}_{0}(w)$ cannot appear from the (\ref{finalfinal}).
 Furthermore, the exponent of the second current
 implies the weight $h_1+h_2-2-h=2-h=-1$ from the presence of $U$.
 Moreover, the weight $h$ is given by $h=0$
 or $h=2$
from the summation. In this case also, the current
$U(w)$ cannot appear from the (\ref{finalfinal}). Therefore,
there appear the extra terms for the currents
having the weights we are considering
which do not satisfy the above constraints.}.

In Appendix $G$, the remaining
(anti) commutator relations are given.
Therefore, the (anti)commutator relations between the
${\cal N}=4$ multiplets are summarized by (\ref{Onerel}),
(\ref{Tworel}), (\ref{Threerel}), (\ref{Fourrel}) and (\ref{finalfinal})
in addition to Appendices
(\ref{sixcase}), (\ref{sevencase}), (\ref{eightcase}),
(\ref{ninecase}), (\ref{tencase}), (\ref{elevencase}), (\ref{twelvecase}),
(\ref{thirteencase}), (\ref{fourteencase}) and (\ref{fifteencase}).
Among these, the more fundamental  (anti)commutator relations
are given by  (\ref{Onerel}),
(\ref{Tworel}), (\ref{Threerel}), Appendices
(\ref{sixcase}),
and (\ref{tencase}) in the sense that
the field contents on the right hand sides of
the remaining ones can be seen from the ones of
these fundamental relations up to
signs.
As mentioned at the end of previous section $3$,
by using (\ref{Onerel}), (\ref{finalfinal}) and
Appendix (\ref{fifteencase})
together with (\ref{rel1}),
the $w_{1+\infty}$ algebra can be obtained by taking the proper limit
on the parameter $q$ at $\la=0$ with the contractions of the currents.

%%%%%%%%%%%%%%%%%%%%%%%%%%%%%%%%%%%%%%%%%%%%%%%%%%%%%%%%%%%%%%%%%%%%%
%%%%%%%%%%%%%%%%%%%%%%%%%%%%%%%%%%%%%%%%%%%%%%%%%%%%%%%%%%%%%%%%%%%%%%
%\section{ Conclusions and outlook}
%4%%%%%%%%%%%%%%%%%%%%%%%%%%%%%%%%%%%%%%%%%%%%%%%%%%%%%%%%%%%%%%%%%%%%%
%%%%%%%%%%%%%%%%%%%%%%%%%%%%%%%%%%%%%%%%%%%%%%%%%%%%%%%%%%%%%%%%%%%%%

%$\bullet$

\vspace{.7cm}

%%%%%%%%%%%%%%%%%%%%%%%%%%%%%%%%%%%%%%%%%%%%%%%%%%%%%%%%%%%%%%
%%%%%%%%%%%%%%%%%%%%%%%%%%%%%%%%%%%%%%%%%%%%%%%%%%%%%%%%%%%%%%%
\centerline{\bf Acknowledgments}
%%%%%%%%%%%%%%%%%%%%%%%%%%%%%%%%%%%%%%%%%%%%%%%%%%%%%%%%%%%%%%%
%%%%%%%%%%%%%%%%%%%%%%%%%%%%%%%%%%%%%%%%%%%%%%%%%%%%%%%%%%%%%%%

%We
%would like to
%thank M.H. Kim for the intensive discussions. 
%This work was supported by
%a National Research Foundation of Korea (NRF) grant
%funded by the Korean government (MSIT)
This work was supported by the National
Research Foundation of Korea(NRF) grant funded by the
Korea government(MSIT) 
(No. 2023R1A2C1003750).

\newpage

\appendix

\renewcommand{\theequation}{\Alph{section}\mbox{.}\arabic{equation}}

%%%%%%%%%%%%%%%%%%%%%%%%%%%%%%%%%%%%%%%%%%%%%%%%%%%%%%%
%%%%%%%%%%%%%%%%%%%%%%%%%%%%%
\section{The structure constants }
%%%%%%%%%%%%%%%%%%%%%%%%%%%%%
%%%%%%AAA%%%%%%%%%%%%%%%%%%%%%%%%%%%%%%%%%%%%%%%%%%%%%%%%%

By introducing the generalized hypergeometric function
\bea
\phi_{r}^{h_1 ,h_2}(\Lambda,a)  \equiv \ _4F_3\left[
\begin{array}{c|}
\frac{1}{2} + \Lambda \ ,  \frac{1}{2} - \Lambda  \ , \frac{1+a-r}{ 2}\ , \frac{a-r}{2}\\
\frac{3}{2}-h_1 \ , \frac{3}{2} -h_2\ , \frac{1}{2}+ h_1+h_2-r
\end{array}  \ 1\right],
\label{phi}
\eea
and mode dependent function
\bea
N^{h_1,h_2}_{h}(m,n)
\!&
\equiv \!&
\sum_{l=0 }^{h+1}(-1)^l
%\binom{h+1}{l}
\left(\begin{array}{c}
h+1 \\  l \\
\end{array}\right)
[h_1-1+m]_{h+1-l}[h_1-1-m]_l
\nonu \\
\!& \times \!& [h_2-1+n]_l [h_2-1-n]_{h+1-l},
\label{Ndef}
\eea
the three kinds of structure constants in \cite{AK2009}
can be summarized, together with (\ref{phi}) and (\ref{Ndef}),  by
\bea
\mathrm{BB}^{h_1,h_2}_{r,\,\pm}(m,n; \mu )
&\equiv&
 -\frac{1  }{ (r-1)!} N_{r-2}^{h_1, h_2}(m,n) \Bigg[
\phi_{r}^{h_1 ,h_2}(\mu,1)  \pm \phi_{r}^{h_1 ,h_2}(1-\mu,1)  
\Bigg],
\nonu\\
\mathrm{BF}^{h_1,h_2+\frac{1}{2}}_{r,\,\pm}(m,\rho; \mu )
&\equiv&
 -\frac{1  }{ (r-1)!} N_{r-2}^{h_1, h_2+\frac{1}{2}}(m,\rho) \Bigg[
 \phi_{r+1}^{h_1 ,h_2+1}(\mu,\frac{3\pm1}{2})
 \nonu \\
 & \pm & \phi_{r+1}^{h_1 ,h_2+1}(1-\mu,\frac{3\pm1}{2})  
\Bigg],
\nonu\\
\mathrm{FF}^{h_1+\frac{1}{2},h_2+\frac{1}{2}}_{r,\,\pm}(\rho,\omega; \mu )
&\equiv&
-\frac{1  }{ (r-1)!}N_{r-2}^{h_1+\frac{1}{2}, h_2+\frac{1}{2}}(\rho,\omega)
\Bigg[
 \phi_{r+1}^{h_1+1 ,h_2+1}(\mu,\frac{3\pm1}{2})  \nonu \\
 & \pm & \phi_{r+1}^{h_1+1 ,h_2+1}(1-\mu,\frac{3\pm1}{2})  
 \Bigg].
\label{3struct}
\eea

%\bea
%\mathrm{BB}^{h_1,h_2}_{r,\,\pm}(m,n; \mu ) & = &
%\pm
%\mathrm{BB}^{h_1,h_2}_{r,\,\pm}(m,n; 1-\mu ),
%\nonu \\
%\mathrm{BF}^{h_1,h_2+\frac{1}{2}}_{r,\,\pm}(m,\rho; \mu ) &=&
%\pm
%\mathrm{BF}^{h_1,h_2+\frac{1}{2}}_{r,\,\pm}(m,\rho; 1-\mu ),
%\nonu \\
%\mathrm{FF}^{h_1+\frac{1}{2},h_2+\frac{1}{2}}_{r,\,\pm}(\rho,\omega; \mu% )
%& = &
%\pm 
%\mathrm{FF}^{h_1+\frac{1}{2},h_2+\frac{1}{2}}_{r,\,\pm}(\rho,\omega; 1-\%mu ).
%\nonu
%\label{symmetry}
%\eea
Then the structure constants in section $3$, from (\ref{3struct}),
are given by
\bea
p_{F,h}^{h_1,h_2}(m,n,\la) & = & -\frac{1}{4} \Bigg[
  \mathrm{BB}^{h_1,h_2}_{h+2,\,+} +
   \mathrm{BB}^{h_1,h_2}_{h+2,\,-} \Bigg]_{\mu = 2 \la},
\nonu \\
p_{B,h}^{h_1,h_2}(m,n,\la) & = & -\frac{1}{4} \Bigg[
  \mathrm{BB}^{h_1,h_2}_{h+2,\,+} -
   \mathrm{BB}^{h_1,h_2}_{h+2,\,-} \Bigg]_{\mu = 2 \la},
\nonu \\
q_{F,2h}^{h_1,h_2+\frac{1}{2}}(m,n,\la) & = &  \Bigg[
 -\frac{1}{8} \mathrm{BF}^{h_1,h_2+\frac{1}{2}}_{2h+2,\,+} +
 \frac{(2h_1-2h-3)}{16(h+1)}\,
 \mathrm{BF}^{h_1,h_2+\frac{1}{2}}_{2h+2,\,-} \Bigg]_{\mu = 2 \la},
\nonu \\
q_{F,2h+1}^{h_1,h_2+\frac{1}{2}}(m,n,\la) & = &  \Bigg[
 \frac{1}{8} \mathrm{BF}^{h_1,h_2+\frac{1}{2}}_{2h+3,\,+} -
 \frac{(h_1-h-2)}{4(2h+3)}\,
 \mathrm{BF}^{h_1,h_2+\frac{1}{2}}_{2h+3,\,-} \Bigg]_{\mu = 2 \la},
\nonu \\
q_{B,2h}^{h_1,h_2+\frac{1}{2}}(m,n,\la) & = &  \Bigg[
 -\frac{1}{8} \mathrm{BF}^{h_1,h_2+\frac{1}{2}}_{2h+2,\,+} -
 \frac{(2h_1-2h-3)}{16(h+1)}\,
 \mathrm{BF}^{h_1,h_2+\frac{1}{2}}_{2h+2,\,-} \Bigg]_{\mu = 2 \la},
\nonu \\
q_{B,2h+1}^{h_1,h_2+\frac{1}{2}}(m,n,\la) & = &  \Bigg[
 -\frac{1}{8} \mathrm{BF}^{h_1,h_2+\frac{1}{2}}_{2h+3,\,+} -
 \frac{(h_1-h-2)}{4(2h+3)}\,
 \mathrm{BF}^{h_1,h_2+\frac{1}{2}}_{2h+3,\,-} \Bigg]_{\mu = 2 \la},
\nonu \\
o_{F,2h}^{h_1+\frac{1}{2},h_2+\frac{1}{2}}(m,n,\la) & = &  \Bigg[
 -\mathrm{FF}^{h_1+\frac{1}{2},h_2+\frac{1}{2}}_{2h+1,\,+} -
 \frac{2(h_1+h_2-h)}{(2h+1)}\,
 \mathrm{FF}^{h_1+\frac{1}{2},h_2+\frac{1}{2}}_{2h+1,\,-} \Bigg]_{\mu = 2 \la},
\nonu \\
o_{F,2h+1}^{h_1+\frac{1}{2},h_2+\frac{1}{2}}(m,n,\la) & = &  \Bigg[
 \mathrm{FF}^{h_1+\frac{1}{2},h_2+\frac{1}{2}}_{2h+2,\,+} +
 \frac{2(h_1+h_2-h)-1}{2(h+1)}\,
 \mathrm{FF}^{h_1+\frac{1}{2},h_2+\frac{1}{2}}_{2h+2,\,-} \Bigg]_{\mu = 2 \la},
\nonu \\
o_{B,2h}^{h_1+\frac{1}{2},h_2+\frac{1}{2}}(m,n,\la) & = &  \Bigg[
 -\mathrm{FF}^{h_1+\frac{1}{2},h_2+\frac{1}{2}}_{2h+1,\,+} +
 \frac{2(h_1+h_2-h)}{(2h+1)}\,
 \mathrm{FF}^{h_1+\frac{1}{2},h_2+\frac{1}{2}}_{2h+1,\,-} \Bigg]_{\mu = 2 \la},
\nonu \\
o_{B,2h+1}^{h_1+\frac{1}{2},h_2+\frac{1}{2}}(m,n,\la) & = &  \Bigg[
- \mathrm{FF}^{h_1+\frac{1}{2},h_2+\frac{1}{2}}_{2h+2,\,+} +
 \frac{2(h_1+h_2-h)-1}{(2h+2)}\,
 \mathrm{FF}^{h_1+\frac{1}{2},h_2+\frac{1}{2}}_{2h+2,\,-} \Bigg]_{\mu = 2 \la}.
\label{structla}
\eea
At $\la=0$, the above structure constants reduce to
the ones in \cite{Odake,AKK1910}.

%%%%%%%%%%%%%%%%%%%%%%%%%%%%%%%%%%%%%%%%%%%%%%%%%%%%%%%
%%%%%%%%%%%%%%%%%%%%%%%%%%%%%
\section{The other central terms }
%%%%%%%%%%%%%%%%%%%%%%%%%%%%%
%%%%%%AAA%%%%%%%%%%%%%%%%%%%%%%%%%%%%%%%%%%%%%%%%%%%%%%%%%

As done in (\ref{++}), we obtain the following
central terms 
\bea
&& V_{\la, \bar{a} b}^{(h_1)+}(z) \, V_{\la, \bar{c} d}^{(h_2)-}(w)\Bigg|_{\frac{1}{
    (z-w)^{h_1+h_2}}} = N\, \de_{b \bar{c}}\, \de_{d \bar{a}}\,
\sum_{j=0}^{h_1-1}\, \sum_{i=0}^{h_2-1}\, \sum_{t=0}^{j+1}\,
\nonu \\
&& \times 
\Bigg( -\frac{(h_2-1+2\la)}{(2h_2-1)} \,
a^j(h_1,\la)\, a^i(h_2,\la)-
\frac{(h_2-2\la)}{(2h_2-1)} \, a^{j}(h_1,\la+\frac{1}{2})\,
a^i(h_2,\la+\frac{1}{2})\Bigg)
\nonu \\
&& \times \frac{ j! \,(t+i)!}{t! \, (t+i+1)!}
\,(-1)^{h_1+t}\, (j+1-t)_{h_1-1-j}\, (h_1-t)_{t+1+i}\,
(h_1+1+i)_{h_2-1-i} \, ,
\nonu \\
&& V_{\la, \bar{a} b}^{(h_1)-}(z) \, V_{\la, \bar{c} d}^{(h_2)-}(w)\Bigg|_{\frac{1}{
    (z-w)^{h_1+h_2}}} = N\, \de_{b \bar{c}}\, \de_{d \bar{a}}\,
\sum_{j=0}^{h_1-1}\, \sum_{i=0}^{h_2-1}\, \sum_{t=0}^{j+1}\,
\nonu \\
&& \times 
\Bigg( \frac{(h_1-1+2\la)}{(2h_1-1)}\, \frac{(h_2-1+2\la)}{(2h_2-1)} \,
a^j(h_1,\la)\, a^i(h_2,\la) \nonu \\
&& -
\frac{(h_1-2\la)}{(2h_1-1)} \,
\frac{(h_2-2\la)}{(2h_2-1)} \, a^{j}(h_1,\la+\frac{1}{2})\,
a^i(h_2,\la+\frac{1}{2})\Bigg)
\nonu \\
&& \times \frac{ j! \,(t+i)!}{t! \, (t+i+1)!}
\,(-1)^{h_1+t}\, (j+1-t)_{h_1-1-j}\, (h_1-t)_{t+1+i}\,
(h_1+1+i)_{h_2-1-i} \, ,
\label{+---}
\eea
where the previous relations in (\ref{coeff}) are used.

%%%%%%%%%%%%%%%%%%%%%%%%%%%%%%%%%%%%%%%%%%%%%%%%%%%%%%%
%%%%%%%%%%%%%%%%%%%%%%%%%%%%%
\section{The OPEs between the nonsinglet currents }
%%%%%%%%%%%%%%%%%%%%%%%%%%%%%
%%%%%%BBB%%%%%%%%%%%%%%%%%%%%%%%%%%%%%%%%%%%%%%%%%%%%%%%%%

In this Appendix, the three kinds of OPEs corresponding to
the ones in section $3$ are given explicitly.

%%%%%%%%%%%%%%%%%%%%%
\subsection{ The OPE between the bosonic currents}
%%%%%%%%%%%%%%%%%%%%%

We present the explicit check for (\ref{first})
and (\ref{BB}) 
with the weights $h_1=h_2=4$. 

%%%%%%%%%%%%%%%%%%%%%
\subsubsection{ The OPE between
$W_{F,4}^{\la,\hat{A}=1}$ and itself}
%%%%%%%%%%%%%%%%%%%%%

For the same indices $\hat{A}=\hat{B}=1$ in (\ref{first}),
we obtain the following result
\bea
&& (W_{F,4}^{\la,12} +W_{F,4}^{\la,21})({z}) \,
(W_{F,4}^{\la,12}+W_{F,4}^{\la,21})({w})  = 
\nonu \\
&& \frac{1}{({z}-{w})^8}\, \Bigg[
  -\frac{1536}{5}  (112 \lambda ^6-280 \lambda ^4+147 \lambda ^2-9
  )  \Bigg]\nonu \\
&& +
  \frac{1}{({z}-{w})^6}\, \Bigg[
    \frac{2048}{5} (\lambda -1) (\lambda +1) (2 \lambda -3)
    (2 \lambda +3) \Bigg] \,
  (W_{F,2}^{\la,11}+W_{F,2}^{\la,22})({w})
  \nonu \\
  & &+ \frac{1}{({z}-{w})^5}\, \frac{1}{2}\, \Bigg[
    \frac{2048}{5} (\lambda -1) (\lambda +1) (2 \lambda -3)
    (2 \lambda +3) \Bigg] \,  {\pa}\,
  (W_{F,2}^{\la,11}+W_{F,2}^{\la,22})({w})
  \nonu \\
    & &+ \frac{1}{({z}-{w})^4}\, \Bigg[ \frac{3}{20}\, 
    \frac{2048}{5} (\lambda -1) (\lambda +1) (2 \lambda -3)
    (2 \lambda +3)  \,  {\pa}^2\, (W_{F,2}^{\la,11}+W_{F,2}^{\la,22})
    \nonu \\
    && - \frac{96}{5}  (4 \lambda ^2-19) \,
  (W_{F,4}^{\la,11}+ W_{F,4}^{\la,22}) \Bigg]({w})
  \nonu \\
    & &+ \frac{1}{({z}-{w})^3}\, \Bigg[ \frac{1}{30}\, 
    \frac{2048}{5} (\lambda -1) (\lambda +1) (2 \lambda -3)
    (2 \lambda +3)  \,  {\pa}^3\, (W_{F,2}^{\la,11}+W_{F,2}^{\la,22})
    \nonu \\
    && - \frac{1}{2}\, \frac{96}{5}  (4 \lambda ^2-19) \,
 {\pa}\,  (W_{F,4}^{\la,11}+ W_{F,4}^{\la,22}) \Bigg]({w})
  \nonu \\
     & &+ \frac{1}{({z}-{w})^2}\, \Bigg[ \frac{1}{168}\, 
    \frac{2048}{5} (\lambda -1) (\lambda +1) (2 \lambda -3)
    (2 \lambda +3)  \,  {\pa}^4\, (W_{F,2}^{\la,11}+W_{F,2}^{\la,22})
    \nonu \\
    && - \frac{5}{36}\, \frac{96}{5}  (4 \lambda ^2-19) \,
    {\pa}^2\,  (W_{F,4}^{\la,11}+W_{F,4}^{\la,22})
    + 6 \,  (W_{F,6}^{\la,11}+W_{F,6}^{\la,22}) \Bigg]({w})
  \nonu \\
  && + \frac{1}{({z}-{w})}\, \Bigg[ \frac{1}{1120}\, 
    \frac{2048}{5} (\lambda -1) (\lambda +1) (2 \lambda -3)
    (2 \lambda +3)  \,  {\pa}^5\, (W_{F,2}^{\la,11}+W_{F,2}^{\la,22})
  \nonu \\
 & &- \frac{1}{36}\, \frac{96}{5}  (4 \lambda ^2-19) \,
  {\pa}^3\,  (W_{F,4}^{\la,11}+W_{F,4}^{\la,22})
  + \frac{1}{2} \, 6 \,   {\pa}\,
  (W_{F,6}^{\la,11}+W_{F,6}^{\la,22}) \Bigg]({w}) + \cdots
  \nonu \\
  & &=  \frac{1}{({z}-{w})^8}\,
  \Bigg[
  -\frac{1536}{5}  (112 \lambda ^6-280 \lambda ^4+147 \lambda ^2-9
  )  \Bigg]
  \nonu \\
  && -
 % \frac{1}{({z}-{w})^6}\,
  p_{F,4}^{4,4}({\pa}_{{z}},{\pa}_{{w}},\la)
  \Bigg[ \frac{ (W_{F,2}^{\la,11}+W_{F,2}^{\la,22})({w})}{
      ({z}-{w})} \Bigg]
  \nonu \\
  && -
  %\frac{1}{({z}-{w})^4}\,
  p_{F,2}^{4,4}({\pa}_{{z}},{\pa}_{{w}},\la)
  \Bigg[ \frac{ (W_{F,4}^{\la,11}+W_{F,4}^{\la,22})({w})}{
      ({z}-{w})} \Bigg]
 -  p_{F,0}^{4,4}({\pa}_{{z}},{\pa}_{{w}},\la)
 \Bigg[ \frac{ (W_{F,6}^{\la,11}+W_{F,6}^{\la,22})({w})}{
     ({z}-{w})} \Bigg]
+ \cdots \, ,
\nonu \\
&& =
 \frac{1}{({z}-{w})^8}\,
  c_F(4,4,\la)\, \de^{11}\, q^4
  \nonu \\
  && + \sum_{h=0,\mbox{\footnotesize even}}^{4}\, q^h\, (-1)^{h-1}\,
  p_{F,h}^{4,4}({\pa}_{{z}},{\pa}_{{w}},\la)
  \Bigg[ \frac{ (W_{F,6-h}^{\la,11}+W_{F,6-h}^{\la,22})({w})}{
      ({z}-{w})} \Bigg] + \cdots \,.
\label{ope-one}
  \eea
  The right hand side of (\ref{ope-one}) can be seen from
  the last two terms in (\ref{first}) by changing the commutator
  to the corresponding OPE. Note that there is an  additional factor
  $(-1)^{h-1}$ in the above.
  
%%%%%%%%%%%%%%%%%%%%%
\subsubsection{ The OPE between
$W_{F,4}^{\la,\hat{A}=1}$ and $W_{F,4}^{\la,\hat{B}=2}$}
%%%%%%%%%%%%%%%%%%%%%

For the different indices $\hat{A}=1$ and $\hat{B}=2$ in (\ref{first}),
the following result is satisfied
\bea
&& (W_{F,4}^{\la,12} +W_{F,4}^{\la,21})({z}) \,
i\, (W_{F,4}^{\la,12}-W_{F,4}^{\la,21})({w})  = 
\nonu \\
&& \frac{1}{({z}-{w})^7}\, \Bigg[
  \frac{2048}{5}(\la-1)(\la+1)(2\la-3)(2\la-1)(2\la+1)(2\la+3)
  \Bigg]
 \,
  i\, (W_{F,1}^{\la,11}-W_{F,1}^{\la,22})({w})
\nonu \\
&& +
  \frac{1}{({z}-{w})^6} \Bigg[
    \frac{1}{2}
    \frac{2048}{5} (\lambda -1) (\lambda +1)(2 \lambda -3) (2\la-1)(2\la+1)
    (2 \lambda +3) \Bigg] 
i  \pa    (W_{F,1}^{\la,11}-W_{F,1}^{\la,22})({w})
  \nonu \\
  & &+ \frac{1}{({z}-{w})^5}\, \Bigg[
  \frac{1}{6}\,   \frac{2048}{5} (\lambda -1) (\lambda +1) (2 \lambda -3)
   (2\la-1)(2\la+1) (2 \lambda +3) \,  i \, {\pa}^2\,
  (W_{F,1}^{\la,11}-W_{F,1}^{\la,22}) \nonu \\
  && -\frac{256}{25}\,
(2\la-3)(2\la+3)(2\la^2-17)\,  i\, (W_{F,3}^{\la,11}-W_{F,3}^{\la,22})
  \Bigg]({w})
  \nonu \\
  & &+ \frac{1}{({z}-{w})^4}\, \Bigg[
 \frac{1}{24}\,   \frac{2048}{5} (\lambda -1) (\lambda +1) (2 \lambda -3)
    (2\la-1)(2\la+1)  (2 \lambda +3) \,  i \, {\pa}^3\,
  (W_{F,1}^{\la,11}-W_{F,1}^{\la,22}) \nonu \\
  && -\frac{1}{2}\, \frac{256}{25}\,
(2\la-3)(2\la+3)(2\la^2-17)\,  i\, \pa\, (W_{F,3}^{\la,11}-W_{F,3}^{\la,22})
    \Bigg]({w})
  \nonu \\
    & &+ \frac{1}{({z}-{w})^3} \Bigg[ \frac{1}{120} 
    \frac{2048}{5} (\lambda -1) (\lambda +1) (2 \lambda -3)
      (2\la-1)(2\la+1)
    (2 \lambda +3)   i {\pa}^4 (W_{F,1}^{\la,11}-W_{F,1}^{\la,22})
    \nonu \\
    &&
-\frac{1}{7} \frac{256}{25}
(2\la-3)(2\la+3)(2\la^2-17)  i \pa^2 (W_{F,3}^{\la,11}-W_{F,3}^{\la,22})
    +\frac{4}{5}(4\la^2-69) i (W_{F,5}^{\la,11}-W_{F,5}^{\la,22})
\Bigg]
  \nonu \\
  & &+ \frac{1}{({z}-{w})^2}\, \Bigg[
\frac{1}{720}
    \frac{2048}{5} (\lambda -1) (\lambda +1) (2 \lambda -3)
      (2\la-1)(2\la+1)
    (2 \lambda +3)   i {\pa}^5 (W_{F,1}^{\la,11}-W_{F,1}^{\la,22})
    \nonu \\
    &&
-\frac{5}{168}\, \frac{256}{25}\,
(2\la-3)(2\la+3)(2\la^2-17)\,  i\, \pa^3\, (W_{F,3}^{\la,11}-W_{F,3}^{\la,22})
\nonu \\
&& + \frac{1}{2}\,
\frac{4}{5}(4\la^2-69)\, i\, \pa \, (W_{F,5}^{\la,11}-W_{F,5}^{\la,22})
    \Bigg]({w})
  \nonu \\
  && + \frac{1}{({z}-{w})}\, \Bigg[
\frac{1}{5040}
    \frac{2048}{5} (\lambda -1) (\lambda +1) (2 \lambda -3)
      (2\la-1)(2\la+1)
    (2 \lambda +3)   i {\pa}^6 (W_{F,1}^{\la,11}-W_{F,1}^{\la,22})
    \nonu \\
    &&
-\frac{5}{1008}\, \frac{256}{25}\,
(2\la-3)(2\la+3)(2\la^2-17)\,  i\, \pa^4\, (W_{F,3}^{\la,11}-W_{F,3}^{\la,22})
\nonu \\
&& + 
\frac{3}{22}\, \frac{4}{5}\,
  (4\la^2-69)\, i\, \pa^2 \, (W_{F,5}^{\la,11}-W_{F,5}^{\la,22})
-\frac{1}{2}\,i\, (W_{F,7}^{\la,11}-W_{F,7}^{\la,22}) 
\Bigg]({w}) + \cdots
  \nonu \\
  & &= 
  -  p_{F,5}^{4,4}({\pa}_{{z}},{\pa}_{{w}},\la)
  \Bigg[ \frac{i\, (W_{F,1}^{\la,11}-W_{F,1}^{\la,22})({w})}{
      ({z}-{w})} \Bigg]
 % \frac{1}{({z}-{w})^6}\,
-  p_{F,3}^{4,4}({\pa}_{{z}},{\pa}_{{w}},\la)
  \Bigg[ \frac{ i\, (W_{F,3}^{\la,11}-W_{F,3}^{\la,22})({w})}{
      ({z}-{w})} \Bigg]
  \nonu \\
  && -
  %\frac{1}{({z}-{w})^4}\,
  p_{F,1}^{4,4}({\pa}_{{z}},{\pa}_{{w}},\la)
  \Bigg[ \frac{ i\, (W_{F,5}^{\la,11}-W_{F,5}^{\la,22})({w})}{
      ({z}-{w})} \Bigg]
 -  p_{F,-1}^{4,4}({\pa}_{{z}},{\pa}_{{w}},\la)
 \Bigg[ \frac{i\, (W_{F,7}^{\la,11}-W_{F,7}^{\la,22})({w})}{
     ({z}-{w})} \Bigg]
+ \cdots \, 
\nonu \\
&& =- \sum_{h=-1,\mbox{\footnotesize odd}}^{5}\, q^h\, (-1)^{h-1}\,
p_{F,h}^{4,4}({\pa}_{{z}},{\pa}_{{w}},\la)
i \, f^{123} \,
  \Bigg[ \frac{ W_{F,6-h}^{\la,\hat{A}=3}({w})}{
      ({z}-{w})} \Bigg] + \cdots \,.
\label{ope-two}
  \eea
  Therefore, we observe that
  this can be seen from the first term in (\ref{first}).
  It turns out that
  the three elements of $ W_{F,4}^{\la,\hat{A}}$ are normalized
  correctly. Here the structure constant is given by
  $f^{123}=1$.
  We can also check other cases of (\ref{first}) by choosing
  the different indices and expect to have similar relations to
  (\ref{ope-one}) and (\ref{ope-two}).
  We have checked that the relation (\ref{first})
  satisfies for the case of the arbitrary weight
  $h_1$ with the restricted weight $h_2$ ($h_1=h_2, h_2\pm 1$).
  
%%%%%%%%%%%%%%%%%  
\subsubsection{The OPE between $W_{B,4}^{\la,\hat{A}=1}$ and itself}
%%%%%%%%%%%%%%%%

We can
consider the second case described by (\ref{BB}).
It turns out that we obtain the result 
\bea
&& (W_{B,4}^{\la,12}+W_{B,4}^{\la,21})({z}) \,
    (W_{B,4}^{\la,12}+W_{B,4}^{\la,21})({w})  = 
  \nonu \\
  && \frac{1}{({z}-{w})^8}\, \Bigg[
\frac{6144}{5} (28 \lambda ^6-84 \lambda ^5+35 \lambda ^4+70 \lambda ^3-42 \lambda ^2-7 \lambda +3)
\Bigg]\nonu \\
&& +
  \frac{1}{({z}-{w})^6}\, \Bigg[
 \frac{2048}{5} (\lambda -2) (\lambda +1) (2 \lambda -3) (2 \lambda +1)
    \Bigg] \,  (W_{B,2}^{\la,11}+W_{B,2}^{\la,22})({w})
  \nonu \\
  & &+ \frac{1}{({z}-{w})^5}\, \frac{1}{2}\, \Bigg[
     \frac{2048}{5} (\lambda -2) (\lambda +1) (2 \lambda -3) (2 \lambda +1)
     \Bigg] \,  {\pa}\, (W_{B,2}^{\la,11}+W_{B,2}^{\la,22})({w})
  \nonu \\
    & &+ \frac{1}{({z}-{w})^4}\, \Bigg[ \frac{3}{20}
    \frac{2048}{5} (\lambda -2) (\lambda +1) (2 \lambda -3)
    (2 \lambda +1)     {\pa}^2\, (W_{B,2}^{\la,11}+W_{B,2}^{\la,22})
    \nonu \\
    && -  \frac{192}{5}  (2 \lambda ^2-2 \lambda -9)
  (W_{B,4}^{\la,11}+W_{B,4}^{\la,22})  \Bigg]({w})
  \nonu \\
    & &+ \frac{1}{({z}-{w})^3} \Bigg[ \frac{1}{30} 
    \frac{2048}{5} (\lambda -2) (\lambda +1) (2 \lambda -3)
    (2 \lambda +1)      {\pa}^3 (W_{B,2}^{\la,11}+W_{B,2}^{\la,22})
    \nonu \\
    && - \frac{1}{2}\frac{192}{5}  (2 \lambda ^2-2 \lambda -9) 
 {\pa}  (W_{B,4}^{\la,11}+W_{B,4}^{\la,22})  \Bigg]({w})
  \nonu \\
     & &+ \frac{1}{({z}-{w})^2}\, \Bigg[ \frac{1}{168}\, 
 \frac{2048}{5} (\lambda -2) (\lambda +1) (2 \lambda -3) (2 \lambda +1)    \,  {\pa}^4\, (W_{B,2}^{\la,11}+W_{B,2}^{\la,22})
 \nonu \\
 && - \frac{5}{36}\,\frac{192}{5}  (2 \lambda ^2-2 \lambda -9) \,
       {\pa}^2\,  (W_{B,4}^{\la,11}+W_{B,4}^{\la,22}) +
       6 \,  (W_{B,6}^{\la,11}+W_{B,6}^{\la,22}) \Bigg]({w})
  \nonu \\
  &&
 + \frac{1}{({z}-{w})}\, \Bigg[ \frac{1}{1120}\, 
   \frac{2048}{5} (\lambda -2) (\lambda +1) (2 \lambda -3) (2 \lambda +1)  \,  {\pa}^5\, (W_{B,2}^{\la,11}+W_{B,2}^{\la,22})
  \nonu \\
 & &- \frac{1}{36}\,\frac{192}{5}  (2 \lambda ^2-2 \lambda -9) \,
        {\pa}^3\,  (W_{B,4}^{\la,11}+W_{B,4}^{\la,22})  +
        \frac{1}{2} \, 6 \,   {\pa}\,
  (W_{B,6}^{\la,11}+W_{B,6}^{\la,22}) \Bigg]({w}) + \cdots
  \nonu \\
  & &=  \frac{1}{({z}-{w})^8}\,
  \Bigg[
    \frac{6144}{5} (28 \lambda ^6-
    84 \lambda ^5+35 \lambda ^4+70 \lambda ^3-42 \lambda ^2-7 \lambda +3)
\Bigg]\nonu \\
  && -
 % \frac{1}{({z}-{w})^6}\,
  p_{B,4}^{4,4}({\pa}_{{z}},{\pa}_{{w}},\la)
  \Bigg[ \frac{ (W_{B,2}^{\la,11}+W_{B,2}^{\la,22})({w})}{({z}-{w})} \Bigg]
-
  %\frac{1}{({z}-{w})^4}\,
  p_{B,2}^{4,4}({\pa}_{{z}},{\pa}_{{w}},\la)
  \Bigg[ \frac{ (W_{B,4}^{\la,11}+W_{B,4}^{\la,22})({w})}{({z}-{w})} \Bigg]
  \nonu \\
  && -  p_{B,0}^{4,4}({\pa}_{{z}},{\pa}_{{w}},\la)
\Bigg[ \frac{ (W_{B,6}^{\la,11}+W_{B,6}^{\la,22})({w})}{({z}-{w})} \Bigg]
+ \cdots \, 
\nonu \\
&& =
 \frac{1}{({z}-{w})^8}\,
  c_B(4,4,\la)\, \de^{11}\, q^4
  \nonu \\
  && + \sum_{h=0,\mbox{\footnotesize even}}^{4}\, q^h\, (-1)^{h-1}\,
  p_{B,h}^{4,4}({\pa}_{{z}},{\pa}_{{w}},\la)
  \Bigg[ \frac{ (W_{B,6-h}^{\la,11}+W_{B,6-h}^{\la,22})({w})}{
      ({z}-{w})} \Bigg] + \cdots \,,
\label{ope-three}
\eea  
which looks similar to the previous one in (\ref{ope-one}).
The corresponding central charge, the structure constants and
the currents occur.

%%%%%%%%%%%%%%%%%  
\subsubsection{The OPE between $W_{B,4}^{\la,\hat{A}=1}$ and
 $W_{B,4}^{\la,\hat{B}=2}$}
%%%%%%%%%%%%%%%%

By taking
the different indices $\hat{A}=1$ and $\hat{B}=2$ in (\ref{BB}),
the following result is obtained
\bea
&& (W_{B,4}^{\la,12}+W_{B,4}^{\la,21})({z}) \,
    i \, (W_{B,4}^{\la,12}-W_{B,4}^{\la,21})({w})  = 
  \nonu \\
  && \frac{1}{({z}-{w})^7}\, \Bigg[
\frac{8192}{5} (\la-2)(\la-1)\la(\la+1)(2\la-3)(2\la+1)
\Bigg]
  \,  i\, (W_{B,1}^{\la,11}-W_{B,1}^{\la,22})({w})\nonu \\
  & &+ \frac{1}{({z}-{w})^6}\, \frac{1}{2}\, \Bigg[
    \frac{8192}{5} (\la-2)(\la-1)\la(\la+1)(2\la-3)(2\la+1)
     \Bigg] \,  i\, {\pa}\, (W_{B,1}^{\la,11}-W_{B,1}^{\la,22})({w})
  \nonu \\
    & &+ \frac{1}{({z}-{w})^5}\, \Bigg[ \frac{1}{6}
    \frac{8192}{5} (\la-2)(\la-1)\la(\la+1)(2\la-3)(2\la+1)
    i\,  {\pa}^2 \, (W_{B,1}^{\la,11}-W_{B,1}^{\la,22})
    \nonu \\
    && - \frac{512}{25}(\la-2)(\la+1)(4\la^2-4\la-33)\,
  i\, (W_{B,3}^{\la,11}-W_{B,3}^{\la,22})  \Bigg]({w})
  \nonu \\
&& + \frac{1}{({z}-{w})^4}\, \Bigg[ \frac{1}{24}
    \frac{8192}{5} (\la-2)(\la-1)\la(\la+1)(2\la-3)(2\la+1)
    i\,  {\pa}^3 \, (W_{B,1}^{\la,11}-W_{B,1}^{\la,22})
    \nonu \\
    && - \frac{1}{2}\, \frac{512}{25}(\la-2)(\la+1)(4\la^2-4\la-33)\,
  i\, \pa \, (W_{B,3}^{\la,11}-W_{B,3}^{\la,22})  \Bigg]({w})
  \nonu \\
&&
 + \frac{1}{({z}-{w})^3}\, \Bigg[ \frac{1}{120}
    \frac{8192}{5} (\la-2)(\la-1)\la(\la+1)(2\la-3)(2\la+1)
    i\,  {\pa}^4 \, (W_{B,1}^{\la,11}-W_{B,1}^{\la,22})
    \nonu \\
    && - \frac{1}{7}\, \frac{512}{25}(\la-2)(\la+1)(4\la^2-4\la-33)\,
    i\, \pa^2 \, (W_{B,3}^{\la,11}-W_{B,3}^{\la,22})  \nonu \\
    && + \frac{16}{5} (\la^2-\la-17)\,
    i \, (W_{B,5}^{\la,11}-W_{B,5}^{\la,22}) \Bigg]({w})  
 \nonu \\
&&
 + \frac{1}{({z}-{w})^2}\, \Bigg[ \frac{1}{720}
    \frac{8192}{5} (\la-2)(\la-1)\la(\la+1)(2\la-3)(2\la+1)
    i\,  {\pa}^5 \, (W_{B,1}^{\la,11}-W_{B,1}^{\la,22})
    \nonu \\
    && - \frac{5}{168}\, \frac{512}{25}(\la-2)(\la+1)(4\la^2-4\la-33)\,
    i\, \pa^3 \, (W_{B,3}^{\la,11}-W_{B,3}^{\la,22})  \nonu \\
    && +\frac{1}{2}\, \frac{16}{5} (\la^2-\la-17)\,
    i \, \pa \, (W_{B,5}^{\la,11}-W_{B,5}^{\la,22}) \Bigg]({w})  
 \nonu \\
&&
 + \frac{1}{({z}-{w})}\, \Bigg[ \frac{1}{5040}
    \frac{8192}{5} (\la-2)(\la-1)\la(\la+1)(2\la-3)(2\la+1)
    i\,  {\pa}^6 \, (W_{B,1}^{\la,11}-W_{B,1}^{\la,22})
    \nonu \\
    && - \frac{5}{1008}\, \frac{512}{25}(\la-2)(\la+1)(4\la^2-4\la-33)\,
    i\, \pa^4 \, (W_{B,3}^{\la,11}-W_{B,3}^{\la,22})  \nonu \\
    && +\frac{3}{22}\, \frac{16}{5} (\la^2-\la-17)\,
    i \, \pa^2 \, (W_{B,5}^{\la,11}-W_{B,5}^{\la,22})
-\frac{1}{2}\, i \, (W_{B,7}^{\la,11}-W_{B,7}^{\la,22})
    \Bigg]({w}) + \cdots
\nonu \\
& &=
-
 % \frac{1}{({z}-{w})^6}\,
  p_{B,5}^{4,4}({\pa}_{{z}},{\pa}_{{w}},\la)
  \Bigg[ \frac{ i \,
      (W_{B,1}^{\la,11}-W_{B,1}^{\la,22})({w})}{({z}-{w})} \Bigg]
\nonu \\
  && -
 % \frac{1}{({z}-{w})^6}\,
  p_{B,3}^{4,4}({\pa}_{{z}},{\pa}_{{w}},\la)
  \Bigg[ \frac{ i\, (W_{B,3}^{\la,11}-W_{B,3}^{\la,22})({w})}{
      ({z}-{w})} \Bigg]
-
  %\frac{1}{({z}-{w})^4}\,
  p_{B,1}^{4,4}({\pa}_{{z}},{\pa}_{{w}},\la)
  \Bigg[ \frac{i\, (W_{B,5}^{\la,11}-W_{B,5}^{\la,22})({w})}{({z}-{w})} \Bigg]
  \nonu \\
  && -  p_{B,-1}^{4,4}({\pa}_{{z}},{\pa}_{{w}},\la)
\Bigg[ \frac{ i \, (W_{B,7}^{\la,11}-W_{B,7}^{\la,22})({w})}{({z}-{w})} \Bigg]
+ \cdots \, 
\nonu \\
&& = -
  \sum_{h=-1,\mbox{\footnotesize odd}}^{5}\, q^h\, (-1)^{h-1}\,
  p_{B,h}^{4,4}({\pa}_{{z}},{\pa}_{{w}},\la)
  i \, f^{123}\,
  \Bigg[ \frac{  W_{B,6-h}^{\la,\hat{A}=3}({w})}{
      ({z}-{w})} \Bigg] + \cdots \,.
\label{ope-four}
\eea  
Note that there is a $\la$ factor in the structure
constants appearing the current 
$(W_{B,1}^{\la,11}-W_{B,1}^{\la,22})$ and its descendants.
It is obvious that they vanish at $\la=0$.

%%%%%%%%%%%%%%%%%  
\subsection{The OPE between the bosonic currents and
the fermionic currents}
%%%%%%%%%%%%%%%%

We present the explicit check for (\ref{WQ})
with the weights $h_1=4$ and $h_2=3$.

%%%%%%%%%%%%%%%%%  
\subsubsection{The OPE between $W_{F,4}^{\la,\hat{A}=1}$ and
 $Q_{\frac{7}{2}}^{\la,\hat{B}=1}$}
%%%%%%%%%%%%%%%%

For the same indices $\hat{A}=\hat{B}=1$ in
the first equation of (\ref{WQ}),
the following result holds
\bea
&& (W_{F,4}^{\la,12}+W_{F,4}^{\la,21})({z}) \,
(Q_{\frac{7}{2}}^{\la,12}+Q_{\frac{7}{2}}^{\la,21})({w})  =
\nonu \\
&& 
  \frac{1}{({z}-{w})^6}\, \Bigg[
\frac{256}{5} (\lambda -1) (\lambda +1) (2 \lambda -3) (2 \lambda +1) (2 \lambda +3)
    \Bigg] \,  (Q_{\frac{3}{2}}^{\la,11}+Q_{\frac{3}{2}}^{\la,22})({w})
  \nonu \\
  && +
 \frac{1}{({z}-{w})^5}\, \Bigg[
   \frac{2}{3}
   \frac{256}{5} (\lambda -1) (\lambda +1) (2 \lambda -3) (2 \lambda +1) (2 \lambda +3) \,  {\pa}\, (Q_{\frac{3}{2}}^{\la,11}+Q_{\frac{3}{2}}^{\la,22})
   \nonu \\
   && -\frac{256}{25} (\lambda -4) (\lambda +1) (2 \lambda -3)
   (2 \lambda +3)\,  (Q_{\frac{5}{2}}^{\la,11}+Q_{\frac{5}{2}}^{\la,22}) \Bigg]({w})
 \nonu \\
 && +
 \frac{1}{({z}-{w})^4}\, \Bigg[
   \frac{1}{4}
   \frac{256}{5} (\lambda -1) (\lambda +1) (2 \lambda -3) (2 \lambda +1) (2 \lambda +3) \,  {\pa}^2\, (Q_{\frac{3}{2}}^{\la,11}+Q_{\frac{3}{2}}^{\la,22})
   \nonu \\
   && -\frac{3}{5} \,
   \frac{256}{25} (\lambda -4) (\lambda +1) (2 \lambda -3)
   (2 \lambda +3)\,  {\pa}\, (Q_{\frac{5}{2}}^{\la,11}+Q_{\frac{5}{2}}^{\la,22})
   \nonu \\
   && -
   \frac{16}{25} (2 \lambda +3) (4 \lambda ^2+18 \lambda -61)
   \, (Q_{\frac{7}{2}}^{\la,11}+ Q_{\frac{7}{2}}^{\la,22})
   \Bigg]({w})
 \nonu \\
  && +
 \frac{1}{({z}-{w})^3}\, \Bigg[
   \frac{1}{15}
   \frac{256}{5} (\lambda -1) (\lambda +1) (2 \lambda -3) (2 \lambda +1) (2 \lambda +3) \,  {\pa}^3\, (Q_{\frac{3}{2}}^{\la,11}+Q_{\frac{3}{2}}^{\la,22})
   \nonu \\
   && -\frac{1}{5} \,
   \frac{256}{25} (\lambda -4) (\lambda +1) (2 \lambda -3)
   (2 \lambda +3)\,  {\pa}^2\,
   (Q_{\frac{5}{2}}^{\la,11}+Q_{\frac{5}{2}}^{\la,22}) \nonu \\
   && -
   \frac{4}{7}\,
   \frac{16}{25} (2 \lambda +3) (4 \lambda ^2+18 \lambda -61)
   \, {\pa}\, (Q_{\frac{7}{2}}^{\la,11}+ Q_{\frac{7}{2}}^{\la,22}) 
   \nonu \\
   && +
   \frac{4}{5} (2 \lambda ^2-5 \lambda -27)\,  (Q_{\frac{9}{2}}^{\la,11}+
   Q_{\frac{9}{2}}^{\la,22})
   \Bigg]({w})
 \nonu \\
  && +
 \frac{1}{({z}-{w})^2}\, \Bigg[
   \frac{1}{72}
   \frac{256}{5} (\lambda -1) (\lambda +1) (2 \lambda -3) (2 \lambda +1) (2 \lambda +3) \,  {\pa}^4\, (Q_{\frac{3}{2}}^{\la,11}+Q_{\frac{3}{2}}^{\la,22})
   \nonu \\
   && -\frac{1}{21} \,
   \frac{256}{25} (\lambda -4) (\lambda +1) (2 \lambda -3)
   (2 \lambda +3)\,  {\pa}^3\,
   (Q_{\frac{5}{2}}^{\la,11}+Q_{\frac{5}{2}}^{\la,22})
   \nonu \\
   && -
   \frac{5}{28}\,
   \frac{16}{25} (2 \lambda +3) (4 \lambda ^2+18 \lambda -61)
   \, {\pa}^2\, (Q_{\frac{7}{2}}^{\la,11}+ Q_{\frac{7}{2}}^{\la,22})
   \nonu \\
   && +\frac{5}{9}\,
   \frac{4}{5} (2 \lambda ^2-5 \lambda -27)\, {\pa}\, (Q_{\frac{9}{2}}^{\la,11}
   +
   Q_{\frac{9}{2}}^{\la,22})
+
\frac{1}{10} (2 \lambda +27) \, (Q_{\frac{11}{2}}^{\la,11}+
   Q_{\frac{11}{2}}^{\la,22})  
\Bigg]({w})
 \nonu \\
 && +
 \frac{1}{({z}-{w})}\, \Bigg[
   \frac{1}{420}
   \frac{256}{5} (\lambda -1) (\lambda +1) (2 \lambda -3) (2 \lambda +1) (2 \lambda +3) \,  {\pa}^5\, (Q_{\frac{3}{2}}^{\la,11}+Q_{\frac{3}{2}}^{\la,22})
   \nonu \\
   && -\frac{1}{112} \,
   \frac{256}{25} (\lambda -4) (\lambda +1) (2 \lambda -3)
   (2 \lambda +3)\,  {\pa}^4\, (Q_{\frac{5}{2}}^{\la,11}+Q_{\frac{5}{2}}^{\la,22})
   \nonu \\
   && -
   \frac{5}{126}\,
   \frac{16}{25} (2 \lambda +3) (4 \lambda ^2+18 \lambda -61)
   \, {\pa}^3\, (Q_{\frac{7}{2}}^{\la,11}+ Q_{\frac{7}{2}}^{\la,22}) 
   \nonu \\
   && +\frac{1}{6}\,
   \frac{4}{5} (2 \lambda ^2-5 \lambda -27)\, {\pa}^2\,
   (Q_{\frac{9}{2}}^{\la,11}+
   Q_{\frac{9}{2}}^{\la,22}) 
+\frac{6}{11}\,
\frac{1}{10} (2 \lambda +27) \, {\pa}\, (Q_{\frac{11}{2}}^{\la,11}
+
   Q_{\frac{11}{2}}^{\la,22})
   \nonu \\
   && -\frac{1}{4} \, (Q_{\frac{13}{2}}^{\la,11}+
   Q_{\frac{13}{2}}^{\la,22}) \Bigg]({w})+\cdots
 \nonu  \\
  & &=  -
 % \frac{1}{({z}-{w})^6}\,
  q_{F,4}^{4,\frac{7}{2}}({\pa}_{{z}},{\pa}_{{w}},\la)
  \Bigg[ \frac{ (Q_{\frac{3}{2}}^{\la,11}+ Q_{\frac{3}{2}}^{\la,22})({w})}{({z}-{w})}  \Bigg]
+
  %\frac{1}{({z}-{w})^4}\,
  q_{F,3}^{4,\frac{7}{2}}({\pa}_{{z}},{\pa}_{{w}},\la)
  \Bigg[ \frac{ (Q_{\frac{5}{2}}^{\la,11}+ Q_{\frac{5}{2}}^{\la,22})({w})}{({z}-{w})} \Bigg]
  \nonu \\
  && -  q_{F,2}^{4,\frac{7}{2}}({\pa}_{{z}},{\pa}_{{w}},\la)
  \Bigg[ \frac{ (Q_{\frac{7}{2}}^{\la,11}+ Q_{\frac{7}{2}}^{\la,22})({w})}{({z}-{w})} \Bigg]
 +  q_{F,1}^{4,\frac{7}{2}}({\pa}_{{z}},{\pa}_{{w}},\la)
  \Bigg[ \frac{ (Q_{\frac{9}{2}}^{\la,11}+
   Q_{\frac{9}{2}}^{\la,22})({w})}{({z}-{w})} \Bigg]
  \nonu \\
  && -
  q_{F,0}^{4,\frac{7}{2}}({\pa}_{{z}},{\pa}_{{w}},\la)
  \Bigg[ \frac{ (Q_{\frac{11}{2}}^{\la,11}+
   Q_{\frac{11}{2}}^{\la,22})({w})}{({z}-{w})} \Bigg]
  +q_{F,-1}^{4,\frac{7}{2}}({\pa}_{{z}},{\pa}_{{w}},\la)
  \Bigg[ \frac{ (Q_{\frac{13}{2}}^{\la,11}+
      Q_{\frac{13}{2}}^{\la,22})({w})}{({z}-{w})} \Bigg]
  + \cdots
  \nonu \\
  && =
 \sum_{h=-1}^{4}\, q^h\, (-1)^{h-1}\,
  q_{F,h}^{4,\frac{7}{2}}({\pa}_{{z}},{\pa}_{{w}},\la)
  \, \de^{11} \,
  \Bigg[ \frac{  Q_{\frac{11}{2}-h}^{\la,\bar{a}a}({w})}{
      ({z}-{w})} \Bigg] + \cdots \,.  
\label{ope-five}
  \eea
 The right hand side of (\ref{ope-five}) can be seen from
 the last term in the first relation of
 (\ref{WQ}) by changing the commutator
  to the corresponding OPE. Note that there is an  additional factor
  $(-1)^{h-1}$ in the above.
  
%%%%%%%%%%%%%%%%%%  
  \subsubsection{
  The OPE between $W_{F,4}^{\la,\hat{A}=1}$ and
 $Q_{\frac{7}{2}}^{\la,\hat{B}=2}$}
  %%%%%%%%%%%%%%%%%%

  The following result is obtained
  by taking
  the different indices $\hat{A}=1$ and $\hat{B}=2$ in
  the first equation of (\ref{WQ}),
  \bea
&& (W_{F,4}^{\la,12}+W_{F,4}^{\la,21})({z}) \,
i \, (Q_{\frac{7}{2}}^{\la,12}-Q_{\frac{7}{2}}^{\la,21})({w})  =
\nonu \\
&& 
  \frac{1}{({z}-{w})^6}\, \Bigg[
\frac{256}{5} (\lambda -1) (\lambda +1) (2 \lambda -3) (2 \lambda +1) (2 \lambda +3)
    \Bigg] \,  i\, (Q_{\frac{3}{2}}^{\la,11}-Q_{\frac{3}{2}}^{\la,22})({w})
  \nonu \\
  && +
 \frac{1}{({z}-{w})^5}\, \Bigg[
   \frac{2}{3}
   \frac{256}{5} (\lambda -1) (\lambda +1) (2 \lambda -3) (2 \lambda +1) (2 \lambda +3) \,  i \, {\pa}\, (Q_{\frac{3}{2}}^{\la,11}-Q_{\frac{3}{2}}^{\la,22})
   \nonu \\
   && -\frac{256}{25} (\lambda -4) (\lambda +1) (2 \lambda -3)
   (2 \lambda +3)\, i\,
   (Q_{\frac{5}{2}}^{\la,11}-Q_{\frac{5}{2}}^{\la,22}) \Bigg]({w})
 \nonu \\
 && +
 \frac{1}{({z}-{w})^4}\, \Bigg[
   \frac{1}{4}
   \frac{256}{5} (\lambda -1) (\lambda +1) (2 \lambda -3) (2 \lambda +1) (2 \lambda +3) \, i\, {\pa}^2\, (Q_{\frac{3}{2}}^{\la,11}-Q_{\frac{3}{2}}^{\la,22})
   \nonu \\
   && -\frac{3}{5} \,
   \frac{256}{25} (\lambda -4) (\lambda +1) (2 \lambda -3)
   (2 \lambda +3)\,  i\,
        {\pa}\, (Q_{\frac{5}{2}}^{\la,11}-Q_{\frac{5}{2}}^{\la,22})
   \nonu \\
   && -
   \frac{16}{25} (2 \lambda +3) (4 \lambda ^2+18 \lambda -61)
   \, i \, (Q_{\frac{7}{2}}^{\la,11}- Q_{\frac{7}{2}}^{\la,22})
   \Bigg]({w})
 \nonu \\
  && +
 \frac{1}{({z}-{w})^3}\, \Bigg[
   \frac{1}{15}
   \frac{256}{5} (\lambda -1) (\lambda +1) (2 \lambda -3) (2 \lambda +1) (2 \lambda +3) \,  i\,
        {\pa}^3\, (Q_{\frac{3}{2}}^{\la,11}-Q_{\frac{3}{2}}^{\la,22})
   \nonu \\
   && -\frac{1}{5} \,
   \frac{256}{25} (\lambda -4) (\lambda +1) (2 \lambda -3)
   (2 \lambda +3)\,  i\, {\pa}^2\,
   (Q_{\frac{5}{2}}^{\la,11}-Q_{\frac{5}{2}}^{\la,22}) \nonu \\
   && -
   \frac{4}{7}\,
   \frac{16}{25} (2 \lambda +3) (4 \lambda ^2+18 \lambda -61)
   \, i\, {\pa}\, (Q_{\frac{7}{2}}^{\la,11}- Q_{\frac{7}{2}}^{\la,22}) 
   \nonu \\
   && +
   \frac{4}{5} (2 \lambda ^2-5 \lambda -27)\,  i\,
   (Q_{\frac{9}{2}}^{\la,11}-
   Q_{\frac{9}{2}}^{\la,22})
   \Bigg]({w})
 \nonu \\
  && +
 \frac{1}{({z}-{w})^2}\, \Bigg[
   \frac{1}{72}
   \frac{256}{5} (\lambda -1) (\lambda +1) (2 \lambda -3) (2 \lambda +1) (2 \lambda +3) \,  i\,
        {\pa}^4\, (Q_{\frac{3}{2}}^{\la,11}-Q_{\frac{3}{2}}^{\la,22})
   \nonu \\
   && -\frac{1}{21} \,
   \frac{256}{25} (\lambda -4) (\lambda +1) (2 \lambda -3)
   (2 \lambda +3)\, i\,  {\pa}^3\,
   (Q_{\frac{5}{2}}^{\la,11}-Q_{\frac{5}{2}}^{\la,22})
   \nonu \\
   && -
   \frac{5}{28}\,
   \frac{16}{25} (2 \lambda +3) (4 \lambda ^2+18 \lambda -61)
   \, i\, {\pa}^2\, (Q_{\frac{7}{2}}^{\la,11}- Q_{\frac{7}{2}}^{\la,22})
   \nonu \\
   && +\frac{5}{9}\,
   \frac{4}{5} (2 \lambda ^2-5 \lambda -27)\, i\,
        {\pa}\, (Q_{\frac{9}{2}}^{\la,11}
   -
   Q_{\frac{9}{2}}^{\la,22})
+
\frac{1}{10} (2 \lambda +27) \, i\, (Q_{\frac{11}{2}}^{\la,11}-
   Q_{\frac{11}{2}}^{\la,22})  
\Bigg]({w})
 \nonu \\
 && +
 \frac{1}{({z}-{w})}\, \Bigg[
   \frac{1}{420}
   \frac{256}{5} (\lambda -1) (\lambda +1) (2 \lambda -3) (2 \lambda +1) (2 \lambda +3) \,  i \, {\pa}^5\, (Q_{\frac{3}{2}}^{\la,11}-Q_{\frac{3}{2}}^{\la,22})
   \nonu \\
   && -\frac{1}{112} \,
   \frac{256}{25} (\lambda -4) (\lambda +1) (2 \lambda -3)
   (2 \lambda +3)\,  i\,
        {\pa}^4\, (Q_{\frac{5}{2}}^{\la,11}-Q_{\frac{5}{2}}^{\la,22})
   \nonu \\
   && -
   \frac{5}{126}\,
   \frac{16}{25} (2 \lambda +3) (4 \lambda ^2+18 \lambda -61)
   \, i\, {\pa}^3\, (Q_{\frac{7}{2}}^{\la,11}- Q_{\frac{7}{2}}^{\la,22}) 
   \nonu \\
   && +\frac{1}{6}\,
   \frac{4}{5} (2 \lambda ^2-5 \lambda -27)\, i\, {\pa}^2\,
   (Q_{\frac{9}{2}}^{\la,11}-
   Q_{\frac{9}{2}}^{\la,22}) 
+\frac{6}{11}\,
\frac{1}{10} (2 \lambda +27) \, i\, {\pa}\, (Q_{\frac{11}{2}}^{\la,11}
-
   Q_{\frac{11}{2}}^{\la,22})
   \nonu \\
   && -\frac{1}{4} \, i\, (Q_{\frac{13}{2}}^{\la,11}-
   Q_{\frac{13}{2}}^{\la,22}) \Bigg]({w})+\cdots
 \nonu  \\
  & &=  -
 % \frac{1}{({z}-{w})^6}\,
  q_{F,4}^{4,\frac{7}{2}}({\pa}_{{z}},{\pa}_{{w}},\la)
  \Bigg[ \frac{i\, (Q_{\frac{3}{2}}^{\la,11}- Q_{\frac{3}{2}}^{\la,22})({w})}{
      ({z}-{w})}  \Bigg]
+
  %\frac{1}{({z}-{w})^4}\,
  q_{F,3}^{4,\frac{7}{2}}({\pa}_{{z}},{\pa}_{{w}},\la)
  \Bigg[ \frac{i\,  (Q_{\frac{5}{2}}^{\la,11}- Q_{\frac{5}{2}}^{\la,22})({w})}{({z}-{w})} \Bigg]
  \nonu \\
  && -  q_{F,2}^{4,\frac{7}{2}}({\pa}_{{z}},{\pa}_{{w}},\la)
  \Bigg[ \frac{i\, (Q_{\frac{7}{2}}^{\la,11}- Q_{\frac{7}{2}}^{\la,22})({w})}{({z}-{w})} \Bigg]
 +  q_{F,1}^{4,\frac{7}{2}}({\pa}_{{z}},{\pa}_{{w}},\la)
  \Bigg[ \frac{ i\, (Q_{\frac{9}{2}}^{\la,11}-
   Q_{\frac{9}{2}}^{\la,22})({w})}{({z}-{w})} \Bigg]
  \nonu \\
  && -
  q_{F,0}^{4,\frac{7}{2}}({\pa}_{{z}},{\pa}_{{w}},\la)
  \Bigg[ \frac{ i\, (Q_{\frac{11}{2}}^{\la,11}-
   Q_{\frac{11}{2}}^{\la,22})({w})}{({z}-{w})} \Bigg]
  +q_{F,-1}^{4,\frac{7}{2}}({\pa}_{{z}},{\pa}_{{w}},\la)
  \Bigg[ \frac{ i\,
      (Q_{\frac{13}{2}}^{\la,11}-Q_{\frac{13}{2}}^{\la,22})({w})}{({z}-{w})} \Bigg]
  + \cdots
  \nonu \\
  && =
 \sum_{h=-1}^{4}\, q^h\, (-1)^{h-1}\,
  q_{F,h}^{4,\frac{7}{2}}({\pa}_{{z}},{\pa}_{{w}},\la)
  \, i \, f^{123}\,
  \Bigg[ \frac{  Q_{\frac{11}{2}-h}^{\la,\hat{A}=3}({w})}{
      ({z}-{w})} \Bigg] + \cdots \,.  
\label{ope-six}
  \eea
  Then we observe the first term of the first equation of (\ref{WQ})
  and  the first relation of (\ref{WQ})
  satisfies for the case of the arbitrary weight
  $h_1$ with the restricted weight $h_2$ ($h_1=h_2, h_2+ 1$).
  
%%%%%%%%%%%%%%%%%  
\subsubsection{The OPE between $W_{B,4}^{\la,\hat{A}=1}$ and
 $Q_{\frac{7}{2}}^{\la,\hat{B}=1}$}
%%%%%%%%%%%%%%%%

We obtain the following result for the same indices $\hat{A}=\hat{B}=1$
  \bea
  && (W_{B,4}^{\la,12}+W_{B,4}^{\la,21})({z}) \,
  (Q_{\frac{7}{2}}^{\la,12}+Q_{\frac{7}{2}}^{\la,21})({w})  = 
  \nonu \\
  && \frac{1}{({z}-{w})^6}\, \Bigg[
-  \frac{512}{5}  (\lambda -2) (\lambda -1) (\lambda +1) (2 \lambda -3) (2 \lambda +1)
    \Bigg] \,  (Q_{\frac{3}{2}}^{\la,11}+Q_{\frac{3}{2}}^{\la,22})({w})
  \nonu \\
  && +
 \frac{1}{({z}-{w})^5}\, \Bigg[
  - \frac{2}{3}
  \frac{512}{5}  (\lambda -2) (\lambda -1) (\lambda +1) (2 \lambda -3) (2 \lambda +1)  \,  {\pa}\, (Q_{\frac{3}{2}}^{\la,11}+Q_{\frac{3}{2}}^{\la,22})
   \nonu \\
   && +\frac{256}{25} (\lambda -2) (\lambda +1) (2 \lambda -3)
   (2 \lambda +7) \,  (Q_{\frac{5}{2}}^{\la,11}+Q_{\frac{5}{2}}^{\la,22}) \Bigg]({w})
 \nonu \\
 && +
 \frac{1}{({z}-{w})^4}\, \Bigg[
-   \frac{1}{4}
\frac{512}{5}  (\lambda -2) (\lambda -1) (\lambda +1) (2 \lambda -3) (2 \lambda +1)  \,  {\pa}^2\,
(Q_{\frac{3}{2}}^{\la,11}+Q_{\frac{3}{2}}^{\la,22})
   \nonu \\
   && +\frac{3}{5} \,\frac{256}{25} (\lambda -2) (\lambda +1) (2 \lambda -3) (2 \lambda +7)
   \,  {\pa}\,
   (Q_{\frac{5}{2}}^{\la,11}+Q_{\frac{5}{2}}^{\la,22})
   \nonu \\
   &&
+\frac{32}{25} (\lambda -2) \left(4 \lambda ^2-22 \lambda -51\right)
   \, (Q_{\frac{7}{2}}^{\la,11}+Q_{\frac{7}{2}}^{\la,22}) 
   \Bigg]({w})
 \nonu \\
  && +
 \frac{1}{({z}-{w})^3}\, \Bigg[
 -  \frac{1}{15}
 \frac{512}{5}  (\lambda -2) (\lambda -1) (\lambda +1) (2 \lambda -3) (2 \lambda +1)   \,  {\pa}^3\,
(Q_{\frac{3}{2}}^{\la,11}+Q_{\frac{3}{2}}^{\la,22})
   \nonu \\
   && +\frac{1}{5} \,
\frac{256}{25} (\lambda -2) (\lambda +1) (2 \lambda -3) (2 \lambda +7)  \,
     {\pa}^2\,
(Q_{\frac{5}{2}}^{\la,11}+Q_{\frac{5}{2}}^{\la,22})
     \nonu \\
     &&
     +\frac{4}{7} \, \frac{32}{25} (\lambda -2) (4 \lambda ^2-22 \lambda -51)
\, {\pa}\,
(Q_{\frac{7}{2}}^{\la,11}+Q_{\frac{7}{2}}^{\la,22}) 
   \nonu \\
   && - \frac{4}{5}  (2 \lambda ^2+3 \lambda -29)
\,  (Q_{\frac{9}{2}}^{\la,11}+Q_{\frac{9}{2}}^{\la,22}) 
   \Bigg]({w})
 \nonu \\
  && +
 \frac{1}{({z}-{w})^2}\, \Bigg[
 -  \frac{1}{72}  \frac{512}{5}  (\lambda -2) (\lambda -1) (\lambda +1) (2 \lambda -3) (2 \lambda +1) 
 \,  {\pa}^4\,
(Q_{\frac{3}{2}}^{\la,11}+Q_{\frac{3}{2}}^{\la,22})
   \nonu \\
   && +\frac{1}{21} \,
   \frac{256}{25} (\lambda -2) (\lambda +1) (2 \lambda -3) (2 \lambda +7)
   \, {\pa}^3 \,
(Q_{\frac{5}{2}}^{\la,11}+Q_{\frac{5}{2}}^{\la,22})
   \nonu \\
   && +
   \frac{5}{28}\,
  \frac{32}{25} (\lambda -2) (4 \lambda ^2-22 \lambda -51)
  \, {\pa}^2\,
(Q_{\frac{7}{2}}^{\la,11}+Q_{\frac{7}{2}}^{\la,22}) 
   \nonu \\
   && -\frac{5}{9}\,
   \frac{4}{5}  (2 \lambda ^2+3 \lambda -29)\, {\pa}\,
 (Q_{\frac{9}{2}}^{\la,11}+Q_{\frac{9}{2}}^{\la,22}) 
+
\frac{1}{5} (- \lambda +14) \, (Q_{\frac{11}{2}}^{\la,11}+
Q_{\frac{11}{2}}^{\la,22})  
\Bigg]({w})
 \nonu \\
 && +
 \frac{1}{({z}-{w})}\, \Bigg[
-   \frac{1}{420}
\frac{512}{5}  (\lambda -2) (\lambda -1) (\lambda +1) (2 \lambda -3) (2 \lambda +1)    \,  {\pa}^5\,
(Q_{\frac{3}{2}}^{\la,11}+Q_{\frac{3}{2}}^{\la,22})
   \nonu \\
   && +\frac{1}{112} \,
   \frac{256}{25} (\lambda -2) (\lambda +1) (2 \lambda -3) (2 \lambda +7)  \,  {\pa}^4\,
   (Q_{\frac{5}{2}}^{\la,11}+Q_{\frac{5}{2}}^{\la,22})\nonu \\
   && +
   \frac{5}{126}\,
 \frac{32}{25} (\lambda -2) (4 \lambda ^2-22 \lambda -51)
 \, {\pa}^3\,
(Q_{\frac{7}{2}}^{\la,11}+Q_{\frac{7}{2}}^{\la,22}) 
   \nonu \\
   && -\frac{1}{6}\,
 \frac{4}{5}  (2 \lambda ^2+3 \lambda -29)
 \, {\pa}^2 \,
 (Q_{\frac{9}{2}}^{\la,11}+Q_{\frac{9}{2}}^{\la,22}) 
+\frac{6}{11}\,
\frac{1}{5} (- \lambda +14)
\, {\pa}\,
 (Q_{\frac{11}{2}}^{\la,11}+
Q_{\frac{11}{2}}^{\la,22})  
\nonu \\
&& +\frac{1}{4} \, (Q_{\frac{13}{2}}^{\la,11}+
Q_{\frac{13}{2}}^{\la,22}) \Bigg]({w})+\cdots
 \nonu  \\
  & &=  -
 % \frac{1}{({z}-{w})^6}\,
  q_{B,4}^{4,\frac{7}{2}}({\pa}_{{z}},{\pa}_{{w}},\la)
  \Bigg[ \frac{ Q_{\frac{3}{2}}^{\la,\bar{a}a}({w})}{({z}-{w})} \Bigg]
+
  %\frac{1}{({z}-{w})^4}\,
  q_{B,3}^{4,\frac{7}{2}}({\pa}_{{z}},{\pa}_{{w}},\la)
  \Bigg[ \frac{ Q_{\frac{5}{2}}^{\la,\bar{a}a}({w})}{({z}-{w})} \Bigg]
   -  q_{B,2}^{4,\frac{7}{2}}({\pa}_{{z}},{\pa}_{{w}},\la)
  \Bigg[ \frac{ Q_{\frac{7}{2}}^{\la,\bar{a}a}({w})}{({z}-{w})} \Bigg]
  \nonu \\
  && +  q_{B,1}^{4,\frac{7}{2}}({\pa}_{{z}},{\pa}_{{w}},\la)
  \Bigg[ \frac{ Q_{\frac{9}{2}}^{\la,\bar{a}a}({w})}{({z}-{w})} \Bigg]
 -
  q_{B,0}^{4,\frac{7}{2}}({\pa}_{{z}},{\pa}_{{w}},\la)
  \Bigg[ \frac{ Q_{\frac{11}{2}}^{\la,\bar{a}a}({w})}{({z}-{w})} \Bigg]
  +q_{B,-1}^{4,\frac{7}{2}}({\pa}_{{z}},{\pa}_{{w}},\la)
  \Bigg[ \frac{ Q_{\frac{13}{2}}^{\la,\bar{a}a}({w})}{({z}-{w})}
    \Bigg] \nonu \\
  && + \cdots
  \nonu \\
  && =
 \sum_{h=-1}^{4}\, q^h\, (-1)^{h-1}\,
  q_{B,h}^{4,\frac{7}{2}}({\pa}_{{z}},{\pa}_{{w}},\la)
  \, \de^{11} \,
  \Bigg[ \frac{  Q_{\frac{11}{2}-h}^{\la,\bar{a}a}({w})}{
      ({z}-{w})} \Bigg] + \cdots \,.    
\label{ope-seven}
  \eea
  We observe that the second relation of (\ref{WQ})
  can be seen from (\ref{ope-seven}).
  
  %%%%%%%%%%%%%%%%%  
\subsubsection{The OPE between $W_{B,4}^{\la,\hat{A}=1}$ and
 $Q_{\frac{7}{2}}^{\la,\hat{B}=2}$}
%%%%%%%%%%%%%%%%

Similarly, we can calculate the following OPE
for different indices $\hat{A}=1$ and $\hat{B}=2$
  \bea
  && (W_{B,4}^{\la,12}+W_{B,4}^{\la,21})({z}) \,
i\,  (Q_{\frac{7}{2}}^{\la,12}-Q_{\frac{7}{2}}^{\la,21})({w})  = 
  \nonu \\
  && \frac{1}{({z}-{w})^6}\, \Bigg[
  \frac{512}{5}  (\lambda -2) (\lambda -1) (\lambda +1) (2 \lambda -3) (2 \lambda +1)
    \Bigg] \,  i\, (Q_{\frac{3}{2}}^{\la,11}-Q_{\frac{3}{2}}^{\la,22})({w})
  \nonu \\
  && +
 \frac{1}{({z}-{w})^5}\, \Bigg[
   \frac{2}{3}
  \frac{512}{5}  (\lambda -2) (\lambda -1) (\lambda +1) (2 \lambda -3) (2 \lambda +1)  \, i\, {\pa}\, (Q_{\frac{3}{2}}^{\la,11}-Q_{\frac{3}{2}}^{\la,22})
   \nonu \\
   && -\frac{256}{25} (\lambda -2) (\lambda +1) (2 \lambda -3)
   (2 \lambda +7) \,  i\,
   (Q_{\frac{5}{2}}^{\la,11}-Q_{\frac{5}{2}}^{\la,22}) \Bigg]({w})
 \nonu \\
 && +
 \frac{1}{({z}-{w})^4}\, \Bigg[
   \frac{1}{4}
\frac{512}{5}  (\lambda -2) (\lambda -1) (\lambda +1) (2 \lambda -3) (2 \lambda +1)  \,  i\, {\pa}^2\,
(Q_{\frac{3}{2}}^{\la,11}-Q_{\frac{3}{2}}^{\la,22})
   \nonu \\
   && -\frac{3}{5} \,\frac{256}{25} (\lambda -2) (\lambda +1) (2 \lambda -3) (2 \lambda +7)
   \,  i\, {\pa}\,
   (Q_{\frac{5}{2}}^{\la,11}-Q_{\frac{5}{2}}^{\la,22})
   \nonu \\
   &&
-\frac{32}{25} (\lambda -2) \left(4 \lambda ^2-22 \lambda -51\right)
   \, i\, (Q_{\frac{7}{2}}^{\la,11}-Q_{\frac{7}{2}}^{\la,22}) 
   \Bigg]({w})
 \nonu \\
  && +
 \frac{1}{({z}-{w})^3}\, \Bigg[
   \frac{1}{15}
 \frac{512}{5}  (\lambda -2) (\lambda -1) (\lambda +1) (2 \lambda -3) (2 \lambda +1)   \,  i\, {\pa}^3\,
(Q_{\frac{3}{2}}^{\la,11}-Q_{\frac{3}{2}}^{\la,22})
   \nonu \\
   && -\frac{1}{5} \,
\frac{256}{25} (\lambda -2) (\lambda +1) (2 \lambda -3) (2 \lambda +7)  \,
    i\, {\pa}^2\,
(Q_{\frac{5}{2}}^{\la,11}-Q_{\frac{5}{2}}^{\la,22})
     \nonu \\
     &&
     -\frac{4}{7} \, \frac{32}{25} (\lambda -2) (4 \lambda ^2-22 \lambda -51)
\, i \, {\pa}\,
(Q_{\frac{7}{2}}^{\la,11}-Q_{\frac{7}{2}}^{\la,22}) 
   \nonu \\
   && + \frac{4}{5}  (2 \lambda ^2+3 \lambda -29)
\,  i\, (Q_{\frac{9}{2}}^{\la,11}-Q_{\frac{9}{2}}^{\la,22}) 
   \Bigg]({w})
 \nonu \\
  && +
 \frac{1}{({z}-{w})^2}\, \Bigg[
   \frac{1}{72}  \frac{512}{5}  (\lambda -2) (\lambda -1) (\lambda +1) (2 \lambda -3) (2 \lambda +1) 
 \, i\,  {\pa}^4\,
(Q_{\frac{3}{2}}^{\la,11}-Q_{\frac{3}{2}}^{\la,22})
   \nonu \\
   && -\frac{1}{21} \,
   \frac{256}{25} (\lambda -2) (\lambda +1) (2 \lambda -3) (2 \lambda +7)
   \, i\, {\pa}^3 \,
(Q_{\frac{5}{2}}^{\la,11}-Q_{\frac{5}{2}}^{\la,22})
   \nonu \\
   && -
   \frac{5}{28}\,
  \frac{32}{25} (\lambda -2) (4 \lambda ^2-22 \lambda -51)
  \, i\, {\pa}^2\,
(Q_{\frac{7}{2}}^{\la,11}-Q_{\frac{7}{2}}^{\la,22}) 
   \nonu \\
   && +\frac{5}{9}\,
   \frac{4}{5}  (2 \lambda ^2+3 \lambda -29)\, i\, {\pa}\,
 (Q_{\frac{9}{2}}^{\la,11}-Q_{\frac{9}{2}}^{\la,22}) 
-
\frac{1}{5} (- \lambda +14) \, i\, (Q_{\frac{11}{2}}^{\la,11}-
Q_{\frac{11}{2}}^{\la,22})  
\Bigg]({w})
 \nonu \\
 && +
 \frac{1}{({z}-{w})}\, \Bigg[
   \frac{1}{420}
\frac{512}{5}  (\lambda -2) (\lambda -1) (\lambda +1) (2 \lambda -3) (2 \lambda +1)    \, i\, {\pa}^5\,
(Q_{\frac{3}{2}}^{\la,11}-Q_{\frac{3}{2}}^{\la,22})
   \nonu \\
   && -\frac{1}{112} \,
   \frac{256}{25} (\lambda -2) (\lambda +1) (2 \lambda -3) (2 \lambda +7)
   i\, \,  {\pa}^4\,
   (Q_{\frac{5}{2}}^{\la,11}-Q_{\frac{5}{2}}^{\la,22})\nonu \\
   && -
   \frac{5}{126}\,
 \frac{32}{25} (\lambda -2) (4 \lambda ^2-22 \lambda -51)
 \, i\, {\pa}^3\,
(Q_{\frac{7}{2}}^{\la,11}-Q_{\frac{7}{2}}^{\la,22}) 
   \nonu \\
   && +\frac{1}{6}\,
 \frac{4}{5}  (2 \lambda ^2+3 \lambda -29)
 \, i\, {\pa}^2 \,
 (Q_{\frac{9}{2}}^{\la,11}-Q_{\frac{9}{2}}^{\la,22}) 
-\frac{6}{11}\,
\frac{1}{5} (- \lambda +14)
\, i\, {\pa}\,
 (Q_{\frac{11}{2}}^{\la,11}-
Q_{\frac{11}{2}}^{\la,22})  
\nonu \\
&& -\frac{1}{4} \, i\, (Q_{\frac{13}{2}}^{\la,11}-
Q_{\frac{13}{2}}^{\la,22}) \Bigg]({w})+\cdots
 \nonu  \\
  & &=  
 % \frac{1}{({z}-{w})^6}\,
  q_{B,4}^{4,\frac{7}{2}}({\pa}_{{z}},{\pa}_{{w}},\la)
  \Bigg[ \frac{ i\, (Q_{\frac{3}{2}}^{\la,11}-
      Q_{\frac{3}{2}}^{\la,22})({w})}{({z}-{w})} \Bigg]
-
  %\frac{1}{({z}-{w})^4}\,
  q_{B,3}^{4,\frac{7}{2}}({\pa}_{{z}},{\pa}_{{w}},\la)
  \Bigg[ \frac{ i\, (Q_{\frac{5}{2}}^{\la,11}-
      Q_{\frac{5}{2}}^{\la,22})({w})}{({z}-{w})} \Bigg]
  \nonu \\
  && +  q_{B,2}^{4,\frac{7}{2}}({\pa}_{{z}},{\pa}_{{w}},\la)
   \Bigg[ \frac{ i\,
 (Q_{\frac{7}{2}}^{\la,11}-
      Q_{\frac{7}{2}}^{\la,22})({w})}{({z}-{w})} \Bigg]
  -  q_{B,1}^{4,\frac{7}{2}}({\pa}_{{z}},{\pa}_{{w}},\la)
  \Bigg[ \frac{i\, (Q_{\frac{9}{2}}^{\la,11}-
      Q_{\frac{9}{2}}^{\la,22})({w})}{({z}-{w})} \Bigg]
  \nonu \\
  && +
  q_{B,0}^{4,\frac{7}{2}}({\pa}_{{z}},{\pa}_{{w}},\la)
  \Bigg[ \frac{ i\, (Q_{\frac{11}{2}}^{\la,11}-
      Q_{\frac{11}{2}}^{\la,22})({w})}{({z}-{w})} \Bigg]
  -q_{B,-1}^{4,\frac{7}{2}}({\pa}_{{z}},{\pa}_{{w}},\la)
  \Bigg[ \frac{ i\, (Q_{\frac{13}{2}}^{\la,11}-
      Q_{\frac{13}{2}}^{\la,22})({w})}{({z}-{w})}
    \Bigg] + \cdots \, 
  \nonu \\
  && =
- \sum_{h=-1}^{4}\, q^h\, (-1)^{h-1}\,
  q_{B,h}^{4,\frac{7}{2}}({\pa}_{{z}},{\pa}_{{w}},\la)
  \, i \, f^{123} \,
  \Bigg[ \frac{  Q_{\frac{11}{2}-h}^{\la,\hat{A}=3}({w})}{
      ({z}-{w})} \Bigg] + \cdots \,.    
\label{ope-eight}
  \eea
 We observe that the first term of second relation of (\ref{WQ})
 can be seen from (\ref{ope-eight}).
 
  %%%%%%%%%%%%%%%%%  
\subsubsection{The OPE between $W_{F,4}^{\la,\hat{A}=1}$ and
 $\bar{Q}_{\frac{7}{2}}^{\la,\hat{B}=1}$}
%%%%%%%%%%%%%%%%

We can consider the third equation of (\ref{WQ})
and the following result can be determined
\bea
&& (W_{F,4}^{\la,12}+W_{F,4}^{\la,21})({z}) \,
(\bar{Q}_{\frac{7}{2}}^{\la,12}+\bar{Q}_{\frac{7}{2}}^{\la,21})({w})  =
\nonu \\
&& \frac{1}{({z}-{w})^7}\,\Bigg[
  -\frac{2048}{5} \, (\lambda -1) \lambda  (\lambda +1) (2 \lambda -3) (2 \lambda -1) (2 \lambda +1)\Bigg]\,
(\bar{Q}_{\frac{1}{2}}^{\la,11}+\bar{Q}_{\frac{1}{2}}^{\la,22})({w})
\nonu \\ 
&&+  \frac{1}{({z}-{w})^6}\, \Bigg[
  -\frac{2048}{5} \, (\lambda -1) \lambda  (\lambda +1) (2 \lambda -3) (2 \lambda -1) (2 \lambda +1)\,  {\pa}\,
(\bar{Q}_{\frac{1}{2}}^{\la,11}+\bar{Q}_{\frac{1}{2}}^{\la,22})({w})
  \nonu \\
  && +\frac{256}{5} (\lambda -1) (\lambda +1) (2 \lambda -3) (2 \lambda +1) (2 \lambda +3)\,
 (\bar{Q}_{\frac{3}{2}}^{\la,11}+\bar{Q}_{\frac{3}{2}}^{\la,22})
    \Bigg]({w})
  \nonu \\
  && +
  \frac{1}{({z}-{w})^5}\, \Bigg[
    -\frac{1}{2} \,
    \frac{2048}{5} \, (\lambda -1) \lambda  (\lambda +1) (2 \lambda -3) (2 \lambda -1) (2 \lambda +1)\,  {\pa}^2 \,
(\bar{Q}_{\frac{1}{2}}^{\la,11}+\bar{Q}_{\frac{1}{2}}^{\la,22})({w})
    \nonu \\
 &&+  \frac{2}{3}
    \frac{256}{5} (\lambda -1) (\lambda +1) (2 \lambda -3) (2 \lambda +1) (2 \lambda +3)  \,  {\pa}\,
(\bar{Q}_{\frac{3}{2}}^{\la,11}+\bar{Q}_{\frac{3}{2}}^{\la,22})
   \nonu \\
   && +\frac{256}{25} (\lambda -4) (\lambda +1) (2 \lambda -3)
   (2 \lambda +3)\,  (\bar{Q}_{\frac{5}{2}}^{\la,11}+
   \bar{Q}_{\frac{5}{2}}^{\la,22}) \Bigg]({w})
 \nonu \\
 && +
 \frac{1}{({z}-{w})^4}\, \Bigg[
 -\frac{1}{6} \,
 \frac{2048}{5} \, (\lambda -1) \lambda  (\lambda +1) (2 \lambda -3) (2 \lambda -1) (2 \lambda +1)\,  {\pa}^3 \,
(\bar{Q}_{\frac{1}{2}}^{\la,11}+\bar{Q}_{\frac{1}{2}}^{\la,22})({w})
 \nonu \\
 &&  \frac{1}{4}
\frac{256}{5} (\lambda -1) (\lambda +1) (2 \lambda -3) (2 \lambda +1) (2 \lambda +3)
\,  {\pa}^2\,
(\bar{Q}_{\frac{3}{2}}^{\la,11}+\bar{Q}_{\frac{3}{2}}^{\la,22})
   \nonu \\
   && +\frac{3}{5} \,
   \frac{256}{25} (\lambda -4) (\lambda +1) (2 \lambda -3)
   (2 \lambda +3)\,  {\pa}\,
 (\bar{Q}_{\frac{5}{2}}^{\la,11}+
   \bar{Q}_{\frac{5}{2}}^{\la,22})\nonu \\
   && -
   \frac{16}{25} (2 \lambda +3) (4 \lambda ^2+18 \lambda -61)
   \, (\bar{Q}_{\frac{7}{2}}^{\la,11}+\bar{Q}_{\frac{7}{2}}^{\la,22}) 
   \Bigg]({w})
 \nonu \\
  && +
 \frac{1}{({z}-{w})^3}\, \Bigg[
   -  \frac{1}{24}\,
   \frac{2048}{5} \, (\lambda -1) \lambda  (\lambda +1) (2 \lambda -3) (2 \lambda -1) (2 \lambda +1)\,  {\pa}^4 \,
(\bar{Q}_{\frac{1}{2}}^{\la,11}+\bar{Q}_{\frac{1}{2}}^{\la,22})({w})
   \nonu \\
   && +
   \frac{1}{15}
\frac{256}{5} (\lambda -1) (\lambda +1) (2 \lambda -3) (2 \lambda +1) (2 \lambda +3)
\,  {\pa}^3\,
(\bar{Q}_{\frac{3}{2}}^{\la,11}+\bar{Q}_{\frac{3}{2}}^{\la,22})
   \nonu \\
   && +\frac{1}{5} \,
  \frac{256}{25} (\lambda -4) (\lambda +1) (2 \lambda -3)
   (2 \lambda +3)
  \,  {\pa}^2\,
 (\bar{Q}_{\frac{5}{2}}^{\la,11}+
  \bar{Q}_{\frac{5}{2}}^{\la,22}) \nonu \\
  && -
   \frac{4}{7}\,
  \frac{16}{25} (2 \lambda +3) (4 \lambda ^2+18 \lambda -61)  
  \, {\pa}\,
 (\bar{Q}_{\frac{7}{2}}^{\la,11}+\bar{Q}_{\frac{7}{2}}^{\la,22})
   \nonu \\
   && -
   \frac{4}{5} (2 \lambda ^2-5 \lambda -27)\,
   (\bar{Q}_{\frac{9}{2}}^{\la,11}+\bar{Q}_{\frac{9}{2}}^{\la,22}) 
   \Bigg]({w})
 \nonu \\
  && +
 \frac{1}{({z}-{w})^2}\, \Bigg[
   - \frac{1}{120} \,
   \frac{2048}{5} \, (\lambda -1) \lambda  (\lambda +1) (2 \lambda -3) (2 \lambda -1) (2 \lambda +1)\,  {\pa}^5 \,
(\bar{Q}_{\frac{1}{2}}^{\la,11}+\bar{Q}_{\frac{1}{2}}^{\la,22})({w})
   \nonu \\
   && +
   \frac{1}{72}
\frac{256}{5} (\lambda -1) (\lambda +1) (2 \lambda -3) (2 \lambda +1) (2 \lambda +3)
\,  {\pa}^4\,
(\bar{Q}_{\frac{3}{2}}^{\la,11}+\bar{Q}_{\frac{3}{2}}^{\la,22})
   \nonu \\
   && +\frac{1}{21} \,
 \frac{256}{25} (\lambda -4) (\lambda +1) (2 \lambda -3)
   (2 \lambda +3)
 \,  {\pa}^3\,
 (\bar{Q}_{\frac{5}{2}}^{\la,11}+
 \bar{Q}_{\frac{5}{2}}^{\la,22})\nonu \\
 && -
   \frac{5}{28}\,
   \frac{16}{25} (2 \lambda +3) (4 \lambda ^2+18 \lambda -61)
   \, {\pa}^2\,
 (\bar{Q}_{\frac{7}{2}}^{\la,11}+\bar{Q}_{\frac{7}{2}}^{\la,22})
   \nonu \\
   && -\frac{5}{9}\,
   \frac{4}{5} (2 \lambda ^2-5 \lambda -27)\, {\pa}\,
 (\bar{Q}_{\frac{9}{2}}^{\la,11}+\bar{Q}_{\frac{9}{2}}^{\la,22}) 
+
\frac{1}{10} (2 \lambda +27) \, (\bar{Q}_{\frac{11}{2}}^{\la,11}+
\bar{Q}_{\frac{11}{2}}^{\la,22})  
\Bigg]({w})
 \nonu \\
 && +
 \frac{1}{({z}-{w})}\, \Bigg[
   -\frac{1}{720}\,
   \frac{2048}{5} \, (\lambda -1) \lambda  (\lambda +1) (2 \lambda -3) (2 \lambda -1) (2 \lambda +1)\,  {\pa}^6 \,
(\bar{Q}_{\frac{1}{2}}^{\la,11}+\bar{Q}_{\frac{1}{2}}^{\la,22})({w})
   \nonu \\
   &&+
   \frac{1}{420}
\frac{256}{5} (\lambda -1) (\lambda +1) (2 \lambda -3) (2 \lambda +1) (2 \lambda +3)
\,  {\pa}^5\,
(\bar{Q}_{\frac{3}{2}}^{\la,11}+\bar{Q}_{\frac{3}{2}}^{\la,22})
   \nonu \\
   && +\frac{1}{112} \,
 \frac{256}{25} (\lambda -4) (\lambda +1) (2 \lambda -3)
   (2 \lambda +3)
 \,  {\pa}^4\,
 (\bar{Q}_{\frac{5}{2}}^{\la,11}+
   \bar{Q}_{\frac{5}{2}}^{\la,22})
   \nonu \\
   && -
   \frac{5}{126}\,
   \frac{16}{25} (2 \lambda +3) (4 \lambda ^2+18 \lambda -61)
   \, {\pa}^3\,
 (\bar{Q}_{\frac{7}{2}}^{\la,11}+\bar{Q}_{\frac{7}{2}}^{\la,22})
   \nonu \\
   && -\frac{1}{6}\,
   \frac{4}{5} (2 \lambda ^2-5 \lambda -27)\, {\pa}^2\,
 (\bar{Q}_{\frac{9}{2}}^{\la,11}+\bar{Q}_{\frac{9}{2}}^{\la,22}) 
+\frac{6}{11}\,
\frac{1}{10} (2 \lambda +27) \, {\pa}\,
(\bar{Q}_{\frac{11}{2}}^{\la,11}+
\bar{Q}_{\frac{11}{2}}^{\la,22})   
\nonu \\
&& +\frac{1}{4} \, (\bar{Q}_{\frac{13}{2}}^{\la,11}+
\bar{Q}_{\frac{13}{2}}^{\la,22}) \Bigg]({w})+\cdots
 \nonu  \\
 & &=
 -
  q_{F,5}^{4,\frac{7}{2}}({\pa}_{{z}},{\pa}_{{w}},\la)
  \Bigg[ \frac{ \bar{Q}_{\frac{1}{2}}^{\la,\bar{a}a}({w})}{({z}-{w})} \Bigg]
 -
 % \frac{1}{({z}-{w})^6}\,
  q_{F,4}^{4,\frac{7}{2}}({\pa}_{{z}},{\pa}_{{w}},\la)
  \Bigg[ \frac{ \bar{Q}_{\frac{3}{2}}^{\la,\bar{a}a}({w})}{({z}-{w})} \Bigg]
-
  %\frac{1}{({z}-{w})^4}\,
  q_{F,3}^{4,\frac{7}{2}}({\pa}_{{z}},{\pa}_{{w}},\la)
  \Bigg[ \frac{ \bar{Q}_{\frac{5}{2}}^{\la,\bar{a}a}({w})}{({z}-{w})} \Bigg]
  \nonu \\
  && -  q_{F,2}^{4,\frac{7}{2}}({\pa}_{{z}},{\pa}_{{w}},\la)
  \Bigg[ \frac{ \bar{Q}_{\frac{7}{2}}^{\la,\bar{a}a}({w})}{({z}-{w})} \Bigg]
 -  q_{F,1}^{4,\frac{7}{2}}({\pa}_{{z}},{\pa}_{{w}},\la)
  \Bigg[ \frac{ \bar{Q}_{\frac{9}{2}}^{\la,\bar{a}a}({w})}{({z}-{w})} \Bigg]
 -
  q_{F,0}^{4,\frac{7}{2}}({\pa}_{{z}},{\pa}_{{w}},\la)
  \Bigg[ \frac{ \bar{Q}_{\frac{11}{2}}^{\la,\bar{a}a}({w})}{({z}-{w})}
    \Bigg] \nonu \\
  &&
  -q_{F,-1}^{4,\frac{7}{2}}({\pa}_{{z}},{\pa}_{{w}},\la)
  \Bigg[ \frac{ \bar{Q}_{\frac{13}{2}}^{\la,\bar{a}a}({w})}{({z}-{w})} \Bigg] + \cdots
  \nonu \\
    && =- \sum_{h=-1}^{5}\, q^h \,
  q_{F,h}^{4,\frac{7}{2}}({\pa}_{{z}},{\pa}_{{w}},\la)
  \, \de^{11} \,
  \Bigg[ \frac{  \bar{Q}_{\frac{11}{2}-h}^{\la,\bar{a}a}({w})}{
      ({z}-{w})} \Bigg] + \cdots \,.    
\label{ope-nine}
  \eea
  There appears the $\la$ factor in
  the coefficients appearing in the current
  $ \bar{Q}_{\frac{1}{2}}^{\la,\bar{a}a}({w})$ and its descendants.
  
    %%%%%%%%%%%%%%%%%  
\subsubsection{The OPE between $W_{F,4}^{\la,\hat{A}=1}$ and
 $\bar{Q}_{\frac{7}{2}}^{\la,\hat{B}=2}$}
%%%%%%%%%%%%%%%%

For different indices, we obtain the 
following result
\bea
&& (W_{F,4}^{\la,12}+W_{F,4}^{\la,21})({z}) \,
(-i)\,(\bar{Q}_{\frac{7}{2}}^{\la,12}-\bar{Q}_{\frac{7}{2}}^{\la,21})({w})  =
\nonu \\
&& \frac{1}{({z}-{w})^7}\,\Bigg[
  -\frac{2048}{5} \, (\lambda -1) \lambda  (\lambda +1) (2 \lambda -3) (2 \lambda -1) (2 \lambda +1)\Bigg]\,
(-i)\,(\bar{Q}_{\frac{1}{2}}^{\la,11}-\bar{Q}_{\frac{1}{2}}^{\la,22})({w})
\nonu \\ 
&&+  \frac{1}{({z}-{w})^6}\, \Bigg[
  -\frac{2048}{5}  (\lambda -1) \lambda  (\lambda +1) (2 \lambda -3) (2 \lambda -1) (2 \lambda +1)  (-i) {\pa}
(\bar{Q}_{\frac{1}{2}}^{\la,11}-\bar{Q}_{\frac{1}{2}}^{\la,22})({w})
  \nonu \\
  && +\frac{256}{5} (\lambda -1) (\lambda +1) (2 \lambda -3) (2 \lambda +1) (2 \lambda +3) (-i)
 (\bar{Q}_{\frac{3}{2}}^{\la,11}-\bar{Q}_{\frac{3}{2}}^{\la,22})
    \Bigg]({w})
  \nonu \\
  && +
  \frac{1}{({z}-{w})^5}\, \Bigg[
    -\frac{1}{2} 
    \frac{2048}{5}  (\lambda -1) \lambda  (\lambda +1) (2 \lambda -3) (2 \lambda -1) (2 \lambda +1)  (-i) {\pa}^2 
(\bar{Q}_{\frac{1}{2}}^{\la,11}-\bar{Q}_{\frac{1}{2}}^{\la,22})({w})
    \nonu \\
 &&+  \frac{2}{3}
    \frac{256}{5} (\lambda -1) (\lambda +1) (2 \lambda -3) (2 \lambda +1) (2 \lambda +3)  \,  (-i)\, {\pa}\,
(\bar{Q}_{\frac{3}{2}}^{\la,11}-\bar{Q}_{\frac{3}{2}}^{\la,22})
   \nonu \\
   && +\frac{256}{25} (\lambda -4) (\lambda +1) (2 \lambda -3)
   (2 \lambda +3)\,  (-i) \, (\bar{Q}_{\frac{5}{2}}^{\la,11}-
   \bar{Q}_{\frac{5}{2}}^{\la,22}) \Bigg]({w})
 \nonu \\
 && +
 \frac{1}{({z}-{w})^4}\, \Bigg[
 -\frac{1}{6} 
 \frac{2048}{5}  (\lambda -1) \lambda  (\lambda +1) (2 \lambda -3) (2 \lambda -1) (2 \lambda +1)  (-i) {\pa}^3 
(\bar{Q}_{\frac{1}{2}}^{\la,11}-\bar{Q}_{\frac{1}{2}}^{\la,22})({w})
 \nonu \\
 &&  \frac{1}{4}
\frac{256}{5} (\lambda -1) (\lambda +1) (2 \lambda -3) (2 \lambda +1) (2 \lambda +3) (-i)\,
\,  {\pa}^2\,
(\bar{Q}_{\frac{3}{2}}^{\la,11}-\bar{Q}_{\frac{3}{2}}^{\la,22})
   \nonu \\
   && +\frac{3}{5} \,
   \frac{256}{25} (\lambda -4) (\lambda +1) (2 \lambda -3)
   (2 \lambda +3)\,  (-i) \, {\pa}\,
 (\bar{Q}_{\frac{5}{2}}^{\la,11}-
   \bar{Q}_{\frac{5}{2}}^{\la,22})\nonu \\
   && -
   \frac{16}{25} (2 \lambda +3) (4 \lambda ^2+18 \lambda -61)
   \, (-i)\,
   (\bar{Q}_{\frac{7}{2}}^{\la,11}-\bar{Q}_{\frac{7}{2}}^{\la,22}) 
   \Bigg]({w})
 \nonu \\
  && +
 \frac{1}{({z}-{w})^3}\, \Bigg[
   -  \frac{1}{24}
   \frac{2048}{5}  (\lambda -1) \lambda  (\lambda +1) (2 \lambda -3) (2 \lambda -1) (2 \lambda +1)  (-i)  {\pa}^4 
(\bar{Q}_{\frac{1}{2}}^{\la,11}-\bar{Q}_{\frac{1}{2}}^{\la,22})({w})
   \nonu \\
   && +
   \frac{1}{15}
\frac{256}{5} (\lambda -1) (\lambda +1) (2 \lambda -3) (2 \lambda +1) (2 \lambda +3)
\,  (-i) \, {\pa}^3\,
(\bar{Q}_{\frac{3}{2}}^{\la,11}-\bar{Q}_{\frac{3}{2}}^{\la,22})
   \nonu \\
   && +\frac{1}{5} \,
  \frac{256}{25} (\lambda -4) (\lambda +1) (2 \lambda -3)
   (2 \lambda +3)
  \,  (-i) \, {\pa}^2\,
 (\bar{Q}_{\frac{5}{2}}^{\la,11}-
  \bar{Q}_{\frac{5}{2}}^{\la,22})\nonu \\
  && -
   \frac{4}{7}\,
  \frac{16}{25} (2 \lambda +3) (4 \lambda ^2+18 \lambda -61)  
  \, (-i) \, {\pa}\,
 (\bar{Q}_{\frac{7}{2}}^{\la,11}-\bar{Q}_{\frac{7}{2}}^{\la,22})
   \nonu \\
   && -
   \frac{4}{5} (2 \lambda ^2-5 \lambda -27)\,
   (-i)\, (\bar{Q}_{\frac{9}{2}}^{\la,11}-\bar{Q}_{\frac{9}{2}}^{\la,22}) 
   \Bigg]({w})
 \nonu \\
  && +
 \frac{1}{({z}-{w})^2}\, \Bigg[
   - \frac{1}{120} 
   \frac{2048}{5}  (\lambda -1) \lambda  (\lambda +1) (2 \lambda -3) (2 \lambda -1) (2 \lambda +1)  \nonu \\
   && \times (-i)  {\pa}^5 
(\bar{Q}_{\frac{1}{2}}^{\la,11}-\bar{Q}_{\frac{1}{2}}^{\la,22})({w})
   \nonu \\
   && +
   \frac{1}{72}
\frac{256}{5} (\lambda -1) (\lambda +1) (2 \lambda -3) (2 \lambda +1) (2 \lambda +3)
\,  (-i) \, {\pa}^4\,
(\bar{Q}_{\frac{3}{2}}^{\la,11}-\bar{Q}_{\frac{3}{2}}^{\la,22})
   \nonu \\
   && +\frac{1}{21} \,
 \frac{256}{25} (\lambda -4) (\lambda +1) (2 \lambda -3)
   (2 \lambda +3)
 \,  (-i) \, {\pa}^3\,
 (\bar{Q}_{\frac{5}{2}}^{\la,11}-
 \bar{Q}_{\frac{5}{2}}^{\la,22}) \nonu \\
 && -
   \frac{5}{28}\,
   \frac{16}{25} (2 \lambda +3) (4 \lambda ^2+18 \lambda -61)
   \, (-i) \, {\pa}^2\,
 (\bar{Q}_{\frac{7}{2}}^{\la,11}-\bar{Q}_{\frac{7}{2}}^{\la,22})
   \nonu \\
   && -\frac{5}{9}\,
   \frac{4}{5} (2 \lambda ^2-5 \lambda -27)\, (-i)\, {\pa}\,
 (\bar{Q}_{\frac{9}{2}}^{\la,11}-\bar{Q}_{\frac{9}{2}}^{\la,22}) 
+
\frac{1}{10} (2 \lambda +27) \, (-i)\, (\bar{Q}_{\frac{11}{2}}^{\la,11}-
\bar{Q}_{\frac{11}{2}}^{\la,22})  
\Bigg]({w})
 \nonu \\
 && +
 \frac{1}{({z}-{w})}\, \Bigg[
   -\frac{1}{720}
   \frac{2048}{5}  (\lambda -1) \lambda  (\lambda +1) (2 \lambda -3) (2 \lambda -1) (2 \lambda +1) \nonu \\
   && \times (-i) {\pa}^6 
(\bar{Q}_{\frac{1}{2}}^{\la,11}-\bar{Q}_{\frac{1}{2}}^{\la,22})({w})
   \nonu \\
   &&+
   \frac{1}{420}
\frac{256}{5} (\lambda -1) (\lambda +1) (2 \lambda -3) (2 \lambda +1) (2 \lambda +3)
\,  (-i)\, {\pa}^5\,
(\bar{Q}_{\frac{3}{2}}^{\la,11}-\bar{Q}_{\frac{3}{2}}^{\la,22})
   \nonu \\
   && +\frac{1}{112} \,
 \frac{256}{25} (\lambda -4) (\lambda +1) (2 \lambda -3)
   (2 \lambda +3)
 \,  (-i) \, {\pa}^4\,
 (\bar{Q}_{\frac{5}{2}}^{\la,11}-
   \bar{Q}_{\frac{5}{2}}^{\la,22})
   \nonu \\
   && -
   \frac{5}{126}\,
   \frac{16}{25} (2 \lambda +3) (4 \lambda ^2+18 \lambda -61)
   \, (-i) \, {\pa}^3\,
 (\bar{Q}_{\frac{7}{2}}^{\la,11}-\bar{Q}_{\frac{7}{2}}^{\la,22})
   \nonu \\
   && -\frac{1}{6}\,
   \frac{4}{5} (2 \lambda ^2-5 \lambda -27)\, (-i) \, {\pa}^2\,
 (\bar{Q}_{\frac{9}{2}}^{\la,11}-\bar{Q}_{\frac{9}{2}}^{\la,22}) 
+\frac{6}{11}\,
\frac{1}{10} (2 \lambda +27) \, (-i) \, {\pa}\,
(\bar{Q}_{\frac{11}{2}}^{\la,11}-
\bar{Q}_{\frac{11}{2}}^{\la,22})   
\nonu \\
&& +\frac{1}{4} \, (-i) \, (\bar{Q}_{\frac{13}{2}}^{\la,11}-
\bar{Q}_{\frac{13}{2}}^{\la,22}) \Bigg]({w})+\cdots
 \nonu  \\
 & &=
 -
  q_{F,5}^{4,\frac{7}{2}}({\pa}_{{z}},{\pa}_{{w}},\la)
  \Bigg[ \frac{
      (-i) (\bar{Q}_{\frac{1}{2}}^{\la,11}-
      \bar{Q}_{\frac{1}{2}}^{\la,22})({w})}{({z}-{w})} \Bigg]
 -
 % \frac{1}{({z}-{w})^6}\,
  q_{F,4}^{4,\frac{7}{2}}({\pa}_{{z}},{\pa}_{{w}},\la)
  \Bigg[ \frac{  (-i) (\bar{Q}_{\frac{3}{2}}^{\la,11}-
      \bar{Q}_{\frac{3}{2}}^{\la,22})({w})}{({z}-{w})} \Bigg]
  \nonu \\
  && -
  %\frac{1}{({z}-{w})^4}\,
  q_{F,3}^{4,\frac{7}{2}}({\pa}_{{z}},{\pa}_{{w}},\la)
  \Bigg[ \frac{ (-i)\, (\bar{Q}_{\frac{5}{2}}^{\la,11}-\bar{Q}_{\frac{5}{2}}^{\la,22})({w})}{({z}-{w})} \Bigg]
 -  q_{F,2}^{4,\frac{7}{2}}({\pa}_{{z}},{\pa}_{{w}},\la)
  \Bigg[ \frac{ (-i)\, (\bar{Q}_{\frac{7}{2}}^{\la,11}-\bar{Q}_{\frac{7}{2}}^{\la,22})({w})}{({z}-{w})} \Bigg]
  \nonu \\
  && -  q_{F,1}^{4,\frac{7}{2}}({\pa}_{{z}},{\pa}_{{w}},\la)
  \Bigg[ \frac{ (-i)\, (\bar{Q}_{\frac{9}{2}}^{\la,11}-\bar{Q}_{\frac{9}{2}}^{\la,22})({w})}{({z}-{w})} \Bigg]
 -
  q_{F,0}^{4,\frac{7}{2}}({\pa}_{{z}},{\pa}_{{w}},\la)
  \Bigg[ \frac{ (-i)\, (\bar{Q}_{\frac{11}{2}}^{\la,11}-\bar{Q}_{\frac{11}{2}}^{\la,22})({w})}{({z}-{w})}
    \Bigg] \nonu \\
  &&-
  q_{F,-1}^{4,\frac{7}{2}}({\pa}_{{z}},{\pa}_{{w}},\la)
  \Bigg[ \frac{ (-i)\, (\bar{Q}_{\frac{13}{2}}^{\la,11}-\bar{Q}_{\frac{13}{2}}^{\la,22})({w})}{({z}-{w})} \Bigg] + \cdots
  \nonu \\
    && =\sum_{h=-1}^{5}\, q^h \,
  q_{F,h}^{4,\frac{7}{2}}({\pa}_{{z}},{\pa}_{{w}},\la)
  \, i \, f^{123}\, 
  \Bigg[ \frac{  \bar{Q}_{\frac{11}{2}-h}^{\la,\hat{A}=3}({w})}{
      ({z}-{w})} \Bigg] + \cdots \,.
\label{ope-ten}
  \eea
  The $\la$ factor appears in the
  structure constant appearing in the current
  $ (\bar{Q}_{\frac{1}{2}}^{\la,11}-
      \bar{Q}_{\frac{1}{2}}^{\la,22})({w})$ and its descendants.
  
   %%%%%%%%%%%%%%%%%  
\subsubsection{The OPE between $W_{B,4}^{\la,\hat{A}=1}$ and
 $\bar{Q}_{\frac{7}{2}}^{\la,\hat{B}=1}$}
%%%%%%%%%%%%%%%%

By considering the last equation of (\ref{WQ}),
the following result is obtained
for the same indices
  \bea
  && (W_{B,4}^{\la,12}+W_{B,4}^{\la,21})({z}) \,
  (\bar{Q}_{\frac{7}{2}}^{\la,12}+\bar{Q}_{\frac{7}{2}}^{\la,21})({w})  =
  \nonu \\
  && \frac{1}{({z}-{w})^7}\,\Bigg[
  \frac{2048}{5} \, (\lambda -1) \lambda  (\lambda +1) (2 \lambda -3) (2 \lambda -1) (2 \lambda +1)\Bigg]\,
(\bar{Q}_{\frac{1}{2}}^{\la,11}+\bar{Q}_{\frac{1}{2}}^{\la,22})({w})
\nonu \\ 
&&+  \frac{1}{({z}-{w})^6}\, \Bigg[
  \frac{2048}{5} \, (\lambda -1) \lambda  (\lambda +1) (2 \lambda -3) (2 \lambda -1) (2 \lambda +1)\,  {\pa}\,
(\bar{Q}_{\frac{1}{2}}^{\la,11}+\bar{Q}_{\frac{1}{2}}^{\la,22})
  \nonu \\
  &&-\frac{512}{5} \,
  (\lambda -2) (\lambda -1) (\lambda +1) (2 \lambda -3) (2 \lambda +1)\,
 (\bar{Q}_{\frac{3}{2}}^{\la,11} + \bar{Q}_{\frac{3}{2}}^{\la,22})
    \Bigg]({w})
  \nonu \\
  && +
  \frac{1}{({z}-{w})^5}\, \Bigg[
    \frac{1}{2} \,
    \frac{2048}{5} \, (\lambda -1) \lambda  (\lambda +1) (2 \lambda -3) (2 \lambda -1) (2 \lambda +1)\,  {\pa}^2 \,
(\bar{Q}_{\frac{1}{2}}^{\la,11}+\bar{Q}_{\frac{1}{2}}^{\la,22})
    \nonu \\
 &&-  \frac{2}{3}
 \frac{512}{5} \,
 (\lambda -2) (\lambda -1) (\lambda +1) (2 \lambda -3) (2 \lambda +1) \,  {\pa}\,
 (\bar{Q}_{\frac{3}{2}}^{\la,11} + \bar{Q}_{\frac{3}{2}}^{\la,22})
   \nonu \\
   &&-\frac{256}{25}\, (\lambda -2) (\lambda +1) (2 \lambda -3) (2 \lambda +7)\,  (\bar{Q}_{\frac{5}{2}}^{\la,11}+\bar{Q}_{\frac{5}{2}}^{\la,22}) \Bigg]({w})
 \nonu \\
 && +
 \frac{1}{({z}-{w})^4}\, \Bigg[
 \frac{1}{6} \,
 \frac{2048}{5} \, (\lambda -1) \lambda  (\lambda +1) (2 \lambda -3) (2 \lambda -1) (2 \lambda +1)\,  {\pa}^3 \,
(\bar{Q}_{\frac{1}{2}}^{\la,11}+\bar{Q}_{\frac{1}{2}}^{\la,22})
 \nonu \\
 &&  -\frac{1}{4}
\frac{512}{5} \,
  (\lambda -2) (\lambda -1) (\lambda +1) (2 \lambda -3) (2 \lambda +1)
\,  {\pa}^2\,
 (\bar{Q}_{\frac{3}{2}}^{\la,11} + \bar{Q}_{\frac{3}{2}}^{\la,22})
   \nonu \\
   && -\frac{3}{5} \,
   \frac{256}{25}\, (\lambda -2) (\lambda +1) (2 \lambda -3) (2 \lambda +7) \,  {\pa}\,
 (\bar{Q}_{\frac{5}{2}}^{\la,11}+\bar{Q}_{\frac{5}{2}}^{\la,22})
   \nonu \\
   && +\frac{32}{25} (\lambda -2) \left(4 \lambda ^2-22 \lambda -51\right)
   \, (\bar{Q}_{\frac{7}{2}}^{\la,11}+ \bar{Q}_{\frac{7}{2}}^{\la,22})
   \Bigg]({w})
 \nonu \\
  && +
 \frac{1}{({z}-{w})^3}\, \Bigg[
     \frac{1}{24}\,
     \frac{2048}{5} \, (\lambda -1) \lambda  (\lambda +1) (2 \lambda -3) (2 \lambda -1) (2 \lambda +1)\,  {\pa}^4 \,
(\bar{Q}_{\frac{1}{2}}^{\la,11}+\bar{Q}_{\frac{1}{2}}^{\la,22})
   \nonu \\
   && -
   \frac{1}{15}
\frac{512}{5} \,
  (\lambda -2) (\lambda -1) (\lambda +1) (2 \lambda -3) (2 \lambda +1)
\,  {\pa}^3\,
 (\bar{Q}_{\frac{3}{2}}^{\la,11} + \bar{Q}_{\frac{3}{2}}^{\la,22})
   \nonu \\
   && -\frac{1}{5} \,
  \frac{256}{25}\, (\lambda -2) (\lambda +1) (2 \lambda -3) (2 \lambda +7)
  \,  {\pa}^2\,
 (\bar{Q}_{\frac{5}{2}}^{\la,11}+\bar{Q}_{\frac{5}{2}}^{\la,22})
  \nonu \\
  && +
   \frac{4}{7}\,
 \frac{32}{25} (\lambda -2) (4 \lambda ^2-22 \lambda -51)
 \, {\pa}\,
(\bar{Q}_{\frac{7}{2}}^{\la,11}+ \bar{Q}_{\frac{7}{2}}^{\la,22})
   \nonu \\
   && +
\frac{4}{5} (2 \lambda ^2+3 \lambda -29)
   \,  (\bar{Q}_{\frac{9}{2}}^{\la,11}+\bar{Q}_{\frac{9}{2}}^{\la,22}) 
   \Bigg]({w})
 \nonu \\
  && +
 \frac{1}{({z}-{w})^2}\, \Bigg[
    \frac{1}{120} \,
    \frac{2048}{5} \, (\lambda -1) \lambda  (\lambda +1) (2 \lambda -3) (2 \lambda -1) (2 \lambda +1)\,  {\pa}^5 \,
(\bar{Q}_{\frac{1}{2}}^{\la,11}+\bar{Q}_{\frac{1}{2}}^{\la,22})
   \nonu \\
   && -
   \frac{1}{72}
\frac{512}{5} \,
  (\lambda -2) (\lambda -1) (\lambda +1) (2 \lambda -3) (2 \lambda +1)
\,  {\pa}^4\,
 (\bar{Q}_{\frac{3}{2}}^{\la,11} + \bar{Q}_{\frac{3}{2}}^{\la,22})
   \nonu \\
   && -\frac{1}{21} \,
  \frac{256}{25}\, (\lambda -2) (\lambda +1) (2 \lambda -3) (2 \lambda +7) 
  \,  {\pa}^3\,
 (\bar{Q}_{\frac{5}{2}}^{\la,11}+\bar{Q}_{\frac{5}{2}}^{\la,22})
  \nonu \\
  && +
   \frac{5}{28}\,
\frac{32}{25} (\lambda -2) (4 \lambda ^2-22 \lambda -51)
\, {\pa}^2\,
(\bar{Q}_{\frac{7}{2}}^{\la,11}+ \bar{Q}_{\frac{7}{2}}^{\la,22})
   \nonu \\
   && +\frac{5}{9}\,
\frac{4}{5} (2 \lambda ^2+3 \lambda -29)
\, {\pa}\,
 (\bar{Q}_{\frac{9}{2}}^{\la,11}+\bar{Q}_{\frac{9}{2}}^{\la,22}) 
+
\frac{1}{5} (- \lambda +14) \, (\bar{Q}_{\frac{11}{2}}^{\la,11}+
\bar{Q}_{\frac{11}{2}}^{\la,22})
\Bigg]({w})
 \nonu \\
 && +
 \frac{1}{({z}-{w})}\, \Bigg[
   \frac{1}{720}\,
   \frac{2048}{5} \, (\lambda -1) \lambda  (\lambda +1) (2 \lambda -3) (2 \lambda -1) (2 \lambda +1)\,  {\pa}^6 \,
(\bar{Q}_{\frac{1}{2}}^{\la,11}+\bar{Q}_{\frac{1}{2}}^{\la,22})
   \nonu \\
   &&-
   \frac{1}{420}
\frac{512}{5} \,
  (\lambda -2) (\lambda -1) (\lambda +1) (2 \lambda -3) (2 \lambda +1)
\,  {\pa}^5\,
 (\bar{Q}_{\frac{3}{2}}^{\la,11} + \bar{Q}_{\frac{3}{2}}^{\la,22})
   \nonu \\
   && -\frac{1}{112} \,
 \frac{256}{25}\, (\lambda -2) (\lambda +1) (2 \lambda -3) (2 \lambda +7) 
 \,  {\pa}^4\,
 (\bar{Q}_{\frac{5}{2}}^{\la,11}+\bar{Q}_{\frac{5}{2}}^{\la,22})
 \nonu \\
 && +
   \frac{5}{126}\,
\frac{32}{25} (\lambda -2) (4 \lambda ^2-22 \lambda -51)
\, {\pa}^3\,
(\bar{Q}_{\frac{7}{2}}^{\la,11}+ \bar{Q}_{\frac{7}{2}}^{\la,22})
   \nonu \\
   && +\frac{1}{6}\,
\frac{4}{5} (2 \lambda ^2+3 \lambda -29)
\, {\pa}^2\,
 (\bar{Q}_{\frac{9}{2}}^{\la,11}+\bar{Q}_{\frac{9}{2}}^{\la,22})  
+\frac{6}{11}\,
\frac{1}{5} (- \lambda +14) 
\, {\pa}\,
(\bar{Q}_{\frac{11}{2}}^{\la,11}+
\bar{Q}_{\frac{11}{2}}^{\la,22})\nonu \\
&& 
-\frac{1}{4} \, (\bar{Q}_{\frac{13}{2}}^{\la,11}+
\bar{Q}_{\frac{13}{2}}^{\la,22}) \Bigg]({w})+\cdots
 \nonu  \\
 & &=
 -
  q_{B,5}^{4,\frac{7}{2}}({\pa}_{{z}},{\pa}_{{w}},\la)
  \Bigg[ \frac{ (\bar{Q}_{\frac{1}{2}}^{\la,11}+\bar{Q}_{\frac{1}{2}}^{\la,22})({w})}{({z}-{w})} \Bigg]
 -
 % \frac{1}{({z}-{w})^6}\,
  q_{B,4}^{4,\frac{7}{2}}({\pa}_{{z}},{\pa}_{{w}},\la)
  \Bigg[ \frac{(\bar{Q}_{\frac{3}{2}}^{\la,11}+\bar{Q}_{\frac{3}{2}}^{\la,22})({w})}{({z}-{w})} \Bigg]
  \nonu \\
  && -
  %\frac{1}{({z}-{w})^4}\,
  q_{B,3}^{4,\frac{7}{2}}({\pa}_{{z}},{\pa}_{{w}},\la)
  \Bigg[ \frac{(\bar{Q}_{\frac{5}{2}}^{\la,11}+\bar{Q}_{\frac{5}{2}}^{\la,22})({w})}{({z}-{w})} \Bigg]
 -  q_{B,2}^{4,\frac{7}{2}}({\pa}_{{z}},{\pa}_{{w}},\la)
  \Bigg[ \frac{(\bar{Q}_{\frac{7}{2}}^{\la,11}+\bar{Q}_{\frac{7}{2}}^{\la,22})({w})}{({z}-{w})} \Bigg]
  \nonu \\
  && -  q_{B,1}^{4,\frac{7}{2}}({\pa}_{{z}},{\pa}_{{w}},\la)
  \Bigg[ \frac{(\bar{Q}_{\frac{9}{2}}^{\la,11}+\bar{Q}_{\frac{9}{2}}^{\la,22})({w})}{({z}-{w})} \Bigg]
 -
  q_{B,0}^{4,\frac{7}{2}}({\pa}_{{z}},{\pa}_{{w}},\la)
  \Bigg[ \frac{(\bar{Q}_{\frac{11}{2}}^{\la,11}+\bar{Q}_{\frac{11}{2}}^{\la,22})({w})}{({z}-{w})}
    \Bigg] \nonu \\
  &&
  -q_{B,-1}^{4,\frac{7}{2}}({\pa}_{{z}},{\pa}_{{w}},\la)
  \Bigg[ \frac{(\bar{Q}_{\frac{13}{2}}^{\la,11}+\bar{Q}_{\frac{13}{2}}^{\la,22})({w})}{({z}-{w})} \Bigg] + \cdots\, 
  \nonu \\
   && =
- \sum_{h=-1}^{5}\, q^h\,
  q_{B,h}^{4,\frac{7}{2}}({\pa}_{{z}},{\pa}_{{w}},\la)
   \, \de^{11}\,
  \Bigg[ \frac{  \bar{Q}_{\frac{11}{2}-h}^{\la,\bar{a}a}({w})}{
      ({z}-{w})} \Bigg] + \cdots \,.  
  \label{ope-eleven}
  \eea
  There appears the $\la$ factor in
  the coefficients of the current
$ \bar{Q}_{\frac{1}{2}}^{\la,\bar{a}a}({w})$ and its descendants.

   %%%%%%%%%%%%%%%%%  
\subsubsection{The OPE between $W_{B,4}^{\la,\hat{A}=1}$ and
 $\bar{Q}_{\frac{7}{2}}^{\la,\hat{B}=2}$}
%%%%%%%%%%%%%%%%

For different indices we obtain the following result
  \bea
  && (W_{B,4}^{\la,12}+W_{B,4}^{\la,21})({z}) \,
  (-i)\, (\bar{Q}_{\frac{7}{2}}^{\la,12}-\bar{Q}_{\frac{7}{2}}^{\la,21})({w})  =
  \nonu \\
  && \frac{1}{({z}-{w})^7}\,\Bigg[
  -\frac{2048}{5} \, (\lambda -1) \lambda  (\lambda +1) (2 \lambda -3) (2 \lambda -1) (2 \lambda +1)\Bigg]\,  (-i)\,
(\bar{Q}_{\frac{1}{2}}^{\la,11}-\bar{Q}_{\frac{1}{2}}^{\la,22})({w})
\nonu \\ 
&&+  \frac{1}{({z}-{w})^6}\, \Bigg[
  -\frac{2048}{5} \, (\lambda -1) \lambda  (\lambda +1) (2 \lambda -3) (2 \lambda -1) (2 \lambda +1)\,  (-i)\, {\pa}\,
(\bar{Q}_{\frac{1}{2}}^{\la,11}-\bar{Q}_{\frac{1}{2}}^{\la,22})
  \nonu \\
  &&+\frac{512}{5} \,
  (\lambda -2) (\lambda -1) (\lambda +1) (2 \lambda -3) (2 \lambda +1)\,
 (-i)\, (\bar{Q}_{\frac{3}{2}}^{\la,11} - \bar{Q}_{\frac{3}{2}}^{\la,22})
    \Bigg]({w})
  \nonu \\
  && +
  \frac{1}{({z}-{w})^5}\, \Bigg[
   - \frac{1}{2} \,
    \frac{2048}{5} \, (\lambda -1) \lambda  (\lambda +1) (2 \lambda -3) (2 \lambda -1) (2 \lambda +1)\,  (-i)\, {\pa}^2 \,
(\bar{Q}_{\frac{1}{2}}^{\la,11}-\bar{Q}_{\frac{1}{2}}^{\la,22})
    \nonu \\
 &&+  \frac{2}{3}
 \frac{512}{5} \,
 (\lambda -2) (\lambda -1) (\lambda +1) (2 \lambda -3) (2 \lambda +1)
 \, (-i)\,  {\pa}\,
 (\bar{Q}_{\frac{3}{2}}^{\la,11} - \bar{Q}_{\frac{3}{2}}^{\la,22})
   \nonu \\
   &&+\frac{256}{25}\, (\lambda -2) (\lambda +1) (2 \lambda -3) (2 \lambda +7)\,  (-i)\, (\bar{Q}_{\frac{5}{2}}^{\la,11}-
   \bar{Q}_{\frac{5}{2}}^{\la,22}) \Bigg]({w})
 \nonu \\
 && +
 \frac{1}{({z}-{w})^4}\, \Bigg[
 -\frac{1}{6} \,
 \frac{2048}{5} \, (\lambda -1) \lambda  (\lambda +1) (2 \lambda -3) (2 \lambda -1) (2 \lambda +1)\, (-i)\,  {\pa}^3 \,
(\bar{Q}_{\frac{1}{2}}^{\la,11}-\bar{Q}_{\frac{1}{2}}^{\la,22})
 \nonu \\
 &&  +\frac{1}{4}
\frac{512}{5} \,
  (\lambda -2) (\lambda -1) (\lambda +1) (2 \lambda -3) (2 \lambda +1)
\,   (-i)\,{\pa}^2\,
 (\bar{Q}_{\frac{3}{2}}^{\la,11} - \bar{Q}_{\frac{3}{2}}^{\la,22})
   \nonu \\
   && +\frac{3}{5} \,
   \frac{256}{25}\, (\lambda -2) (\lambda +1) (2 \lambda -3) (2 \lambda +7) \,  (-i)\, {\pa}\,
 (\bar{Q}_{\frac{5}{2}}^{\la,11}-\bar{Q}_{\frac{5}{2}}^{\la,22})
   \nonu \\
   && -\frac{32}{25} (\lambda -2) \left(4 \lambda ^2-22 \lambda -51\right)
   \,  (-i)\,(\bar{Q}_{\frac{7}{2}}^{\la,11}- \bar{Q}_{\frac{7}{2}}^{\la,22})
   \Bigg]({w})
 \nonu \\
  && +
 \frac{1}{({z}-{w})^3}\, \Bigg[
     -\frac{1}{24}\,
     \frac{2048}{5} \, (\lambda -1) \lambda  (\lambda +1) (2 \lambda -3) (2 \lambda -1) (2 \lambda +1)\, (-i)\,  {\pa}^4 \,
(\bar{Q}_{\frac{1}{2}}^{\la,11}-\bar{Q}_{\frac{1}{2}}^{\la,22})
   \nonu \\
   && +
   \frac{1}{15}
\frac{512}{5} \,
  (\lambda -2) (\lambda -1) (\lambda +1) (2 \lambda -3) (2 \lambda +1)
\,  (-i)\, {\pa}^3\,
 (\bar{Q}_{\frac{3}{2}}^{\la,11} - \bar{Q}_{\frac{3}{2}}^{\la,22})
   \nonu \\
   && +\frac{1}{5} \,
  \frac{256}{25}\, (\lambda -2) (\lambda +1) (2 \lambda -3) (2 \lambda +7)
  \, (-i)\,  {\pa}^2\,
 (\bar{Q}_{\frac{5}{2}}^{\la,11}-\bar{Q}_{\frac{5}{2}}^{\la,22})
  \nonu \\
  && -
   \frac{4}{7}\,
 \frac{32}{25} (\lambda -2) (4 \lambda ^2-22 \lambda -51)
 \, (-i)\, {\pa}\,
(\bar{Q}_{\frac{7}{2}}^{\la,11}- \bar{Q}_{\frac{7}{2}}^{\la,22})
   \nonu \\
   && -
\frac{4}{5} (2 \lambda ^2+3 \lambda -29)
   \,  (-i)\, (\bar{Q}_{\frac{9}{2}}^{\la,11}-\bar{Q}_{\frac{9}{2}}^{\la,22}) 
   \Bigg]({w})
 \nonu \\
  && +
 \frac{1}{({z}-{w})^2}\, \Bigg[
    -\frac{1}{120} \,
    \frac{2048}{5} \, (\lambda -1) \lambda  (\lambda +1) (2 \lambda -3) (2 \lambda -1) (2 \lambda +1)\,  (-i)\, {\pa}^5 \,
(\bar{Q}_{\frac{1}{2}}^{\la,11}-\bar{Q}_{\frac{1}{2}}^{\la,22})
   \nonu \\
   && +
   \frac{1}{72}
\frac{512}{5} \,
  (\lambda -2) (\lambda -1) (\lambda +1) (2 \lambda -3) (2 \lambda +1)
\,  (-i)\, {\pa}^4\,
 (\bar{Q}_{\frac{3}{2}}^{\la,11} - \bar{Q}_{\frac{3}{2}}^{\la,22})
   \nonu \\
   && +\frac{1}{21} \,
  \frac{256}{25}\, (\lambda -2) (\lambda +1) (2 \lambda -3) (2 \lambda +7) 
  \, (-i)\,  {\pa}^3\,
 (\bar{Q}_{\frac{5}{2}}^{\la,11}-\bar{Q}_{\frac{5}{2}}^{\la,22})
  \nonu \\
  && -
   \frac{5}{28}\,
\frac{32}{25} (\lambda -2) (4 \lambda ^2-22 \lambda -51)
\, (-i)\, {\pa}^2\,
(\bar{Q}_{\frac{7}{2}}^{\la,11}- \bar{Q}_{\frac{7}{2}}^{\la,22})
   \nonu \\
   && -\frac{5}{9}\,
\frac{4}{5} (2 \lambda ^2+3 \lambda -29)
\, (-i)\, {\pa}\,
 (\bar{Q}_{\frac{9}{2}}^{\la,11}-\bar{Q}_{\frac{9}{2}}^{\la,22}) 
-
\frac{1}{5} (- \lambda +14) \, (-i)\, (\bar{Q}_{\frac{11}{2}}^{\la,11}-
\bar{Q}_{\frac{11}{2}}^{\la,22})
\Bigg]({w})
 \nonu \\
 && +
 \frac{1}{({z}-{w})}\, \Bigg[
  - \frac{1}{720}\,
   \frac{2048}{5} \, (\lambda -1) \lambda  (\lambda +1) (2 \lambda -3) (2 \lambda -1) (2 \lambda +1)\,  (-i)\, {\pa}^6 \,
(\bar{Q}_{\frac{1}{2}}^{\la,11}-\bar{Q}_{\frac{1}{2}}^{\la,22})
   \nonu \\
   &&+
   \frac{1}{420}
\frac{512}{5} \,
  (\lambda -2) (\lambda -1) (\lambda +1) (2 \lambda -3) (2 \lambda +1)
\,  (-i)\, {\pa}^5\,
 (\bar{Q}_{\frac{3}{2}}^{\la,11} - \bar{Q}_{\frac{3}{2}}^{\la,22})
   \nonu \\
   && +\frac{1}{112} \,
 \frac{256}{25}\, (\lambda -2) (\lambda +1) (2 \lambda -3) (2 \lambda +7) 
 \,  (-i)\, {\pa}^4\,
 (\bar{Q}_{\frac{5}{2}}^{\la,11}-\bar{Q}_{\frac{5}{2}}^{\la,22})
 \nonu \\
 && -
   \frac{5}{126}\,
\frac{32}{25} (\lambda -2) (4 \lambda ^2-22 \lambda -51)
\, (-i)\, {\pa}^3\,
(\bar{Q}_{\frac{7}{2}}^{\la,11}- \bar{Q}_{\frac{7}{2}}^{\la,22})
   \nonu \\
   && -\frac{1}{6}\,
\frac{4}{5} (2 \lambda ^2+3 \lambda -29)
\, (-i)\, {\pa}^2\,
 (\bar{Q}_{\frac{9}{2}}^{\la,11}-\bar{Q}_{\frac{9}{2}}^{\la,22})  
-\frac{6}{11}\,
\frac{1}{5} (- \lambda +14) 
\, (-i)\, {\pa}\,
(\bar{Q}_{\frac{11}{2}}^{\la,11}-
\bar{Q}_{\frac{11}{2}}^{\la,22})\nonu \\
&& 
+\frac{1}{4} \, (-i)\, (\bar{Q}_{\frac{13}{2}}^{\la,11}-
\bar{Q}_{\frac{13}{2}}^{\la,22}) \Bigg]({w})+\cdots
 \nonu  \\
 & &=
  q_{B,5}^{4,\frac{7}{2}}({\pa}_{{z}},{\pa}_{{w}},\la)
  \Bigg[ \frac{ (-i)\, (\bar{Q}_{\frac{1}{2}}^{\la,11}-\bar{Q}_{\frac{1}{2}}^{\la,22})({w})}{({z}-{w})} \Bigg]
 +
 % \frac{1}{({z}-{w})^6}\,
  q_{B,4}^{4,\frac{7}{2}}({\pa}_{{z}},{\pa}_{{w}},\la)
  \Bigg[ \frac{ (-i)\,(\bar{Q}_{\frac{3}{2}}^{\la,11}-\bar{Q}_{\frac{3}{2}}^{\la,22})({w})}{({z}-{w})} \Bigg]
  \nonu \\
  && +
  %\frac{1}{({z}-{w})^4}\,
  q_{B,3}^{4,\frac{7}{2}}({\pa}_{{z}},{\pa}_{{w}},\la)
  \Bigg[ \frac{ (-i)\,(\bar{Q}_{\frac{5}{2}}^{\la,11}-\bar{Q}_{\frac{5}{2}}^{\la,22})({w})}{({z}-{w})} \Bigg]
 +  q_{B,2}^{4,\frac{7}{2}}({\pa}_{{z}},{\pa}_{{w}},\la)
  \Bigg[ \frac{ (-i)\,(\bar{Q}_{\frac{7}{2}}^{\la,11}-\bar{Q}_{\frac{7}{2}}^{\la,22})({w})}{({z}-{w})} \Bigg]
  \nonu \\
  && +  q_{B,1}^{4,\frac{7}{2}}({\pa}_{{z}},{\pa}_{{w}},\la)
  \Bigg[ \frac{ (-i)\,(\bar{Q}_{\frac{9}{2}}^{\la,11}-\bar{Q}_{\frac{9}{2}}^{\la,22})({w})}{({z}-{w})} \Bigg]
 +
  q_{B,0}^{4,\frac{7}{2}}({\pa}_{{z}},{\pa}_{{w}},\la)
  \Bigg[ \frac{ (-i)\,(\bar{Q}_{\frac{11}{2}}^{\la,11}-\bar{Q}_{\frac{11}{2}}^{\la,22})({w})}{({z}-{w})}
    \Bigg] \nonu \\
  &&
  +q_{B,-1}^{4,\frac{7}{2}}({\pa}_{{z}},{\pa}_{{w}},\la)
  \Bigg[ \frac{ (-i)\,(\bar{Q}_{\frac{13}{2}}^{\la,11}-\bar{Q}_{\frac{13}{2}}^{\la,22})({w})}{({z}-{w})} \Bigg] + \cdots\, 
  \nonu \\
   && =-
\sum_{h=-1}^{5}\, q^h\,
  q_{B,h}^{4,\frac{7}{2}}({\pa}_{{z}},{\pa}_{{w}},\la)
   \, i \, f^{123}\,
  \Bigg[ \frac{  \bar{Q}_{\frac{11}{2}-h}^{\la,\hat{A}=3}({w})}{
      ({z}-{w})} \Bigg] + \cdots \,.  
  \label{ope-twelve}
  \eea
  Similarly,
  we observe that
  the last equation of (\ref{WQ})
  can be seen from the above (\ref{ope-twelve}).
  
   %%%%%%%%%%%%%%%%%  
\subsection{The OPE between the fermionic currents}
%%%%%%%%%%%%%%%%

We check (\ref{Final}) with the weights $h_1=h_2=3$.

  %%%%%%%%%%%%%%%%%  
\subsubsection{The OPE between $Q_{\frac{7}{2}}^{\la,\hat{A}=1}$ and
 $\bar{Q}_{\frac{7}{2}}^{\la,\hat{B}=1}$}
%%%%%%%%%%%%%%%%

When the indices $\hat{A}$ and $\hat{B}$ ($\hat{A}=\hat{B}=1$)
are the same, 
then we obtain the following expression 
\bea
&& (Q_{\frac{7}{2}}^{\la,12}+Q_{\frac{7}{2}}^{\la,21})({z}) \,
(\bar{Q}_{\frac{7}{2}}^{\la,12}+\bar{Q}_{\frac{7}{2}}^{\la,21})({w})=
\nonu \\
&& \frac{1}{({z}-{w})^7} \Bigg[
 - \frac{6144}{5}  (4 \lambda -1)
  (12 \lambda ^4-12 \lambda ^3-13 \lambda ^2+8 \lambda +3)
  \Bigg]
\nonu \\
&& +
\frac{1}{({z}-{w})^6}\, \Bigg[
  \frac{4096}{5} (\lambda -1) (\lambda +1) (2 \lambda -3)
  (2 \lambda +1) (2\la-1)\,
 W_{F,1}^{\la,\bar{a}a} \nonu \\
 && +
 \frac{8192}{5} (\lambda -1) \la
 (\lambda +1) (2 \lambda -3) (2 \lambda +1)\,
 W_{B,1}^{\la,\bar{a}a}
  \Bigg]({w})
\nonu \\
&& +
\frac{1}{({z}-{w})^5}\, \Bigg[
 \frac{1}{2}\, \frac{4096}{5} (\lambda -1) (\lambda +1) (2 \lambda -3)
  (2 \lambda +1) (2\la-1)\,
 {\pa} \, W_{F,1}^{\la,\bar{a}a} \nonu \\
 && + \frac{1}{2} \,
 \frac{8192}{5} (\lambda -1) \la
 (\lambda +1) (2 \lambda -3) (2 \lambda +1)\,
 {\pa} \,  W_{B,1}^{\la,\bar{a}a} \nonu \\
 && +
\frac{1024}{25} (\lambda -1) (\lambda +1) (2 \lambda -3) (2 \lambda +17)
 \, W_{F,2}^{\la,\bar{a}a}
 \nonu \\
 && + \frac{1024}{25} (\lambda -9) (\lambda +1) (2 \lambda -3) (2 \lambda +1)
 \,  W_{B,2}^{\la,\bar{a}a}
  \Bigg]({w})
\nonu \\
&& +
\frac{1}{({z}-{w})^4}\, \Bigg[
 \frac{1}{6}\, \frac{4096}{5} (\lambda -1) (\lambda +1) (2 \lambda -3)
  (2 \lambda +1) (2\la-1)\,
 {\pa}^2 \, W_{F,1}^{\la,\bar{a}a} \nonu \\
 && + \frac{1}{6} \,
 \frac{8192}{5} (\lambda -1) \la
 (\lambda +1) (2 \lambda -3) (2 \lambda +1)\,
 {\pa}^2 \,  W_{B,1}^{\la,\bar{a}a} \nonu \\
 && + \frac{1}{2}\,
\frac{1024}{25} (\lambda -1) (\lambda +1) (2 \lambda -3) (2 \lambda +17)
 \,
 {\pa} \, W_{F,2}^{\la,\bar{a}a}
 \nonu \\
 && +\frac{1}{2}\,
 \frac{1024}{25} (\lambda -9) (\lambda +1) (2 \lambda -3) (2 \lambda +1)
 \,  {\pa} \, W_{B,2}^{\la,\bar{a}a}
 \nonu \\
 &&-\frac{256}{25}  (2 \lambda -3)
 (4 \lambda ^2-6 \lambda -25)\, W_{F,3}^{\la,\bar{a}a}
   - \frac{512}{25}  (\lambda +1)
 (4 \lambda ^2+2 \lambda -27)
 W_{B,3}^{\la,\bar{a}a}
 \Bigg]({w})
\nonu \\
&& +
\frac{1}{({z}-{w})^3}\, \Bigg[
 \frac{1}{24}\, \frac{4096}{5} (\lambda -1) (\lambda +1) (2 \lambda -3)
  (2 \lambda +1) (2\la-1)\,
 {\pa}^3 \, W_{F,1}^{\la,\bar{a}a} \nonu \\
 && + \frac{1}{24} \,
 \frac{8192}{5} (\lambda -1) \la
 (\lambda +1) (2 \lambda -3) (2 \lambda +1)\,
 {\pa}^3 \,  W_{B,1}^{\la,\bar{a}a} \nonu \\
 && + \frac{3}{20}\,
\frac{1024}{25} (\lambda -1) (\lambda +1) (2 \lambda -3) (2 \lambda +17)
 \,
 {\pa}^2 \, W_{F,2}^{\la,\bar{a}a}
 \nonu \\
 && +\frac{3}{20}\,
 \frac{1024}{25} (\lambda -9) (\lambda +1) (2 \lambda -3) (2 \lambda +1)
 \,  {\pa}^2 \, W_{B,2}^{\la,\bar{a}a}
 \nonu \\
 &&-\frac{1}{2}\,
 \frac{256}{25}  (2 \lambda -3) (4 \lambda ^2-6 \lambda -
 25)\, {\pa} \, W_{F,3}^{\la,\bar{a}a}
- \frac{1}{2}\, \frac{512}{25}  (\lambda +1)
 (4 \lambda ^2+2 \lambda -27)
 \, {\pa} \, W_{B,3}^{\la,\bar{a}a} \nonu \\
&&
- \frac{64}{25}  (4 \lambda ^2+18 \lambda -61)\,
 W_{F,4}^{\la,\bar{a}a} 
-\frac{64}{25}  (4 \lambda ^2-22 \lambda -51)\,
  W_{B,4}^{\la,\bar{a}a}
 \Bigg]({w})
\nonu \\
&& +
\frac{1}{({z}-{w})^2}\, \Bigg[
 \frac{1}{120}\, \frac{4096}{5} (\lambda -1) (\lambda +1) (2 \lambda -3)
  (2 \lambda +1) (2\la-1)\,
 {\pa}^4 \, W_{F,1}^{\la,\bar{a}a} \nonu \\
 && + \frac{1}{120} \,
 \frac{8192}{5} (\lambda -1) \la
 (\lambda +1) (2 \lambda -3) (2 \lambda +1)\,
 {\pa}^4 \,  W_{B,1}^{\la,\bar{a}a} \nonu \\
 && + \frac{1}{30}\,
\frac{1024}{25} (\lambda -1) (\lambda +1) (2 \lambda -3) (2 \lambda +17)
 \,
 {\pa}^3 \, W_{F,2}^{\la,\bar{a}a}
 \nonu \\
 && +\frac{1}{30}\,
 \frac{1024}{25} (\lambda -9) (\lambda +1) (2 \lambda -3) (2 \lambda +1)
 \,  {\pa}^3 \, W_{B,2}^{\la,\bar{a}a}
 \nonu \\
 &&-\frac{1}{7}\,
 \frac{256}{25}  (2 \lambda -3) (4 \lambda ^2-6 \lambda -
 25)\, {\pa}^2 \, W_{F,3}^{\la,\bar{a}a}
- \frac{1}{7}\, \frac{512}{25}  (\lambda +1)
 (4 \lambda ^2+2 \lambda -27)
 \, {\pa}^2 \, W_{B,3}^{\la,\bar{a}a} \nonu \\
&&
- \frac{1}{2} \, \frac{64}{25}  (4 \lambda ^2+18 \lambda -61)\,
 {\pa} \, W_{F,4}^{\la,\bar{a}a} 
-\frac{1}{2} \, \frac{64}{25}  (4 \lambda ^2-22 \lambda -51)\,
 {\pa} \,  W_{B,4}^{\la,\bar{a}a}
 \nonu \\
 && + \frac{8}{5} (2 \lambda -13) \,W_{F,5}^{\la,\bar{a}a} 
+\frac{16 (\lambda +6)}{5}\, 
 W_{B,5}^{\la,\bar{a}a}
 \Bigg]({w})
\nonu \\
&& +
\frac{1}{({z}-{w})}\, \Bigg[
 \frac{1}{720}\, \frac{4096}{5} (\lambda -1) (\lambda +1) (2 \lambda -3)
  (2 \lambda +1) (2\la-1)\,
 {\pa}^5 \, W_{F,1}^{\la,\bar{a}a} \nonu \\
 && + \frac{1}{720} \,
 \frac{8192}{5} (\lambda -1) \la
 (\lambda +1) (2 \lambda -3) (2 \lambda +1)\,
 {\pa}^5 \,  W_{B,1}^{\la,\bar{a}a} \nonu \\
 && + \frac{1}{168}\,
\frac{1024}{25} (\lambda -1) (\lambda +1) (2 \lambda -3) (2 \lambda +17)
 \,
 {\pa}^4 \, W_{F,2}^{\la,\bar{a}a}
 \nonu \\
 && +\frac{1}{168}\,
 \frac{1024}{25} (\lambda -9) (\lambda +1) (2 \lambda -3) (2 \lambda +1)
 \,  {\pa}^4 \, W_{B,2}^{\la,\bar{a}a}
 \nonu \\
 &&-\frac{5}{168}\,
 \frac{256}{25}  (2 \lambda -3) (4 \lambda ^2-6 \lambda -
 25)\, {\pa}^3 \, W_{F,3}^{\la,\bar{a}a}
- \frac{5}{168}\, \frac{512}{25}  (\lambda +1)
 (4 \lambda ^2+2 \lambda -27)
 \, {\pa}^3 \, W_{B,3}^{\la,\bar{a}a} \nonu \\
&&
- \frac{5}{36} \, \frac{64}{25}  (4 \lambda ^2+18 \lambda -61)\,
 {\pa}^2 \, W_{F,4}^{\la,\bar{a}a} 
-\frac{5}{36} \, \frac{64}{25}  (4 \lambda ^2-22 \lambda -51)\,
 {\pa}^2 \,  W_{B,4}^{\la,\bar{a}a}
 \nonu \\
 && + \frac{1}{2}\, \frac{8}{5} (2 \lambda -13) \,
 {\pa} \, W_{F,5}^{\la,\bar{a}a} 
+\frac{1}{2} \, \frac{16 (\lambda +6)}{5}\, 
 {\pa} \, W_{B,5}^{\la,\bar{a}a}
 + 2\, W_{F,6}^{\la,\bar{a}a} + 2\,  W_{B,6}^{\la,\bar{a}a}\Bigg]({w})
+\cdots
\nonu \\
&& =
\frac{1}{({z}-{w})^7} \Bigg[
  -\frac{6144}{5}  (4 \lambda -1)
  (12 \lambda ^4-12 \lambda ^3-13 \lambda ^2+8 \lambda +3)
  \Bigg]\nonu \\
&&
-
  o_{F,5}^{\frac{7}{2},\frac{7}{2}}({\pa}_{{z}},{\pa}_{{w}},\la)
  \Bigg[ \frac{
      W_{F,1}^{\la,\bar{a}a}({w})}{({z}-{w})} \Bigg]
  -
  o_{B,5}^{\frac{7}{2},\frac{7}{2}}({\pa}_{{z}},{\pa}_{{w}},\la)
  \Bigg[ \frac{
      W_{B,1}^{\la,\bar{a}a}({w})}{({z}-{w})} \Bigg]
  +o_{F,4}^{\frac{7}{2},\frac{7}{2}}({\pa}_{{z}},{\pa}_{{w}},\la)
  \Bigg[ \frac{
      W_{F,2}^{\la,\bar{a}a}({w})}{({z}-{w})} \Bigg]
  \nonu \\
  && +
  o_{B,4}^{\frac{7}{2},\frac{7}{2}}({\pa}_{{z}},{\pa}_{{w}},\la)
  \Bigg[ \frac{
      W_{B,2}^{\la,\bar{a}a}({w})}{({z}-{w})} \Bigg]
   -o_{F,3}^{\frac{7}{2},\frac{7}{2}}({\pa}_{{z}},{\pa}_{{w}},\la)
  \Bigg[ \frac{
      W_{F,3}^{\la,\bar{a}a}({w})}{({z}-{w})} \Bigg]
  -
  o_{B,3}^{\frac{7}{2},\frac{7}{2}}({\pa}_{{z}},{\pa}_{{w}},\la)
  \Bigg[ \frac{
      W_{B,3}^{\la,\bar{a}a}({w})}{({z}-{w})} \Bigg]
  \nonu \\
&&  +
 o_{F,2}^{\frac{7}{2},\frac{7}{2}}({\pa}_{{z}},{\pa}_{{w}},\la)
  \Bigg[ \frac{
      W_{F,4}^{\la,\bar{a}a}({w})}{({z}-{w})} \Bigg]
  +
  o_{B,2}^{\frac{7}{2},\frac{7}{2}}({\pa}_{{z}},{\pa}_{{w}},\la)
  \Bigg[ \frac{
      W_{B,4}^{\la,\bar{a}a}({w})}{({z}-{w})} \Bigg]
  -
 o_{F,1}^{\frac{7}{2},\frac{7}{2}}({\pa}_{{z}},{\pa}_{{w}},\la)
  \Bigg[ \frac{
      W_{F,5}^{\la,\bar{a}a}({w})}{({z}-{w})} \Bigg]
  \nonu \\
  && -
  o_{B,1}^{\frac{7}{2},\frac{7}{2}}({\pa}_{{z}},{\pa}_{{w}},\la)
  \Bigg[ \frac{
      W_{B,5}^{\la,\bar{a}a}({w})}{({z}-{w})} \Bigg]
   +
 o_{F,0}^{\frac{7}{2},\frac{7}{2}}({\pa}_{{z}},{\pa}_{{w}},\la)
  \Bigg[ \frac{
      W_{F,6}^{\la,\bar{a}a}({w})}{({z}-{w})} \Bigg]
  +
  o_{B,0}^{\frac{7}{2},\frac{7}{2}}({\pa}_{{z}},{\pa}_{{w}},\la)
  \Bigg[ \frac{
      W_{B,6}^{\la,\bar{a}a}({w})}{({z}-{w})} \Bigg]
  \nonu \\
  && + \cdots\, 
  \nonu \\
 && =
\frac{1}{({z}-{w})^7}\,c_Q(3,3,\la) \, \de^{11}\, q^4
  + \sum_{h=0}^{5}\, q^h\, (-1)^h\,
  o_{F,h}^{\frac{7}{2},\frac{7}{2}}({\pa}_{{z}},{\pa}_{{w}},\la)
   \, \de^{11} \,
  \Bigg[ \frac{  W_{F,6-h}^{\la,\bar{a}a}({w})}{
      ({z}-{w})} \Bigg] \nonu \\
  && +
\sum_{h=0}^{5}\, q^h\, (-1)^h\,
  o_{B,h}^{\frac{7}{2},\frac{7}{2}}({\pa}_{{z}},{\pa}_{{w}},\la)
   \, \de^{11} \,
  \Bigg[ \frac{  W_{B,6-h}^{\la,\bar{a}a}({w})}{
      ({z}-{w})} \Bigg]
  +\cdots \,.  
  \label{ope-thirteen}
  \eea
  We can check that the corresponding terms
  in (\ref{Final}) can be seen from the above (\ref{ope-thirteen})
  and note that there is an additional factor $(-1)^h$
  coming from the anticommutator to the OPE.
  
%%%%%%%%%%%%%%%%%  
\subsubsection{The OPE between $Q_{\frac{7}{2}}^{\la,\hat{A}=1}$ and
 $\bar{Q}_{\frac{7}{2}}^{\la,\hat{B}=2}$}
%%%%%%%%%%%%%%%%

By considering the different indices as before,
we determine the following result
\bea
&& (Q_{\frac{7}{2}}^{\la,12}+Q_{\frac{7}{2}}^{\la,21})({z}) \,
(-i)\, (\bar{Q}_{\frac{7}{2}}^{\la,12}-\bar{Q}_{\frac{7}{2}}^{\la,21})({w})=
\nonu \\
&& 
\frac{1}{({z}-{w})^6}\, \Bigg[
  -\frac{4096}{5} (\lambda -1) (\lambda +1) (2 \lambda -3)
  (2 \lambda +1) (2\la-1)\,
 (-i)\, (W_{F,1}^{\la,11}-W_{F,1}^{\la,22}) \nonu \\
 && +
 \frac{8192}{5} (\lambda -1) \la
 (\lambda +1) (2 \lambda -3) (2 \lambda +1)\,
 (-i) \, (W_{B,1}^{\la,11}-W_{B,1}^{\la,22})
  \Bigg]({w})
\nonu \\
&& +
\frac{1}{({z}-{w})^5}\, \Bigg[
 -\frac{1}{2}\, \frac{4096}{5} (\lambda -1) (\lambda +1) (2 \lambda -3)
  (2 \lambda +1) (2\la-1)\,
   (-i)\,   {\pa} \,
 (W_{F,1}^{\la,11}-W_{F,1}^{\la,22}) 
       \nonu \\
 && + \frac{1}{2} \,
 \frac{8192}{5} (\lambda -1) \la
 (\lambda +1) (2 \lambda -3) (2 \lambda +1)\,
     (-i)\,  {\pa} \,
(W_{B,1}^{\la,11}-W_{B,1}^{\la,22})
       \nonu \\
 && -
\frac{1024}{25} (\lambda -1) (\lambda +1) (2 \lambda -3) (2 \lambda +17)
\,(-i)\,
(W_{F,2}^{\la,11}-W_{F,2}^{\la,22})
 \nonu \\
 && + \frac{1024}{25} (\lambda -9) (\lambda +1) (2 \lambda -3) (2 \lambda +1) \,
 (-i) \,  (W_{B,2}^{\la,11}-W_{B,2}^{\la,22})
  \Bigg]({w})
\nonu \\
&& +
\frac{1}{({z}-{w})^4}\, \Bigg[
 -\frac{1}{6}\, \frac{4096}{5} (\lambda -1) (\lambda +1) (2 \lambda -3)
  (2 \lambda +1) (2\la-1)\,
    (-i)\,  {\pa}^2 \,
 (W_{F,1}^{\la,11}-W_{F,1}^{\la,22}) 
       \nonu \\
 && + \frac{1}{6} \,
 \frac{8192}{5} (\lambda -1) \la
 (\lambda +1) (2 \lambda -3) (2 \lambda +1)\,
      (-i)\, {\pa}^2 \,
(W_{B,1}^{\la,11}-W_{B,1}^{\la,22})
       \nonu \\
 && - \frac{1}{2}\,
\frac{1024}{25} (\lambda -1) (\lambda +1) (2 \lambda -3) (2 \lambda +17)
 \, (-i)\,
   {\pa} \,
(W_{F,2}^{\la,11}-W_{F,2}^{\la,22})
 \nonu \\
 && +\frac{1}{2}\,
 \frac{1024}{25} (\lambda -9) (\lambda +1) (2 \lambda -3) (2 \lambda +1)
 \,  (-i) \, {\pa} \,
 (W_{B,2}^{\la,11}-W_{B,2}^{\la,22})
 \nonu \\
 &&+\frac{256}{25}  (2 \lambda -3)
 (4 \lambda ^2-6 \lambda -25)\,
(-i)\,
 (W_{F,3}^{\la,11}-W_{F,3}^{\la,22})
 \nonu \\
 && - \frac{512}{25}  (\lambda +1)
 (4 \lambda ^2+2 \lambda -27)
 (-i) \, (W_{B,3}^{\la,11}-W_{B,3}^{\la,22})
 \Bigg]({w})
\nonu \\
&& +
\frac{1}{({z}-{w})^3}\, \Bigg[
 -\frac{1}{24}\, \frac{4096}{5} (\lambda -1) (\lambda +1) (2 \lambda -3)
  (2 \lambda +1) (2\la-1)\,
   (-i)\,   {\pa}^3 \,
(W_{F,1}^{\la,11}-W_{F,1}^{\la,22}) 
      \nonu \\
 && + \frac{1}{24} \,
 \frac{8192}{5} (\lambda -1) \la
 (\lambda +1) (2 \lambda -3) (2 \lambda +1)\,
   (-i)\,   {\pa}^3 \,
(W_{B,1}^{\la,11}-W_{B,1}^{\la,22})
       \nonu \\
 && - \frac{3}{20}\,
\frac{1024}{25} (\lambda -1) (\lambda +1) (2 \lambda -3) (2 \lambda +17)
 \,(-i)\,
   {\pa}^2 \,
(W_{F,2}^{\la,11}-W_{F,2}^{\la,22})
 \nonu \\
 && +\frac{3}{20}\,
 \frac{1024}{25} (\lambda -9) (\lambda +1) (2 \lambda -3) (2 \lambda +1)
 \,  (-i) \, {\pa}^2 \,
 (W_{B,2}^{\la,11}-W_{B,2}^{\la,22})
 \nonu \\
 &&+\frac{1}{2}\,
 \frac{256}{25}  (2 \lambda -3) (4 \lambda ^2-6 \lambda -
 25)\, (-i) {\pa} \,
(W_{F,3}^{\la,11}-W_{F,3}^{\la,22})
 \nonu \\
 && - \frac{1}{2}\, \frac{512}{25}  (\lambda +1)
 (4 \lambda ^2+2 \lambda -27)
\, (-i) \,{\pa} \,
 (W_{B,3}^{\la,11}-W_{B,3}^{\la,22})
 \nonu \\
&&
+ \frac{64}{25}  (4 \lambda ^2+18 \lambda -61)\,
 (-i)\, (W_{F,4}^{\la,11}-W_{F,4}^{\la,22}) 
\nonu \\
&& -\frac{64}{25}  (4 \lambda ^2-22 \lambda -51)\,
  (-i) \, (W_{B,4}^{\la,11}-W_{B,4}^{\la,22})
 \Bigg]({w})
\nonu \\
&& +
\frac{1}{({z}-{w})^2}\, \Bigg[
 -\frac{1}{120}\, \frac{4096}{5} (\lambda -1) (\lambda +1) (2 \lambda -3)
  (2 \lambda +1) (2\la-1)\,
      (-i) \, {\pa}^4 \,
(W_{F,1}^{\la,11}-W_{F,1}^{\la,22}) 
       \nonu \\
 && + \frac{1}{120} \,
 \frac{8192}{5} (\lambda -1) \la
 (\lambda +1) (2 \lambda -3) (2 \lambda +1)\,
      (-i)\, {\pa}^4 \,
(W_{B,1}^{\la,11}-W_{B,1}^{\la,22})
       \nonu \\
 && - \frac{1}{30}\,
\frac{1024}{25} (\lambda -1) (\lambda +1) (2 \lambda -3) (2 \lambda +17)
 \,
 (-i)\,  {\pa}^3 \,
(W_{F,2}^{\la,11}-W_{F,2}^{\la,22})
 \nonu \\
 && +\frac{1}{30}\,
 \frac{1024}{25} (\lambda -9) (\lambda +1) (2 \lambda -3) (2 \lambda +1)
 \,  (-i) \, {\pa}^3 \,
 (W_{B,2}^{\la,11}-W_{B,2}^{\la,22})
 \nonu \\
 &&+\frac{1}{7}\,
 \frac{256}{25}  (2 \lambda -3) (4 \lambda ^2-6 \lambda -
 25)\, (-i)\, {\pa}^2 \,
(W_{F,3}^{\la,11}-W_{F,3}^{\la,22})
 \nonu \\
 && - \frac{1}{7}\, \frac{512}{25}  (\lambda +1)
 (4 \lambda ^2+2 \lambda -27)
\, (-i) \, {\pa}^2 \,
 (W_{B,3}^{\la,11}-W_{B,3}^{\la,22})
 \nonu \\
&&
+ \frac{1}{2}  \frac{64}{25}  (4 \lambda ^2+18 \lambda -61)
(-i) {\pa} 
(W_{F,4}^{\la,11}-W_{F,4}^{\la,22})
\nonu \\
&& -\frac{1}{2}  \frac{64}{25}  (4 \lambda ^2-22 \lambda -51)
(-i) {\pa} 
(W_{B,4}^{\la,11}-W_{B,4}^{\la,22})
 \nonu \\
 && - \frac{8}{5} (2 \lambda -13) \,
 (-i) \, (W_{F,5}^{\la,11} -W_{F,5}^{\la,22})
+\frac{16 (\lambda +6)}{5}\, 
(-i)\,  (W_{B,5}^{\la,11}-W_{B,5}^{\la,22})
 \Bigg]({w})
\nonu \\
&& +
\frac{1}{({z}-{w})}\, \Bigg[
 -\frac{1}{720}\, \frac{4096}{5} (\lambda -1) (\lambda +1) (2 \lambda -3)
  (2 \lambda +1) (2\la-1)\,
    (-i)\,   {\pa}^5 \,
(W_{F,1}^{\la,11}-W_{F,1}^{\la,22}) 
       \nonu \\
 && + \frac{1}{720} \,
 \frac{8192}{5} (\lambda -1) \la
 (\lambda +1) (2 \lambda -3) (2 \lambda +1)\,
      (-i)\, {\pa}^5 \,
(W_{B,1}^{\la,11}-W_{B,1}^{\la,22})
       \nonu \\
 && - \frac{1}{168}\,
\frac{1024}{25} (\lambda -1) (\lambda +1) (2 \lambda -3) (2 \lambda +17)
 \, (-i)\,
   {\pa}^4 \,
(W_{F,2}^{\la,11}-W_{F,2}^{\la,22})
 \nonu \\
 && +\frac{1}{168}\,
 \frac{1024}{25} (\lambda -9) (\lambda +1) (2 \lambda -3) (2 \lambda +1)
 \,  (-i) \, {\pa}^4 \,
 (W_{B,2}^{\la,11}-W_{B,2}^{\la,22})
 \nonu \\
 &&+\frac{5}{168}\,
 \frac{256}{25}  (2 \lambda -3) (4 \lambda ^2-6 \lambda -
 25)\, (-i) \, {\pa}^3 \,
(W_{F,3}^{\la,11}-W_{F,3}^{\la,22})
 \nonu \\
 && - \frac{5}{168}\, \frac{512}{25}  (\lambda +1)
 (4 \lambda ^2+2 \lambda -27)
\, (-i) \, {\pa}^3 \,
 (W_{B,3}^{\la,11}-W_{B,3}^{\la,22})
 \nonu \\
&&
+ \frac{5}{36} \, \frac{64}{25}  (4 \lambda ^2+18 \lambda -61)\,
(-i) \, {\pa}^2 \,
 (W_{F,4}^{\la,11}-W_{F,4}^{\la,22}) 
\nonu \\
&& -\frac{5}{36} \, \frac{64}{25}  (4 \lambda ^2-22 \lambda -51)\,
(-i) \, {\pa}^2 \,
(W_{B,4}^{\la,11}-W_{B,4}^{\la,22})
 \nonu \\
 && - \frac{1}{2}\, \frac{8}{5} (2 \lambda -13) \,
      (-i) \,  {\pa} \,
 (W_{F,5}^{\la,11} -W_{F,5}^{\la,22})
+\frac{1}{2} \, \frac{16 (\lambda +6)}{5}\, 
(-i) \, {\pa} \,
 (W_{B,5}^{\la,11}-W_{B,5}^{\la,22})
\nonu \\
&& - 2\, (-i)\,
(W_{F,6}^{\la,11}-W_{F,6}^{\la,22}) + 2\, (-i)\,
(W_{B,6}^{\la,11}-W_{B,6}^{\la,22}) \Bigg]({w})
+\cdots
\nonu \\
&& =
\nonu \\
&&
  o_{F,5}^{\frac{7}{2},\frac{7}{2}}({\pa}_{{z}},{\pa}_{{w}},\la)
  \Bigg[ (-i)\, \frac{
      (W_{F,1}^{\la,11}-W_{F,1}^{\la,22})({w})}{({z}-{w})} \Bigg]
  -
  o_{B,5}^{\frac{7}{2},\frac{7}{2}}({\pa}_{{z}},{\pa}_{{w}},\la)
  \Bigg[ (-i)\, \frac{(W_{B,1}^{\la,11}-W_{B,1}^{\la,22})
      ({w})}{({z}-{w})} \Bigg]
  \nonu \\
  && -o_{F,4}^{\frac{7}{2},\frac{7}{2}}({\pa}_{{z}},{\pa}_{{w}},\la)
  \Bigg[(-i)\, \frac{
      (W_{F,2}^{\la,11}-W_{F,2}^{\la,22})
      ({w})}{({z}-{w})} \Bigg]
  +
  o_{B,4}^{\frac{7}{2},\frac{7}{2}}({\pa}_{{z}},{\pa}_{{w}},\la)
  \Bigg[(-i)\, \frac{(W_{B,2}^{\la,11}-W_{B,2}^{\la,22})
      ({w})}{({z}-{w})} \Bigg]
  \nonu \\
  && +o_{F,3}^{\frac{7}{2},\frac{7}{2}}({\pa}_{{z}},{\pa}_{{w}},\la)
  \Bigg[(-i)\, \frac{(W_{F,3}^{\la,11}-W_{F,3}^{\la,22})
      ({w})}{({z}-{w})} \Bigg]
  -
  o_{B,3}^{\frac{7}{2},\frac{7}{2}}({\pa}_{{z}},{\pa}_{{w}},\la)
  \Bigg[ (-i)\, \frac{(W_{B,3}^{\la,11}-W_{B,3}^{\la,22})
      ({w})}{({z}-{w})} \Bigg]
  \nonu \\
&&  -
 o_{F,2}^{\frac{7}{2},\frac{7}{2}}({\pa}_{{z}},{\pa}_{{w}},\la)
  \Bigg[(-i)\, \frac{(W_{F,4}^{\la,11}-W_{F,4}^{\la,22})
      ({w})}{({z}-{w})} \Bigg]
  +
  o_{B,2}^{\frac{7}{2},\frac{7}{2}}({\pa}_{{z}},{\pa}_{{w}},\la)
  \Bigg[ (-i)\, \frac{(W_{B,4}^{\la,11}-W_{B,4}^{\la,22})
      ({w})}{({z}-{w})} \Bigg]
  \nonu \\
  && +
 o_{F,1}^{\frac{7}{2},\frac{7}{2}}({\pa}_{{z}},{\pa}_{{w}},\la)
  \Bigg[ (-i) \, \frac{(W_{F,5}^{\la,11}-W_{F,5}^{\la,22})
      ({w})}{({z}-{w})} \Bigg]
  -
  o_{B,1}^{\frac{7}{2},\frac{7}{2}}({\pa}_{{z}},{\pa}_{{w}},\la)
  \Bigg[(-i)\, \frac{(W_{B,5}^{\la,11}-W_{B,5}^{\la,22})
      ({w})}{({z}-{w})} \Bigg]
  \nonu \\
  && -
 o_{F,0}^{\frac{7}{2},\frac{7}{2}}({\pa}_{{z}},{\pa}_{{w}},\la)
  \Bigg[ (-i)\, \frac{(W_{F,6}^{\la,11}-W_{F,6}^{\la,22})
      ({w})}{({z}-{w})} \Bigg]
  +
  o_{B,0}^{\frac{7}{2},\frac{7}{2}}({\pa}_{{z}},{\pa}_{{w}},\la)
  \Bigg[ (-i)\, \frac{(W_{B,6}^{\la,11}-W_{B,6}^{\la,22})
      ({w})}{({z}-{w})} \Bigg]
  \nonu \\
  && + \cdots\, 
  \nonu \\
 && =
\sum_{h=0}^{5}\, q^h\, (-1)^h\,
  o_{F,h}^{\frac{7}{2},\frac{7}{2}}({\pa}_{{z}},{\pa}_{{w}},\la)
   \, i\, f^{123} \,
  \Bigg[ \frac{  W_{F,6-h}^{\la,\hat{A}=3}({w})}{
      ({z}-{w})} \Bigg] \nonu \\
  && -
\sum_{h=0}^{5}\, q^h\, (-1)^h\,
  o_{B,h}^{\frac{7}{2},\frac{7}{2}}({\pa}_{{z}},{\pa}_{{w}},\la)
   \, i\, f^{123} \,
  \Bigg[ \frac{  W_{B,6-h}^{\la,\hat{A}=3}({w})}{
      ({z}-{w})} \Bigg]
  +\cdots \,.  
  \label{ope-fourteen}
  \eea
  As before, the $\la$ factor appears in the
  structure constants appearing in the
  current $(W_{B,1}^{\la,11}-W_{B,1}^{\la,22})$ and its descendants.
  
%%%%%%%%%%%%%%%%%%%%%%%%%%%%%%%%%%%%%%%%%%%%%%%%%%%%%%%
%%%%%%%%%%%%%%%%%%%%%%%%%%%%%
  \section{The other central terms in the
  OPEs between the fermionic currents }
%%%%%%%%%%%%%%%%%%%%%%%%%%%%%
%%%%%%AAA%%%%%%%%%%%%%%%%%%%%%%%%%%%%%%%%%%%%%%%%%%%%%%%%%  

%%%%%%%%%%%%%%%%%%%%%
%\subsubsection{The central term}
%%%%%%%%%%%%%%%%%%%%%

  As in (\ref{++}) and (\ref{+---}), we can calculate the
  following OPEs for the highest order poles
\bea
&& Q_{\la, \bar{a} b}^{(h_1)+}(z) \,
Q_{\la, \bar{c} d}^{(h_2)+}(w)\Bigg|_{\frac{1}{
    (z-w)^{h_1+h_2-1}}} = N\, \de_{b \bar{c}}\, \de_{d \bar{a}}\,
\Bigg(\sum_{j=0}^{h_1-2}\, \sum_{i=0}^{h_2-1}\, \sum_{t=0}^{j+1}\,
 \beta^j(h_1,\la)\, \al^i(h_2,\la)
\nonu \\
&& \times \frac{ j! \,(t+i)!}{t! \, (t+i+1)!}
\,(-1)^{h_1-1+t}\, (j+1-t)_{h_1-2-j}\, (h_1-1-t)_{t+1+i}\,
(h_1+i)_{h_2-1-i} \,
\nonu \\
&& - \sum_{j=0}^{h_1-1}\, \sum_{i=0}^{h_2-2}\, \sum_{t=0}^{j+1}\,
\al^{j}(h_1,\la)\,
\beta^i(h_2,\la)
\nonu \\
&& \times \frac{ j! \,(t+i)!}{t! \, (t+i+1)!}
\,(-1)^{h_1+t}\, (j+1-t)_{h_1-1-j}\, (h_1-t)_{t+1+i}\,
(h_1+1+i)_{h_2-2-i} \Bigg)\, ,
\nonu \\
%%%%%%%%%%%%%%%%%%%%%%%%%%%%%%%%%%%%%%%%%%%%%%%%%%%%%%%%
&& Q_{\la, \bar{a} b}^{(h_1)+}(z) \,
Q_{\la, \bar{c} d}^{(h_2)-}(w)\Bigg|_{\frac{1}{
    (z-w)^{h_1+h_2-1}}} = N\, \de_{b \bar{c}}\, \de_{d \bar{a}}\,
\Bigg(\sum_{j=0}^{h_1-2}\, \sum_{i=0}^{h_2-1}\, \sum_{t=0}^{j+1}\,
 \beta^j(h_1,\la)\, \al^i(h_2,\la)
\nonu \\
&& \times \frac{ j! \,(t+i)!}{t! \, (t+i+1)!}
\,(-1)^{h_1-1+t}\, (j+1-t)_{h_1-2-j}\, (h_1-1-t)_{t+1+i}\,
(h_1+i)_{h_2-1-i} \,
\nonu \\
&& + \sum_{j=0}^{h_1-1}\, \sum_{i=0}^{h_2-2}\, \sum_{t=0}^{j+1}\,
\al^{j}(h_1,\la)\,
\beta^i(h_2,\la)
\nonu \\
&& \times \frac{ j! \,(t+i)!}{t! \, (t+i+1)!}
\,(-1)^{h_1+t}\, (j+1-t)_{h_1-1-j}\, (h_1-t)_{t+1+i}\,
(h_1+1+i)_{h_2-2-i} \Bigg)\, ,
\nonu \\
%%%%%%%%%%%%%%%%%%%%%%%%%%%%%%%%%%%%%%%%%%%%%%%%%%%%%%%%%%%%%%
&& Q_{\la, \bar{a} b}^{(h_1)-}(z) \,
Q_{\la, \bar{c} d}^{(h_2)-}(w)\Bigg|_{\frac{1}{
    (z-w)^{h_1+h_2-1}}} = N\, \de_{b \bar{c}}\, \de_{d \bar{a}}\,
\Bigg(-\sum_{j=0}^{h_1-2}\, \sum_{i=0}^{h_2-1}\, \sum_{t=0}^{j+1}\,
 \beta^j(h_1,\la)\, \al^i(h_2,\la)
\nonu \\
&& \times \frac{ j! \,(t+i)!}{t! \, (t+i+1)!}
\,(-1)^{h_1-1+t}\, (j+1-t)_{h_1-2-j}\, (h_1-1-t)_{t+1+i}\,
(h_1+i)_{h_2-1-i} \,
\nonu \\
&& + \sum_{j=0}^{h_1-1}\, \sum_{i=0}^{h_2-2}\, \sum_{t=0}^{j+1}\,
\al^{j}(h_1,\la)\,
\beta^i(h_2,\la)
\nonu \\
&& \times \frac{ j! \,(t+i)!}{t! \, (t+i+1)!}
\,(-1)^{h_1+t}\, (j+1-t)_{h_1-1-j}\, (h_1-t)_{t+1+i}\,
(h_1+1+i)_{h_2-2-i} \Bigg)\, .
\label{CEN}
\eea
These results will be used in (\ref{Final}) and (\ref{cQ}).
  
%%%%%%%%%%%%%%%%%%%%%%%%%%%%%%%%%%%%%%%%%%%%%%%%%%%%%%%
%%%%%%%%%%%%%%%%%%%%%%%%%%%%%
\section{The other (anti)commutator relations }
%%%%%%%%%%%%%%%%%%%%%%%%%%%%%
%%%%%%CCC%%%%%%%%%%%%%%%%%%%%%%%%%%%%%%%%%%%%%%%%%%%%%%%%%  

We present the remaining (anti)commutator relations
discussed in section $3$.

%%%%%%%%%%%%%%%%%%
\subsection{ The commutators between the currents consisting of $\beta\,
  \ga$ system  with $h_1=h_2, h_2\pm 1$
for nonzero $\la$}
%%%%%%%%%%%%%%%%%%

As done in section $3$, we present the final result for
the commutators between the currents consisting of $\beta\,
\ga$ system as follows:
\bea
\big[(W^{\la,21}_{\mathrm{B},h_1})_m,(W^{\la,12}_{\mathrm{B},h_2})_n\big] 
\!&=& \!
\frac{1}{2}\, \sum^{h_1+h_2-3}_{h= 0,
\mbox{\footnotesize even}} \, q^h\,
p_{\mathrm{B}}^{h_1,h_2, h}(m,n,\la)
 \, ( W^{\la,11}_{\mathrm{B},h_1+h_2-2-h} + W^{\la,22}_{\mathrm{B},h_1+h_2-2-h} )_{m+n}
\nonu \\
\!& + \!&
\frac{1}{2}\,
\left(\begin{array}{c}
m+h_1-1 \\  h_1+h_2-1 \\
 \end{array}\right) \,
c_{B}(h_1,h_2,\la) \,
%\delta^{\hat{A} \hat{B}}\, \delta^{h_1 h_2}\,
q^{h_1+h_2-4}\,\delta_{m+n}
\nonu \\
\!&-\!& \frac{1}{2}\,
\sum^{h_1+h_2-3}_{h= -1, \mbox{\footnotesize odd}} \, q^h\,
p_{\mathrm{B}}^{h_1,h_2, h}(m,n,\la)
\,
(   W^{\la,11}_{\mathrm{B},h_1+h_2-2-h}- W^{\la,22}_{\mathrm{B},h_1+h_2-2-h} )_{m+n}
\, ,
\nonu \\
\big[(W^{\la,12}_{\mathrm{B},h_1})_m,(W^{\la,21}_{\mathrm{B},h_2})_n\big] 
\!&=& \!
\frac{1}{2}\, \sum^{h_1+h_2-3}_{h= 0,
\mbox{\footnotesize even}} \, q^h\,
p_{\mathrm{B}}^{h_1,h_2, h}(m,n,\la)
 \, ( W^{\la,11}_{\mathrm{B},h_1+h_2-2-h} + W^{\la,22}_{\mathrm{B},h_1+h_2-2-h} )_{m+n}
\nonu \\
\!& + \!&
\frac{1}{2}\,
\left(\begin{array}{c}
m+h_1-1 \\  h_1+h_2-1 \\
 \end{array}\right) \,
c_{B}(h_1,h_2,\la) \,
%\delta^{\hat{A} \hat{B}}\, \delta^{h_1 h_2}\,
q^{h_1+h_2-4}\,\delta_{m+n}
\nonu \\
\!&+\!& \frac{1}{2}\,
\sum^{h_1+h_2-3}_{h= -1, \mbox{\footnotesize odd}} \, q^h\,
p_{\mathrm{B}}^{h_1,h_2, h}(m,n,\la)
\,
(   W^{\la,11}_{\mathrm{B},h_1+h_2-2-h}- W^{\la,22}_{\mathrm{B},h_1+h_2-2-h} )_{m+n}
\, ,
\nonu \\
\big[(W^{\la,12}_{\mathrm{B},h_1})_m,(W^{\la,11}_{\mathrm{B},h_2})_n\big] 
\!&=& \!
\frac{1}{2}\, (\sum^{h_1+h_2-3}_{h= 0,
  \mbox{\footnotesize even}} -\sum^{h_1+h_2-3}_{h= -1,
  \mbox{\footnotesize odd}})
\, q^h\,
p_{\mathrm{B}}^{h_1,h_2, h}(m,n,\la)
 \, ( W^{\la,12}_{\mathrm{B},h_1+h_2-2-h} )_{m+n} \, ,
 \nonu \\
\big[(W^{\la,12}_{\mathrm{B},h_1})_m,(W^{\la,22}_{\mathrm{B},h_2})_n\big] 
\!&=& \!
\frac{1}{2}\, (\sum^{h_1+h_2-3}_{h= 0,
  \mbox{\footnotesize even}} +\sum^{h_1+h_2-3}_{h= -1,
  \mbox{\footnotesize odd}})
\, q^h\,
p_{\mathrm{B}}^{h_1,h_2, h}(m,n,\la)
 \, ( W^{\la,12}_{\mathrm{B},h_1+h_2-2-h} )_{m+n} \, ,
 \nonu \\
 \big[(W^{\la,21}_{\mathrm{B},h_1})_m,(W^{\la,11}_{\mathrm{B},h_2})_n\big] 
\!&=& \!
\frac{1}{2}\, (\sum^{h_1+h_2-3}_{h= 0,
  \mbox{\footnotesize even}} +\sum^{h_1+h_2-3}_{h= -1,
  \mbox{\footnotesize odd}})
\, q^h\,
p_{\mathrm{B}}^{h_1,h_2, h}(m,n,\la)
 \, ( W^{\la,21}_{\mathrm{B},h_1+h_2-2-h} )_{m+n} \, ,
 \nonu \\
 \big[(W^{\la,21}_{\mathrm{B},h_1})_m,(W^{\la,22}_{\mathrm{B},h_2})_n\big] 
\!&=& \!
\frac{1}{2}\, (\sum^{h_1+h_2-3}_{h= 0,
  \mbox{\footnotesize even}} -\sum^{h_1+h_2-3}_{h= -1,
  \mbox{\footnotesize odd}})
\, q^h\,
p_{\mathrm{B}}^{h_1,h_2, h}(m,n,\la)
 \, ( W^{\la,21}_{\mathrm{B},h_1+h_2-2-h} )_{m+n} \, ,
 \nonu \\
 \big[(W^{\la,11}_{\mathrm{B},h_1})_m,(W^{\la,12}_{\mathrm{B},h_2})_n\big] 
\!&=& \!
\frac{1}{2}\, (\sum^{h_1+h_2-3}_{h= 0,
  \mbox{\footnotesize even}} +\sum^{h_1+h_2-3}_{h= -1,
  \mbox{\footnotesize odd}})
\, q^h\,
p_{\mathrm{B}}^{h_1,h_2, h}(m,n,\la)
 \, ( W^{\la,12}_{\mathrm{B},h_1+h_2-2-h} )_{m+n} \, ,
 \nonu \\
 \big[(W^{\la,22}_{\mathrm{B},h_1})_m,(W^{\la,12}_{\mathrm{B},h_2})_n\big] 
\!&=& \!
\frac{1}{2}\, (\sum^{h_1+h_2-3}_{h= 0,
  \mbox{\footnotesize even}} -\sum^{h_1+h_2-3}_{h= -1,
  \mbox{\footnotesize odd}})
\, q^h\,
p_{\mathrm{B}}^{h_1,h_2, h}(m,n,\la)
 \, ( W^{\la,12}_{\mathrm{B},h_1+h_2-2-h} )_{m+n} \, ,
 \nonu \\
 \big[(W^{\la,11}_{\mathrm{B},h_1})_m,(W^{\la,21}_{\mathrm{B},h_2})_n\big] 
\!&=& \!
\frac{1}{2}\, (\sum^{h_1+h_2-3}_{h= 0,
  \mbox{\footnotesize even}} -\sum^{h_1+h_2-3}_{h= -1,
  \mbox{\footnotesize odd}})
\, q^h\,
p_{\mathrm{B}}^{h_1,h_2, h}(m,n,\la)
 \, ( W^{\la,21}_{\mathrm{B},h_1+h_2-2-h} )_{m+n} \, ,
 \nonu \\
 \big[(W^{\la,22}_{\mathrm{B},h_1})_m,(W^{\la,21}_{\mathrm{B},h_2})_n\big] 
\!&=& \!
\frac{1}{2}\, (\sum^{h_1+h_2-3}_{h= 0,
  \mbox{\footnotesize even}} +\sum^{h_1+h_2-3}_{h= -1,
  \mbox{\footnotesize odd}})
\, q^h\,
p_{\mathrm{B}}^{h_1,h_2, h}(m,n,\la)
 \, ( W^{\la,21}_{\mathrm{B},h_1+h_2-2-h} )_{m+n} \, ,
 \nonu \\
 \big[(W^{\la,11}_{\mathrm{B},h_1})_m,(W^{\la,11}_{\mathrm{B},h_2})_n\big] 
\!&=& \!
\sum^{h_1+h_2-3}_{h= 0,
  \mbox{\footnotesize even}} 
\, q^h\,
p_{\mathrm{B}}^{h_1,h_2, h}(m,n,\la)
\, ( W^{\la,11}_{\mathrm{B},h_1+h_2-2-h} )_{m+n} \, \nonu \\
& + &
\frac{1}{2}\,
\left(\begin{array}{c}
m+h_1-1 \\  h_1+h_2-1 \\
 \end{array}\right) \,
c_{B}(h_1,h_2,\la) \,
%\delta^{\hat{A} \hat{B}}\, \delta^{h_1 h_2}\,
q^{h_1+h_2-4}\,\delta_{m+n} \, ,
\nonu \\
 \big[(W^{\la,22}_{\mathrm{B},h_1})_m,(W^{\la,22}_{\mathrm{B},h_2})_n\big] 
\!&=& \!
\sum^{h_1+h_2-3}_{h= 0,
  \mbox{\footnotesize even}} 
\, q^h\,
p_{\mathrm{B}}^{h_1,h_2, h}(m,n,\la)
\, ( W^{\la,22}_{\mathrm{B},h_1+h_2-2-h} )_{m+n} \, \nonu \\
& + &
\frac{1}{2}\,
\left(\begin{array}{c}
m+h_1-1 \\  h_1+h_2-1 \\
 \end{array}\right) \,
c_{B}(h_1,h_2,\la) \,
%\delta^{\hat{A} \hat{B}}\, \delta^{h_1 h_2}\,
q^{h_1+h_2-4}\,\delta_{m+n} \, ,
\label{res1}
\eea
where the central term (\ref{cB}) can be substituted.

%%%%%%%%%%%%%%%%%%
\subsection{ The commutators between the currents consisting of $b \,
  c$ system and the currents consisting of $ \ga \, b$ system
 with $h_1=h_2, h_2+ 1$
for nonzero $\la$}
%%%%%%%%%%%%%%%%%%

By analyzing the first equation of (\ref{WQ}),
we obtain the following
commutators between the currents consisting of $b \,
  c$ system and the currents consisting of $ \ga \, b$ system
\bea
\big[(W^{\la,21}_{\mathrm{F},h_1})_m,(Q^{\la,12}_{h_2+\frac{1}{2}})_r\big] 
\!&=& \!
\sum^{h_1+h_2-3}_{h= -1} \, q^h\,
q_{\mathrm{F}}^{h_1,h_2+\frac{1}{2}, h}(m,r,\la)
 \, ( Q^{\la,11}_{h_1+h_2-\frac{3}{2}-h})_{m+r} \, ,
 \nonu \\
 \big[(W^{\la,12}_{\mathrm{F},h_1})_m,(Q^{\la,21}_{h_2+\frac{1}{2}})_r\big] 
\!&=& \!
\sum^{h_1+h_2-3}_{h= -1} \, q^h\,
q_{\mathrm{F}}^{h_1,h_2+\frac{1}{2}, h}(m,r,\la)
 \, ( Q^{\la,22}_{h_1+h_2-\frac{3}{2}-h})_{m+r} \, ,
 \nonu \\
 \big[(W^{\la,12}_{\mathrm{F},h_1})_m,(Q^{\la,11}_{h_2+\frac{1}{2}})_r\big] 
\!&=& \!
\sum^{h_1+h_2-3}_{h= -1} \, q^h\,
q_{\mathrm{F}}^{h_1,h_2+\frac{1}{2}, h}(m,r,\la)
 \, ( Q^{\la,12}_{h_1+h_2-\frac{3}{2}-h})_{m+r} \, ,
 \nonu \\
 \big[(W^{\la,21}_{\mathrm{F},h_1})_m,(Q^{\la,22}_{h_2+\frac{1}{2}})_r\big] 
\!&=& \!
\sum^{h_1+h_2-3}_{h= -1} \, q^h\,
q_{\mathrm{F}}^{h_1,h_2+\frac{1}{2}, h}(m,r,\la)
 \, ( Q^{\la,21}_{h_1+h_2-\frac{3}{2}-h})_{m+r} \, ,
 \nonu \\
 \big[(W^{\la,22}_{\mathrm{F},h_1})_m,(Q^{\la,12}_{h_2+\frac{1}{2}})_r\big] 
\!&=& \!
\sum^{h_1+h_2-3}_{h= -1} \, q^h\,
q_{\mathrm{F}}^{h_1,h_2+\frac{1}{2}, h}(m,r,\la)
 \, ( Q^{\la,12}_{h_1+h_2-\frac{3}{2}-h})_{m+r} \, ,
 \nonu \\
 \big[(W^{\la,11}_{\mathrm{F},h_1})_m,(Q^{\la,21}_{h_2+\frac{1}{2}})_r\big] 
\!&=& \!
\sum^{h_1+h_2-3}_{h= -1} \, q^h\,
q_{\mathrm{F}}^{h_1,h_2+\frac{1}{2}, h}(m,r,\la)
 \, ( Q^{\la,21}_{h_1+h_2-\frac{3}{2}-h})_{m+r} \, ,
 \nonu \\
 \big[(W^{\la,11}_{\mathrm{F},h_1})_m,(Q^{\la,11}_{h_2+\frac{1}{2}})_r\big] 
\!&=& \!
\sum^{h_1+h_2-3}_{h= -1} \, q^h\,
q_{\mathrm{F}}^{h_1,h_2+\frac{1}{2}, h}(m,r,\la)
 \, ( Q^{\la,11}_{h_1+h_2-\frac{3}{2}-h})_{m+r} \, ,
 \nonu \\
 \big[(W^{\la,22}_{\mathrm{F},h_1})_m,(Q^{\la,22}_{h_2+\frac{1}{2}})_r\big] 
\!&=& \!
\sum^{h_1+h_2-3}_{h= -1} \, q^h\,
q_{\mathrm{F}}^{h_1,h_2+\frac{1}{2}, h}(m,r,\la)
 \, ( Q^{\la,22}_{h_1+h_2-\frac{3}{2}-h})_{m+r} \, .
\label{res2}
 \eea
As in previous footnote \ref{foot}, the OPE between
  $W^{\la,\hat{A}=1}_{\mathrm{F},h_1=6}(z)$ and
$Q^{\la,\hat{B}=1}_{\mathrm{B},h_2+\frac{1}{2}=\frac{7}{2}}(w)$
where the weights satisfy $h_1=h_2+3$ can be
obtained and the seventh order
  pole of this OPE contains the structure constant
  $-\frac{2048}{105}  (\la+1) (\la+2) (2 \la-3)
  (2 \la+3) (2 \la+5) (11 \la-17)$
  appearing in the current $ Q^{\la}_{h_1+h_2-\frac{3}{2}-h=\frac{5}{2}}(w)$
  with weight $h=5$.
  On the other hand, the 
  structure constant $ q_{\mathrm{F}}^{6,
    \frac{7}{2},5}(m,r,\la)$
  contains
  $\frac{4}{945} (\la+1) (2 \la-3) (4 \la^4-28 \la^3-
  13 \la^2+49 \la+51)$.
  By subtracting  the contribution
  $\frac{2}{225} (\la-1) \la (\la+1) (2 \la-3)
  (2 \la-1) (2 \la+1)$
  coming from the 
  structure constant $ q_{\mathrm{F}}^{4,
    \frac{7}{2},5}(m,r,\la)$ from this,
  we obtain
  $-\frac{2}{4725} (\la+1) (\la+2) (2 \la-3) (2 \la+3)
  (2 \la+5) (11 \la-17)$. The weight
  $h_1=6$ is replaced by the weight $(h_1-2)=4$.
  Note that the additional term
  appearing in the  $ q_{\mathrm{F}}^{4,
    \frac{7}{2},5}(m,r,\la)$ contains the factor
  $\la$.
  By considering the numerical factor $46080$
  when we move from the modes to the differential operator in the OPE
  and multiplying this into the above factor,
  we obtain the above structure constant in the current
   $ Q^{\la}_{\frac{5}{2}}(w)$.
 
%%%%%%%%%%%%%%%%%%
\subsection{ The commutators between the currents consisting of $\beta \,
  \ga$ system and the currents consisting of $\ga\, b$ system
 with $h_1=h_2, h_2+ 1$
for nonzero $\la$}
%%%%%%%%%%%%%%%%%%

In this case, by using the second equation of (\ref{WQ}),
we can write down the following
 commutators between the currents consisting of $\beta \,
  \ga$ system and the currents consisting of $\ga\, b$ system
\bea
\big[(W^{\la,21}_{\mathrm{B},h_1})_m,(Q^{\la,12}_{h_2+\frac{1}{2}})_r\big] 
\!&=& \!
\sum^{h_1+h_2-3}_{h= -1} \, q^h\,
q_{\mathrm{B}}^{h_1,h_2+\frac{1}{2}, h}(m,r,\la)
 \, ( Q^{\la,22}_{h_1+h_2-\frac{3}{2}-h})_{m+r} \, ,
 \nonu \\
 \big[(W^{\la,12}_{\mathrm{B},h_1})_m,(Q^{\la,21}_{h_2+\frac{1}{2}})_r\big] 
\!&=& \!
\sum^{h_1+h_2-3}_{h= -1} \, q^h\,
q_{\mathrm{B}}^{h_1,h_2+\frac{1}{2}, h}(m,r,\la)
 \, ( Q^{\la,11}_{h_1+h_2-\frac{3}{2}-h})_{m+r} \, ,
 \nonu \\
 \big[(W^{\la,12}_{\mathrm{B},h_1})_m,(Q^{\la,22}_{h_2+\frac{1}{2}})_r\big] 
\!&=& \!
\sum^{h_1+h_2-3}_{h= -1} \, q^h\,
q_{\mathrm{B}}^{h_1,h_2+\frac{1}{2}, h}(m,r,\la)
 \, ( Q^{\la,12}_{h_1+h_2-\frac{3}{2}-h})_{m+r} \, ,
 \nonu \\
 \big[(W^{\la,21}_{\mathrm{B},h_1})_m,(Q^{\la,11}_{h_2+\frac{1}{2}})_r\big] 
\!&=& \!
\sum^{h_1+h_2-3}_{h= -1} \, q^h\,
q_{\mathrm{B}}^{h_1,h_2+\frac{1}{2}, h}(m,r,\la)
 \, ( Q^{\la,21}_{h_1+h_2-\frac{3}{2}-h})_{m+r} \, ,
 \nonu \\
 \big[(W^{\la,11}_{\mathrm{B},h_1})_m,(Q^{\la,12}_{h_2+\frac{1}{2}})_r\big] 
\!&=& \!
\sum^{h_1+h_2-3}_{h= -1} \, q^h\,
q_{\mathrm{B}}^{h_1,h_2+\frac{1}{2}, h}(m,r,\la)
 \, ( Q^{\la,12}_{h_1+h_2-\frac{3}{2}-h})_{m+r} \, ,
 \nonu \\
 \big[(W^{\la,22}_{\mathrm{B},h_1})_m,(Q^{\la,21}_{h_2+\frac{1}{2}})_r\big] 
\!&=& \!
\sum^{h_1+h_2-3}_{h= -1} \, q^h\,
q_{\mathrm{B}}^{h_1,h_2+\frac{1}{2}, h}(m,r,\la)
 \, ( Q^{\la,21}_{h_1+h_2-\frac{3}{2}-h})_{m+r} \, ,
 \nonu \\
 \big[(W^{\la,11}_{\mathrm{B},h_1})_m,(Q^{\la,11}_{h_2+\frac{1}{2}})_r\big] 
\!&=& \!
\sum^{h_1+h_2-3}_{h= -1} \, q^h\,
q_{\mathrm{B}}^{h_1,h_2+\frac{1}{2}, h}(m,r,\la)
 \, ( Q^{\la,11}_{h_1+h_2-\frac{3}{2}-h})_{m+r} \, ,
 \nonu \\
 \big[(W^{\la,22}_{\mathrm{B},h_1})_m,(Q^{\la,22}_{h_2+\frac{1}{2}})_r\big] 
\!&=& \!
\sum^{h_1+h_2-3}_{h= -1} \, q^h\,
q_{\mathrm{B}}^{h_1,h_2+\frac{1}{2}, h}(m,r,\la)
 \, ( Q^{\la,22}_{h_1+h_2-\frac{3}{2}-h})_{m+r} \, ,
 \label{res3}
 \eea
which look similar to the previous relations in (\ref{res2}).
 
%%%%%%%%%%%%%%%%%%
\subsection{The commutators between the currents consisting of $b \,
  c$ system and the currents consisting of $ \beta\, c$ system
 with $h_1=h_2, h_2+ 1$
for nonzero $\la$
}
%%%%%%%%%%%%%%%%%%

Similarly, from the third equation of (\ref{WQ}),
we can write down the 
commutators between the currents consisting of $b \,
c$ system and the currents consisting of $ \beta\, c$ system
\bea
\big[(W^{\la,21}_{\mathrm{F},h_1})_m,(\bar{Q}^{\la,21}_{h_2+\frac{1}{2}})_r\big] 
\!&=& \!
\sum^{h_1+h_2-2}_{h= -1} \, q^h\, (-1)^h\,
q_{\mathrm{F}}^{h_1,h_2+\frac{1}{2}, h}(m,r,\la)
 \, ( \bar{Q}^{\la,22}_{h_1+h_2-\frac{3}{2}-h})_{m+r} \, ,
 \nonu \\
\big[(W^{\la,12}_{\mathrm{F},h_1})_m,(\bar{Q}^{\la,12}_{h_2+\frac{1}{2}})_r\big] 
\!&=& \!
\sum^{h_1+h_2-2}_{h= -1} \, q^h\, (-1)^h\,
q_{\mathrm{F}}^{h_1,h_2+\frac{1}{2}, h}(m,r,\la)
 \, ( \bar{Q}^{\la,11}_{h_1+h_2-\frac{3}{2}-h})_{m+r} \, ,
 \nonu \\
 \big[(W^{\la,12}_{\mathrm{F},h_1})_m,(\bar{Q}^{\la,22}_{h_2+\frac{1}{2}})_r\big] 
\!&=& \!
\sum^{h_1+h_2-2}_{h= -1} \, q^h\, (-1)^h\,
q_{\mathrm{F}}^{h_1,h_2+\frac{1}{2}, h}(m,r,\la)
 \, ( \bar{Q}^{\la,21}_{h_1+h_2-\frac{3}{2}-h})_{m+r} \, ,
 \nonu \\
 \big[(W^{\la,21}_{\mathrm{F},h_1})_m,(\bar{Q}^{\la,11}_{h_2+\frac{1}{2}})_r\big] 
\!&=& \!
\sum^{h_1+h_2-2}_{h= -1} \, q^h\, (-1)^h\,
q_{\mathrm{F}}^{h_1,h_2+\frac{1}{2}, h}(m,r,\la)
 \, ( \bar{Q}^{\la,12}_{h_1+h_2-\frac{3}{2}-h})_{m+r} \, ,
 \nonu \\
\big[(W^{\la,11}_{\mathrm{F},h_1})_m,(\bar{Q}^{\la,21}_{h_2+\frac{1}{2}})_r\big] 
\!&=& \!
\sum^{h_1+h_2-2}_{h= -1} \, q^h\, (-1)^h\,
q_{\mathrm{F}}^{h_1,h_2+\frac{1}{2}, h}(m,r,\la)
 \, ( \bar{Q}^{\la,21}_{h_1+h_2-\frac{3}{2}-h})_{m+r} \, ,
 \nonu \\
 \big[(W^{\la,22}_{\mathrm{F},h_1})_m,(\bar{Q}^{\la,12}_{h_2+\frac{1}{2}})_r\big] 
\!&=& \!
\sum^{h_1+h_2-2}_{h= -1} \, q^h\, (-1)^h\,
q_{\mathrm{F}}^{h_1,h_2+\frac{1}{2}, h}(m,r,\la)
 \, ( \bar{Q}^{\la,12}_{h_1+h_2-\frac{3}{2}-h})_{m+r} \, ,
 \nonu \\
 \big[(W^{\la,11}_{\mathrm{F},h_1})_m,(\bar{Q}^{\la,11}_{h_2+\frac{1}{2}})_r\big] 
\!&=& \!
\sum^{h_1+h_2-2}_{h= -1} \, q^h\, (-1)^h\,
q_{\mathrm{F}}^{h_1,h_2+\frac{1}{2}, h}(m,r,\la)
 \, ( \bar{Q}^{\la,11}_{h_1+h_2-\frac{3}{2}-h})_{m+r} \, ,
 \nonu \\
 \big[(W^{\la,22}_{\mathrm{F},h_1})_m,(\bar{Q}^{\la,22}_{h_2+\frac{1}{2}})_r\big] 
\!&=& \!
\sum^{h_1+h_2-2}_{h= -1} \, q^h\, (-1)^h\,
q_{\mathrm{F}}^{h_1,h_2+\frac{1}{2}, h}(m,r,\la)
 \, ( \bar{Q}^{\la,22}_{h_1+h_2-\frac{3}{2}-h})_{m+r} \, .
 \label{res4}
 \eea
 Note that
 there exists the factor $(-1)^h$ in (\ref{res4})
 which appears in the (anti)commutators described in section $4$.
 
%%%%%%%%%%%%%%%%%%
 \subsection{The commutators between the currents consisting of
     $\beta \,
   \ga$ system and the currents consisting of $ \beta \, c$ system
  with $h_1=h_2, h_2+ 1$
for nonzero $\la$}
%%%%%%%%%%%%%%%%%%

 Finally, from the analysis of the last equation of (\ref{WQ}),
 we obtain the following result
 for the commutators between the currents consisting of
     $\beta \,
   \ga$ system and the currents consisting of $ \beta \, c$ system
\bea
\big[(W^{\la,21}_{\mathrm{B},h_1})_m,(\bar{Q}^{\la,21}_{h_2+\frac{1}{2}})_r\big] 
\!&=& \!
\sum^{h_1+h_2-2}_{h= -1} \, q^h\, (-1)^h\,
q_{\mathrm{B}}^{h_1,h_2+\frac{1}{2}, h}(m,r,\la)
 \, ( \bar{Q}^{\la,11}_{h_1+h_2-\frac{3}{2}-h})_{m+r} \, ,
 \nonu \\
 \big[(W^{\la,12}_{\mathrm{B},h_1})_m,(\bar{Q}^{\la,12}_{h_2+\frac{1}{2}})_r\big] 
\!&=& \!
\sum^{h_1+h_2-2}_{h= -1} \, q^h\, (-1)^h\,
q_{\mathrm{B}}^{h_1,h_2+\frac{1}{2}, h}(m,r,\la)
 \, ( \bar{Q}^{\la,22}_{h_1+h_2-\frac{3}{2}-h})_{m+r} \, ,
 \nonu \\
 \big[(W^{\la,12}_{\mathrm{B},h_1})_m,(\bar{Q}^{\la,11}_{h_2+\frac{1}{2}})_r\big] 
\!&=& \!
\sum^{h_1+h_2-2}_{h= -1} \, q^h\, (-1)^h\,
q_{\mathrm{B}}^{h_1,h_2+\frac{1}{2}, h}(m,r,\la)
 \, ( \bar{Q}^{\la,21}_{h_1+h_2-\frac{3}{2}-h})_{m+r} \, ,
 \nonu \\
 \big[(W^{\la,21}_{\mathrm{B},h_1})_m,(\bar{Q}^{\la,22}_{h_2+\frac{1}{2}})_r\big] 
\!&=& \!
\sum^{h_1+h_2-2}_{h= -1} \, q^h\, (-1)^h\,
q_{\mathrm{B}}^{h_1,h_2+\frac{1}{2}, h}(m,r,\la)
 \, ( \bar{Q}^{\la,12}_{h_1+h_2-\frac{3}{2}-h})_{m+r} \, ,
 \nonu \\
 \big[(W^{\la,22}_{\mathrm{B},h_1})_m,(\bar{Q}^{\la,21}_{h_2+\frac{1}{2}})_r\big] 
\!&=& \!
\sum^{h_1+h_2-2}_{h= -1} \, q^h\, (-1)^h\,
q_{\mathrm{B}}^{h_1,h_2+\frac{1}{2}, h}(m,r,\la)
 \, ( \bar{Q}^{\la,21}_{h_1+h_2-\frac{3}{2}-h})_{m+r} \, ,
 \nonu \\
 \big[(W^{\la,11}_{\mathrm{B},h_1})_m,(\bar{Q}^{\la,12}_{h_2+\frac{1}{2}})_r\big] 
\!&=& \!
\sum^{h_1+h_2-2}_{h= -1} \, q^h\, (-1)^h\,
q_{\mathrm{B}}^{h_1,h_2+\frac{1}{2}, h}(m,r,\la)
 \, ( \bar{Q}^{\la,12}_{h_1+h_2-\frac{3}{2}-h})_{m+r} \, ,
 \nonu \\
 \big[(W^{\la,11}_{\mathrm{B},h_1})_m,(\bar{Q}^{\la,11}_{h_2+\frac{1}{2}})_r\big] 
\!&=& \!
\sum^{h_1+h_2-2}_{h= -1} \, q^h\, (-1)^h\,
q_{\mathrm{B}}^{h_1,h_2+\frac{1}{2}, h}(m,r,\la)
 \, ( \bar{Q}^{\la,11}_{h_1+h_2-\frac{3}{2}-h})_{m+r} \, ,
 \nonu \\
 \big[(W^{\la,22}_{\mathrm{B},h_1})_m,(\bar{Q}^{\la,22}_{h_2+\frac{1}{2}})_r\big] 
\!&=& \!
\sum^{h_1+h_2-2}_{h= -1} \, q^h\, (-1)^h\,
q_{\mathrm{B}}^{h_1,h_2+\frac{1}{2}, h}(m,r,\la)
 \, ( \bar{Q}^{\la,22}_{h_1+h_2-\frac{3}{2}-h})_{m+r} \, .
 \label{res5}
 \eea
Similar to the previous footnote \ref{foot},  the OPE between
$W^{\la,\hat{A}=1}_{\mathrm{B},h_1=7}(z)$ and
$\bar{Q}^{\la,\hat{B}=1}_{\mathrm{B},h_2+\frac{1}{2}=\frac{7}{2}}(w)$ provides
the seventh order with the structure constant
$\frac{2048}{1155} (\la+2) (2 \la+3) (2 \la+5)
(662 \la^3-3165 \la^2+5434 \la-3339)$
appearing in the current $ \bar{Q}^{\la}_{\frac{7}{2}}(w)$.
This can be obtained by adding the
extra contribution from  $ q_{\mathrm{B}}^{4,
  \frac{7}{2},5}(m,r,\la)$ where $h_1$ is replaced by
$(h_1-3)$ in addition to
the one from  $ q_{\mathrm{B}}^{7,
  \frac{7}{2},5}(m,r,\la)$.
Note that 
the $ \bar{Q}^{\la}_{\frac{1}{2}}(w)$
term on the right hand side of the tenth order pole of the OPE can be
determined by the contribution from the
structure constant $ q_{\mathrm{B}}^{5,
  \frac{7}{2},8}(m,r,\la)$ only.

 Note that
 there exists the factor $(-1)^h$ in (\ref{res5})
 which appears in the (anti)commutators described in section $4$
 and Appendix $G$.

 %%%%%%%%%%%%%%%%%%%%%%%%%%%
 \subsection{The anticommutators between the currents consisting of
     $
   \ga\, b$ system and the currents consisting of
   $\beta\, c$ system  with $h_1=h_2$
for nonzero $\la$}
%%%%%%%%%%%%%%%%%%%%%%%%%%%%%%

 By analyzing the equation of (\ref{Final}),
 the following result satisfies 
\bea
\{(Q^{\la,21}_{h_1+\frac{1}{2}})_r,(\bar{Q}^{\la,21}_{h_2+\frac{1}{2}})_s\} 
\!&=& \!
\sum^{h_1+h_2-1}_{h= 0} \, q^h \, \Bigg( o_{\mathrm{F}}^{h_1+\frac{1}{2},h_2+
\frac{1}{2}, h}(r,s,\la)
\,
(W^{\la,11}_{F,h_1+h_2-h})_{r+s} \nonu \\
&+& 
o_{\mathrm{B}}^{h_1+\frac{1}{2},h_2+\frac{1}{2}, h}(r,s,\la)
\,
 (W^{\la,22}_{B,h_1+h_2-h} )_{r+s} \Bigg)
\nonu \\
\!&+\!&  \frac{1}{2} \,
\left(\begin{array}{c}
r+h_1-\frac{1}{2} \\  h_1+h_2 \\
 \end{array}\right) \,
c_{Q}(h_1,h_2,\la)
%\, \delta^{\hat{A} \hat{B}}
%\, \delta^{h_1 h_2}
\, q^{h_1+h_2-2}
\delta_{r+s} \, ,
\nonu \\
\{(Q^{\la,12}_{h_1+\frac{1}{2}})_r,(\bar{Q}^{\la,12}_{h_2+\frac{1}{2}})_s\} 
\!&=& \!
\sum^{h_1+h_2-1}_{h= 0} \, q^h \, \Bigg( o_{\mathrm{F}}^{h_1+\frac{1}{2},h_2+
\frac{1}{2}, h}(r,s,\la)
\,
(W^{\la,22}_{F,h_1+h_2-h})_{r+s} \nonu \\
&+& 
o_{\mathrm{B}}^{h_1+\frac{1}{2},h_2+\frac{1}{2}, h}(r,s,\la)
\,
 (W^{\la,11}_{B,h_1+h_2-h} )_{r+s} \Bigg)
\nonu \\
\!&+\!&  \frac{1}{2} \,
\,
\left(\begin{array}{c}
r+h_1-\frac{1}{2} \\  h_1+h_2 \\
 \end{array}\right) \,
c_{Q}(h_1,h_2,\la)
%\, \delta^{\hat{A} \hat{B}}
%\, \delta^{h_1 h_2}
\, q^{h_1+h_2-2}
\delta_{r+s} \, ,
\nonu \\
\{(Q^{\la,12}_{h_1+\frac{1}{2}})_r,(\bar{Q}^{\la,11}_{h_2+\frac{1}{2}})_s\} 
\!&=& \!
\sum^{h_1+h_2-1}_{h= 0} \, q^h \,  o_{\mathrm{F}}^{h_1+\frac{1}{2},h_2+
\frac{1}{2}, h}(r,s,\la)
\,
(W^{\la,12}_{F,h_1+h_2-h})_{r+s} \, ,
\nonu \\
\{(Q^{\la,12}_{h_1+\frac{1}{2}})_r,(\bar{Q}^{\la,22}_{h_2+\frac{1}{2}})_s\} 
\!&=& \!
\sum^{h_1+h_2-1}_{h= 0} \, q^h \,  o_{\mathrm{B}}^{h_1+\frac{1}{2},h_2+
\frac{1}{2}, h}(r,s,\la)
\,
(W^{\la,12}_{B,h_1+h_2-h})_{r+s} \, ,
\nonu \\
\{(Q^{\la,21}_{h_1+\frac{1}{2}})_r,(\bar{Q}^{\la,11}_{h_2+\frac{1}{2}})_s\} 
\!&=& \!
\sum^{h_1+h_2-1}_{h= 0} \, q^h \,  o_{\mathrm{B}}^{h_1+\frac{1}{2},h_2+
\frac{1}{2}, h}(r,s,\la)
\,
(W^{\la,21}_{B,h_1+h_2-h})_{r+s} \, ,
\nonu \\
\{(Q^{\la,21}_{h_1+\frac{1}{2}})_r,(\bar{Q}^{\la,22}_{h_2+\frac{1}{2}})_s\} 
\!&=& \!
\sum^{h_1+h_2-1}_{h= 0} \, q^h \,  o_{\mathrm{F}}^{h_1+\frac{1}{2},h_2+
\frac{1}{2}, h}(r,s,\la)
\,
(W^{\la,21}_{F,h_1+h_2-h})_{r+s} \, ,
\nonu \\
\{(Q^{\la,11}_{h_1+\frac{1}{2}})_r,(\bar{Q}^{\la,21}_{h_2+\frac{1}{2}})_s\} 
\!&=& \!
\sum^{h_1+h_2-1}_{h= 0} \, q^h \,  o_{\mathrm{B}}^{h_1+\frac{1}{2},h_2+
\frac{1}{2}, h}(r,s,\la)
\,
(W^{\la,12}_{B,h_1+h_2-h})_{r+s} \, ,
\nonu \\
\{(Q^{\la,22}_{h_1+\frac{1}{2}})_r,(\bar{Q}^{\la,21}_{h_2+\frac{1}{2}})_s\} 
\!&=& \!
\sum^{h_1+h_2-1}_{h= 0} \, q^h \,  o_{\mathrm{F}}^{h_1+\frac{1}{2},h_2+
\frac{1}{2}, h}(r,s,\la)
\,
(W^{\la,12}_{F,h_1+h_2-h})_{r+s} \, ,
\nonu \\
\{(Q^{\la,11}_{h_1+\frac{1}{2}})_r,(\bar{Q}^{\la,12}_{h_2+\frac{1}{2}})_s\} 
\!&=& \!
\sum^{h_1+h_2-1}_{h= 0} \, q^h \,  o_{\mathrm{F}}^{h_1+\frac{1}{2},h_2+
\frac{1}{2}, h}(r,s,\la)
\,
(W^{\la,21}_{F,h_1+h_2-h})_{r+s} \, ,
\nonu \\
\{(Q^{\la,22}_{h_1+\frac{1}{2}})_r,(\bar{Q}^{\la,12}_{h_2+\frac{1}{2}})_s\} 
\!&=& \!
\sum^{h_1+h_2-1}_{h= 0} \, q^h \,  o_{\mathrm{B}}^{h_1+\frac{1}{2},h_2+
\frac{1}{2}, h}(r,s,\la)
\,
(W^{\la,21}_{B,h_1+h_2-h})_{r+s} \, ,
\nonu \\
\{(Q^{\la,11}_{h_1+\frac{1}{2}})_r,(\bar{Q}^{\la,11}_{h_2+\frac{1}{2}})_s\} 
\!&=& \!
\sum^{h_1+h_2-1}_{h= 0} \, q^h \, \Bigg( o_{\mathrm{F}}^{h_1+\frac{1}{2},h_2+
\frac{1}{2}, h}(r,s,\la)
\,
(W^{\la,11}_{F,h_1+h_2-h})_{r+s} \nonu \\
&+& 
o_{\mathrm{B}}^{h_1+\frac{1}{2},h_2+\frac{1}{2}, h}(r,s,\la)
\,
 (W^{\la,11}_{B,h_1+h_2-h} )_{r+s} \Bigg)
\nonu \\
\!&+\!&  \frac{1}{2} \,
\,
\left(\begin{array}{c}
r+h_1-\frac{1}{2} \\  h_1+h_2 \\
 \end{array}\right) \,
c_{Q}(h_1,h_2,\la)
\, q^{h_1+h_2-2}
\delta_{r+s} \, ,
\nonu \\
\{(Q^{\la,22}_{h_1+\frac{1}{2}})_r,(\bar{Q}^{\la,22}_{h_2+\frac{1}{2}})_s\} 
\!&=& \!
\sum^{h_1+h_2-1}_{h= 0} \, q^h \, \Bigg( o_{\mathrm{F}}^{h_1+\frac{1}{2},h_2+
\frac{1}{2}, h}(r,s,\la)
\,
(W^{\la,22}_{F,h_1+h_2-h})_{r+s} \nonu \\
&+& 
o_{\mathrm{B}}^{h_1+\frac{1}{2},h_2+\frac{1}{2}, h}(r,s,\la)
\,
 (W^{\la,22}_{B,h_1+h_2-h} )_{r+s} \Bigg)
\nonu \\
\!&+\!&  \frac{1}{2} \,
\,
\left(\begin{array}{c}
r+h_1-\frac{1}{2} \\  h_1+h_2 \\
 \end{array}\right) \,
c_{Q}(h_1,h_2,\la)
\, q^{h_1+h_2-2}
\delta_{r+s} \, .
\label{res6}
\eea
There are the relations when we interchange
$h_1$ and $h_2$ as 
$o_{\mathrm{F}}^{h_2+\frac{1}{2},h_1+
\frac{1}{2}, h}=(-1)^h\, o_{\mathrm{F}}^{h_1+\frac{1}{2},h_2+
\frac{1}{2}, h}$ and
$o_{\mathrm{B}}^{h_2+\frac{1}{2},h_1+\frac{1}{2}, h}= (-1)^h\,
o_{\mathrm{B}}^{h_1+\frac{1}{2},h_2+\frac{1}{2}, h}$.
These will be used in Appendix $G$.

%%%%%%%%%%%%%%%%%%%%%%%%%%%%%%%%%%%%%%%%%%%%%%%%%%%%%%%
%%%%%%%%%%%%%%%%%%%%%%%%%%%%%
\section{Some relations which can be used in section $4$  }
%%%%%%%%%%%%%%%%%%%%%%%%%%%%%
%%%%%%BBB%%%%%%%%%%%%%%%%%%%%%%%%%%%%%%%%%%%%%%%%%%%%%%%%%

Other component result related to (\ref{twotwo})
can be summarized by
\bea
-\frac{i}{2}\,\Big(
Q^{\la,11}_{h+\frac{1}{2}}
+2i\sqrt{2} \,Q^{\la,21}_{h+\frac{1}{2}}
-2 \,Q^{\la,22}_{h+\frac{1}{2}} \Big) & = &
\frac{1}{2}\, \Bigg[ \frac{1}{4(-4)^{h-4}}\, \Phi^{(h),2}_{\frac{1}{2}}
  - \frac{1}{4(-4)^{h-5}}\, \tilde{\Phi}^{(h-1),2}_{\frac{3}{2}} \Bigg]\, ,
\nonu \\
-\frac{i}{2}\,\Big(2\, \bar{Q}^{\la,11}_{h+\frac{1}{2}}
+2i\sqrt{2} \, \bar{Q}^{\la,12}_{h+\frac{1}{2}}
-\bar{Q}^{\la,22}_{h+\frac{1}{2}}\Big) & = &
\frac{1}{2}\, \Bigg[ \frac{1}{4(-4)^{h-4}}\, \Phi^{(h),2}_{\frac{1}{2}}
  + \frac{1}{4(-4)^{h-5}}\, \tilde{\Phi}^{(h-1),2}_{\frac{3}{2}} \Bigg]\, ,
\nonu \\
-\frac{i}{2}\,\Big(
Q^{\la,11}_{h+\frac{1}{2}}
+i\sqrt{2} \,Q^{\la,12}_{h+\frac{1}{2}}
-2\,Q^{\la,22}_{h+\frac{1}{2}} \Big) & = &
\frac{1}{2}\, \Bigg[ \frac{1}{4(-4)^{h-4}}\, \Phi^{(h),3}_{\frac{1}{2}}
  - \frac{1}{4(-4)^{h-5}}\, \tilde{\Phi}^{(h-1),3}_{\frac{3}{2}} \Bigg]\, ,
\nonu \\
-\frac{i}{2}\,\Big(2\, \bar{Q}^{\la,11}_{h+\frac{1}{2}}
+i \sqrt{2} \, \bar{Q}^{\la,21}_{h+\frac{1}{2}}
-\bar{Q}^{\la,22}_{h+\frac{1}{2}}
\Big) & = &
\frac{1}{2}\, \Bigg[ \frac{1}{4(-4)^{h-4}}\, \Phi^{(h),3}_{\frac{1}{2}}
  + \frac{1}{4(-4)^{h-5}}\, \tilde{\Phi}^{(h-1),3}_{\frac{3}{2}} \Bigg]\, ,
\nonu \\
-\frac{1}{2}\,Q^{\la,11}_{h+\frac{1}{2}}
-Q^{\la,22}_{h+\frac{1}{2}} & = &
\frac{1}{2}\, \Bigg[ \frac{1}{4(-4)^{h-4}}\, \Phi^{(h),4}_{\frac{1}{2}}
  - \frac{1}{4(-4)^{h-5}}\, \tilde{\Phi}^{(h-1),4}_{\frac{3}{2}} \Bigg]\, , 
\nonu \\
\bar{Q}^{\la,11}_{h+\frac{1}{2}}
+\frac{1}{2}\,\bar{Q}^{\la,22}_{h+\frac{1}{2}}
& = &
\frac{1}{2}\, \Bigg[ \frac{1}{4(-4)^{h-4}}\, \Phi^{(h),4}_{\frac{1}{2}}
  + \frac{1}{4(-4)^{h-5}}\, \tilde{\Phi}^{(h-1),4}_{\frac{3}{2}} \Bigg]\, .
\label{RELL}
\eea
Similarly, other relations associated with (\ref{rell})
can be described as
\bea
-2i\,W^{\la,11}_{\mathrm{B},h+1}
+4\sqrt{2}\,W^{\la,21}_{\mathrm{B},h+1}
+2i\,\,W^{\la,22}_{\mathrm{B},h+1} & = &
\frac{1}{8(-4)^{h-4}} \Bigg[ \Phi_1^{(h),13}+  \Phi_1^{(h),24}\Bigg]\, ,
\nonu \\
-  2i\,W^{\la,11}_{\mathrm{F},h+1}
+2\sqrt{2}\,W^{\la,21}_{\mathrm{F},h+1}
+2i\,W^{\la,22}_{\mathrm{F}+h+1} & = &
\frac{1}{8(-4)^{h-4}} \Bigg[ \Phi_1^{(h),13}-  \Phi_1^{(h),24}\Bigg] \, ,
\nonu \\
2\,W^{\la,11}_{\mathrm{B},h+1}
+i\sqrt{2}\,W^{\la,12}_{\mathrm{B},h+1}
+4i\sqrt{2}\,\,W^{\la,21}_{\mathrm{B},h+1}
-2\,W^{\la,22}_{\mathrm{B},h+1} & = &
\frac{1}{8(-4)^{h-4}} \Bigg[ \Phi_1^{(h),14}-  \Phi_1^{(h),23}\Bigg]\, ,
\nonu \\
-  2W^{\la,11}_{\mathrm{F},h+1}
-2i\sqrt{2}W^{\la,12}_{\mathrm{F},h+1}
 -  2i\sqrt{2}W^{\la,21}_{\mathrm{F},h+1}
 +2W^{\la,22}_{\mathrm{F},h+1} & = &
 \frac{1}{8(-4)^{h-4}} \Bigg[ \Phi_1^{(h),14}+  \Phi_1^{(h),23}\Bigg]  .
 \label{RELL1}
\eea
Both relations in (\ref{RELL}) and (\ref{RELL1})
are used in section $4$ and Appendix $G$.

%%%%%%%%%%%%%%%%%%%%%%%%%%%%%%%%%%%%%%%%%%%%%%%%%%%%%%%
%%%%%%%%%%%%%%%%%%%%%%%%%%%%%
\section{The remaining (anti)commutator relations between the
${\cal N}=4$ multiplets}
%%%%%%%%%%%%%%%%%%%%%%%%%%%%%
%%%%%%BBB%%%%%%%%%%%%%%%%%%%%%%%%%%%%%%%%%%%%%%%%%%%%%%%%%

In this Appendix, the remaining ten (anti)commutators
with the particular examples for the specific weights
$h_1$ and $h_2$ showing that the extra structures on the right
hand sides of these (anti)commutators are given
explicitly. 

%%%%%%%%%%%%%%%%%
\subsection{The anticommutator relation between the second components
with $h_1=h_2$  for nonzero $\la$}
%%%%%%%%%%%%%%%%%

By using the expressions in (\ref{two}) and (\ref{res6})
together with (\ref{rel1}), (\ref{rell}) and (\ref{RELL1}),
the following anticommutator can be obtained
\bea
&& \big\{(\Phi^{(h_1),i}_{\frac{1}{2}})_r,
  (\Phi^{(h_2),j}_{\frac{1}{2}})_s \big\}  = 
16(-4)^{h_1-4} \, (-4)^{h_2-4}
\Bigg[ \nonu \\
  & &
\left(\begin{array}{c}
r+h_1-\frac{1}{2} \\  h_1+h_2 \\
 \end{array}\right) \, q^{h_1+h_2-2} \, c_Q \,
 \de^{ij}\,  \de_{r+s}
  \nonu \\
  && +  \de^{ij}\, \sum_{h=0}^{h_1+h_2-1} \,
 q^h\, 
  (1+ (-1)^h)\,
  \Bigg(  o_{\mathrm{F}}^{h_1+\frac{1}{2},h_2+\frac{1}{2}, h}(r,s,\la)
 - 
 o_{\mathrm{B}}^{h_1+\frac{1}{2},h_2+\frac{1}{2}, h}(r,s,\la)
 \Bigg)\nonu \\
 && \times \frac{1}{2(-4)^{h_1+h_2-h-2}}\,
(\Phi_{0}^{(h_1+h_2-h)})_{r+s}
\nonu \\
&& + \de^{ij}\, \sum_{h=0}^{h_1+h_2-1} \, \Bigg(
(h_1+h_2-h-1+2\la)\, q^h\, 
(1+(-1)^h) \, o_{\mathrm{F}}^{h_1+\frac{1}{2},h_2+\frac{1}{2}, h}(r,s,\la)
\nonu \\
&& +
(h_1+h_2-h-2\la)\, q^h\, 
(1+(-1)^h) \,
o_{\mathrm{B}}^{h_1+\frac{1}{2},h_2+\frac{1}{2}, h}(r,s,\la)
\,
\Bigg)\nonu \\
&& \times \frac{1}{
  16(2h_1+2h_2-2h-1)(-4)^{h_1+h_2-h-6}}
\, (\tilde{\Phi}_{2}^{(h_1+h_2-h-2)})_{r+s}
\nonu \\
&& -
  \sum_{h=0}^{h_1+h_2-1} \, q^h\, 
  (1-(-1)^h)\, o_{\mathrm{F}}^{h_1+\frac{1}{2},h_2+\frac{1}{2}, h}(r,s,\la)
  \nonu \\
  && \times
 \frac{1}{32(-4)^{h_1+h_2-h-5}}\,
  \Bigg(\Phi_{1}^{(h_1+h_2-h-1),ij} + \frac{1}{2}\,
  \varepsilon^{ijkl} \,\Phi_{1}^{(h_1+h_2-h-1),kl} \Bigg)_{r+s}
\nonu \\
 &&   +
  \sum_{h=0}^{h_1+h_2-1} \, q^h\, 
 (1-(-1)^h)\,  o_{\mathrm{B}}^{h_1+\frac{1}{2},h_2+\frac{1}{2}, h}(r,s,\la)
  \nonu \\
  && \times
\frac{1}{32(-4)^{h_1+h_2-h-5}}\,
  \Bigg(\Phi_{1}^{(h_1+h_2-h-1),ij} - \frac{1}{2}\,
  \varepsilon^{ijkl} \,\Phi_{1}^{(h_1+h_2-h-1),kl}\Bigg)_{r+s} \,
\Bigg] \, .
\label{sixcase}
\eea
The central term is given by (\ref{cQ}).
Due to the factor $(1\pm (-1)^h)$ in the above,
the two kinds of the currents, the $SO(4)$ singlets and the $SO(4)$
adjoints appear alternatively.
When the weight $h$ is equal to its maximum value
$h=h_1+h_2-1$, then the currents $\Phi_0^{(1)}$ and $\tilde{\Phi}_2^{(-1)}$
terms in the $SO(4)$ singlets appear on the right hand side
while the currents
$\Phi^{(0),ij}_1$ and $\frac{1}{2}\, \varepsilon^{ijkl}\,\Phi^{(0),kl}_1$
appear similarly.

%%%%%%%%%%%%%%%%%
\subsection{The commutator relation between the second component and the
third component with $h_1 =h_2, h_2+1$  for nonzero $\la$}
%%%%%%%%%%%%%%%%%

By using the relation (\ref{two}), (\ref{three})
together with  (\ref{res2}), (\ref{res3}), (\ref{res4}),
(\ref{res5}), (\ref{twotwo}), and (\ref{RELL}),
It turns out that we obtain
\bea
&& \big[(\Phi^{(h_1),i}_{\frac{1}{2}})_r,
  (\Phi^{(h_2),jk}_{1})_m \big]  = 
 4(-4)^{h_1-4}\, 4(-4)^{h_2-4}
\Bigg[ 
 \de^{ij} \, \Bigg( -
 \sum_{h=-1}^{h_1+h_2-2} \, q^h\, 
  q_{\mathrm{F}}^{h_2+1,h_1+\frac{1}{2}, h}(m,r,\la)
  \nonu \\
  && +  \sum_{h=-1}^{h_1+h_2-1} \,
  q^h\, (-1)^h\, 
q_{\mathrm{F}}^{h_2+1,h_1+\frac{1}{2}, h}(m,r,\la)
 +  
\sum_{h=-1}^{h_1+h_2-2} \,
 q^h\, 
q_{\mathrm{B}}^{h_2+1,h_1+\frac{1}{2}, h}(m,r,\la)
\nonu \\
&& - 
\sum_{h=-1}^{h_1+h_2-1} \,
 q^h\, (-1)^h\,
q_{\mathrm{B}}^{h_2+1,h_1+\frac{1}{2}, h}(m,r,\la)
\Bigg)\,
 \frac{1}{4(-4)^{h_1+h_2-h-5}}\,
(\Phi_{\frac{1}{2}}^{(h_1+h_2-1-h),k})_{m+r}
\nonu \\
 &&   +
\de^{ij}\,  \Bigg(  
  \sum_{h=-1}^{h_1+h_2-2} \, q^h\, 
  q_{\mathrm{F}}^{h_2+1,h_1+\frac{1}{2}, h}(m,r,\la)
  \nonu \\
  && +  \sum_{h=-1}^{h_1+h_2-1} \,
  q^h\, (-1)^h\, 
q_{\mathrm{F}}^{h_2+1,h_1+\frac{1}{2}, h}(m,r,\la)
 -  
\sum_{h=-1}^{h_1+h_2-2} \,
 q^h\, 
q_{\mathrm{B}}^{h_2+1,h_1+\frac{1}{2}, h}(m,r,\la)
\nonu \\
&& -  
\sum_{h=-1}^{h_1+h_2-1} \,
 q^h\, (-1)^h\,
q_{\mathrm{B}}^{h_2+1,h_1+\frac{1}{2}, h}(m,r,\la)
\Bigg)\,
\frac{1}{4(-4)^{h_1+h_2-h-6}}\,
(\tilde{\Phi}_{\frac{3}{2}}^{(h_1+h_2-2-h),k})_{m+r} \,
\nonu \\
%%%%%%%%%%%%%%%%%%%%%%%%%%%%%%%%%%%%%%%%%%%%% 
 && -\de^{ik} \, \Bigg( -
 \sum_{h=-1}^{h_1+h_2-2} \, q^h\, 
  q_{\mathrm{F}}^{h_2+1,h_1+\frac{1}{2}, h}(m,r,\la)
  \nonu \\
  && +  \sum_{h=-1}^{h_1+h_2-1} \,
  q^h\, (-1)^h\, 
q_{\mathrm{F}}^{h_2+1,h_1+\frac{1}{2}, h}(m,r,\la)
 +  
\sum_{h=-1}^{h_1+h_2-2} \,
 q^h\, 
q_{\mathrm{B}}^{h_2+1,h_1+\frac{1}{2}, h}(m,r,\la)
\nonu \\
&& - 
\sum_{h=-1}^{h_1+h_2-1} \,
 q^h\, (-1)^h\,
q_{\mathrm{B}}^{h_2+1,h_1+\frac{1}{2}, h}(m,r,\la)
\Bigg)\,
 \frac{1}{4(-4)^{h_1+h_2-h-5}}\,
(\Phi_{\frac{1}{2}}^{(h_1+h_2-1-h),j})_{m+r}
\nonu \\
 &&   -
\de^{ik}\,  \Bigg(  
  \sum_{h=-1}^{h_1+h_2-2} \, q^h\, 
  q_{\mathrm{F}}^{h_2+1,h_1+\frac{1}{2}, h}(m,r,\la)
  \nonu \\
  && +  \sum_{h=-1}^{h_1+h_2-1} \,
  q^h\, (-1)^h\, 
q_{\mathrm{F}}^{h_2+1,h_1+\frac{1}{2}, h}(m,r,\la)
 -  
\sum_{h=-1}^{h_1+h_2-2} \,
 q^h\, 
q_{\mathrm{B}}^{h_2+1,h_1+\frac{1}{2}, h}(m,r,\la)
\nonu \\
&& -  
\sum_{h=-1}^{h_1+h_2-1} \,
 q^h\, (-1)^h\,
q_{\mathrm{B}}^{h_2+1,h_1+\frac{1}{2}, h}(m,r,\la)
\Bigg)\,
\frac{1}{4(-4)^{h_1+h_2-h-6}}\,
(\tilde{\Phi}_{\frac{3}{2}}^{(h_1+h_2-2-h),j})_{m+r} \,
\nonu \\
%%%%%%%%%%%%%%%%%%%%%%%%%%%%%%%%%%%%%%%%%%%%%%%%%%%%%%%
&&
+ \varepsilon^{ijkl} \, \Bigg( -
 \sum_{h=-1}^{h_1+h_2-2} \, q^h\, 
  q_{\mathrm{F}}^{h_2+1,h_1+\frac{1}{2}, h}(m,r,\la)
  \nonu \\
  && +  \sum_{h=-1}^{h_1+h_2-1} \,
  q^h\, (-1)^h\, 
q_{\mathrm{F}}^{h_2+1,h_1+\frac{1}{2}, h}(m,r,\la)
 -  
\sum_{h=-1}^{h_1+h_2-2} \,
 q^h\, 
q_{\mathrm{B}}^{h_2+1,h_1+\frac{1}{2}, h}(m,r,\la)
\nonu \\
&& + 
\sum_{h=-1}^{h_1+h_2-1} \,
 q^h\, (-1)^h\,
q_{\mathrm{B}}^{h_2+1,h_1+\frac{1}{2}, h}(m,r,\la)
\Bigg)\,
 \frac{1}{4(-4)^{h_1+h_2-h-5}}\,
(\Phi_{\frac{1}{2}}^{(h_1+h_2-1-h),l})_{m+r}
\nonu \\
 &&   +
\varepsilon^{ijkl}\,  \Bigg(  
  \sum_{h=-1}^{h_1+h_2-2} \, q^h\, 
  q_{\mathrm{F}}^{h_2+1,h_1+\frac{1}{2}, h}(m,r,\la)
  \nonu \\
  && +  \sum_{h=-1}^{h_1+h_2-1} \,
  q^h\, (-1)^h\, 
q_{\mathrm{F}}^{h_2+1,h_1+\frac{1}{2}, h}(m,r,\la)
 +  
\sum_{h=-1}^{h_1+h_2-2} \,
 q^h\, 
q_{\mathrm{B}}^{h_2+1,h_1+\frac{1}{2}, h}(m,r,\la)
\nonu \\
&& +  
\sum_{h=-1}^{h_1+h_2-1} \,
 q^h\, (-1)^h\,
q_{\mathrm{B}}^{h_2+1,h_1+\frac{1}{2}, h}(m,r,\la)
\Bigg)\,
\frac{1}{4(-4)^{h_1+h_2-h-6}}\,
(\tilde{\Phi}_{\frac{3}{2}}^{(h_1+h_2-2-h),l})_{m+r} \,
\Bigg]\, .
\label{sevencase}
\eea
Due to the antisymmetric property in the indices
$j$ and $k$ on the left hand side, we observe that there exists
antisymmetric property for the interchange of those indices
on the right hand side.
The field contents in (\ref{sevencase}) look like as
the ones in (\ref{Tworel}).
Note that the ordering of first two upper elements in the
structure constants in the above is opposite to the one in (\ref{Tworel}).
As before, in the summation over the dummy variable of the weight $h$,
there are current terms for the case of $h=h_1+h_2-2$
where we see the presence of the current $\Phi^{(1),k}_{\frac{1}{2}}$
and $\tilde{\Phi}_{\frac{3}{2}}^{(0),k}$ terms with Kronecker delta
or epsilon tensor
of $SO(4)$
and for the weights $h \leq h_1+h_2-3$, there are some cancellations
in the current terms due to the factor $(-1)^h$ depending on the even or
odd property of the weight $h$.

Let us consider the OPE
  between
$\Phi^{(h_1=3),i}_{\frac{1}{2}}(z)$ and $
  \Phi^{(h_2=4),jk}_{1}(w)$ with $i=j$ where $h_1=h_2-1$
  as in the footnotes of section $4$.
  The seventh order pole of this OPE gives us
  the structure constant $\frac{131072}{35} (\la-2) (\la-1)
  (\la+1) (2 \la-3) (2 \la+1) (2 \la+3)$ appearing in the current
  $\Phi_{\frac{1}{2}}^{(h_1+h_2-1-h=1),k}(w)$ with weight $h=5$.
  By substituting the various expressions in the corresponding terms of
  (\ref{sevencase}),
  we can check that we obtain the above structure constant correctly
  where $q_F^{5,\frac{7}{2},5}$ term corresponds to
  $\frac{4096}{35} (\la-1) (\la+1) (2 \la-3)
  (2 \la+1) (4 \la^2-14 \la-9)$
  while  $q_B^{5,\frac{7}{2},5}$ term
  corresponds to $-\frac{4096}{35}  (\la-1)
    (\la+1) (2 \la-3) (2 \la+1)
 (4 \la^2+10 \la-15)$.
  There appears
  the current $\tilde{\Phi}_{\frac{3}{2}}^{(0),i}(w)$ term
with nonzero structure constant having the $\la$ factor 
  in the seventh order pole of this OPE
  and this is not consistent with (\ref{sevencase}) because
  all the coefficients vanish for the odd weight $h=5$.
  This implies that there is additional term on the right hand side
  of the OPE for
  the weights $h_1=3$ and $h_2=4$.
 
%%%%%%%%%%%%%%%%%
\subsection{The anticommutator relation between the second component and the
fourth component with $h_1=h_2+1$  for nonzero $\la$}
%%%%%%%%%%%%%%%%%

By using the relations (\ref{two}) and (\ref{four}),
we can rewrite it in terms of the corresponding anticommutators
in (\ref{res6}) and using the relations (\ref{rel1}), (\ref{rell})
and (\ref{RELL1}),
the following result can be obtained
\bea
&& \big\{(\Phi^{(h_1),i}_{\frac{1}{2}})_r,
  (\tilde{\Phi}^{(h_2),j}_{\frac{3}{2}})_s \big\}  = 
16(-4)^{h_1-4} \, (-4)^{h_2-4}
\Bigg[ \nonu \\
  & &
-\left(\begin{array}{c}
r+h_1-\frac{1}{2} \\  h_1+h_2+1 \\
 \end{array}\right) \, q^{h_1+h_2-1} \, c'_Q \,
 \de^{ij}\,  \de_{r+s}
  \nonu \\
  && +  \de^{ij}\, \sum_{h=0}^{h_1+h_2} \,
 q^h\, 
  (1- (-1)^h)\,
  \Bigg(  o_{\mathrm{F}}^{h_1+\frac{1}{2},h_2+\frac{3}{2}, h}(r,s,\la)
 - 
 o_{\mathrm{B}}^{h_1+\frac{1}{2},h_2+\frac{3}{2}, h}(r,s,\la)
 \Bigg)\nonu \\
 && \times \frac{1}{2(-4)^{h_1+h_2-h-1}}\,
(\Phi_{0}^{(h_1+h_2+1-h)})_{r+s}
\nonu \\
&& + \de^{ij}\, \sum_{h=0}^{h_1+h_2} \, \Bigg(
(h_1+h_2-h+2\la)\, q^h\, 
(1-(-1)^h) \, o_{\mathrm{F}}^{h_1+\frac{1}{2},h_2+\frac{3}{2}, h}(r,s,\la)
\nonu \\
&& +
(h_1+h_2+1-h-2\la)\, q^h\, 
(1-(-1)^h) \,
o_{\mathrm{B}}^{h_1+\frac{1}{2},h_2+\frac{3}{2}, h}(r,s,\la)
\,
\Bigg)\nonu \\
&& \times \frac{1}{
  16(2h_1+2h_2-2h+1)(-4)^{h_1+h_2-h-5}}
\, (\tilde{\Phi}_{2}^{(h_1+h_2-h-1)})_{r+s}
\nonu \\
&& -
  \sum_{h=0}^{h_1+h_2} \, q^h\, 
  (1+(-1)^h)\, o_{\mathrm{F}}^{h_1+\frac{1}{2},h_2+\frac{3}{2}, h}(r,s,\la)
  \nonu \\
  && \times
 \frac{1}{32(-4)^{h_1+h_2-h-4}}\,
  \Bigg(\Phi_{1}^{(h_1+h_2-h),ij} + \frac{1}{2}\,
  \varepsilon^{ijkl} \,\Phi_{1}^{(h_1+h_2-h),kl} \Bigg)_{r+s}
\nonu \\
 &&   +
  \sum_{h=0}^{h_1+h_2} \, q^h\, 
 (1+(-1)^h)\,  o_{\mathrm{B}}^{h_1+\frac{1}{2},h_2+\frac{3}{2}, h}(r,s,\la)
  \nonu \\
  && \times
\frac{1}{32(-4)^{h_1+h_2-h-4}}\,
  \Bigg(\Phi_{1}^{(h_1+h_2-h),ij} - \frac{1}{2}\,
  \varepsilon^{ijkl} \,\Phi_{1}^{(h_1+h_2-h),kl}\Bigg)_{r+s} \,
\Bigg] \, .
\label{eightcase}
\eea
The central charge will be given later soon.
The right hand side of (\ref{eightcase})
is similar to the one in (\ref{sixcase}) in the sense that
the field contents are the same and the relative signs are different
from each other.
It is obvious to see that some expressions having $h_2$ in (\ref{sixcase})
are replaced by $(h_2+1)$.
As mentioned before, the two kinds of currents occur alternatively
depending on the even or odd property of the weight $h$.

The OPE
  between
$\Phi^{(h_1=4),i}_{\frac{1}{2}}(z)$ and $
  \Phi^{(h_2=1),j}_{\frac{3}{2}}(w)$ with $i=j$ where the weights
  satisfy $h_1=h_2+3$ can be calculated.
  The fifth order pole of this OPE gives us
  the vanishing structure constant appearing in the current
  $\Phi_{0}^{(h_1+h_2+1-h=2)}(w)$ with weight $h=4$ from our calculation.
  There appears
  the current $\tilde{\Phi}_{2}^{(0)}(w)$ term
  with nonzero structure constant
  $\frac{256}{3} (\la-1) \la (2 \la-1) (2 \la+1)$
  in the fifth order pole of this OPE
  and this is not consistent with (\ref{eightcase}) because
  all the coefficients vanish for the even weight $h=4$.
  This implies that there is additional term on the right hand side
  of the OPE for
  the weights $h_1=4$ and $h_2=1$ which are outside of the
  allowed region we consider.

%%%%%%%%%%%%%%%%%%%%
%\subsubsection{The central term}
%%%%%%%%%%%%%%%%%%%%
Here the central term is given by
\bea
c'_Q & = &  8(-4)^{h_1+h_2-4} \,
\de_{b \bar{a}} \, \de_{d \bar{c}} \, \Bigg[
-Q_{\la, \bar{a} b}^{(h_1+1)+}(z) \, Q_{\la, \bar{c} d}^{(h_2+2)+}(w)  + 
Q_{\la, \bar{a} b}^{(h_1+1)-}(z) \, Q_{\la, \bar{c} d}^{(h_2+2)-}(w)
\nonu \\
& - &
Q_{\la, \bar{a} b}^{(h_1+1)+}(z) \, Q_{\la, \bar{c} d}^{(h_2+2)-}(w)
+ Q_{\la, \bar{a} b}^{(h_1+1)-}(z) \, Q_{\la, \bar{c} d}^{(h_2+2)+}(w)
\label{cQ'}
 \\
& + & (-1)^{h_1+h_2}\,
Q_{\la, \bar{c} d}^{(h_2+2)+}(z) \, Q_{\la, \bar{a} b}^{(h_1+1)+}(w)  -
 (-1)^{h_1+h_2}\,Q_{\la, \bar{c} d}^{(h_2+2)-}(z) \, Q_{\la, \bar{a} b}^{(h_1+1)-}(w)
\nonu \\
& + & (-1)^{h_1+h_2}\,
Q_{\la, \bar{c} d}^{(h_2+2)+}(z) \, Q_{\la, \bar{a} b}^{(h_1+1)-}(w)
-  (-1)^{h_1+h_2}\,
Q_{\la, \bar{c} d}^{(h_2+2)-}(z) \, Q_{\la, \bar{a} b}^{(h_1+1)+}(w)
\Bigg]_{\frac{1}{
    (z-w)^{h_1+h_2+2}}}\, .
\nonu
\eea
Note that from the inside of the bracket in (\ref{cQ}),
after the $h_2$ is replaced with $(h_2+1)$, we obtain the
above result (\ref{cQ'}).

%%%%%%%%%%%%%%%%%
\subsection{The commutator relation between the second component and the
last component with $h_1=h_2+1,h_2+2$  for nonzero $\la$}
%%%%%%%%%%%%%%%%%

By using the relations (\ref{two}) and (\ref{last}),
we can rewrite it in terms of the corresponding commutators
in (\ref{res2}), (\ref{res3}), (\ref{res4}), and
(\ref{res5}) and using the relations (\ref{twotwo})
and (\ref{RELL}),
the following result can be obtained
\bea
&& \big[(\Phi^{(h_1),i}_{\frac{1}{2}})_r,
  (\tilde{\Phi}^{(h_2)}_{2})_m \big]  = 
4(-4)^{h_1-4} \, 4(-4)^{h_2-4}
\Bigg[ 
\Bigg( 
 \sum_{h=-1}^{h_1+h_2-1} \, q^h\, 
  q_{\mathrm{F}}^{h_2+2,h_1+\frac{1}{2}, h}(m,r,\la)
  \nonu \\
  && +  \sum_{h=-1}^{h_1+h_2} \,
  q^h\, (-1)^h\, 
q_{\mathrm{F}}^{h_2+2,h_1+\frac{1}{2}, h}(m,r,\la)
 +  
\sum_{h=-1}^{h_1+h_2-1} \,
 q^h\, 
q_{\mathrm{B}}^{h_2+2,h_1+\frac{1}{2}, h}(m,r,\la)
\nonu \\
&& + 
\sum_{h=-1}^{h_1+h_2} \,
 q^h\, (-1)^h\,
q_{\mathrm{B}}^{h_2+2,h_1+\frac{1}{2}, h}(m,r,\la)
\Bigg)\,
 \frac{1}{4(-4)^{h_1+h_2-h-4}}\,
(\Phi_{\frac{1}{2}}^{(h_1+h_2-h),i})_{m+r}
\nonu \\
 &&   +
\Bigg(-  
  \sum_{h=-1}^{h_1+h_2-1} \, q^h\, 
  q_{\mathrm{F}}^{h_2+2,h_1+\frac{1}{2}, h}(m,r,\la)
  \nonu \\
  && +  \sum_{h=-1}^{h_1+h_2} \,
  q^h\, (-1)^h\, 
q_{\mathrm{F}}^{h_2+2,h_1+\frac{1}{2}, h}(m,r,\la)
 -  
\sum_{h=-1}^{h_1+h_2-1} \,
 q^h\, 
q_{\mathrm{B}}^{h_2+2,h_1+\frac{1}{2}, h}(m,r,\la)
\nonu \\
&& +  
\sum_{h=-1}^{h_1+h_2} \,
 q^h\, (-1)^h\,
q_{\mathrm{B}}^{h_2+2,h_1+\frac{1}{2}, h}(m,r,\la)
\Bigg)\,
\frac{1}{4(-4)^{h_1+h_2-h-5}}\,
(\tilde{\Phi}_{\frac{3}{2}}^{(h_1+h_2-1-h),i})_{m+r} \,
\Bigg]\, .
\label{ninecase}
\eea
The field contents on the right hand side
in (\ref{ninecase}) look similar to
the ones in (\ref{Tworel}).
Except the four terms having $(-1)^h$ factor for the weight $h=h_1+h_2$,
there are some cancellation between the currents
due to the factor $(-1)^h$.
In other words, the two kinds of currents appear alternatively
depending on the even or odd property of the weight $h$.
Note the presence of different ordering in the elements of the structure
constants.

Let us consider the OPE
  between
$\Phi^{(h_1=5),i}_{\frac{1}{2}}(z)$ and $
  \Phi^{(h_2=1)}_{2}(w)$ where the weights satisfy $h_1=h_2+4$.
  The seventh order pole of this OPE gives us
  the structure constant $\frac{16384}{21} (\la-1) (2 \la+1)
  (4 \la-1) (6 \la^2-3 \la+5)$
  appearing in the current
  $G^{i}(w)$ with weight $h=5$.
  By substituting the various expressions in
  the corresponding terms of (\ref{ninecase}),
  we can check that we obtain the above structure constant correctly
  where $q_F^{3,\frac{7}{2},5}$ term corresponds to
  $-\frac{4096}{21}  (\la-1) (\la+1)
  (2 \la+1) (8 \la^3+8 \la^2+24 \la-25)$
  while  $q_B^{3,\frac{7}{2},5}$ term
  corresponds to $\frac{4096}{21} (\la-1) (2 \la-3) (2 \la+1)
  (4 \la^3-10 \la^2+19 \la+5)$.
  There appears
  the current $\Phi_{\frac{1}{2}}^{(1),i}(w)$ term
with nonzero structure constant having the $\la$ factor 
  in the seventh order pole of the OPE
  and this is not consistent with (\ref{ninecase}) because
  all the coefficients vanish for the odd weight $h=5$.
  There is additional term on the right hand side
  of the OPE for
  the weights $h_1=5$ and $h_2=1$ which are outside of the above
  allowed region.

%%%%%%%%%%%%%%%%%
\subsection{The commutator between the third components
with $h_1= h_2-1,h_2, h_2+1$  for nonzero $\la$}
%%%%%%%%%%%%%%%%%

By using the relations (\ref{three}), we can rewrite them in terms of
the commutators in (\ref{FIRST}), (\ref{SECOND}),
(\ref{THIRD}), (\ref{FOURTH}), (\ref{FIFTH}), and
(\ref{res1}). Then we can use (\ref{rel1}), (\ref{rell})
and (\ref{RELL1}), we obtain the following commutator
\bea
&& \big[(\Phi^{(h_1),ij}_{1})_m,
  (\Phi^{(h_2),kl}_{1})_n \big]  = 
16(-4)^{h_1-4} \, (-4)^{h_2-4}
\Bigg[ \nonu \\
  & &
\left(\begin{array}{c}
m+h_1 \\  h_1+h_2+1 \\
 \end{array}\right) \, q^{h_1+h_2-2} \, \Bigg( 
-4\, ( \de^{ik} \, \de^{jl} - \de^{jk}\, \de^{il}) \, (c_F'+c_B')
- 4 \, \varepsilon^{ijkl} \, (c_F'-c_B') \Bigg)
\,  \de_{m+n}
  \nonu \\
  && +  (\de^{ik} \, \de^{jl}-\de^{il}\, \de^{jk}) \,
  \sum_{h=0,\mbox{\footnotesize even}}^{h_1+h_2-1} \,
 q^h\, 
  \Bigg(  p_{\mathrm{F}}^{h_1+1,h_2+1, h}(m,n,\la)
 -
 p_{\mathrm{B}}^{h_1+1,h_2+1, h}(m,n,\la)
 \Bigg)\nonu \\
 && \times \frac{4}{(-4)^{h_1+h_2-h-2}}\,
(\Phi_{0}^{(h_1+h_2-h)})_{m+n}
\nonu \\
&& + (\de^{ik} \,
\de^{jl} -\de^{il}\, \de^{jk}) \,
\sum_{h=0,\mbox{\footnotesize even}}^{h_1+h_2-1} \, \Bigg(
(h_1+h_2-h-1+2\la)\, q^h\, 
 p_{\mathrm{F}}^{h_1+1,h_2+1, h}(m,n,\la)
\nonu \\
&& +
(h_1+h_2-h-2\la)\, q^h\, 
p_{\mathrm{B}}^{h_1+1,h_2+1, h}(m,n,\la)
\,
\Bigg)\nonu \\
&& \times \frac{1}{
  2(2h_1+2h_2-2h-1)(-4)^{h_1+h_2-h-6}}
\, (\tilde{\Phi}_{2}^{(h_1+h_2-h-2)})_{m+n}
\nonu \\
  && +  \varepsilon^{ijkl} \,
  \sum_{h=0,\mbox{\footnotesize even}}^{h_1+h_2-1} \,
 q^h\, 
  \Bigg(  p_{\mathrm{F}}^{h_1+1,h_2+1, h}(m,n,\la)
 +
 p_{\mathrm{B}}^{h_1+1,h_2+1, h}(m,n,\la)
 \Bigg)\nonu \\
 && \times \frac{4}{(-4)^{h_1+h_2-h-2}}\,
(\Phi_{0}^{(h_1+h_2-h)})_{m+n}
\nonu \\
&& + 
\varepsilon^{ijkl} \, \sum_{h=0,\mbox{\footnotesize even}}^{h_1+h_2-1} \, \Bigg(
(h_1+h_2-h-1+2\la)\, q^h\, 
 p_{\mathrm{F}}^{h_1+1,h_2+1, h}(m,n,\la)
\nonu \\
&& -
(h_1+h_2-h-2\la)\, q^h\, 
p_{\mathrm{B}}^{h_1+1,h_2+1, h}(m,n,\la)
\,
\Bigg)\nonu \\
&& \times \frac{1}{
  2(2h_1+2h_2-2h-1)(-4)^{h_1+h_2-h-6}}
\, (\tilde{\Phi}_{2}^{(h_1+h_2-h-2)})_{m+n}
\nonu \\
%%%%%%%%%%%%%%%%%%%%%%%%%%%%%%%%%%%%%%%%%%%%%
&& + \de^{ik}\,
  \sum_{h=-1,\mbox{\footnotesize odd}}^{h_1+h_2-1} \, q^h\, 
  p_{\mathrm{F}}^{h_1+1,h_2+1, h}(m,n,\la)
  \nonu \\
  && \times
 \frac{1}{4(-4)^{h_1+h_2-h-5}}\,
  \Bigg(\Phi_{1}^{(h_1+h_2-h-1),jl} + \frac{1}{2}\,
  \varepsilon^{jlm_1n_1} \,\Phi_{1}^{(h_1+h_2-h-1),m_1n_1} \Bigg)_{m+n}
\nonu \\
 &&   +  \de^{ik}\,
  \sum_{h=-1,\mbox{\footnotesize odd}}^{h_1+h_2-1} \, q^h\, 
  p_{\mathrm{B}}^{h_1+1,h_2+1, h}(m,n,\la)
  \nonu \\
  && \times
\frac{1}{4(-4)^{h_1+h_2-h-5}}\,
  \Bigg(\Phi_{1}^{(h_1+h_2-h-1),jl} - \frac{1}{2}\,
  \varepsilon^{jlm_1n_1} \,\Phi_{1}^{(h_1+h_2-h-1),m_1n_1}\Bigg)_{m+n} \,
  \nonu \\
%%%%%%%%%%%%%%%%%%%%%%%%%
  && - \de^{il}\,
  \sum_{h=-1,\mbox{\footnotesize odd}}^{h_1+h_2-1} \, q^h\, 
  p_{\mathrm{F}}^{h_1+1,h_2+1, h}(m,n,\la)
  \nonu \\
  && \times
 \frac{1}{4(-4)^{h_1+h_2-h-5}}\,
  \Bigg(\Phi_{1}^{(h_1+h_2-h-1),jk} + \frac{1}{2}\,
  \varepsilon^{jkm_1n_1} \,\Phi_{1}^{(h_1+h_2-h-1),m_1n_1} \Bigg)_{m+n}
\nonu \\
 &&   -  \de^{il}\,
  \sum_{h=-1,\mbox{\footnotesize odd}}^{h_1+h_2-1} \, q^h\, 
  p_{\mathrm{B}}^{h_1+1,h_2+1, h}(m,n,\la)
  \nonu \\
  && \times
\frac{1}{4(-4)^{h_1+h_2-h-5}}\,
  \Bigg(\Phi_{1}^{(h_1+h_2-h-1),jk} - \frac{1}{2}\,
  \varepsilon^{jkm_1n_1} \,\Phi_{1}^{(h_1+h_2-h-1),m_1n_1}\Bigg)_{m+n} \, 
  \nonu \\
%%%%%%%%%%%%%%%%%%%%%%%%%%%%%%%%%%%%%%%%%%
  && - \de^{jk}\,
  \sum_{h=-1,\mbox{\footnotesize odd}}^{h_1+h_2-1} \, q^h\, 
  p_{\mathrm{F}}^{h_1+1,h_2+1, h}(m,n,\la)
  \nonu \\
  && \times
 \frac{1}{4(-4)^{h_1+h_2-h-5}}\,
  \Bigg(\Phi_{1}^{(h_1+h_2-h-1),il} + \frac{1}{2}\,
  \varepsilon^{ilm_1n_1} \,\Phi_{1}^{(h_1+h_2-h-1),m_1n_1} \Bigg)_{m+n}
\nonu \\
 &&   -  \de^{jk}\,
  \sum_{h=-1,\mbox{\footnotesize odd}}^{h_1+h_2-1} \, q^h\, 
  p_{\mathrm{B}}^{h_1+1,h_2+1, h}(m,n,\la)
  \nonu \\
  && \times
\frac{1}{4(-4)^{h_1+h_2-h-5}}\,
  \Bigg(\Phi_{1}^{(h_1+h_2-h-1),il} - \frac{1}{2}\,
  \varepsilon^{ilm_1n_1} \,\Phi_{1}^{(h_1+h_2-h-1),m_1n_1}\Bigg)_{m+n} \,
  \nonu \\
%%%%%%%%%%%%%%%%%%%%%%%%%%%%%%%%%%
  && + \de^{jl}\,
  \sum_{h=-1,\mbox{\footnotesize odd}}^{h_1+h_2-1} \, q^h\, 
  p_{\mathrm{F}}^{h_1+1,h_2+1, h}(m,n,\la)
  \nonu \\
  && \times
 \frac{1}{4(-4)^{h_1+h_2-h-5}}\,
  \Bigg(\Phi_{1}^{(h_1+h_2-h-1),ik} + \frac{1}{2}\,
  \varepsilon^{ikm_1n_1} \,\Phi_{1}^{(h_1+h_2-h-1),m_1n_1} \Bigg)_{m+n}
\nonu \\
 &&   +  \de^{jl}\,
  \sum_{h=-1,\mbox{\footnotesize odd}}^{h_1+h_2-1} \, q^h\, 
  p_{\mathrm{B}}^{h_1+1,h_2+1, h}(m,n,\la)
  \nonu \\
  && \times
\frac{1}{4(-4)^{h_1+h_2-h-5}}\,
  \Bigg(\Phi_{1}^{(h_1+h_2-h-1),ik} - \frac{1}{2}\,
  \varepsilon^{ikm_1n_1} \,\Phi_{1}^{(h_1+h_2-h-1),m_1n_1}\Bigg)_{m+n} \,
  \Bigg] \, .
\label{tencase}
\eea
The central terms  will be given later.
There are antisymmetric properties both in the indices $i$ and $j$
and in the indices $k$ and $l$ on the left hand side above.
This will give us rather complicated expressions on the right hand side.
We do not use any simplified notations.

Let us consider the OPE
  between
$\Phi^{(h_1=3),ij}_{1}(z)$ and $
  \Phi^{(h_2=1),kl}_{1}(w)$ where the weights satisfy $h_1=h_2+2$.
  The fifth order pole of this OPE gives us
  the nonvanishing structure constants
  in the current 
  $\Phi^{(h_1+h_2-h=1)}_{0}(w)$ and
 $\tilde{\Phi}^{(h_1+h_2-2-h=-1)}_{2}(w)$
  with weight $h=3$.  
This is not consistent with (\ref{tencase}) because
the dummy variable $h$ can appear as even number.
There is additional term on the right hand side
  of the OPE for
  the weights $h_1=3$ and $h_2=1$ which do not satisfy the above
  constraint between the weights.

%%%%%%%%%%%%%%%%%%%%%
%\subsubsection{The central term}
%%%%%%%%%%%%%%%%%%%%%
The central term can be written in terms of
\bea
c'_F & = &  (-4)^{h_1+h_2-4}\, \de_{b \bar{a}} \, \de_{d \bar{c}} \, \Bigg[
  \frac{(h_1+2\la)}{(2h_1+1)} \, \frac{(h_2+2\la)}{(2h_2+1)}\,
V_{\la, \bar{a} b}^{(h_1+1)+}(z) \, V_{\la, \bar{c} d}^{(h_2+1)+}(w) \nonu \\
& + &
V_{\la, \bar{a} b}^{(h_1+1)-}(z) \, V_{\la, \bar{c} d}^{(h_2+1)-}(w)+
\frac{(h_1+2\la)}{(2h_1+1)}\,
V_{\la, \bar{a} b}^{(h_1+1)+}(z) \, V_{\la, \bar{c} d}^{(h_2+1)-}(w)
\nonu \\
&+& \frac{(h_2+2\la)}{(2h_2+1)}\,
V_{\la, \bar{a} b}^{(h_1+1)-}(z) \, V_{\la, \bar{c} d}^{(h_2+1)+}(w) \Bigg]_{\frac{1}{
    (z-w)^{h_1+h_2+2}}}\, ,
\label{cF'}
\eea
and 
\bea
c'_B & = &  (-4)^{h_1+h_2-4}\, \de_{b \bar{a}} \, \de_{d \bar{c}} \, \Bigg[
  \frac{(h_1+1-2\la)}{(2h_1+1)} \, \frac{(h_2+1-2\la)}{(2h_2+1)}\,
V_{\la, \bar{a} b}^{(h_1+1)+}(z) \, V_{\la, \bar{c} d}^{(h_2+1)+}(w) \nonu \\
& + &
V_{\la, \bar{a} b}^{(h_1+1)-}(z) \, V_{\la, \bar{c} d}^{(h_2+1)-}(w)-
\frac{(h_1+1-2\la)}{(2h_1+1)}\,
V_{\la, \bar{a} b}^{(h_1+1)+}(z) \, V_{\la, \bar{c} d}^{(h_2+1)-}(w)
\nonu \\
&-& \frac{(h_2+1-2\la)}{(2h_2+1)}\,
V_{\la, \bar{a} b}^{(h_1+1)-}(z) \, V_{\la, \bar{c} d}^{(h_2+1)+}(w)
\Bigg]_{\frac{1}{
    (z-w)^{h_1+h_2+2}}}\, .
\label{cB'}
\eea
We can easily see that after $h_1$ inside the bracket of (\ref{cF})
and $h_2$ inside the bracket of (\ref{cB}) are
replaced by $(h_1+1)$ and $(h_2+1)$ respectively,
the central terms in (\ref{cF'}) and (\ref{cB'})
can be obtained.

%%%%%%%%%%%%%%%%%
\subsection{The commutator relation between the third component and the
fourth component with $h_1=h_2, h_2+1$  for nonzero $\la$}
%%%%%%%%%%%%%%%%%

By using the relations (\ref{three}) and (\ref{four})
we can have the commutators in (\ref{res2}), (\ref{res3}), (\ref{res4})
and (\ref{res5}). Then by using (\ref{twotwo}) and (\ref{RELL}),
the following result can be determined
\bea
&& \big[(\Phi^{(h_1),ij}_{1})_r,
  (\tilde{\Phi}^{(h_2),k}_{\frac{3}{2}})_m \big]  = 
4(-4)^{h_1-4} \, 4(-4)^{h_2-4}
\Bigg[ 
\de^{ik} \, \Bigg( 
 \sum_{h=-1}^{h_1+h_2-1} \, q^h\, 
  q_{\mathrm{F}}^{h_1+1,h_2+\frac{3}{2}, h}(m,r,\la)
  \nonu \\
  && +  \sum_{h=-1}^{h_1+h_2} \,
  q^h\, (-1)^h\, 
q_{\mathrm{F}}^{h_1+1,h_2+\frac{3}{2}, h}(m,r,\la)
 -  
\sum_{h=-1}^{h_1+h_2-1} \,
 q^h\, 
q_{\mathrm{B}}^{h_1+1,h_2+\frac{3}{2}, h}(m,r,\la)
\nonu \\
&& - 
\sum_{h=-1}^{h_1+h_2} \,
 q^h\, (-1)^h\,
q_{\mathrm{B}}^{h_1+1,h_2+\frac{3}{2}, h}(m,r,\la)
\Bigg)\,
 \frac{1}{4(-4)^{h_1+h_2-h-4}}\,
(\Phi_{\frac{1}{2}}^{(h_1+h_2-h),j})_{m+r}
\nonu \\
 &&   +
\de^{ik} \,\Bigg(-  
  \sum_{h=-1}^{h_1+h_2-1} \, q^h\, 
  q_{\mathrm{F}}^{h_1+1,h_2+\frac{3}{2}, h}(m,r,\la)
  \nonu \\
  && +  \sum_{h=-1}^{h_1+h_2} \,
  q^h\, (-1)^h\, 
q_{\mathrm{F}}^{h_1+1,h_2+\frac{3}{2}, h}(m,r,\la)
 +  
\sum_{h=-1}^{h_1+h_2-1} \,
 q^h\, 
q_{\mathrm{B}}^{h_1+1,h_2+\frac{3}{2}, h}(m,r,\la)
\nonu \\
&& -  
\sum_{h=-1}^{h_1+h_2} \,
 q^h\, (-1)^h\,
q_{\mathrm{B}}^{h_1+1,h_2+\frac{3}{2}, h}(m,r,\la)
\Bigg)\,
\frac{1}{4(-4)^{h_1+h_2-h-5}}\,
(\tilde{\Phi}_{\frac{3}{2}}^{(h_1+h_2-1-h),j})_{m+r} \,
\nonu \\
%%%%%%%%%%%%%%%%%%%%%%%%%%%%%%%%%%%%%%%%%%%%%%%%
&& -\de^{jk} \, \Bigg( 
 \sum_{h=-1}^{h_1+h_2-1} \, q^h\, 
  q_{\mathrm{F}}^{h_1+1,h_2+\frac{3}{2}, h}(m,r,\la)
  \nonu \\
  && +  \sum_{h=-1}^{h_1+h_2} \,
  q^h\, (-1)^h\, 
q_{\mathrm{F}}^{h_1+1,h_2+\frac{3}{2}, h}(m,r,\la)
 -  
\sum_{h=-1}^{h_1+h_2-1} \,
 q^h\, 
q_{\mathrm{B}}^{h_1+1,h_2+\frac{3}{2}, h}(m,r,\la)
\nonu \\
&& - 
\sum_{h=-1}^{h_1+h_2} \,
 q^h\, (-1)^h\,
q_{\mathrm{B}}^{h_1+1,h_2+\frac{3}{2}, h}(m,r,\la)
\Bigg)\,
 \frac{1}{4(-4)^{h_1+h_2-h-4}}\,
(\Phi_{\frac{1}{2}}^{(h_1+h_2-h),i})_{m+r}
\nonu \\
 &&   -
\de^{jk} \,\Bigg(-  
  \sum_{h=-1}^{h_1+h_2-1} \, q^h\, 
  q_{\mathrm{F}}^{h_1+1,h_2+\frac{3}{2}, h}(m,r,\la)
  \nonu \\
  && +  \sum_{h=-1}^{h_1+h_2} \,
  q^h\, (-1)^h\, 
q_{\mathrm{F}}^{h_1+1,h_2+\frac{3}{2}, h}(m,r,\la)
 +  
\sum_{h=-1}^{h_1+h_2-1} \,
 q^h\, 
q_{\mathrm{B}}^{h_1+1,h_2+\frac{3}{2}, h}(m,r,\la)
\nonu \\
&& -  
\sum_{h=-1}^{h_1+h_2} \,
 q^h\, (-1)^h\,
q_{\mathrm{B}}^{h_1+1,h_2+\frac{3}{2}, h}(m,r,\la)
\Bigg)\,
\frac{1}{4(-4)^{h_1+h_2-h-5}}\,
(\tilde{\Phi}_{\frac{3}{2}}^{(h_1+h_2-1-h),i})_{m+r} \,
\nonu \\
%%%%%%%%%%%%%%%%%%%%%%%%%%%%%%%%%%%%%%%%%%%%%%%%%%%
&& + \varepsilon^{ijkl} \, \Bigg( 
 \sum_{h=-1}^{h_1+h_2-1} \, q^h\, 
  q_{\mathrm{F}}^{h_2+2,h_1+\frac{1}{2}, h}(m,r,\la)
  \nonu \\
  && +  \sum_{h=-1}^{h_1+h_2} \,
  q^h\, (-1)^h\, 
q_{\mathrm{F}}^{h_2+2,h_1+\frac{1}{2}, h}(m,r,\la)
 +  
\sum_{h=-1}^{h_1+h_2-1} \,
 q^h\, 
q_{\mathrm{B}}^{h_2+2,h_1+\frac{1}{2}, h}(m,r,\la)
\nonu \\
&& + 
\sum_{h=-1}^{h_1+h_2} \,
 q^h\, (-1)^h\,
q_{\mathrm{B}}^{h_2+2,h_1+\frac{1}{2}, h}(m,r,\la)
\Bigg)\,
 \frac{1}{4(-4)^{h_1+h_2-h-4}}\,
(\Phi_{\frac{1}{2}}^{(h_1+h_2-h),l})_{m+r}
\nonu \\
 &&   +
\varepsilon^{ijkl} \,\Bigg(-  
  \sum_{h=-1}^{h_1+h_2-1} \, q^h\, 
  q_{\mathrm{F}}^{h_2+2,h_1+\frac{1}{2}, h}(m,r,\la)
  \nonu \\
  && +  \sum_{h=-1}^{h_1+h_2} \,
  q^h\, (-1)^h\, 
q_{\mathrm{F}}^{h_2+2,h_1+\frac{1}{2}, h}(m,r,\la)
 - 
\sum_{h=-1}^{h_1+h_2-1} \,
 q^h\, 
q_{\mathrm{B}}^{h_2+2,h_1+\frac{1}{2}, h}(m,r,\la)
\nonu \\
&& +  
\sum_{h=-1}^{h_1+h_2} \,
 q^h\, (-1)^h\,
q_{\mathrm{B}}^{h_2+2,h_1+\frac{1}{2}, h}(m,r,\la)
\Bigg)\,
\frac{1}{4(-4)^{h_1+h_2-h-5}}\,
(\tilde{\Phi}_{\frac{3}{2}}^{(h_1+h_2-1-h),l})_{m+r} \,
\Bigg]\, .
\label{elevencase}
\eea
The field contents in (\ref{elevencase}) are similar to the ones
in (\ref{sevencase}). The antisymmetric property in the indices
$i$ and $j$ on the left hand side is seen on the right hand side.
Except the case of $h=h_1+h_2$, due to the $(-1)^h$ factor,
the two kinds of currents appear alternatively.

Let us consider the OPE
  between
$\Phi^{(h_1=3),ij}_{1}(z)$ and $
  \tilde{\Phi}^{(h_2=1),k}_{\frac{3}{2}}(w)$ where $h_1=h_2+2$.
  The fifth order pole of this OPE gives us
  the structure constant $\frac{32}{5}
  (\la-1) (\la+1) (2 \la-3) (2 \la+1)$ appearing in the current
  $G^{i}(w)$ of ${\cal N}=4$ stress energy tensor with weight $h=3$.
  By substituting the various expressions in the corresponding terms of
  (\ref{elevencase}),
  we can check that we obtain the above structure constant correctly
  where $q_F^{4,\frac{5}{2},3}$ term corresponds to
  $-\frac{128}{5}  (\la-1) (2 \la+1)
  (2 \la^2-5 \la-2)$
  while  $q_B^{4,\frac{5}{2},3}$ term
  corresponds to $\frac{128}{5} (\la-1) (2 \la+1)
  (2 \la^2+3 \la-4)$.
  On the other hand, there exists the current 
  $\Phi_{\frac{1}{2}}^{(h_1+h_2-h=1),i}(w)$ term with
  the structure constant
  $-\frac{512}{3}  (\la-1) \la (2 \la-1) (2 \la+1)$
  having the $\la$ factor.
  We can check that this structure constant
  is equal to the
  one in the first current terms in (\ref{elevencase})
  where the previous
$q_F^{4,\frac{5}{2},3}$
  and $q_B^{4,\frac{5}{2},3}$ are replaced by
$q_F^{3,\frac{5}{2},3}$
  and $q_B^{3,\frac{5}{2},3}$ respectively.
  Note that there is a replacement of $h_1$ by $(h_1-1)$.
  
%%%%%%%%%%%%%%%%%
\subsection{The commutator relation between the third component and the
last component with $h_1=h_2, h_2+1,h_2+2$  for nonzero $\la$}
%%%%%%%%%%%%%%%%%

By using (\ref{three}) and (\ref{last}) together with (\ref{FIRST}),
(\ref{SECOND}),
(\ref{THIRD}), (\ref{FOURTH}), (\ref{FIFTH}), (\ref{res1}),
(\ref{rel1}), (\ref{rell}), and
(\ref{RELL1}), we determine the following commutator
\bea
&& \big[(\Phi^{(h_1),ij}_{1})_m,
  (\tilde{\Phi}^{(h_2)}_{2})_n \big]  = 
16(-4)^{h_1-4} \, (-4)^{h_2-4}
\Bigg[ \nonu \\
&& -
  \sum_{h=0,\mbox{\footnotesize even}}^{h_1+h_2} \, q^h\, 
 p_{\mathrm{F}}^{h_1+1,h_2+2, h}(m,n,\la)
  \nonu \\
  && \times
 \frac{1}{4(-4)^{h_1+h_2-h-4}}\,
  \Bigg(\Phi_{1}^{(h_1+h_2-h),ij} + \frac{1}{2}\,
  \varepsilon^{ijkl} \,\Phi_{1}^{(h_1+h_2-h),kl} \Bigg)_{m+n}
\nonu \\
 &&   -
  \sum_{h=0,\mbox{\footnotesize even}}^{h_1+h_2} \, q^h\, 
  p_{\mathrm{B}}^{h_1+1,h_2+2, h}(m,n,\la)
  \nonu \\
  && \times
\frac{1}{4(-4)^{h_1+h_2-h-4}}\,
  \Bigg(\Phi_{1}^{(h_1+h_2-h),ij} - \frac{1}{2}\,
  \varepsilon^{ijkl} \,\Phi_{1}^{(h_1+h_2-h),kl}\Bigg)_{m+n} \,
\Bigg] \, .
\label{twelvecase}
\eea
The field contents on the right hand side of (\ref{twelvecase})
are similar to the ones in (\ref{Threerel}).
At the maximum value of the weight $h=h_1+h_2$,
we observe that there exists the current of the weight-$1$ current
having $SO(4)$ indices
of the ${\cal N}=4$ stress energy tensor.

Let us calculate the OPE
  between
$\Phi^{(h_1=5),ij}_{1}(z)$ and $
  \Phi^{(h_2=2)}_{2}(w)$ where the weights satisfy
  $h_1=h_2+3$.
  The ninth order pole of this OPE gives us
  the nonvanishing structure constants
  in the current
  $\Phi^{(h_1+h_2-h=1),ij}_{1}(w)$ and
  $\frac{1}{2}\, \varepsilon^{ijkl}
  \Phi^{(h_1+h_2-2-h=-1),kl}_{1}(w)$
  with weight $h=7$.
  They have the $\la$ factor explicitly.
This is not consistent with (\ref{twelvecase}) because
the dummy variable $h$ can appear as even number.
This implies that
there is additional term on the right hand side
  of the OPE for
  the weights $h_1=5$ and $h_2=2$ which are outside of the
  above allowed region.
  
%%%%%%%%%%%%%%%%%
\subsection{The anticommutator relation between the fourth components
with $h_1=h_2$  for nonzero $\la$}
%%%%%%%%%%%%%%%%%

By using the relations (\ref{four}), we can rewrite in terms of the
in terms of anticommutator in (\ref{res6}).
The relations (\ref{rel1}), (\ref{rell}), and (\ref{RELL1}) can be used further.
We obtain
\bea
&& \big\{(\tilde{\Phi}^{(h_1),i}_{\frac{3}{2}})_r,
  (\tilde{\Phi}^{(h_2),j}_{\frac{3}{2}})_s \big\}  = 
16(-4)^{h_1-4} \, (-4)^{h_2-4}
\Bigg[ \nonu \\
  & &
\left(\begin{array}{c}
r+h_1+\frac{1}{2} \\  h_1+h_2+2 \\
 \end{array}\right) \, q^{h_1+h_2-1} \, c''_Q \,
 \de^{ij}\,  \de_{r+s}
  \nonu \\
  && -  \de^{ij}\, \sum_{h=0}^{h_1+h_2+1} \,
 q^h\, 
  (1+ (-1)^h)\,
  \Bigg(  o_{\mathrm{F}}^{h_1+\frac{3}{2},h_2+\frac{3}{2}, h}(r,s,\la)
 - 
 o_{\mathrm{B}}^{h_1+\frac{3}{2},h_2+\frac{3}{2}, h}(r,s,\la)
 \Bigg)\nonu \\
 && \times \frac{1}{2(-4)^{h_1+h_2-h}}\,
(\Phi_{0}^{(h_1+h_2+2-h)})_{r+s}
\nonu \\
&& - \de^{ij}\, \sum_{h=0}^{h_1+h_2+1} \, \Bigg(
(h_1+h_2+1-h+2\la)\, q^h\, 
(1+(-1)^h) \, o_{\mathrm{F}}^{h_1+\frac{3}{2},h_2+\frac{3}{2}, h}(r,s,\la)
\nonu \\
&& -
(h_1+h_2+2-h-2\la)\, q^h\, 
(1+(-1)^h) \,
o_{\mathrm{B}}^{h_1+\frac{3}{2},h_2+\frac{3}{2}, h}(r,s,\la)
\,
\Bigg)\nonu \\
&& \times \frac{1}{
  8(2h_1+2h_2-2h+3)(-4)^{h_1+h_2-h-4}}
\, (\tilde{\Phi}_{2}^{(h_1+h_2-h)})_{r+s}
\nonu \\
&& +
  \sum_{h=0}^{h_1+h_2+1} \, q^h\, 
  (1-(-1)^h)\, o_{\mathrm{F}}^{h_1+\frac{3}{2},h_2+\frac{3}{2}, h}(r,s,\la)
  \nonu \\
  && \times
 \frac{1}{32(-4)^{h_1+h_2-h-3}}\,
  \Bigg(\Phi_{1}^{(h_1+h_2+1-h),ij} + \frac{1}{2}\,
  \varepsilon^{ijkl} \,\Phi_{1}^{(h_1+h_2-h),kl} \Bigg)_{r+s}
\nonu \\
 &&   -
  \sum_{h=0}^{h_1+h_2+1} \, q^h\, 
 (1-(-1)^h)\,  o_{\mathrm{B}}^{h_1+\frac{3}{2},h_2+\frac{3}{2}, h}(r,s,\la)
  \nonu \\
  && \times
\frac{1}{32(-4)^{h_1+h_2-h-3}}\,
  \Bigg(\Phi_{1}^{(h_1+h_2+1-h),ij} - \frac{1}{2}\,
  \varepsilon^{ijkl} \,\Phi_{1}^{(h_1+h_2+1-h),kl}\Bigg)_{r+s} \,
\Bigg] \, .
\label{thirteencase}
\eea
The central term will be given later.
The structure of (\ref{thirteencase}) is very similar to
the one of (\ref{sixcase}). Once we replace the
weights $h_1$ and $h_2$ with $(h_1+1)$ and $(h_2+1)$ in the latter
respectively,
then we can see most of the structure of (\ref{thirteencase}). 
Even the factor $(1 \pm (-1)^h)$ appears precisely.

Let us calculate the OPE
  between
$\Phi^{(h_1=3),i}_{\frac{3}{2}}(z)$ and $
  \Phi^{(h_2=1),j}_{\frac{3}{2}}(w)$ with $i=j$ where the weights
  satisfy $h_1=h_2+2$.
  The sixth order pole of this OPE gives us
  the nonvanishing structure constants
  in the currents 
  $\Phi^{(h_1+h_2+2-h=1)}_{0}(w)$ and
  $\tilde{\Phi}^{(h_1+h_2-h=-1),kl}_{2}(w)$
  with weight $h=5$.
  They have the $\la$ factor.
This is not consistent with (\ref{thirteencase}) because
all the coefficients are vanishing for odd number $h$.
There is additional term on the right hand side
  of the OPE for
  the weights $h_1=3$ and $h_2=1$ which do not satisfy the above
  constraint between the weights.

%%%%%%%%%%%%%%%%%%%%%%%%
%\subsubsection{The central term}
%%%%%%%%%%%%%%%%%%%%%%%%
The central term contains
\bea
c''_Q & = &  8(-4)^{h_1+h_2-4} \,
\de_{b \bar{a}} \, \de_{d \bar{c}} \, \Bigg[
-Q_{\la, \bar{a} b}^{(h_1+2)+}(z) \, Q_{\la, \bar{c} d}^{(h_2+2)+}(w)  + 
Q_{\la, \bar{a} b}^{(h_1+2)-}(z) \, Q_{\la, \bar{c} d}^{(h_2+2)-}(w)
\nonu \\
& - &
Q_{\la, \bar{a} b}^{(h_1+2)+}(z) \, Q_{\la, \bar{c} d}^{(h_2+2)-}(w)
+ Q_{\la, \bar{a} b}^{(h_1+2)-}(z) \, Q_{\la, \bar{c} d}^{(h_2+2)+}(w)
\label{cQ''}
\\
& - & (-1)^{h_1+h_2}\,
Q_{\la, \bar{c} d}^{(h_2+2)+}(z) \, Q_{\la, \bar{a} b}^{(h_1+2)+}(w)  + 
 (-1)^{h_1+h_2}\,Q_{\la, \bar{c} d}^{(h_2+2)-}(z) \, Q_{\la, \bar{a} b}^{(h_1+2)-}(w)
\nonu \\
& - & (-1)^{h_1+h_2}\,
Q_{\la, \bar{c} d}^{(h_2+2)+}(z) \, Q_{\la, \bar{a} b}^{(h_1+2)-}(w)
+  (-1)^{h_1+h_2}\,
Q_{\la, \bar{c} d}^{(h_2+2)-}(z) \, Q_{\la, \bar{a} b}^{(h_1+2)+}(w)
\Bigg]_{\frac{1}{
    (z-w)^{h_1+h_2+3}}}\, .
\nonu
\eea
Note that from the inside of the bracket in (\ref{cQ}),
after the $h_1$ and the $h_2$ are replaced with $(h_1+1)$ and
$(h_2+1)$ respectively, we obtain the
above result (\ref{cQ''}).

%%%%%%%%%%%%%%%%%
\subsection{The commutator relation between the fourth component and the
last component with $h_1=h_2, h_2+1$  for nonzero $\la$}
%%%%%%%%%%%%%%%%%

By using the relations (\ref{four}) and (\ref{last}),
we can reexpress them in terms of (\ref{res2}), (\ref{res3}),
(\ref{res4}) and (\ref{res5}). After using (\ref{twotwo}) and (\ref{RELL})
the following result can be obtained 
\bea
&& \big[(\tilde{\Phi}^{(h_1),i}_{\frac{3}{2}})_r,
  (\tilde{\Phi}^{(h_2)}_{2})_m \big]  = 
4(-4)^{h_1-4} \, 4(-4)^{h_2-4}
\Bigg[ 
-\Bigg( -
 \sum_{h=-1}^{h_1+h_2} \, q^h\, 
  q_{\mathrm{F}}^{h_2+2,h_1+\frac{3}{2}, h}(m,r,\la)
  \nonu \\
  && +  \sum_{h=-1}^{h_1+h_2+1} \,
  q^h\, (-1)^h\, 
q_{\mathrm{F}}^{h_2+2,h_1+\frac{3}{2}, h}(m,r,\la)
 -  
\sum_{h=-1}^{h_1+h_2} \,
 q^h\, 
q_{\mathrm{B}}^{h_2+2,h_1+\frac{3}{2}, h}(m,r,\la)
\nonu \\
&& + 
\sum_{h=-1}^{h_1+h_2+1} \,
 q^h\, (-1)^h\,
q_{\mathrm{B}}^{h_2+2,h_1+\frac{3}{2}, h}(m,r,\la)
\Bigg)\,
 \frac{1}{4(-4)^{h_1+h_2-h-3}}\,
(\Phi_{\frac{1}{2}}^{(h_1+h_2+1-h),i})_{m+r}
\nonu \\
%%%%%%%%%%%%%%%%%%%%%%%%%%%%%%%%%%%%%%%%%%%%%%
&&   -
\Bigg(  
  \sum_{h=-1}^{h_1+h_2} \, q^h\, 
  q_{\mathrm{F}}^{h_2+2,h_1+\frac{3}{2}, h}(m,r,\la)
  \nonu \\
  && +  \sum_{h=-1}^{h_1+h_2+1} \,
  q^h\, (-1)^h\, 
q_{\mathrm{F}}^{h_2+2,h_1+\frac{3}{2}, h}(m,r,\la)
 +  
\sum_{h=-1}^{h_1+h_2} \,
 q^h\, 
q_{\mathrm{B}}^{h_2+2,h_1+\frac{3}{2}, h}(m,r,\la)
\nonu \\
&& +  
\sum_{h=-1}^{h_1+h_2+1} \,
 q^h\, (-1)^h\,
q_{\mathrm{B}}^{h_2+2,h_1+\frac{3}{2}, h}(m,r,\la)
\Bigg)\,
\frac{1}{4(-4)^{h_1+h_2-h-4}}\,
(\tilde{\Phi}_{\frac{3}{2}}^{(h_1+h_2-h),i})_{m+r} \,
\Bigg]\, .
\label{fourteencase} 
\eea
In this case we can compare the above with the previous
result in (\ref{ninecase}).
They look similar to each other.
The alternating feature between the currents depending on
the property of even or odd in the weight $h$ can be observed.

Let us consider the OPE
  between
$\tilde{\Phi}^{(h_1=1),i}_{\frac{3}{2}}(z)$ and $
  \tilde{\Phi}^{(h_2=2)}_{2}(w)$ where the weights satisfy
  $h_1=h_2-1$.
  The sixth order pole of this OPE gives us
  the nonvanishing structure constant
  having the $\la$ factor 
  in the current
  $\Ga^i(w)$ of the ${\cal N}=4$ stress energy tensor.
This is not consistent with (\ref{fourteencase}) because
the minimum value of the
exponent $(h_1+h_2-h)$ of the first current
is given by $3$ and those of 
the second current is
$2$.
There is additional term on the right hand side
  of the OPE for
  the weights $h_1=1$ and $h_2=2$ which are outside of the allowed
  region between the weights.

%%%%%%%%%%%%%%%%%
\subsection{The commutator relation between the
last components with $h_1=h_2-1,h_2,h_2+1$  for nonzero $\la$}
%%%%%%%%%%%%%%%%%

Finally, by using (\ref{last}) together with (\ref{FIRST}),
(\ref{SECOND}), (\ref{THIRD}), (\ref{FOURTH}), (\ref{FIFTH}),
(\ref{res1}),  (\ref{rel1}), (\ref{rell}), and (\ref{RELL1})
the following result can be determined 
\bea
&& \big[(\tilde{\Phi}^{(h_1)}_{2})_m,
  (\tilde{\Phi}^{(h_2)}_{2})_n \big]  = 
16(-4)^{h_1-4} \, (-4)^{h_2-4}
\Bigg[ \nonu \\
  & &
\left(\begin{array}{c}
m+h_1+1 \\  h_1+h_2+3 \\
 \end{array}\right) \, 4 \, q^{h_1+h_2} \, (c''_F +c''_B) \,
 \de^{ij}\,  \de_{r+s}
  \nonu \\
  && +   \sum_{h=0,\mbox{\footnotesize even}}^{h_1+h_2+1} \,
 q^h\, 
  \Bigg(  -p_{\mathrm{F}}^{h_1+2,h_2+2, h}(m,n,\la)
 +
 p_{\mathrm{B}}^{h_1+2,h_2+2, h}(m,n,\la)
 \Bigg)\nonu \\
 && \times \frac{4}{(-4)^{h_1+h_2-h}}\,
(\Phi_{0}^{(h_1+h_2+2-h)})_{m+n}
\nonu \\
&& +  \sum_{h=0,\mbox{\footnotesize even}}^{h_1+h_2+1} \, \Bigg(
-(h_1+h_2+1-h+2\la)\, q^h\, 
 p_{\mathrm{F}}^{h_1+2,h_2+2, h}(m,n,\la)
\nonu \\
&& -
(h_1+h_2+2-h-2\la)\, q^h\, 
p_{\mathrm{B}}^{h_1+2,h_2+2, h}(m,n,\la)
\,
\Bigg)\nonu \\
&& \times \frac{1}{
  2(2h_1+2h_2-2h+3)(-4)^{h_1+h_2-h-4}}
\, (\tilde{\Phi}_{2}^{(h_1+h_2-h)})_{m+n}
\Bigg] \, .
\label{fifteencase}
\eea
This has the similar structure to the one in (\ref{Onerel}).

Let us calculate the OPE
  between
$\tilde{\Phi}^{(h_1=2)}_{2}(z)$ and $
  \tilde{\Phi}^{(h_2=0)}_{2}(w)$ where the weights satisfy
  $h_1=h_2+2$.
  The fifth order pole of this OPE gives us
  the nonvanishing structure constants
  having the $\la$ factor 
  in the currents
  $\Phi^{(h_1+h_2+2-h=1)}_{0}(w)$ and
    $\tilde{\Phi}^{(h_1+h_2-h=-1)}_{2}(w)$ with weight $h=3$.
This is not consistent with (\ref{fifteencase}) because
the dummy variable $h$ can appear as even number.
This implies that there is additional term on the right hand side
  of the OPE for
  the weights $h_1=2$ and $h_2=0$ we are considering.

%%%%%%%%%%%%%%%%%%%%%%
%\subsubsection{The central term}
%%%%%%%%%%%%%%%%%%%%%%
  The central term has
  \bea
c''_F & = &  (-4)^{h_1+h_2-4}\, \de_{b \bar{a}} \, \de_{d \bar{c}} \, \Bigg[
  \frac{(h_1+1+2\la)}{(2h_1+3)} \, \frac{(h_2+1+2\la)}{(2h_2+3)}\,
V_{\la, \bar{a} b}^{(h_1+2)+}(z) \, V_{\la, \bar{c} d}^{(h_2+2)+}(w) \nonu \\
& + &
V_{\la, \bar{a} b}^{(h_1+2)-}(z) \, V_{\la, \bar{c} d}^{(h_2+2)-}(w)+
\frac{(h_1+1+2\la)}{(2h_1+3)}\,
V_{\la, \bar{a} b}^{(h_1+2)+}(z) \, V_{\la, \bar{c} d}^{(h_2+2)-}(w)
\nonu \\
&+& \frac{(h_2+1+2\la)}{(2h_2+3)}\,
V_{\la, \bar{a} b}^{(h_1+2)-}(z) \, V_{\la, \bar{c} d}^{(h_2+2)+}(w) \Bigg]_{\frac{1}{
    (z-w)^{h_1+h_2+4}}}\, ,
\label{cF''}
\eea

\bea
c''_B & = &  (-4)^{h_1+h_2-4}\, \de_{b \bar{a}} \, \de_{d \bar{c}} \, \Bigg[
  \frac{(h_1+2-2\la)}{(2h_1+3)} \, \frac{(h_2+2-2\la)}{(2h_2+3)}\,
V_{\la, \bar{a} b}^{(h_1+2)+}(z) \, V_{\la, \bar{c} d}^{(h_2+2)+}(w) \nonu \\
& + &
V_{\la, \bar{a} b}^{(h_1+2)-}(z) \, V_{\la, \bar{c} d}^{(h_2+2)-}(w)-
\frac{(h_1+2-2\la)}{(2h_1+3)}\,
V_{\la, \bar{a} b}^{(h_1+2)+}(z) \, V_{\la, \bar{c} d}^{(h_2+2)-}(w)
\nonu \\
&-& \frac{(h_2+2-2\la)}{(2h_2+3)}\,
V_{\la, \bar{a} b}^{(h_1+2)-}(z) \, V_{\la, \bar{c} d}^{(h_2+2)+}(w)
\Bigg]_{\frac{1}{
    (z-w)^{h_1+h_2+4}}}\, .
\label{cB''}
\eea
We can easily see that after $h_1$ inside the bracket of (\ref{cF})
and $h_2$ inside the bracket of (\ref{cB}) are
replaced by $(h_1+2)$ and $(h_2+2)$ respectively,
the central terms in (\ref{cF''}) and (\ref{cB''})
can be obtained.

%%%%%%%%%%%%%%%%%%%%%%%%%%%%%%%%%%%%%%%%%%%%%%%%%%%%%%%%%%%%%%%%%%%%%%%%%%%
%%%%%%%%%%%%%%%%%%%%%%%%%%%%%%%%%%%%%%%%%%%%%%%%%%%%%%%%%%%%%%%%%%%%%%%%%%

\end{document}